\documentclass[a4paper,12pt,twoside]{report}
\usepackage{color}
\usepackage{graphicx}
\usepackage{latexsym}
\usepackage{amsmath}
\usepackage{axodraw}
\usepackage{fancyhdr}
\usepackage{amsfonts}
\def\I{\textrm{i}}
\def\be{\begin{eqnarray}}
\def\ee{\end{eqnarray}}
\def\bc{\begin{center}}
\def\ec{\end{center}}
\def\ba{\begin{array}}
\def\ea{\end{array}}
\def\bi{\begin{itemize}}
\def\ei{\end{itemize}}
\def\um{\frac{1}{2}}

\def\odz*{\oint\frac{d\bar{z}}{2\pi i}}
\def\odz{\oint\frac{dz}{2\pi i}}

\def\Tr{\mbox{Tr}}
\def\bee{\begin{eqnarray}}
\def\eee{\end{eqnarray}}
\def\nn{\nonumber}
\def\bi{\begin{itemize}}
\def\ei{\end{itemize}}

\def\sst{\scriptstyle}

\def\bigK{\mbox{\bf K}}

\def\squm{\frac{1}{\sqrt{2}}}
\def\st{\sigma,\tau}
\def\text{\mbox}
\def\zz{(z,\bar{z})}
\def\ww{(w,\bar{w})}
\def\re{\text{{Re}}}
\def\im{\text{{Im}}}
\def\>{\rightarrow}
\def\<{\leftarrow}
\def\ph{\phantom}

\def\cc{\c c}
\def\vs{\vspace}

\newlength{\dinwidth}
\newlength{\dinmargin}
\author{}
\date{}                 
\title{\Huge \bf Open Descendants at {\em c} = 1}
 
\begin{document}

\maketitle
\thispagestyle{empty}
\mbox{}
\newpage

\setcounter{page}{1}
\pagenumbering{Roman}
\begin{center}
\mbox{}
\vs{2cm}
{\Huge \bf Open Descendants at {\em c} = 1}
\vs{1.5cm}
\begin{tabular}{p{11cm}}
\bc
{Een wetenschappelijke dissertatie op het gebied van Natuurkunde en Wiskunde}
\ec
\end{tabular}

\vs{1.5cm}
{\large PROEFSCHRIFT}
\vs{1.5cm}

\begin{tabular}{p{11cm}}
\bc 
Ter verkrijging van de graad van Doctor aan de Katholieke Universiteit van Nijmegen, volgens besluit van Het College van Decanen in het openbaar te verdedigen op maandag 13 october 2003 des namiddags om 1:30 uur precies
\ec
\end{tabular}

\vs{1.5cm}
DOOR
\vs{1.5cm}

{\large Nuno Miguel Marques de Sousa}

\vs{1.5cm}
Geboren op 17 januari 1973 te Coimbra, Portugal \\
\vs{2cm}

Rozenberg Publishers, Amsterdam
\end{center}
\newpage

\mbox{}
\vs{3cm}

\begin{tabular}{ll}
Promotor:           & {\bf Prof. Dr. Adrianus Norbertus J.J. Schellekens} \\
                    &             \\
Manuscriptcomissie: & {\bf Prof. Dr. Robbert Dijkgraaf} \\
                    & {\bf Prof. Dr. Jan-Willem van Holten} \\  
                    & {\bf Prof. Dr. Christoph Schweigert}
\end{tabular}

\newpage
\mbox{}
\vs{3cm}
\noindent
This thesis is based on the following articles:
\vs{0cm}

\begin{tabular}{lll}
Sections \ref{nimish}, \ref{exceptable}: 
     &\hspace{1cm}& A. Schellekens and N. Sousa, \\
     &\hspace{1cm}& {\em Orientation matters for NIMreps,} \\
     &\hspace{1cm}& Nucl.Phys. B653 (2003) 339, hep-th/0210014. \\
     &\hspace{1cm}& \\
Chapter \ref{freebosons}:  
     &\hspace{1cm}& A. Schellekens and N. Sousa, \\
     &\hspace{1cm}& {\em Open descendants of $U(2N)$ 
                     orbifolds at  rational radii,} \\
     &\hspace{1cm}& Int.J.Mod.Phys. A16 (2001) 3659, hep-th/0009100. \\
     &\hspace{1cm}& \\
Chapter \ref{orbfus}:  
     &\hspace{1cm}& The results of this chapter have not been published before. \\
\end{tabular}

\vs{8cm}
\noindent
This work was performed at the Dutch National Institute for Nuclear and High-Energy Physics, NIKHEF, Amsterdam, Holland. Financial support granted by the Portuguese `Funda\cc\~ao para a Ci\^encia e Tecnologia - FCT', under the reference BD/12770/97 and by the Dutch `Samenwerkingsverband Mathematische Fysica - FOM-SWON', member of the `Nederlandse Organisatie voor Wetenschappelijk Onderzoek - NWO'.

\newpage
\tableofcontents
\listoffigures
\listoftables
\pagestyle{fancy}
\fancyhf{}
\renewcommand{\chaptermark}[1]{\markboth{\thechapter.\ #1}{}}
\renewcommand{\sectionmark}[1]{\markright{\thesection\ #1}{}}
\fancyhead[LO,RE]{\slshape \rightmark}
\fancyhead[RO,LE]{\slshape \leftmark}
\fancyfoot[C]{\thepage}
\renewcommand{\headrulewidth}{0.5pt}
\renewcommand{\footrulewidth}{0pt}

\newpage
\pagenumbering{arabic}
\chapter{Introduction} 
\section{Proposition} 
String Theory is a tribute to human imagination. A line of research in which a few scientists embarked thirty years ago, which appeared to be a promising candidate to solve the intriguing enigma of Quantum Gravity. As of today, this goal still seems far away, but the fascination String Theory exerts on scientists is so great that they just keep going as if the years elapsed did not matter the slightest. Science is after all proliferous in cases where seemingly impossible breakthroughs were the outcome of enduring work on ideas that looked strange at first. As examples of this we can cite General Relativity, Quantum Mechanics or, in a more distant past, the invention of Calculus.

Flashing back to the early 1900s', we realize the phenomena scientists observed at the time must have looked to them very, very strange. Indeed, the results of many experiments blatantly escaped common sense and seemed to have no relation to anything known thus far. What logic could there be in equally prepared experiments giving different results every time? Or, how could someone be convinced that if his brother were to fly in a spaceship, he would return to Earth twenty years younger? Yet these questions have been answered. The key for finding the answers was imagination. Imagination freed the scientists' minds from the clutches of the prejudice of determinism and made possible a rational interpretation to the chaos of contradicting experiments.

The flamboyant examples of Quantum Mechanics and General Relativity are perhaps misleading, for imagination and preserverance more often than not take their time to sort matters out. For instance, it took well over three centuries to prove Fermat's last theorem. We can also cite Apollonius of Perga, who had to wait more than 2000 years before his study of conical sections was used by Kepler to explain the orbits of planets. It is therefore not to say when should the quest of unifying the fundamental forces of Nature be accomplished, for it has barely started.

String Theory is one of the candidates for realizing unification. Despite its eerie formulation and some disheartening technicalities, it still seems the best one. The good perturbative behavior of String Theory and natural appearance of gravitons are nice indicators that we are on the right track. String Theory also stimulated progress in various branches of Mathematics and provided possible explanations for several cosmological observations. Coincidence or not, it is, for instance, possible to derive formulas for black-hole entropy from String Theory. The path it treads is however so widespread that the number of possible scenarios for the Universe is immense. This book tells the story of one of them.

\section{Past, present and future}
\subsection{Strings from hadronic models}
The upsurge of particle accelerators in the early 1960's brought forth many surprises. One of the intriguing facts observed was the mass-squared behavior of hadronic resonances, which seemed to be linear with spin
\be
m^2 = \frac{J}{\alpha'}, \;\;\; \alpha' \sim 1\text{(GeV)}^{-2}.
\ee
This behavior seemed to go on indefinitely, so it was not plausible that all these resonances were fundamental and even if one took them as so, the high spins meant bad high-energy behavior of tree-level amplitudes and certain renormalization trouble. Trying to find a way around these problems, in 1968 Veneziano proposed the hadron scattering amplitude \cite{veneziano}
\be
A(s,t) = \frac{\Gamma \Big( -\alpha(s) \Big)\Gamma\Big( -\alpha(t) \Big)}{\Gamma \Big( -\alpha(s)-\alpha(t) \Big)}, \;\;\; s,t: \text{Mandelstam}, \;\;\; \alpha(x) = \alpha' x + \alpha(0).
\ee
This amplitude is exponentially soft at high energies in all kinematic regimes, even though it includes contributions from particles of all spins\footnote{This can be seen, for example, from the pole structure.}. However,
while exponential decay at high energies matched experimental results in some kinematic regimes, it did not match the observed power-law decay in others. The power-law decay is nowadays interpreted as a signal of partonic structure, and Veneziano's amplitude failure to reproduce this was the chief reason for its demise.

Later in the 1970's a much better description of the strong interactions appeared in the form of QCD, the mass spectrum being interpreted as spin-orbit excitations of the constituent quarks. With this the Veneziano model was finally abandoned as theory of hadronic interactions. It was however noticed that the amplitude $A(s,t)$ wasn't irrelevant. It actually described scattering of quantum strings.

\subsection{Strings as quantum gravity?}
The subject remained dormant for a while until later on some physicists noticed that the pole $s=0,t=0$ of the corresponding Veneziano formula for closed strings signaled a massless spin-2 particle. The only known consistent field theory of massless spin-2 particles is gravity. Could this pole be the graviton?

The Veneziano amplitude is an amplitude for string scattering, so, in this way String Theory became a proposal for a theory of Quantum Gravity. Quantum Gravity arises when radiative corrections to General Relativity due to one-graviton exchange become important. This happens at the so-called Planck scale
\be 
M_{\text{{\scriptsize Planck}}}=\sqrt{\frac{\hbar c}{G_N}}\sim 10^{19}\,\mbox{GeV},\;\;\; L_{\text{{\scriptsize Planck}}}\sim 10^{-16}\,\mbox{cm}. \label{Mpl}
\ee
This length scale is unfortunately too small for accelerator reach. However, recent developments show us that there may be ways around this problem.

The first attempt at a String Theory produced, as we shall see below, more than was desired. For example, the graviton didn't come out of the mass spectrum alone - it came together with tachyons, particles of negative mass-squared, whose phenomenological interpretation was problematic, to say the least. Also, the absence of space-time fermions wasn't that good an omen either, as wasn't the existence of critical dimensions for the quantized theory. Nevertheless, these problems have been successfully dealt with and at the end of the day, five Superstring Theories were obtained, that are believed to be completely consistent in $D=10$ space-time dimensions. But ten space-time dimensions is more than the four we see around us and therefore something had to be done with the six extra dimensions.

\subsection{Compactification and extended objects}
The first proposal to deal with the critical dimension problem was to compactify the extra dimensions via the Kaluza-Klein mechanism. Since one has to curl-up six dimensions, the possibilities for the compactification manifold are more than many. In the middle the plethora of choices, a phenomenologically interesting model arose. This was the compactification of the heterotic $E_8\otimes E_8$ superstring on a Calabi-Yau manifold \cite{candelas}. The compactification process breaks one of the $E_8$ gauge symmetries in such a way that in the four-dimensional world, one gets the Standard Model group $SU(3)\otimes SU(2) \otimes U(1)$, with four generations of fermions. The model had however some problems, like gauge coupling hierarchy \cite{heteroticgaugecoupling}, but it marked nevertheless the beginning of string phenomenology.

In 1995 D-branes, extended objects of String Theory, were discovered \cite{dbranes} and with it a new piece of the puzzle fit in. The D-branes are non-perturbative objects, hyper-surfaces where open strings can end. Their discovery allowed briefly afterwards the establishment of a net of dualities relating the various Superstring Theories, which culminated with the postulation of the existence of an eleven-dimensional unified theory, M-theory, which would in different asymptotic regimes reduce to the five known String Theories. From the net of dualities came also the AdS/CFT correspondence \cite{maldacena}, a interesting duality which relates gauge theories to gravity ones.

The D-branes can accommodate gauge theories in their world-volume. This opened the possibility of the so-called Brane World Scenario \cite{braneworld}. In such a scenario, one assumes that our four-dimensional world is confined to live on the world-volume of a D-brane. This allows the extra dimensions to grow very large. Large extra dimensions lower the string scale, which can eventually drop as low as the TeV scale, giving string phenomenology a whole new perspective, perhaps even an experimental one. Other scenarios featuring D-branes and their close relatives, the orientifold O-planes, can be setup to try and understand cosmological puzzles such as black-hole entropy or Hawking radiation.

\subsection{Prospects}
The recent developments opened many new windows to String Theory, and the topics mentioned above are but a small fraction of what experts are dealing with currently. Today, the subject and its branches undergoes a growing researching frenzy as the start of the running of the Large Hadron Collider at CERN approaches. This accelerator will, amongst many things, probe the energy scales where Supersymmetry, a key feature of String Theory, is belied to be restored. If the existence of Supersymmetry is indeed confirmed, it will forever stand as one of the great achievements of Theoretical Physics, for it was postulated relying solely on theoretical symmetry arguments.

With so many subjects being frantically researched at the same time, the next years promise to be a rich period in intellectual endeavor. For String Theory is indeed a fascinating subject whose details interest mathematicians and physicists alike.

\section{This work}
This thesis focuses on one of the basic ingredients of String Theory: Conformal Field Theory. As the string evolves in space-time, it sweeps out a two-dimensional world-sheet, in which a two-dimensional conformal field theory lives. To first approximation, the field theory living in space-time is a mapping from the world-sheet theory to space-time. Thus, studying the world-sheet theory enables us to understand the structure of space-time.

The conformal symmetry that the world-sheet enjoys at classical level does not get through to the quantum level. In general, it acquires quantum corrections due to the presence of an anomaly. Quantum anomalies are complicated to deal with and normally bar a theory from being consistent. In String Theory, if one interprets matters in the right way, the accursed quantum conformal anomaly can be dispelled and turned into a blessing instead. The point is that in a path-integral quantization procedure, the classical gauge symmetries of the string world-sheet generate Faddeev-Popov ghosts, which contribute to conformal anomaly. Their contribution can however be canceled by adding the contribution of the world-sheet bosonic and fermionic fields. Since one needs to add a precise number of world-sheet fields, this procedure normally leads to the appearing of a very definite critical dimension. The blessing appears now because we are not obliged to add {\em exclusively} world-sheet bosons and fermions. We can add whatever conformal fields we please, as long as we do it in a consistent manner. In this way, what we get in the end of the day is an algebraic alternative to compactification.

Of particular interest for this work are string theories that have open strings. The presence of open strings introduces surfaces with boundaries - the world-sheet boundaries drawn by the end points of the open strings. This makes it necessary to formulate conformal field theory on such surfaces. Two open strings can join to create one closed string, which means one must consider open and closed surfaces alike. Because the end points of open strings can end on D-branes, one has to consider these extended objects as well. Not only that, one must also consider their relatives, the orientifold O-planes. One way to see that they should be included is to note that D-branes carry so-called Ramond-Ramond charges, whose flux lines must end somewhere. If the D-brane fills out the whole of space-time, the flux lines cannot escape and must therefore end on a surface of negative Ramond-Ramond charge. The O-planes carry this sort of charge in String Theory, and therefore they should be present. The O-planes act as mirrors in a sense: they truncate the string spectrum to states invariant under world-sheet orientation reversal. In the presence of O-planes, the string theory becomes unoriented. At the level of the world-sheet, this means crosscaps are added to the array of surfaces drawn out by strings. In the end, an open string theory means one has to consider conformal field theory on four types of surfaces: closed or open, oriented or unoriented. Because of this, throughout this thesis, by the words `open string theory' it will be meant a theory of both open and closed unoriented strings.

In this work we search for open string conformal field theories with conformal anomaly $c=1$ which can be consistently added to the world-sheet theory and take the first steps towards proving consistency of the resulting model to all orders in string perturbation theory. The Feynman diagrams of point-particle quantum field theory have analogs in String Theory. Put simply, in String Theory one replaces the propagator lines of point-particles for tubes or strips representing propagation of closed and open strings respectively. The surfaces obtained can, as in quantum field theory, include loop-like configurations. In open string theory they can also and have boundaries and crosscaps, which allow the open and closed strings to draw surfaces like the annulus, the Klein bottle or the Moebius strip in space-time. Constructing the partition functions for the open string theory at one-loop order, one can write down very simple constraint equations that enable us to write down the spectrum of the theory and do some basic consistency checks. Going further into the details of the conformal field theory we can find its chiral four-point functions, which can be used to extract fusing and braiding matrices, which can in turn be used to test the consistency of the theory to all orders in perturbation theory via the induction mechanism of the so-called sewing constraints \cite{sonoda} \cite{lewellen} \cite{sagnotticross}.

This thesis is organized as follows. In chapter \ref{bosonic} a review of the bosonic string is presented as motivation for String Theory and to clarify and motivate the algebraic approach. In chapter \ref{cftchap} conformal field theory is reviewed. In chapter \ref{opendes} the construction of open strings is presented and an interesting result is found. In chapter \ref{freebosons} the open string construction is used to derive results for $c=1$ orbifold theories. In chapter \ref{sewchap} consistency of string perturbation theory is discussed and in chapter \ref{orbfus} correlators of the orbifold theory are used to derive some of the chiral quantities needed to test perturbative consistency. Finally, in chapter \ref{conclusion} conclusions and prospects for future work are discussed.

\chapter{The Bosonic String}\label{bosonic}
\section{Classical String Theory}
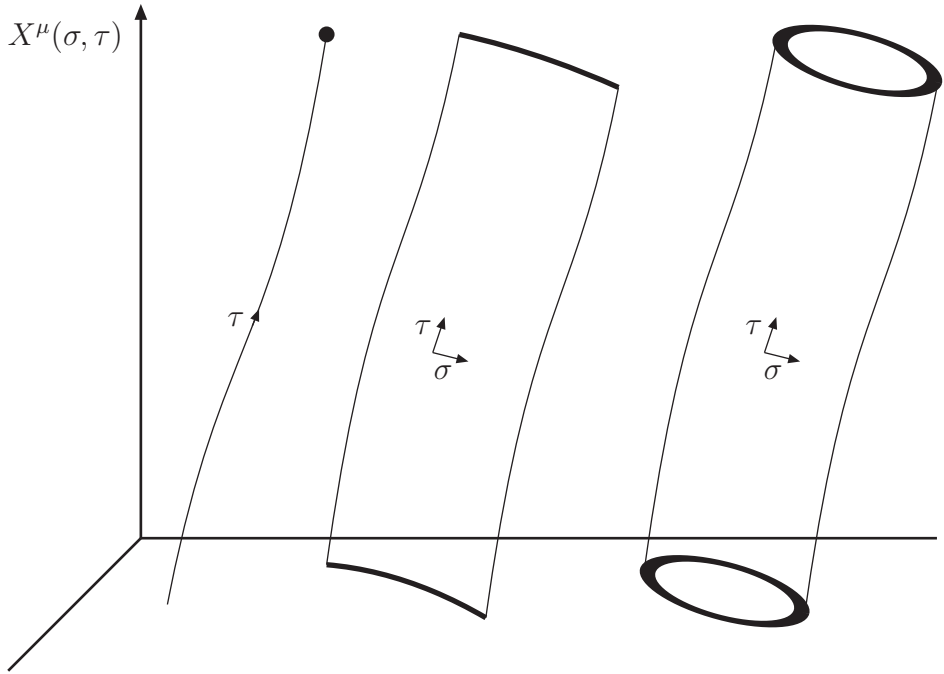
\begin{figure}
\SetScale{1}
\begin{center}
\begin{picture}(400,300)(0,0)
\SetWidth{1}
\Line(70,50)(370,50)
\Line(70,50)(20,0)
\LongArrow(70,50)(70,250)
\Text(20,240)[l]{${X^\mu(\sigma,\tau)}$}
\SetWidth{.5}
\Curve{(80,25)(81,30)(139,234)(140,240)}
\LongArrow(112,130)(114,135)
\Text(103,133)[l]{$\tau$}
\Vertex(140,240){3}
\SetWidth{2}
\Curve{(140,40)(180,30)(200,20)}
\Curve{(190,240)(225,230)(250,220)}
\SetWidth{.5}
\Curve{(140,40)(141,47)(189,235)(190,240)}
\Curve{(200,20)(201,27)(249,215)(250,220)}
\LongArrow(180,120)(184,132)
\LongArrow(180,120)(192,117)
\Text(174,130)[l]{$\tau$}
\Text(181,113)[l]{$\sigma$}
\SetWidth{2}
\Oval(340,230)(10,30)(345)
\Oval(290,30)(10,30)(345)
\SetWidth{.5}
\Curve{(260,40)(261,47)(309,235)(310,240)}
\Curve{(320,20)(321,27)(369,215)(370,220)}
\LongArrow(305,120)(309,132)
\LongArrow(305,120)(317,117)
\Text(299,130)[l]{$\tau$}
\Text(306,113)[l]{$\sigma$}
\end{picture}
\end{center}
\caption{String world-sheet}
\label{stringws}
\end{figure}

The idea of String Theory is the quite simple assumption that elementary particles are not point-like objects but tiny little strings instead. The string can be defined as a series of maps $X^\mu(\sigma,\tau)$ from the world-sheet to space-time, with dynamics governed by an action which is proportional to the area swept out by the world-sheet as it moves through space-time. This can be seen more clearly from figure \ref{stringws}.

From the picture, looking first to the point-particle case on the left, we see the world-line it draws at it goes through space-time is parametrized by the proper time $\tau$. Moving on to the string, we see that the world-sheet parameters $\sigma^a = (\sigma,\tau)$ can be interpreted as the world-sheet space and time respectively. The classical dynamics of the string action is governed by the Nambu-Goto action
\be 
S_{NG}
=-T\int_{A} d^2\sigma \, \sqrt{(\dot{X} \! \cdot \! X^\prime)^2 - \dot{X}^2 X^{\prime 2}}, \;\;\; \dot{X}^\mu = 
{\partial_\tau X^\mu},\;\; X^{\prime \mu} ={\partial_\sigma X^\mu},
\label{SNG} 
\ee
with $T$ the string tension. The integrand is the infinitesimal world-sheet area element, therefore its integral is the total area swept out by the world-sheet. We define also $A\cdot B = A^\mu B^\nu \eta_{\mu\nu}$, the $D$-dimensional Minkowskian signature being $(-+\cdots+)$ and $A^2=A\cdot A$. Minkowski metric can, of course, be replaced by a more general $G_{\mu\nu}(X)$, which would mean the string is propagating in a curved background. Making the space-time metric dynamical would however move us to the realm of String Field Theory, which is beyond the scope of this work. For simplicity we consider only the flat space case, which is in any case the first term in a weak field expansion.

The Nambu-Goto action is the simplest and most intuitive string action one can write, but it is not the best starting point for quantization. Therefore we replace it by a classically equivalent, more symmetrical action, the Polyakov action. On a flat $D$-dimensional Minkowskian background, the Polyakov action takes the form 
\be 
S_P = -\frac{T}{2} \int d^2\sigma \,\sqrt{g} \, g^{ab} \partial_a X^\mu \partial_b X^\nu \eta_{\mu \nu}. \label{SP} 
\ee 
Here $g_{ab}(\sigma,\tau)$ is the metric on the world-sheet and $g=-\det{g_{ab}}$. (Note that $\det{g_{ab}}$ is again the area of the world-sheet swept.) Polyakov's action has two dynamical variables: the coordinate fields $X^\mu$ and the metric $g_{ab}$ (which was absent in the Nambu-Goto action). Vanishing of variations of the action with respect to these variables defines the equations of motion
\be 
\frac{\delta S}{\delta X^\mu} &=& \frac{1}{\sqrt{g}} \partial_a (\sqrt{g} g^{ab} \partial_b X^\mu) = 0, \nn \\ 
-\frac{1}{T}\frac{1}{\sqrt{g}}\frac{\delta S}{\delta g^{ab}} &=& T_{ab} = \partial_a X^\mu \partial_b X_\mu - \um g_{ab}g^{cd} \partial_c X^\mu \partial_d X_\mu = 0, \label{Pmotion} 
\ee
with $T_{ab}$ the system's energy-momentum tensor.
Classical equivalence of Polyakov's action to the Nambu-Goto one is easily seen by splitting $T_{ab}=0$ into two members, taking determinants on both sides and plugging the result back into $S_P$.
 
The string dynamics of (\ref{SP}) can be simplified by using the symmetries available to remove some of the gauge freedom. The symmetries of the Polyakov action are 
\bi 
\item{Poincar\'e invariance:  
\be 
X^\mu(\sigma,\tau) &\rightarrow& \tilde{X}^\mu(\sigma,\tau) = \Lambda^\mu_{\phantom\mu \nu} X^\nu (\sigma,\tau) + a^\mu, \nn \\ 
g_{ab}(\st) &\rightarrow& \tilde{g}_{ab}(\st) = g_{ab}(\st), 
\ee 
with $\Lambda^\mu_{\phantom\mu \nu}$ a Lorentz transformation and $a^\mu$ a translation.} 
\item{Reparametrization invariance: 
\be 
X^\mu(\st) &\rightarrow&  \tilde{X}^\mu(\tilde{\sigma}(\st),\tilde{\tau}(\st)) = X^\mu(\st), \nn \\ 
g_{ab}(\st) &\rightarrow& \tilde{g}_{ab}(\tilde{\sigma},\tilde{\tau}) =  
\frac{\partial \sigma^a}{\partial \tilde{\sigma}^c} 
\frac{\partial \sigma^b}{\partial \tilde{\sigma}^d} g_{ab}(\st). 
\ee} 
\item{Conformal invariance:
\be 
X^\mu(\st) &\rightarrow&  \tilde{X}^\mu(\st) = X^\mu(\st), \nn \\ 
g_{ab}(\st) &\rightarrow& \tilde{g}_{ab}(\st) =  
e^{2\omega({\st})} g_{ab}(\st), 
\ee} 
with $\omega(\st)$ arbitrary. 
\ei 
Poincar\'e invariance is a global space-time symmetry, reflecting the fact that it doesn't matter which space-time reference frame we take. Reparametrization invariance is the statement of general coordinate invariance at the world-sheet level. Conformal invariance is a local scaling symmetry of the world-sheet metric. The world-sheet theory is thus a conformal field theory (see also chapter \ref{cftchap}). Conformal symmetry was not explicitly present in the Nambu-Goto action.

The world-sheet symmetries can now be used to eliminate the degrees of freedom of $g_{ab}$. Heuristically speaking, a general $d$-dimensional object is described by a metric which is a $d\times d$ matrix. The metric is symmetric, so it has $\um d(d+1)$ independent components. Each reparametrization transformation can then be used to eliminate one further degree of freedom of the metric and conformal scaling can eliminate one final degree of freedom. In the end we have $\um d(d+1)-d-1$ dynamical components left in the world-volume metric of a $d$-dimensional object. For $d=2$ (strings) this is zero and therefore any world-sheet metric $g_{ab}$ is essentially equivalent to any other metric we choose\footnote{This is at genus zero. For higher genera finitely many moduli will remain.}. For our purposes, a convenient choice is $\eta_{ab}=(-+)$ everywhere, a metric which has indeed zero dynamical components. Note that for higher dimensional objects like membranes, the world-volume metric cannot be gauged away to a constant. If the world-volume metric remains dynamical, the problem becomes much more complicated, already at the classical level. This is one of the reasons why we only consider strings, rather than going further into theories of higher-dimensional fundamental objects like membranes.

Setting the world-sheet metric to $\eta_{ab}$ simplifies the Polyakov action to
\be 
S_P = \frac{T}{2} \int d^2\sigma \, \left( \dot{X}^2 - X^{\prime 2} \right). \label{SPconf} 
\ee 
In this gauge the equations of motion become very simple. Setting $\sigma \in [0,\pi]$ for the open string and $\sigma \in [0,2\pi]$ for the closed string, the variation of the action with respect to $X^\mu$ yields 
\be 
\delta S = T\int_A d^2\sigma\left( \partial_\sigma^2 X^\mu - \partial_\tau^2 X^\mu \right) \delta X_\mu - T \int_{\tau_0}^{\tau_1} d\tau \, X_\mu^\prime \delta X^\mu \Big|_{\sigma=0}^{\sigma=\pi}=0, 
   \label{SPvar} 
\ee 
where the surface term is absent for closed strings. For open strings this surface term can be set to zero if either $X_\mu^\prime$ or $\delta X^\mu$ are zero (at $\sigma=0,\pi$). The motion is then $\Box X^\mu = 0$, subject to the boundary conditions
\be 
& X^{\mu}(\sigma+2\pi,\tau) = X^\mu(\sigma,\tau) & \mbox{(Closed string)} \nn \\ 
&X^\prime_\mu\Big|_{\sigma=0,\pi} = 0 & \mbox{(Neumann open string)} \label{bcs}\\ 
&\delta X_\mu\Big|_{\sigma=0,\pi} = 0 & \mbox{(Dirichlet open string)}  \nn
\ee 
The Neumann boundary condition can be interpreted as momentum conservation at the freely-moving end points of an open string, while the Dirichlet boundary condition means the open string end points are kept fixed, attached to something.

The general solution for the wave equations, subject to the boundary conditions (\ref{bcs}) is, (define left- and right-movers $\sigma^\pm = \tau \pm \sigma$)
\be 
X^\mu(\st) &=& 
   \um q^\mu + \frac{\alpha'}{2}p^\mu\sigma^- + 
   \frac{\I}{\sqrt{2/\alpha'}} \sum_{n\neq 0} 
      \frac{1}{n} \alpha^\mu_n e^{-\I n \sigma^-}  
\nn \\ &&+ \um q^\mu + \frac{\alpha'}{2}p^\mu\sigma^+ +
   \frac{\I}{\sqrt{2/\alpha'}} \sum_{n\neq 0} \frac{1}{n} 
      \bar{\alpha}^\mu_n e^{-\I n \sigma^+} \;\;\;\;\;\;\;\;\,\;\;\;  
      \text{(Closed string)}  \nn \\
&=& X_L^\mu (\sigma^-) + X_R^\mu (\sigma^+)  \nn \\
X^\mu(\st) &=& 
   q^\mu + 2\alpha' p^\mu \tau + {\I}{\sqrt{2\alpha'}} \sum_n \frac{1}{n} 
      \alpha^\mu_n e^{-\I n\tau} \cos{(n\sigma)} \;\;\;\;\,\;\;\;\; 
      \text{(Neumann open string)}   \nn \\
X^\mu(\st) &=& q_1^\mu\sigma + q_2^\mu(\pi\!-\!\sigma) + 
   {\I}{\sqrt{2\alpha'}} \sum_n \frac{1}{n} \alpha^\mu_n e^{-\I n\tau} 
   \cos{(n\sigma)}  \;\;\; \text{(Dirichlet open string)} \nn 
\label{closedstringmode}
\ee 
Here we replaced the string tension by the Regge slope parameter, $\alpha'=1/2\pi T$. The $q^\mu, p^\mu$ are respectively the position and momentum of the string center-of-mass. The $\alpha_n^\mu,\bar{\alpha}_n^\mu$ are Fourier modes, which in the classical theory are numbers subject to $\alpha_{-n}^\mu = (\alpha_n^\mu)^*,\bar{\alpha}_{-n}^\mu = (\bar{\alpha}_n^\mu)^*$ (because we want $X^\mu$ to be real). In the quantum theory all these quantities get be promoted to operators. For quantization purposes we write the canonical Poisson brackets in terms of $q^\mu,p^\mu$ and the Fourier modes
\be 
\left\{ \alpha_m^\mu,\alpha_n^\nu \right\}= 
\left\{ \bar{\alpha}_m^\mu,\bar{\alpha}_n^\nu \right\}=-\I m \,\eta^{\mu \nu}\, \delta_{m+n,0}, \;\;\; \left\{ q^\mu,p^\mu \right\} =\eta^{\mu \nu} & \text{(Closed string)}\nn \\ 
\left\{ {\alpha}_m^\mu,{\alpha}_n^\nu \right\} = -\I m \,\eta^{\mu \nu}\, \delta_{m+n,0}, \;\;\; \left\{ q^\mu,p^\mu \right\} =\eta^{\mu \nu} & \text{(Open string)},
\ee
with all other Poisson brackets zero.

One must not forget to enforce the equations of motion for the metric, which are $T_{ab}=0$. This is simply
\be
T_{ab}=0 \>  \left( \dot{X} \pm X^{\prime}\right)^2=0.
\ee
It is however more convenient to write these constraints in terms of Fourier modes. For this purpose we define Virasoro operators\footnote{These are conserved charges associated with world-sheet energy-momentum conservation: $\partial_- T_{++}= \partial_+T_{--}=0$, with $T_{\pm\pm}$ two components of the energy-momentum tensor written in terms of left- and right-modes.}
\be 
L_m &=& 2T\int_0^{2\pi} d\sigma\, e^{-\I m\sigma} T_{--} = 
   \um\sum_n \alpha_{m-n} \cdot \alpha_{n},  \nn \\
\overline{L}_m &=& 2T\int_0^{2\pi} d\sigma \,e^{+\I m\sigma} T_{++} = 
   \um\sum_n \bar{\alpha}_{m-n} \cdot \bar{\alpha}_{n} 
   \;\;\;\;\;\;\;\;\;\;\;\;\;\;\;\;\;\;\;\;\;\; \text{(Closed string)} \nn \\ 
L_m &=& T\int_0^{\pi} d\sigma\, \left( e^{-\I m\sigma} T_{--} + 
   e^{+\I m\sigma} T_{++} \right) = 
   \um\sum_n {\alpha}_{m-n} \cdot {\alpha}_{n}\;\;\; \text{(Open string)}.
\ee 
These have Poisson brackets 
\be 
\left\{ L_m,L_n \right\} = -\I (m-n)L_{m+n},\;\;
\left\{ \overline{L}_m,\overline{L}_n \right\} = -\I (m-n)\overline{L}_{m+n} & 
   \text{(Closed string)} \nn \\ 
\left\{ L_m,L_n \right\} = -\I (m-n)L_{m+n} & \text{(Open string)}.
\ee 
This algebra is the classical Virasoro algebra. The constraint equations $T_{ab}=0$ are then equivalent to vanishing of the components: $L_n=\overline{L}_n=0$. For the closed string $L_0-\overline{L}_0$ generates rigid $\sigma$-translations and must therefore be identically zero. The case $L_0=\overline{L}_0=0$ can be used to express the string mass in terms of the Fourier modes by means of the mass-shell condition $m^2=-p^\mu p_\mu$, 
\be 
m^2 &=& \frac{2}{\alpha'}\sum_{n=1}^\infty 
        \left( \alpha_{-n} \cdot \alpha_{n} + 
               \bar{\alpha}_{-n} \cdot \bar{\alpha}_{n} 
        \right)\;\;\; \text{(Closed string)} \nn \\ 
m^2 &=& \frac{1}{\alpha'}\sum_{n=1}^\infty 
        \alpha_{-n} \cdot \alpha_{n}
\;\;\;\;\;\;\;\;\;\;\;\;\;\;\;\;\;\;\;\;\;\;\;\, \text{(Open string)}.
\ee 
As a curiosity we note that the end points of the Neumann open string move at the speed of light. The centrifugal force of this motion compensates the inward-pulling tension and prevents the collapse of the string to a point. In its turn, the Dirichlet open string solution can be applied to as many of the space-like directions $\mu$ as we want. For each direction with such boundary conditions, $q^\mu_{1,2}$ are constants labeling the precise position of the string end points. These are kept fixed and the bulk of the string vibrates, much like a guitar string when fingered.

\section{Light-cone quantization} 
In this section we review a few basic properties of quantized strings, which were the original motivation for studying the subject. These properties are derived via light-cone quantization, a quantization method does not have manifest space-time Lorentz invariance but will provide the energy spectrum and an Hilbert space free of negative-norm states. Lorentz invariance will be imposed `by hand' at the end, and this requirement will restrict the number of space-time dimensions. In the next section we present an alternative quantization method based on path-integrals, which will introduce a few more elements necessary for the ensuing discussion.
 
In light-cone quantization we first define light-cone and transverse coordinates 
\be 
X^\pm = \frac{1}{\sqrt{2}} (X^0 \pm X^1), \;\;\; X^i, \, i=2, \ldots, D-1. 
\ee 
Note that we are treating $X^\pm$ in a non-covariant way. In light-cone coordinates the space-time Minkowskian metric becomes $\eta_{ij}=\delta_{ij},\; \eta_{+-}=\eta_{-+}=-1$, so inner products become $A\cdot B=A^i B^i - A^+B^- -A^-B^+$. Now we make a parametrization choice for $X^+$. With this choice everything is fixed and the theory has no more gauge freedom. We parametrize $X^+$ as
\be 
X^+ = \left\{ \ba{l} q^+ + \alpha' p^+ \tau  \;\;\;\;\, 
\mbox{(Closed string)} \\ 
q^+ + 2\alpha' p^+ \tau \;\;\, \mbox{(Open string)} 
\ea \right.
\ee 
This corresponds classically to setting the Fourier modes $\alpha^+_n=0$ for $n\neq 0$. The Virasoro constraints $(\dot{X}\pm X')^2=0$ become 
\be 
4\alpha'p^+ \left( \dot{X}^- \pm X^{'-}\right) = \left( \dot{X}^i \pm X^{'i} \right)^2, \label{LCvirasoroconstraint} 
\ee 
from which we see that by solving the constraints, the $X^-$ can be written explicitly in terms of $X^i$. With $X^+$ imposed and $X^-$ eliminated, the transverse oscillators $X^i$ are the only dynamical variables left.

\begin{figure}
\SetScale{1}
\begin{center}
\begin{picture}(200,120)(0,00)
\SetWidth{1}
\Line(0,20)(60,20)
\Line(0,40)(60,40)
\Line(0,60)(60,60)
\Line(0,80)(60,80)
\SetWidth{.5}
\LongArrow(30,23)(30,37)
\LongArrow(30,43)(30,57)
\LongArrow(30,63)(30,77)
\Text(30,93)[c]{$\vdots$}
\Text(35,30)[l]{$a^\dagger$}
\Text(35,50)[l]{$a^\dagger$}
\Text(35,70)[l]{$a^\dagger$}
\Text(30,10)[c]{$|0\rangle$}
\Text(30,-5)[c]{Harmonic oscillator}
\Vertex(30,20){2}
\SetWidth{1}
\Line(130,40)(170,40)
\Line(110,80)(190,80)
\Line(120,60)(180,60)
\SetWidth{.5}
\Curve{(100,100)(150,20)(200,100)}
\LongArrow(147,23)(135,37)
\LongArrow(150,23)(155,57)
\LongArrow(153,23)(180,77)
\LongArrow(133,43)(125,57)
\LongArrow(124,63)(120,77)
\LongArrow(156,63)(157,77)
\LongArrow(136,43)(143,77)
\Text(150,10)[c]{$|0,k\rangle$}
\Text(180,30)[l]{${ \alpha_{-1}^i}$}
\Text(190,50)[l]{${ \alpha_{-2}^i}$}
\Text(200,70)[l]{${ \alpha_{-3}^i}$}
\Text(150,-5)[c]{Open string}
\Text(150,93)[c]{$\vdots$}
\Vertex(150,20){2}
\end{picture}
\end{center}
\caption{Hilbert spaces for the harmonic oscillator and the string}
\label{hilbert}
\end{figure}
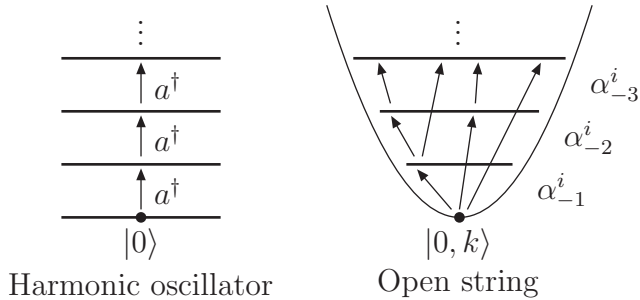

To quantize, we promote classical quantities to operators and impose canonical equal-$\tau$ commutation relations according to the Heisenberg rule: $\{\,,\}\rightarrow -\I \,[\,,]$. In terms of Fourier components this is 
\be 
&&\left[q^-,p^+\right]=\I \eta^{-+}=-\I ,\;\;\;\left[q^i,p^j\right]=\I \delta^{ij},\nn \\
&&\left[\alpha^i_m,\alpha^j_n\right]=m\,\delta^{ij}\delta_{m+n,0}, \;\;\;\text{with} \; \left(\alpha_m^i \right)^\dagger = \alpha_{-m}^i,
\ee 
with similar commutation relations for the $\bar{\alpha}^i_m$ in the closed string case.
The commutation relations of the Fourier oscillators are similar to the harmonic oscillator creation/annihilation commutation relation $[a,a^\dagger]=1$, which suggests interpreting the quantized string theory as set of harmonic oscillators. Therefore $\alpha^i_m,\; m>0$ become annihilation operators, while $\alpha^i_m,\; m<0$ are creation operators. The ground state of the quantum theory is labeled by $|0,k\rangle$, with `0' the excitation level and $k=(k^+,k^i)$ the momentum over the $X^+$ and transverse directions. A general state $|n,k\rangle$ has then excitation level $n$ and momentum $k$. The ground state is defined to be annihilated by the lowering operators $\alpha^i_m,\; m>0$ and to be an eigenstate of the momentum operators 
\be 
&& p^+ |0,k\rangle = 
   k^+ |0,k\rangle, \;\;\; p^i|0,k\rangle = k^i|0,k\rangle, \nn \\ 
&&\alpha^i_m|0,k\rangle = 0, \; m>0. 
\ee
A general state can be built upon the ground state by acting with linear combinations of creation operators of products of $\alpha^i_m,\; m<0$. From this point of view, the quantized string resembles a series of overlapping harmonic oscillators, as we can see from figure \ref{hilbert}. For the closed string we would have the tensor product of two modules: one for the $\alpha$ and another for the $\bar{\alpha}$.

The light-cone Virasoro constraints (\ref{LCvirasoroconstraint}) solve the classical open string $\alpha^-_n$ in terms of bilinear combinations of the $\alpha^i_n$. In the quantum theory, bilinear operator combinations are replaced normally-ordered operators. So we get
\be 
\alpha_n^- = \frac{1}{\sqrt{2\alpha'}p^+} 
             \left( \um\sum_{m=-\infty}^{+\infty}  
            :\alpha^i_{n-m}\alpha_{m}^i: -a\delta_{n,0}\right), 
\label{LCalphan} 
\ee 
with a similar expression for the $\bar{\alpha}_n^-$ in the closed string case. The dots $::$ stand for the usual normal ordering of the creation/annihilation operators and the factor $a\delta_{n,0}$ accounts for possible contributions coming from this normal-ordering. (Cf. also (\ref{NO})) The light-cone mass-shell condition is $m^2= 2p^+ p^- -p^i p^i$, which yields 
\be
\ba{lll}
\displaystyle{m^2 = \frac{1}{\alpha'}(N-a),} &  
\displaystyle{N=\sum_{n=1}^{\infty} \alpha^i_{-n}\alpha_{n}^i} & 
\displaystyle{\text{(Open string)}} \\ 
\displaystyle{m^2 =\frac{2}{\alpha'}(N-\overline{N}-2a),} & 
\displaystyle{\bar{N}=\sum_{n=1}^{\infty} 
              \bar{\alpha}^i_{-n}\overline{\alpha}_{n}^i} & 
\displaystyle{\text{(Closed string)}}
\ea
\ee 
The operators $N,\overline{N}$ are the level number operators, whose eigenvalues tell us the highest excitation level of a state. For closed strings the condition $N=\overline{N}$, coming from $L_0-\overline{L}_0=0$, must also be satisfied. Since the oscillators are integer-moded, each excitation raises the mass by $1/\alpha'$ ($4/\alpha'$ for the closed case) and the quantized string spectrum is thus discrete.
 
Open string excitations are generated by the transverse $\alpha_{-n}^i$ oscillators. At excitation level zero we have a scalar $|0,k\rangle$ of mass $-1/\alpha'$. At the first excited level we have a vector $\sum_i\xi_i\alpha^i|0,k\rangle,\; i=2,\dots, D$. A vector with $D-2$ components in a $D$-dimensional space-time must be massless, otherwise it would gain a longitudinal component under Lorentz transformations. The first excited string state has $N=1$, so, if Lorentz invariance is to be preserved, that state must have $m^2=0$, which means $a=1$. Knowing $a$, we can calculate $D$ explicitly from the $\alpha_0^-$ normal-ordering formula
\be 
\um\sum_{n=-\infty}^{+\infty} \alpha_{-n}^i \alpha_{n}^i =  
\um\sum_{n=-\infty}^{+\infty} :\alpha_{-n}^i \alpha_{n}^i: + \underbrace{\frac{D-2}{2} \sum_{n=1}^{\infty} n}_{=-a=-1}.  \label{NO}
\ee 
The infinite sum $\sum_{n>0} n$ is divergent but can be evaluated using Riemann $\zeta$-function regularization, yielding
\be 
\sum_{n=1}^{\infty} n \rightarrow -\frac{1}{12},\;\; \Rightarrow \;\; D=26. 
\ee 
Here is a remarkable result. It seems that quantized string is Lorentz-invariant only in 26 space-time dimensions. While the regularization argument may not be totally convincing, the result $a\!=\!1, D\!=\!26$ also comes out of other quantization methods, as we will see in the next section, where we will relate the critical dimension $D=26$ not to a space-time symmetry but to a world-sheet one.  
 
The mass of the ground state space-time scalar is now known,
\be 
m^2 = -\frac{1}{\alpha'} \;\;\; \text{(Open string)}, \;\;\;\;\;\; 
m^2 = -\frac{4}{\alpha'} \;\;\; \text{(Closed string)} 
\ee 
The negative mass-squared signals a tachyon, an object that moves faster than light and causes consistency problems at various levels. Later we will see how tachyons can be removed from the theory. At the first excited level we have states
\be 
m^2=0,& \;\; \sum_i\xi_i \alpha^i_{-1}|0,k\rangle \;\;\; \;\;\;\;\;\;\;\text{(Open string)} \nn \\
 m^2=0,& \; 
\; \sum_{ij} \Omega_{ij}\alpha^i_{-1} \bar{\alpha}^j_{-1}|0,k\rangle \;\;\; \text{(Closed string)}. 
\ee 
The open string state is the aforementioned space-time massless spin-1 vector particle, which can be identified with a photon. The closed string state can be decomposed in a space-time symmetric traceless tensor, which can be identified with a graviton; an anti-symmetric tensor and a scalar trace, called the dilaton 
\be 
\Omega_{ij} \alpha^i_{-1} \bar{\alpha}^j_{-1}|0,k\rangle = 
   |\Omega^{ij}\rangle \rightarrow \sum \left\{ \ba{cl} 
   |\Omega^{ij}\rangle = |\Omega^{ji}\rangle & \text{graviton (traceless)} \\ 
|\Omega^{ij}\rangle = -|\Omega^{ji}\rangle & 
    \text{anti-symmetric tensor field}\\ 
\delta_{ij}|\Omega^{ij}\rangle & \text{dilaton} 
\ea 
\right. 
\ee 
The presence of a graviton, a massless spin-2 particle, on the spectrum of a quantized theory was the original motivation for studying strings. At higher excitation levels we find all sorts of massive particles, filling various multiplets of $SO(25)$. All of them turn out to be bosons, though, so space-time fermions are missing in this simple model. 
 
Naturally, the string model exposed above is still quite far from reality. Solving the tachyon problem and obtaining space-time fermions on the spectrum can done by means of introducing world-sheet supersymmetry. The $D=26$ problem can be dealt with in two ways, as we will see below.

\subsection{Remarks on covariant quantization} 
Here we review very briefly some aspects of covariant quantization. The aim is to introduce the quantum Virasoro algebra.

Had we not solved the constraints $L_n=0$ at the classical level, and kept explicit space-time covariance, we would have had to impose extra conditions on states created by action of a full set of independent $\alpha^\mu_m$. If the constraints $L_n=0$ are not solved at the classical level, they would have to be so at the quantum level. In the quantized theory, these identities are promoted to operator identities acting on physical states. Solving these identities is crucial to avoid possible negative-norm states, such as for instance $\alpha_{-1}^0|0,k\rangle$, which has norm $-1$. This state was not present in the light-cone formalism because there only the transverse $\alpha^i_{-n}$ oscillators existed.

The quantum Virasoro $L_n$ operators are defined by normally-ordered classical expressions 
\be 
L_n = \sum_m :\alpha_{n-m}^\mu \alpha_{m\mu}:+a\delta_{n,0}, 
\ee 
where the normal-ordering constant is again present to account for ambiguities in defining $L_0$. From the commutator $[\alpha_{m}^\mu,\alpha_{n}^\nu]=m\delta_{m+n,0}\eta^{\mu\nu}$ we get the Virasoro algebra 
\be 
\left[L_m,L_n\right]=(m-n)L_{m+n} + A(m)\delta_{m+n}, 
\ee 
with $A(m)$ an anomaly term emerging from an eventual normal-ordering ambiguity. From the Jacobi identity we see that the general form of the anomaly is $A(m)=am^3+bm$. Acting with the commutator $[L_2,L_{-2}]$ on the string ground state with momentum $k=0$ we can determine $a$ and $b$, obtaining 
\be 
\left[L_m,L_n\right]=(m-n)L_{m+n} + \frac{1}{12}D(m^3-m)\delta_{m+n}. 
\ee 
This is the quantum version of the Virasoro algebra for $D$ world-sheet scalars $X^\mu$. Note that the anomaly term only appears because of the normal-ordering ambiguities, which are absent in the classical case. The presence of an anomaly term signals breakdown of a classical symmetry at the quantum level; in this case conformal symmetry. In quantum conformal field theory, the `central charge' number $c$ appearing in $A(m)=\frac{1}{12}c(m^3-m)$ is one of the parameters characterizing the theory. 
 
It is precisely because of the anomaly central term in the commutator $[L_n,L_{-n}]$ that we see that $L_n|\psi\rangle=0$, with $|\psi\rangle$ a physical state, is only possible for a subset of the $L_n$ 
\be 
\left( L_n - a \delta_{n,0} \right)|\psi\rangle = 0, \;\;\; n\geq 0. \label{OCQvirasoroconstraints} 
\ee 
This is however enough to ensure decoupling of unphysical states, because, since $L_n|\psi\rangle = 0, \; n>0$ and $\langle \psi| L_n = 0, \; n<0$, then $\langle \psi_1|L_n |\psi_2\rangle =0, \; \forall n$. Enforcing (\ref{OCQvirasoroconstraints}) protects the spectrum from physical negative-norm states if $a=1,D=26$. This is the `no-ghost theorem' \cite{noghost}. It can be shown that the spectrum of light-cone quantization is equivalent to the covariant one by means of this theorem.

\subsection{Unoriented strings and Chan-Paton factors}
The formulation developed for closed and open strings has the discrete symmetry of world-sheet parity, $\Omega$. This symmetry interchanges the left- and right-moving sector of the string and in a geometric description may also have a non-trivial action in space-time. The world-sheet effect of $\Omega$ on open strings is $\sigma \rightarrow \pi-\sigma$, which acts on the oscillators as
\be
\Omega: \; \alpha_n^\mu \rightarrow (-1)^n \alpha_n^\mu \;\;\;\;\;\text{(open string)}
\ee
and swaps the open strings' end-points, which can eventually introduce an extra minus sign.
On the closed string, the effect of $\Omega$ is to change $\sigma \rightarrow -\sigma$, which on the oscillators yields
\be
\Omega: \; \alpha_n^\mu \leftrightarrow \bar{\alpha}_n^\mu 
    \;\;\;\;\;\;\;\;\text{(closed string)}
\ee
Since $\Omega$ is an interchange of sectors, it squares to 1, so its eigenvalues are $\pm 1$. Now we can construct a new theory, the unoriented theory, by projecting out of the spectrum states which are not invariant under the action of $\Omega$. From the table below, we see this removes the open string photon and the closed string anti-symmetric tensor, leaving the tachyons and the graviton and dilaton.
\bc
\begin{table}[h]
\begin{tabular}{|c|c|} \hline
          open string & closed string  \\ \hline
Tachyon $\;\;\; \Omega\left( |k\rangle \right)= +\left( |k\rangle \right)$ &
Tachyon $\;\;\; \Omega\left( |k\rangle \right)= +\left( |k\rangle \right)$ \\ \hline
Photon   $\Omega\left( \alpha_{-1}^\mu |k\rangle\right) = -\left(\alpha_{-1}^\mu |k\rangle\right) $ &
 $\ba{c} \text{Graviton} \\ \text{Asym tensor} \\ \text{Dilaton}  \ea$ 
$\ba{c}
 \Omega \left( \Omega_{\mu\nu} \alpha_{-1}^\mu \bar{\alpha}_{-1}^\nu |k\rangle \right) = +\left( \Omega_{\mu\nu} \alpha_{-1}^\mu \bar{\alpha}_{-1}^\nu |k\rangle \right)\\ \Omega \left( \Omega_{\mu\nu} \alpha_{-1}^\mu \bar{\alpha}_{-1}^\nu |k\rangle \right) = -\left( \Omega_{\mu\nu} \alpha_{-1}^\mu \bar{\alpha}_{-1}^\nu |k\rangle \right)\\ \Omega \left( \delta_{\mu\nu} \alpha_{-1}^\mu \bar{\alpha}_{-1}^\nu |k\rangle \right) = +\left( \delta_{\mu\nu} \alpha_{-1}^\mu \bar{\alpha}_{-1}^\nu |k\rangle \right)
\ea$ \\
\hline
\end{tabular}
\caption{$\Omega$-eigenvalues of the bosonic string spectrum}
\end{table}
\ec
Projecting out the gauge bosons is however not what one wishes to do from a phenomenological point of view. The gauge theory can however be restored using the trick of attaching a label, usually called Chan-Paton factor, to the open string end-points \cite{cpfactors}. Adding these {\em ad hoc} degrees of freedom is compatible with all the symmetries of the string action and enhances the open string spectrum. The open string states are then characterized by two extra labels $|N,k\rangle {\rightarrow} \lambda_{ij}|N,k,ij\rangle$. The action of $\Omega$ when Chan-Paton factors are present is
\be
\Omega \lambda_{ij}|N,k,ij\rangle = \lambda'_{ji} |N,k,ji\rangle, \;\;\; \lambda' = M\lambda^T M^{-1},
\ee
with the prime on the matrix $\lambda'$ reflecting the above mentioned non-trivial action on the end-points. (It can, for instance, change the sign of the Moebius strip projection.) Acting with $\Omega$ twice we see that $M$ can be symmetric or anti-symmetric, leading to $\lambda = \pm \lambda'$ respectively \cite{tasi}. If the number of Chan-Paton factors is $n$, before introducing $\Omega$ there were $n^2$ photonic states, which correspond to the degeneracy of states for a gauge group $U(n)$. For symmetric/anti-symmetric $\lambda$ only $\um n(n\pm 1)$ states survive the $\Omega$-projection and the gauge group becomes $Sp(n)$ or $SO(n)$ respectively.

Chan-Paton factors can in this way be used bring back the gauge theory, giving the unoriented theory a structure similar to the oriented one, but with different gauge groups and without the anti-symmetric tensor field from the closed string sector. Unoriented strings introduce crosscaps on the world-sheet, which can now include surfaces like Klein bottles and Moebius strips.

\section{Path-integral quantization} \label{faddeevpopov}
The presence of the graviton on the quantized spectrum was original motivation for studying strings. The motivation for this thesis is in turn to solve one of the problems that arise in the quantized theory: $D=26$. There is an interesting and elegant way out of the space-time dimension problem. This solution is not obvious from the previous quantization methods, but becomes clear when we introduce an alternative method to quantize strings, path-integral quantization.

The path-integral method is a versatile quantization method, particularly fit to quantize systems with gauge symmetry, which often gives insight that is difficult to get otherwise. When applied to strings, it will introduce the ghost system and with it an idea to deal with the space-time dimension problem. The derivation of the string path-integral presented here is simplified. For a more technical derivation see \cite{polyakov}.
 
The idea of a path-integral is to represent transition amplitudes between states by an integration over all possible classical paths, each of them exponentially weighted by the corresponding classical Euclidean action $e^{-S}$. This Euclidean path-integral takes the form
\be
Z=\int {\cal D}g \, {\cal D}X^\mu \; e^{-S[g,X^\mu]}. \label{naivePI}
\ee
Now, we know that many of the world-sheet metrics $g$ over which we are integrating are actually physically equivalent because they are related via reparametrizations and/or conformal transformations. Therefore the path-integral (\ref{naivePI}) contains a huge over counting which must be eliminated if the integral is to make sense.

The modern way to deal with the gauge freedom in path-integrals is via the Faddeev-Popov procedure. The first step in this procedure is to choose a reference metric on the world-sheet, say, $\hat{g}_{ab}$. We have seen that classically all metrics are essentially equivalent to this one. It is pertinent to ask ourselves whether the same is true in the quantum case. This is an important point and we will come back to it later. For the moment let us assume it is always possible to go from any metric into the reference metric via a combined reparametrization and conformal transformation. Such a general coordinate transformation $t$ is
\be
t:g\rightarrow g^t, \;\;\; g^t_{ab}(\tilde{\sigma}) = e^{2\omega({\sigma})} \frac{\partial \sigma^a}{\partial \tilde{\sigma}^c} \frac{\partial \sigma^b}{\partial \tilde{\sigma}^d}g_{cd}(\sigma).
\ee
The second step in the Faddeev-Popov method is to write `1' as
\be
1=\det{{}_{FP}}(g)\int {\cal D}t\; \delta(g-\hat{g}^t), \label{FPdeterminant}
\ee
where $\det_{FP}(g)$ is the Faddeev-Popov determinant and ${\cal D}t$ is a gauge-invariant measure, whose specific form will not be needed. The argument of the delta-function means we used $t$ to bring $g$ into the reference metric $\hat{g}$. The next step is to insert `1' into the path-integral
\be
Z[\hat{g}] = \int {\cal D}g \,{\cal D}X^\mu {\cal D}t \; \det{{}_{FP}}(g) \; \delta(g-\hat{g}^t) \; e^{-S[g,X^\mu]}. \label{PI1}
\ee
Now we can do the ${\cal D}g$ integration. The delta-function requires that $g=\hat{g}^t$ everywhere, so we get 
\be
Z[\hat{g}] = \int {\cal D}X^\mu \, {\cal D}t \; \det{{}_{FP}}(\hat{g}^t) \; e^{-S[\hat{g}^t,X^\mu]}. \label{PI2}
\ee
The Faddeev-Popov determinant is invariant under gauge transformations and so is the action $S$ and ${\cal D}X^\mu$, so (\ref{PI2}) is actually independent of $t$. Being so, we can perform the ${\cal D}t$ integration, which contributes with an infinite gauge volume factor and drops out as an overall normalization factor. We are then left with
\be
Z[\hat{g}] = \int {\cal D}X^\mu \; \det{{}_{FP}}(\hat{g}) \; e^{-S[\hat{g},X^\mu]}. \label{PI3}
\ee
The final step is to rewrite the Faddeev-Popov determinant as a path-integral over anti-commuting ghost fields $b,c$
\be
\det{{}_{FP}}(\hat{g}) = \int {\cal D}b \, {\cal D}c \; e^{-S_g}.
\ee
Polyakov's action then becomes
\be
Z[\hat{g}]= \int {\cal D}X^\mu \,{\cal D}b \, {\cal D}c \; e^{-S_{X}-S_g}.\label{PI4}
\ee
The construction of the ghost action $S_g$ requires careful differential geometry calculations. In the end it turns out that the $c$ ghost is an anti-commuting vector field (thus it has one index) of conformal weight $-1$, whereas the anti-ghost $b$ is a symmetric traceless tensor (has two indices) of conformal weight $2$. Now we pick our reference metric to be the conformal metric $\hat{g}_{ab}=e^{2\omega}\eta_{ab}$, under which the ghost action is, in terms of the world-sheet left- and right-movers,
\be
S_g = \frac{\I}{\pi}\int d\sigma^+ d\sigma^- \; 
   \left( c^+ \partial_- b_{++} + c^- \partial_+ b_{--}\right). \label{Sg}
\ee

Looking at (\ref{PI4}) one sees that path-integral quantization gives a very interesting insight: getting rid of the gauge freedom in the metric of the string action is equivalent to choosing a particular gauge and adding ghost fields to the world-sheet theory. In conformal field theory language, one says that the world-sheet gauge-fixed theory is the tensor product of $D$ world-sheet bosons $X^\mu$ with the ghost conformal field theory. It can be shown that ghost system acts in such a way that the spectrum of the world-sheet theory of `bosons plus ghosts' exactly matches the spectrum of light-cone and covariant quantization. The Faddeev-Popov path-integral is therefore an alternative and equivalent description of the quantized string.

The world-sheet conformal field theory of bosons plus ghosts is worth exploring. The ghost energy-momentum tensor can be derived as the variation of the ghost action with respect to the metric (before going to reference metric, of course). In terms of $\sigma^\pm$ it simplifies to
\be
T^g_{++}&=& \I \left( 2b_{++} \partial_+ c^+ + (\partial_+ b_{++}) c^+ \right), \nn \\
T^g_{--}&=& \I \left( 2b_{--} \partial_- c^- + 
               (\partial_- b_{--}) c^- \right),
\label{Tg}
\ee
with other components vanishing. The equations of motion for the ghosts are
\be
\partial_- c^+ = \partial_- b_{++} = \partial_+ c^- = \partial_+ b_{--}=0.
\ee
The ghosts, being anti-commuting, are canonically quantized by equal-$\tau$ anti-commutation relations
\be
&& \left\{ b_{++}(\sigma),c^+(\tilde{\sigma})\right\} = 
   2\pi \delta(\sigma -\tilde{\sigma}), \nn \\
&& \left\{ b_{--}(\sigma),c^-(\tilde{\sigma})\right\} = 
   2\pi \delta(\sigma -\tilde{\sigma}), \label{ghostAC}
\ee 
supplemented by periodicity conditions (closed string) or boundary conditions (open string). The explicit open string solution is
\be
   c^+=\sum_n c_n e^{-\I n(\tau+\sigma)}, &&  
   c^-=\sum_n c_n e^{-\I n(\tau-\sigma)}, \nn \\
   b_{++}=\sum_n b_n e^{-\I n(\tau+\sigma)}, &&
   b_{--}=\sum_n b_n e^{-\I n(\tau-\sigma)}, \label{bcopen}
\ee
which, when inserted in (\ref{ghostAC}) yields
\be
   \left\{ c_m,b_n \right\} = \delta_{m+n,0}, \;\;\;
   \left\{ c_m,c_n \right\} = 0, \;\;\; 
   \left\{ b_m,b_n \right\} = 0.
\ee
For closed strings $c^+,b_{++}$ and $c^-,b_{--}$ have independent mode expansions, which leads to a second set of modes $\bar{c}_n, \bar{b}_n$. The ghost Virasoro operators for the open string are defined by $L^g_m=\frac{1}{\pi}\int_{-\pi}^\pi d\sigma (e^{\I m\sigma} T^g_{++} + e^{-\I m\sigma} T^g_{--}),$ at $\tau=0$. Defining the quantum $L^g_m$'s by the normal-ordered expressions we get
\be
L_m^g = \sum_{n=-\infty}^{+\infty} (m-n):b_{m+n}c_{-n}: + b\delta_{m+n,0},
\ee
where we inserted $b$ as the ghost normal-ordering constant. The algebra of the $L_m^g$ is similar to the one for the bosons, but the anomaly derived from the Jacobi identity is different and leads to
\be
\left[L_m^g,L_n^g \right]=(m-n)L^g_{m+n} + A^g(m),\;\;\;  
   A^g(m)=\frac{1}{6}(m-13m^3).
\ee
We can define a complete world-sheet energy momentum tensor by simply adding the energy-momentum tensors coming from the $X^\mu$ bosons and the ghost system. This leads to complete quantum Virasoro generators
\be
L_m = L_m^X + L_m^g - a \, \delta_{m,0}, \;\;\; 
\left[L_m,L_n \right]=(m-n)L_{m+n} + A(m),
\ee
with total conformal anomaly
\be
 A(m)=\frac{1}{12}D(m^3-m) + \frac{1}{6}(m-13m^3) + 2am.
\ee
Again, this vanishes for the magical values $D=26,\, a=1$. If we incorporate the normal-ordering factor $2am$ into the ghost part, we see that the ghost system has a central charge $c^g=-26$.

We turn now to the remark made in the beginning of this section. As we see, existence of a conformal anomaly for $D\neq 26$ signals breakdown of conformal symmetry at the quantum level. Since reparametrizations can only bring an arbitrary metric into the form $e^{2\omega}\eta_{ab}$, conformal transformations are always necessary to eliminate the exponential factor. If conformal symmetry is not exact at the quantum level, the ${\cal D}g$ integration will not decouple. It is only if the total central charge of the system vanishes that the decoupling of ${\cal D}g$ is valid and path-integral calculation holds.

\subsection{Internal theories}
The elegance of the Faddeev-Popov method is that it re-expresses the gauge freedom of the string in terms of a ghost system which contributes $-26$ to the conformal anomaly. This ghost contribution is an expression of the classical string symmetries and does not change with the space-time dimension. To cancel the conformal anomaly, one must add bosons $X^\mu$ to the world-sheet theory. Each of them contributes with $c=+1$ to the total anomaly, and therefore when we add 26 of them the anomaly vanishes and we are done. But adding a boson $X^\mu$ also means adding a space-time dimension, so when we put in 26 of them, we go to $D=26$.

This leads to the following idea. We are adding free bosons $X^\mu$ only. Couldn't we add something else instead? The answer is yes. We can, in fact, tensor the world-sheet bosons and ghosts with whatever conformal field theory we want, as long as it is unitary, i.e. that it doesn't introduce negative-norm states into the theory. The central charge of each theory adds up and, as long as it adds up to zero, the outcome is in principle perfectly consistent. The dimensional problem $D=26$ has therefore a very simple solution. Find so-called `internal' conformal field theories $I$, tensor them with the world-sheet bosons and ghosts in such a way that
\be
c^X+c^g+c^I=0.
\ee
For instance, we can arrive at our favorite number of space-time dimensions $D=4$ by adding four bosons and finding an internal conformal field theory $I$ such that $c^I = 0-4+26 = 22$. Adding an internal conformal field theory to the world-sheet produces changes at various levels, such as the spectrum, transition amplitudes, partition functions and much more. 

\subsection{Superstrings and compactification}
The extra fields added to the world-sheet theory can have non-trivial space-time interpretation. For instance, we can add fermions with space-time indexes $\mu$ to the world-sheet. Historically this was motivated by the desire to have space-time fermions in the spectrum, something that could not be achieved with world-sheet bosons $X^\mu$ alone. By inserting world-sheet fermions and requiring a supersymmetric space-time spectrum \cite{gso} it was also possible to project out the bothersome tachyonic states. In conformal gauge, the supersymmetric world-sheet action for $D$ bosons and fermions is
\be
S=-\frac{1}{2\pi} \int d^2\sigma  \left\{ \partial_a X^\mu \partial^a X_\mu -\I \bar{\psi}^\mu \rho^a \partial_a \psi_\mu  \right\}, \label{SSaction}
\ee
where $X^\mu(\st)$ are the usual world-sheet bosons, $\psi^\mu(\st)$ a world-sheet two-component spinor which transforms in the vector representation of the Lorentz group $SO(D-1,1)$ and $\rho^a$ the two-dimensional gamma matrices. This action can be quantized by the path-integral method. The world-sheet supersymmetry also produces changes in the Faddeev-Popov determinant. As one could expect since the world-sheet theory is supersymmetric, the $b,c$ ghosts gain super partners, the $\beta,\gamma$ ghosts, which turn out to contribute +11 to the central charge. Each fermion contributes $1/2$, so in the end we have
\be
c = c^X + c^\psi + c^g + c^{sg} = D+\frac{1}{2}D -26 + 11,
\ee
from which we get the critical dimension $D=10$. This is better than the purely bosonic $D=26$, but still means we need an internal conformal field theory.

Path-integral quantization is of course but one way to look at the problem of quantizing strings. After all, in covariant or light-cone quantization the ghost formalism is not used and therefore sense must be made of the extra dimensions. The usual way to deal with them is via the Kaluza-Klein mechanism, or compactification. This amounts to assuming that the extra dimensions are compact and small enough to be out of accelerator reach. We then have strings living in the critical dimension, but with some directions highly curled up.

Compactification is a geometric and intuitive approach to the dimensional problem, which is why it has become much more popular than the abstract approach of adding {\em ad hoc} degrees of freedom. The two ways of looking at the problem should in the end be equivalent, although that remains a conjecture. A formal proof would take a titanic effort, but some steps towards that end have been taken in simple cases \cite{geomwzw}. Each method has its advantages and drawbacks. Compactification, while intuitive, easier to deal with and having all the machinery of differential geometry to help characterizing the compactification manifolds, has problems in going beyond simple calculations. It also feels unnatural to impose a specific geometry on which to quantize strings, when the strings themselves create geometry. The algebraic method on the other hand, while being quite abstract and shedding little light as to what is going on, makes however explicit stringy calculations feasible.

It is the purpose of this book to deepen the understanding of the internal conformal field theory by studying examples and checking the strict consistency requirements they must obey.

\chapter{Conformal Field Theory}\label{cftchap}
We have seen that building a string theory outside the critical dimension can be done by means of adding an internal conformal field theory to the world-sheet theory. In order to understand the internal theory better, we review the key concepts of classical and quantum conformal field theory.

\section{Classical conformal field theory}
Consider classical field theory in $D$-dimensions, with action
\be
S= \int d^D x \, {\cal L} 
   \Big(  g_{\mu\nu}(x), \phi(x), \partial_x \phi(x) \Big).
\ee
This action can have a series of symmetries, each of which gives rise to a conserved quantity via the Noether theorem. For instance, invariance under general coordinate transformations $x \rightarrow x'(x)$ is equivalent to $D_\mu T^{\mu\nu}=0$, with $D_\mu$ a covariant derivative.

Another symmetry the action can have is Weyl invariance, which is invariance under local re-scalings of the metric; that is, transformations of the metric of type $g_{\mu\nu}(x)\rightarrow \Lambda(x) g_{\mu\nu}(x)$. Such symmetry implies a traceless energy-momentum tensor: $T_\mu^\mu=0$. A conformal transformation is a coordinate transformation which acts on the metric as a Weyl transformation. If the action is Weyl-invariant, the action of the conformal transformation on the metric can be compensated by a Weyl transformation. The theory is then invariant under conformal transformations and is said to be a conformal field theory.

Since the string world-sheet is two-dimensional, we concentrate on two-dimensional conformal field theories from now on. This is most fortunate since it is precisely in two dimensions that conformal symmetry is especially powerful. As we have seen from the string, two-dimensional metrics are essentially trivial, so we take as reference metric the Euclidean metric\footnote{Going from world-sheet Minkowskian metric to Euclidean is done to make use of the machinery of complex analysis. However, one has to make sure the relevant quantities can be analytically continued to the complex plane.} $(+,+)$. In this case, an infinitesimal coordinate transformation $x^{\mu} = {x'}^{\mu} - \epsilon^\mu(x^\mu)$ is conformal if
\be
\partial_{x^1} \epsilon^1 = \partial_{x^2} \epsilon^2, \;\;\;
\partial_{x^1} \epsilon^2 = -\partial_{x^2} \epsilon^1,
\ee
or, going to complex coordinates $z,\bar{z}=x^1\mp ix^2$,
\be
\partial_z \bar{\epsilon}(z,\bar{z})=0,\;\;\; \partial_{\bar{z}} {\epsilon}(z,\bar{z})=0.
\ee
The global form of such a transformation would be $z\rightarrow f(z), \bar{z}\rightarrow \overline{f}(\bar{z})$. The generators for infinitesimal transformations of a function of $\zz$ are
\be
L_n = -z^{n+1}\partial_z, \;\;\;
\overline{L}_n = -\bar{z}^{n+1}\partial_{\bar{z}},
\ee
which satisfy the classical Virasoro algebra
\be
\left[ L_m,L_n \right]=(m-n)L_{m+n},\;\;\;
\left[ \overline{L}_m,\overline{L}_n \right]=(m-n)\overline{L}_{m+n},\;\;\;
\left[ L_m,\overline{L}_n \right]=0.
\ee
In complex coordinates, energy-momentum conservation is simply $\partial_{\bar{z}}T_{zz}=\partial_{{z}}T_{\bar{z}\bar{z}}=0$. We see that the energy-momentum tensor has a purely holomorphic and a purely anti-holomorphic component, respectively $T(z),\overline{T}(\bar{z})$. The corresponding conserved charges are
\be
Q_\epsilon = \int dx^2 \epsilon(z)T(z), \;\;\;
Q_{\bar{\epsilon}} = \int dx^2 \bar{\epsilon}(\bar{z})\overline{T}(\bar{z}).
\ee
There is an infinity of conserved charges, since we have one for every function $\epsilon$ that is holomorphic. Conformal symmetry in two dimensions has then an infinite-dimensional symmetry group.

\section{Quantum conformal field theory}
To quantize the conformal field theory we go first to the complex plane via the conformal transformation $\omega = e^z$. The closed string Euclidean theory lives on a cylinder of periodic spatial boundary conditions, so the world-sheet fields obey $\phi\zz=\phi(e^{2\pi \I}z,e^{-2\pi \I}\bar{z})$. Upon mapping the cylinder to the plane, time on the plane flows in circles from the center to the exterior and spatial integrals become contour integrals around the origin $z=0$. Time-ordering of fields in correlators on the cylinder becomes radial ordering on the plane. From now on we work on the plane unless otherwise stated and use $\zz$ as coordinates on the plane (i.e. we {renamed} $(\omega,\bar{\omega})$ back into $\zz$). Conformal field theory on the plane describes closed strings. To describe open strings we need conformal field theory on the half-plane, which we review later in chapter \ref{sewchap}.

Going to the Heisenberg picture we see that the commutator $[Q_\epsilon,\phi{\zz}]$ generates quantum conformal transformations. In order for the infinitesimal quantum conformal transformations to be in agreement with the commutator, the radially-ordered product of the energy-momentum tensor with $\phi\zz$ must be
\be
T(z)\phi\ww &\sim& \frac{h}{(z-w)^2}\phi\ww + \frac{1}{z-w}\partial_w\phi\ww + \cdots \nn \\
\overline{T}(\bar{z})\phi\ww &\sim& \frac{\bar{h}}{(\bar{z}-\bar{w})^2}\phi\ww + \frac{1}{\bar{z}-\bar{w}}\partial_{\bar{w}}\phi\ww + 
\cdots \label{TfOPE}
\ee
where the dots stand for an analytic power series in $(z-w)$ or $(\bar{z}-\bar{w})$. The numbers $h,\bar{h}$ are the holomorphic and anti-holomorphic conformal weights of $\phi\zz$. A conformal field, or primary field is a field which has an operator product with the holomorphic and anti-holomorphic components of the energy-momentum tensor of the form (\ref{TfOPE}).

To fully describe a conformal field one has to attach to it a chiral (or holomorphic) and an anti-chiral (or anti-holomorphic) label, which tells us in which what representation of the chiral algebra it transforms (see below). Formulas where two nearby fields are expressed in terms of a sum of single fields, are called operator products. Attaching the chiral and anti-chiral labels, operator products take the general form
\be
\phi_{i\bar{i}}\zz \phi_{j\bar{j}}\ww \sim \sum_{k\bar{k}} C_{i\bar{i},j\bar{j}}^{\ph{i\bar{i},j\bar{j}}k\bar{k}} (z-w)^{h_k-h_i-h_j} 
(\bar{z}-\bar{w})^{\bar{h}_k-\bar{h}_i-\bar{h}_j} \phi_{k\bar{k}}\ww +\cdots
\ee
The coefficient $C_{i\bar{i},j\bar{j}}^{\ph{i\bar{i},j\bar{j}}k\bar{k}}$ is called an operator product coefficient and its value depends on the details of the theory. The energy-momentum has an operator product with itself of the form
\be
T(z)T(w) \sim \frac{c/2}{(z-w)^4} + \frac{2}{(z-w)^2}T(w) + \frac{1}{z-w}\partial_w T(w)+\cdots
\ee
(and similar for $\overline{T}(\bar{z})\overline{T}(\bar{w})$). We see the central charge arising in this operator product. If $c=0$, the energy-momentum tensor is a conformal field of weight 2. The classical $L_m,\overline{L}_m$ generators become in the quantum theory modes of the energy-momentum tensor
\be
L_m = \oint_0 \frac{dz}{2\pi \I} z^{m+1} T(z), \;\;\;
\overline{L}_m = \oint_0 \frac{d\bar{z}}{2\pi \I} \bar{z}^{m+1} \overline{T}(\bar{z}).
\ee
(Remember that the integral over space $\int \!d(x^2)$ becomes in the complex plane a contour integral around the origin.) Doing the integrals, we see that the modes $L_m$ obey the Virasoro algebra
\be
\left[ L_m,L_n \right] = (m-n)L_{m+n} + \frac{c}{12}m(m^3-m)\delta_{m+n,0}\,.
\ee
Being an algebra, its representations can be studied. Of particular importance are the irreducible highest-weight representations, which are annihilated by $L_n,n\geq 1$. If the number of these representations is finite, the conformal field theory is said to be {\em rational}. An highest-weight representation can be related to a conformal field $\phi_{i\bar{i}}$ of weight $h_i,\bar{h}_i$\footnote{The anti-holomorphic weight can also be designated by $h_{\bar{i}}$.} via the isomorphism
\be
|h_i,\bar{h}_i\rangle = 
   \lim_{z\rightarrow 0} \phi_{i\bar{i}}\zz|0\rangle, \;\;\; 
   \langle h_i,\bar{h}_i| = \lim_{z\rightarrow \infty} |z|^{-2(h_i+\bar{h}_i)}
   \langle 0|\phi_{i\bar{i}}\zz. \label{inandout}
\ee
where $|0\rangle$ is the vacuum of the theory, the state with the most symmetries and therefore the state which is annihilated by the most $L_n$'s, in this case $L_n|0\rangle = 0,n\geq -1$. Highest-weight states have eigenvalue $L_0|h,\overline{h}\rangle = h |h,\bar{h}\rangle,\bar{L}_0|h,\bar{h}\rangle = \bar{h} |h,\bar{h}\rangle$ and upon them a whole representation module (the Verma module) can be build by acting with $L_m,m<0$. One can also relate non-highest-weight states with a field. Such field is called a descendant field.

The Hilbert space of a conformal field theory is the tensor product of a holomorphic and an anti-holomorphic {\em chiral algebra}
\be
{\cal H} = {\cal A} \otimes \overline{{\cal A}}.
\ee
For a general state in ${\cal H}$, the holomorphic representation need not coincide with the anti-holomorphic one, which is why the corresponding primary fields, which are sometimes also called bulk fields, have two labels instead of just one. This will be made clearer when we study the partition function of a conformal field theory. The Virasoro algebra is the simplest example of a chiral algebra.

\subsection{Extended chiral algebras}
It can happen that a conformal field theory has more symmetry than just the Virasoro algebra. The chiral algebra algebra spanned by the $L_m$ modes of the energy-momentum tensor can be enlarged by modes of holomorphic currents $J(z)$ such that
\be
\left[ L_m,J_n \right] = (m(h_J-1)-n)J_{m+n},\;\;\;
J_n = \odz z^{n+h_J-1}J(z). \label{LJ}
\ee
There is, of course, also a commutator $[J_m,J_n]$ to close the algebra\footnote{The index $n$ on the current modes need not be integer.}, but its form depends on the explicit form of $J(z)$. The algebra (\ref{LJ}) is an {\em extended} chiral algebra. Extended chiral algebras have in general bigger Verma modules than unextended ones because to generate a whole representation we can act not only with the $L_m, m<0$ modes but also with $J_n, n<0$. Unextended representations are uniquely determined by the values of $(c,h)$, but this is not true for extended ones because $h$-eigenvalues states may behave differently under $J_n$.

In the presence of extended symmetry, the definition of the out-going state in (\ref{inandout}) changes due to non-trivial charge conjugation. It becomes
\be
\langle h_i,\bar{h}_i| = 
\lim_{z\rightarrow \infty} |z|^{-2(h_i+\bar{h}_i)}
\langle 0|\phi_{i\bar{i}}^\dagger\zz,
\ee
with $\phi_{i\bar{i}}^\dagger \equiv \phi_{i^c\bar{i}^c}$, where the label $i^c$ is said to be the charge conjugate label of $i$. Labels for which $i=i^c$ are called real and those with $i\neq i^c$ complex. Chiral labels can be lowered and raised with the charge conjugation matrix $C_{ij}=\delta_{ij^c}$.

\subsection{Correlation functions on the plane}
Correlation or Green's functions are what one wants to compute in a quantum field theory. They are transition amplitudes, with which we can study any scattering process. If the conformal field theory is formulated on the plane, correlators take the form of an expectation value
\be
\langle 0| \phi_{i_1\bar{i}_1}(z_1,\bar{z}_1) \ldots \phi_{i_n\bar{i}_n}(z_n,\bar{z}_n) |0\rangle, \;\;\; |z_{n}|<|z_{n-1}|. \label{generalcorrelator} 
\ee
Requiring invariance of correlators under infinitesimal transformations generated by $L_{-1},L_0,L_1$ imposes constraints on their form. Assuming for simplicity $\bar{i}={i}$ for primary fields, the two-, three- and four-point functions take the form\footnote{One-point functions vanish due to translation invariance and zero-point functions are trivial.}
\be
\langle 0| \phi_{1}(z_1,\bar{z}_1) \phi_{2}(z_2,\bar{z}_2) |0\rangle 
&=& {\delta_{12^c}}{|z_{12}|^{-4h_1}}, \nn \\
\langle 0| \phi_{1}(z_1,\bar{z}_1) \phi_{2}(z_2,\bar{z}_2) 
           \phi_{3}(z_3,\bar{z}_3)|0 \rangle &=&
   C_{12}^{\ph{12}3}|z_{12}|^{h_3-h_2-h_1}|z_{13}|^{h_2-h_1-h_3}
      |z_{23}|^{h_1-h_2-h_3}, \nn \\
\langle 0| \phi_{1}(z_1,\bar{z}_1) \phi_{2}(z_2,\bar{z}_2) 
           \phi_{3}(z_3,\bar{z}_3)\phi_{4}(z_4,\bar{z}_4)|0\rangle &=& 
f(x,\bar{x}) \prod_{i<j} |z_{ij}|^{-h_i-h_j+h/3}, \label{1234corr}
\ee
with $z_{ij}=z_i-z_j$ and $x=z_{12}z_{34}/z_{13}z_{24}$. The two-point function need not be normalized to unity; a redefinition $\phi\rightarrow\lambda\phi$ may change it\footnote{With $\lambda$ real, otherwise the Hilbert space becomes complex.}. By the same token, the operator product coefficient in the three-point function can be redefined. Note that both the form of the two- and the three-point function is completely determined by conformal invariance. The four-point function is partially determined by it, but a general dependence on $f(x,\bar{x})$ remains, whose form depends specifically on the theory at hand.

Correlators involving descendant fields can be derived from correlators with primary fields by applying differential operators \cite{bpz}.

\section{String perturbation theory}
\begin{figure}
\SetScale{1}
\begin{center}
\begin{picture}(400,200)(20,0)
\Text(30,160)[l]{$\langle a|b \rangle\;\;\;=$}
\Line(100,160)(140,160)
\Vertex(100,160){1}
\Vertex(140,160){1}
\Text(155,160)[l]{$+$}
\Line(180,160)(200,160)
\BCirc(210,160){10}
\Line(220,160)(240,160)
\Vertex(180,160){1}
\Vertex(240,160){1}
\Text(255,160)[l]{$+$}
\Line(280,175)(300,175)
\BCirc(310,175){10}
\Line(320,175)(340,175)
\BCirc(350,175){10}
\Line(360,175)(380,175)
\Vertex(280,175){1}
\Vertex(380,175){1}
\Line(280,145)(320,145)
\BCirc(330,145){10}
\Line(340,145)(380,145)
\Line(330,155)(330,135)
\Vertex(280,145){1}
\Vertex(380,145){1}
\Text(395,160)[l]{$+\;\cdots$}
\Text(220,120)[c]{QFT Feynmann diagrams}
\Text(220,10)[c]{Closed string diagrams}
\Text(30,50)[l]{$\langle a|b \rangle\;\;\;=$}
\Text(255,50)[l]{$+$}
\Text(155,50)[l]{$+$}
\Text(395,50)[l]{$+\;\cdots$}
\Line(100,40)(140,40)
\Line(100,60)(140,60)
\Oval(100,50)(10,5)(0)
\Oval(140,50)(10,5)(0)
\Line(180,40)(190,40)
\Line(180,60)(190,60)
\Line(230,40)(240,40)
\Line(230,60)(240,60)
\Oval(180,50)(10,5)(0)
\Oval(240,50)(10,5)(0)
\Curve{(190,60)(210,70)(230,60)}
\Curve{(190,40)(210,30)(230,40)}
\Curve{(200,50)(210,45)(220,50)}
\Curve{(205,47)(210,50)(215,47)}
\Line(280,40)(290,40)
\Line(280,60)(290,60)
\Line(370,40)(380,40)
\Line(370,60)(380,60)
\Line(320,40)(340,40)
\Line(320,60)(340,60)
\Oval(280,50)(10,5)(0)
\Oval(380,50)(10,5)(0)
\Curve{(290,60)(305,70)(320,60)}
\Curve{(340,60)(355,70)(370,60)}
\Curve{(290,40)(305,30)(320,40)}
\Curve{(340,40)(355,30)(370,40)}
\Curve{(295,50)(305,45)(315,50)}
\Curve{(345,50)(355,45)(365,50)}
\Curve{(300,47)(305,50)(310,47)}
\Curve{(350,47)(355,50)(360,47)}
\end{picture}
\end{center}
\caption{String perturbation series}
\label{pert}
\end{figure}
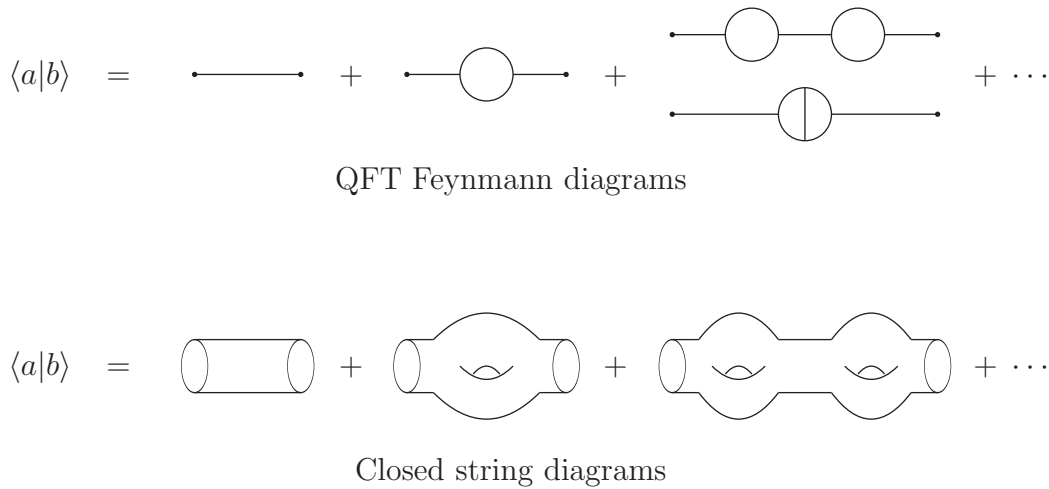

Conformal field theory on the complex plane represents but the first order in string perturbation theory. In field theories like quantum electrodynamics (or chromodynamics in the high-energy regime), one calculates a full transition amplitude from an initial state $|b\rangle$ to a final state $|a\rangle$ by summing up over all possible intermediate processes, each of them represented by a Feynman diagram. In String Theory we can adopt the same principle, this time replacing the point-particles' lines by stringy tubes or strips. The pictorial representation of these two is shown on figure \ref{pert}.

The tree-level closed string diagram is topologically equivalent to a sphere with two insertion points (the initial and final states). This we can map to the complex plane with two field insertions. The sphere is the first order diagram in string perturbation theory, so to study the perturbation series to higher orders one must then study conformal field theory on more complicated surfaces. The general form of the transition amplitude for an $n$-point scattering process would be
\be
\langle 0| V_1(z_1,\bar{z}_1)\ldots V_n(z_n,\bar{z}_n) |0\rangle 
=\sum_{\cal T} \int {\cal D}g \, {\cal D}X^\mu \! \int \! d^2 z_1 \ldots d^2 z_n  V_1(z_1,\bar{z}_1)\ldots V_n(z_n,\bar{z}_n) \, e^{-S[g,X^\mu]}. \label{An} \nn
\ee
This amplitude for the scattering process reads as follows: it is the sum over all possible world-sheet topologies ${\cal T}$ of the path-integral with $n$ vertex operator insertions integrated over. A vertex operator  $V_i\zz$ is some combination of primary fields with a physically meaningful space-time interpretation, which generate the in-going (or outgoing) strings states. For a closed string scattering process, the $n$-point amplitude is the sum over the sphere, the torus, the two-handled torus, etc. with $n$ vertex operator insertions. If the sum is to converge, the contributions from the various topologies must of course decrease as the number of handles increases. If this is the case, we say we are in the string perturbative regime.

Naturally, this crude form of the transition amplitude again over-counts configurations and must be made sensible via the Faddeev-Popov procedure. Each particular topology has its own peculiarities and must be dealt with separately. For example, we have seen that at the tree-level every closed string surface regardless of its metric $g$ could be conformally brought to a sphere and subsequently to a complex plane. For the torus what happens is that we can conformally bring every possible closed, oriented genus one surface to a torus with one complex parameter. Thus, using conformal invariance we avoid having to consider all the infinity of configurations for $g$, which is what actually makes computations possible. This is one of the reasons why quantum conformal invariance is so crucial in String Theory.

\subsection{Open and closed surfaces at one-loop}
Since we will be interested in open string theories, we will have unoriented open and closed strings, which requires us to consider world-sheet surfaces with boundaries and crosscaps. Since this increases our array of possibile surfaces, we need to keep track of the order in perturbation theory, and for that we use the Euler number $\chi$, whose exponential is the inverse string coupling \cite{gsw}. The Euler number counts the number of handles, boundaries and crosscaps of the two-dimensional string world-sheet
\be
\chi =  2(1-h) - b - c.
\ee
The first topologies are listed below

\begin{table}[h]
\begin{center}
\begin{tabular}{l|ccc|c}
   Surface & $\hspace{.1cm}h\hspace{.1cm}$ & 
             $\hspace{.1cm}b\hspace{.1cm}$ & 
             $\hspace{.1cm}c\hspace{.1cm}$ & 
             $\hspace{.3cm}$$\chi$$\hspace{.3cm}$ 
              \\ [.1cm]                         \hline
   Sphere           &  0  &  0  &  0  &  2  \\  \hline
   Disk             &  0  &  1  &  0  &  1  \\
   Projective plane &  0  &  0  &  1  &  1  \\  \hline
   Torus            &  1  &  0  &  0  &  0  \\
   Klein bottle     &  0  &  0  &  2  &  0  \\
   Annulus          &  0  &  2  &  0  &  0  \\ 
   Moebius strip    &  0  &  1  &  1  &  0  
\end{tabular}
\caption{Surfaces of highest Euler number}
\label{surfaces}
\end{center}
\end{table}
\noindent
The sphere and disk are the surfaces for tree-level scattering of closed and open strings respectively. Sphere amplitudes are correlation functions on the plane, which are bilinear combinations of conformal blocks. Disk amplitudes are correlators on the disk, which are linear combinations of conformal blocks (see chapter \ref{orbfus}). The projective plane is like a sphere with a crosscap and can scatter closed strings, the amplitudes being also conformal blocks \cite{usecross}. At Euler number $\chi=0$ come the first loop diagrams of String Theory, to which we now turn to.

\subsection{Torus}
The torus looks like a cylinder whose ends have been sewn together. Having no boundaries, this topology arises from closed string loops only. Any two-dimensional surface with Euler number parameters $(h,b,c)=(1,0,0)$ can be deformed to a torus by means of reparametrizations and conformal transformations. 

A torus can be defined as the complex plane modulo a lattice, as shown in fig. \ref{toruspic}.
\begin{figure}
\SetScale{1}
\begin{center}
\begin{picture}(400,150)(0,10)
\SetWidth{2}
\Oval(100,80)(25,60)(0)
\Curve{(80,80)(100,75)(120,80)}
\Curve{(85,78)(100,82)(115,78)}
\SetWidth{.5}
\LongArrow(240,40)(240,130)
\LongArrow(240,40)(330,40)
\SetWidth{2}
\Line(280,120)(360,120)
\Line(360,120)(320,40)
\Line(240,40)(320,40)
\LongArrow(240,40)(280,120)
\SetWidth{.5}
\DashLine(240,120)(280,120){3}
\Text(235,120)[r]{$\I t$}
\Text(320,30)[c]{$1$}
\Text(280,127)[c]{$\tau$}
\Vertex(240,40){1}
\end{picture}
\end{center}
\caption{Torus}
\label{toruspic}
\end{figure}
The complex parameter $\tau$ is the {\em modulus} of the torus. In the same way one brings the infinity of tree-level closed surfaces to a sphere, one can also bring the infinity of one-loop closed surfaces to a torus. The difference is however that while all spheres are conformally equivalent, the various tori are not - they are distinguished by the modular parameter $\tau$.

The zero-point function on the torus is calculable and has an important physical interpretation. In quantum mechanics, the Euclidean path-integral
\be
Z=\int_{PBC} {\cal D}q \, e^{-S(q)} = \Tr \, e^{-\beta H},
\ee
with $PBC$ standing for periodic boundary conditions, is the thermodynamic partition function. For two-dimensional quantum field theory (like String Theory is) this integral can be generalized. Imposing periodic boundary conditions not only on the spatial direction but also on the time direction is equivalent to saying that the fields of the theory live on the torus, which is precisely the zero-point function we want. Rescaling $\beta=2\pi \im(\tau)$ we get
\be
T = \int_{PBC} {\cal D}X^\mu \, e^{-S[X^\mu]} = 
    \Tr \,  e^{-2\pi \im (\tau) H} \,e^{2\pi \I \re (\tau)P},
\ee
with $H$ and $P$ the Hamiltonian and total momentum operators {\em on the  cylinder}, and the second exponential reflecting the fact that we can twist the torus in the $\sigma$-direction before gluing the edges back together. The operators $H$ and $P$ are \cite{bertnotes}
\be
H=L_0 - \frac{c}{24} + \overline{L}_0 - \frac{\bar{c}}{24}, \;\;\;
P=L_0 - \frac{c}{24} - \overline{L}_0 + \frac{\bar{c}}{24}.
\ee
With this we finally get
\be
T(\tau,\bar{\tau})= \Tr \; e^{2\pi \I 
   \tau (L_0-\frac{c}{24})}\,e^{-2\pi \I \bar{\tau} 
  (\overline{L}_0-\frac{\bar{c}}{24})}. \label{T}
\ee
The trace is over all states in the Hilbert space, null vectors excluded. Note also the path-integral is evaluated for a particular value of the modular parameter $\tau$. To get the complete zero-point function one must in principle integrate over $\tau$. For the bosonic string, which is the space-time part of a more general theory, one has also to do the integral over the space-time momentum $k$ (which comes from the contribution of $k$ to $L_0,\overline{L}_0$). With all this, the bosonic string torus partition function is, for $\tau = s+\I t$,
\be
T(t)= \int_0^{2\pi} \frac{ds}{(2t)^{12}} \; q^{-1}(\bar{q})^{-1} 
      \prod_{n=1}^{\infty}(1-q^n)^{-24} 
      \prod_{n=1}^{\infty}(1-\bar{q}^n)^{-24} = 
         \frac{1}{(2t)^{12}} e^{4\pi t} 
      \sum_{n=0}^{\infty} d_n^2 e^{-4\pi n t}. 
\label{toruspf}
\ee
This reads as follows. For each free boson field $X^\mu$, the $k$-integration yields a factor $1/\sqrt{2t}$ while the oscillators give a Dedekind $\eta$-function. The ghost contribution kills off two oscillators, $X^0$ and $X^1$, to leave us with the physical transverse oscillations only and furthermore introduces a factor $1/t^2$ in the integration measure. Then the integration $\int \! ds$ enforces the condition $L_0=\overline{L}_0$. In the end, the torus degeneracy of states at level $n$ is $d_n^2$. The total torus is obtained by integrating (\ref{toruspf}) over $dt/t^2$, {i.e.} $T=\int \! \frac{dt}{t^2} T(t)$. This leads naively to infinity, but as we shall see below, this integration has a natural cut-off due to modular invariance.

When the theory at hand has an internal part, the torus partition function for the internal sector can be written in terms of primaries and descendants as
\be
T(\tau,\bar{\tau}) = \sum_{ij} \chi_i(\tau) Z_{ij} 
                     \overline{\chi}_j(\bar{\tau}), \;\;\;
   \chi_i (\tau) = \Tr_{\text{rep} \; i} \; 
      e^{2\pi \I \tau (L_0-\frac{c}{24})},
\ee
with $\chi_i(\tau)$ the {\em character} of representation $i$. Its trace is over the representation $i$, modulo null states (zero-norm states), and can be expanded in terms of $\tau$ with the expansion coefficient $d_n$ counting the internal number of states at each excitation level
\be
\chi_i(\tau) = q^{(h_i-\frac{c}{24})}\sum_n d_n^i q^n,\;\;\; 
   q=e^{2\pi \I \tau}.
\label{qdef}
\ee
The matrix $Z_{ij}$ is called the {\em invariant matrix}. It tells us how the holomorphic and anti-holomorphic parts of the chiral algebra combine into the full conformal field theory.

\subsubsection{Modular invariance}
An interesting feature of the torus is modular invariance. One can check the transformations
\be
T: \tau \> \tau + 1, \;\;\; S: \tau \> -\frac{1}{\tau},
\ee
define the same lattice and thus the same torus. Actually, any combination of $S$ and $T$ does this, as
\be
\tau \> \frac{a\tau+b}{c\tau+d}\,,\;\;\ 
ac-bd=1, \;\;\; a,b,c,d \in {\mathbb{Z}}
\ee
describes the same torus. These transformations generate the group $SL(2,{{\mathbb Z}})/{{\mathbb Z}}_2$ and are called modular transformations. The torus partition function should then reflect this symmetry. The characters $\chi_i(\tau)$ transform under modular transformations like
\be
T: \chi_i(\tau+1)=
   \underbrace{e^{2\pi \I (h_i-\frac{c}{24})}}_{T_{ii}} \chi_i(\tau), \;\;\;
S: \chi_i(-1/\tau)=\sum_j S_{ij} \chi_j(\tau).
\ee
If the modular matrices $S_{ij},T_{ij}$ commute with the modular matrix $Z_{ij}$, the theory is invariant under modular transformations and is said to be {\em modular invariant}. It is because of modular invariance that we can leave out of the integration over $\tau$ regions that are related to other via the modular group. Performing the full integration over $\tau$ to get the partition function would lead to over-counting and to a divergent result. This is the above-mentioned natural cut-off of String Theory that controls the ultraviolet divergences that plague quantum field theories. For more details see for instance \cite{bertnotes} and references therein.

\subsubsection{Fusion rules}
The matrix $S$ is symmetric and further satisfies $S^2=C$, $S_{ij^c}=S_{ij}^*=S_i^{\ph{i}j}$. It can be also used to define {\em fusion rules} via the Verlinde formula \cite{verlinde}. Fusion rules are integer numbers $N_{ij}^{\ph{ij}k}$ such that
\be
N_{ij}^{\ph{ij}k} = \sum_m \frac{S_{mi}S_{mj}S_{mk}^*}{S_{m0}}.
\ee
The fusion rules compute couplings. If the chiral algebras ${\cal A}$ and $\overline{\cal A}$ are fully extended, the operator product coefficient $C_{i\bar{i},j\bar{j}}^{\ph{i\bar{i},j\bar{j}}k\bar{k}}$ vanishes if and only if either $N_{ij}^{\ph{ij}k}$ or $N_{\bar{i}\bar{j}}^{\ph{\bar{i}\bar{j}}\bar{k}}$ vanish.  This is the Naturality theorem of \cite{naturality}.

There may exist chiral labels $J$ for which the fusion rules contain only one field on the right-hand-side. E.g. $J\times i = j$ plus nothing else. The label $J$ is called a {\em simple current} \cite{simple} \cite{aligator}. A field $f$ is called a {\em fixed point} of $J$ if $J\times f=f$.

\subsection{Klein bottle}
\begin{figure}
\SetScale{1}
\begin{center}
\begin{picture}(400,100)(0,40)
\SetWidth{2}
\CArc(80,80)(40,0,270)
\CArc(80,80)(20,0,270)
\Line(80,40)(140,40)
\Line(80,60)(140,60)
\Oval(140,50)(10,5)(0)
\Line(100,80)(100,60)
\Line(120,80)(120,60)
\SetWidth{.5}
\DashCurve{(100,60)(110,50)(140,40)}{3}
\DashCurve{(120,60)(130,55)(140,60)}{3}
\Line(240,40)(240,130)
\Line(240,40)(330,40)
\SetWidth{2}
\ArrowLine(240,40)(240,120)
\ArrowLine(240,40)(320,40)
\ArrowLine(320,120)(240,120)
\ArrowLine(320,40)(320,120)
\SetWidth{.5}
\Text(235,120)[r]{$\um \I t$}
\Text(320,30)[c]{$1$}
\Vertex(240,40){1}
\end{picture}
\end{center}
\caption{Klein bottle}
\label{klein1}
\end{figure}
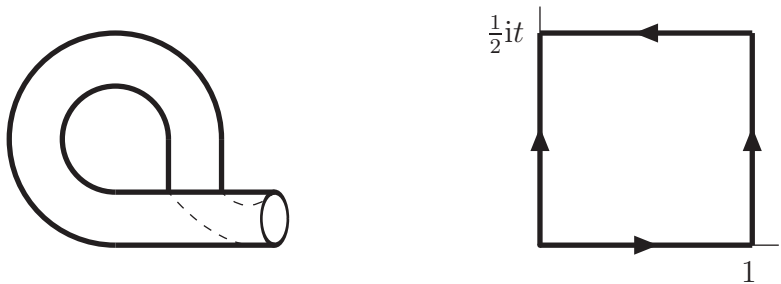

The Klein bottle is a cylinder whose ends have been sewn together but with opposite orientation. This surface can be drawn by unoriented closed strings running in a loop in the world-sheet time-like direction. Surfaces with $\chi=0$ can be obtained from the torus by means of anti-conformal involutions. The Klein bottle for instance be obtained from the a torus of purely imaginary modulus $\I t$ via the identification $z\sim z^* + \tau/2$, as we can see from fig. \ref{klein1}.
Taking the imaginary direction $\I t$ as world-sheet time, from the picture we see that a closed string is propagating in a loop and returning to itself with opposite orientation. 

The Klein bottle partition function can be derived from the torus one by requiring that is projects out states that are not left invariant under the world-sheet parity $\Omega$. At the level of the character this means that instead of $\Tr_{\text{rep}\,i} \, q^{(L_0-c/24)}$, we should evaluate $\Tr_{\text{rep}\,i}(\frac{1+\Omega}{2}) \, q^{(L_0-c/24)}$. To achieve the symmetrization of states at level $n$ we add $d_n$ and average out the result:
\be
\um \bigg( T(t)+K(t) \bigg) = \frac{1}{(2t)^{12}}
   e^{4\pi t}\sum_{n=0}^\infty \um (d_n^2 + d_n) \, e^{-4\pi nt}.
\ee
For the bosonic string, the Klein bottle partition function that achieves this is
\be
K(t) = \frac{1}{(2t)^{12}} \eta(2\I t)^{-24}.
\ee
The Klein bottle, like the torus, has one free parameter, but it is real, not complex. Note that if we do a modular transformation on $T(\tau,\bar{\tau})$ we get the same torus, but this is {\em not} true for the Klein bottle. The $T$-transformation, for instance, shifts the argument of the $\eta$-function. This means there is no analog of modular transformations for the Klein bottle.

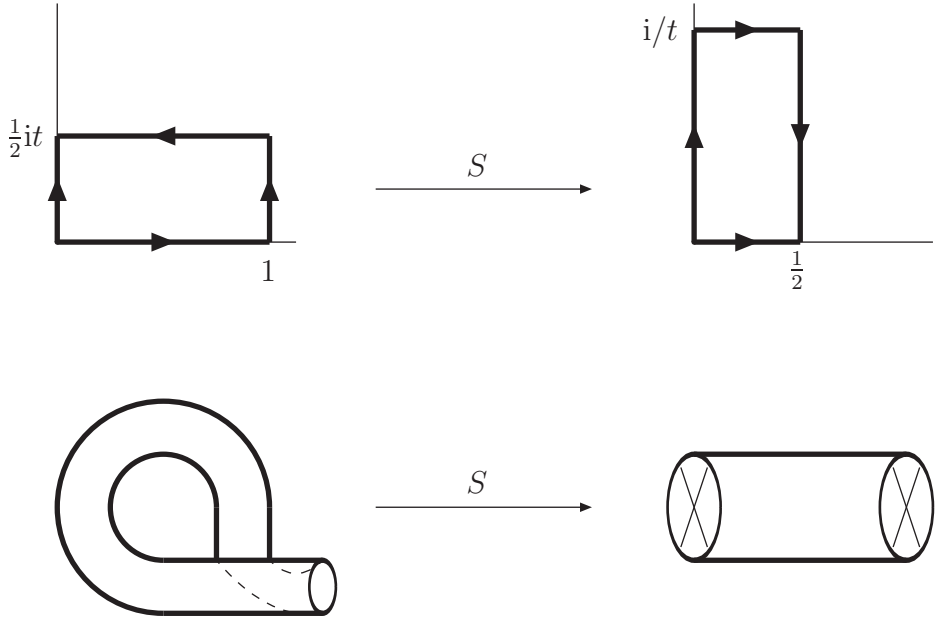
\begin{figure}
\SetScale{1}
\begin{center}
\begin{picture}(400,250)(0,40)
\SetWidth{2}
\CArc(80,80)(40,0,270)
\CArc(80,80)(20,0,270)
\Line(80,40)(140,40)
\Line(80,60)(140,60)
\Oval(140,50)(10,5)(0)
\Line(100,80)(100,60)
\Line(120,80)(120,60)
\SetWidth{.5}
\DashCurve{(100,60)(110,50)(140,40)}{3}
\DashCurve{(120,60)(130,55)(140,60)}{3}
\LongArrow(160,80)(240,80)
\Text(195,90)[l]{$S$}
\SetWidth{2}
\Line(280,60)(360,60)
\Line(280,100)(360,100)
\Oval(280,80)(20,10)(0)
\Oval(360,80)(20,10)(0)
\SetWidth{0.5}
\Line(275,65)(285,95)
\Line(275,95)(285,65)
\Line(355,65)(365,95)
\Line(355,95)(365,65)
\Line(40,220)(40,270)
\Line(40,180)(130,180)
\SetWidth{2}
\ArrowLine(40,180)(40,220)
\ArrowLine(120,220)(40,220)
\ArrowLine(120,180)(120,220)
\ArrowLine(40,180)(120,180)
\SetWidth{.5}
\Text(35,220)[r]{$\um \I t$}
\Text(120,170)[c]{$1$}
\Vertex(40,180){1}
\LongArrow(160,200)(240,200)
\Text(195,210)[l]{$S$}
\Line(280,180)(280,270)
\Line(280,180)(370,180)
\SetWidth{2}
\ArrowLine(280,180)(280,260)
\ArrowLine(280,180)(320,180)
\ArrowLine(280,260)(320,260)
\ArrowLine(320,260)(320,180)
\SetWidth{.5}
\Text(275,260)[r]{${\I /t}$}
\Text(320,170)[c]{$\um$}
\Vertex(280,180){1}
\end{picture}
\end{center}
\caption{Klein bottle channel transformation}
\label{sklein}
\end{figure}

However, modular transformations still play a role. The $S$-transformation, when applied to the Klein bottle interchanges space and time, but does not change its Teichmueller parameter. This transformation has a nice interpretation when we look at the picture \ref{sklein}.
In the picture on the right-hand-side space is periodic and time is cross-identified. The geometric interpretation is then that of a closed string propagating between two crosscaps.

This transformation of the Klein bottle is called {\em channel transformation} and plays a key role in the construction of unoriented string theories. In general when one has a $\chi=0$ surface defined via an identification on the torus, a channel transformation is a modular transformation that does not alter the Teichmueller parameters of the identification. The channel where world-sheet space is real is the {\em direct channel} and the channel where it can be imaginary is the {\em transverse channel}.

\subsection{Annulus}
The annulus is the equivalent of the torus for open strings, in the sense that it is the oriented one-loop diagram. The anti-conformal involution defining it is $z\sim 1-z^*$, which, for a purely imaginary $\tau$ leads to the picture \ref{annulus1}.
\begin{figure}
\SetScale{1}
\begin{center}
\begin{picture}(400,100)(0,40)
\SetWidth{2}
\BCirc(100,80){40}
\BCirc(100,80){20}
\SetWidth{.5}
\Line(240,40)(330,40)
\Line(240,40)(240,130)
\SetWidth{2}
\ArrowLine(240,120)(320,120)
\ArrowLine(240,40)(320,40)
\Line(320,40)(320,120)
\Line(240,40)(240,120)
\SetWidth{.5}
\Text(235,120)[r]{$2 \I t$}
\Text(320,30)[c]{$1$}
\Vertex(240,40){1}
\end{picture}
\end{center}
\caption{Annulus}
\label{annulus1}
\end{figure}
The annulus partition function is similar to the torus, but now using the open string Hamiltonian on the cylinder, which is just equal to $H=L_0-c/24$. The result is
\be
A(t) = \Tr_{\text{rep}\; i} \, q^{(L_0-\frac{c}{24})}.
\ee
For the bosonic string this gives
\be
A(t) = \frac{1}{(2t)^{12}} \eta(\I t /2)^{-24}.
\ee
The different argument for the $\eta$-function is due to the difference between the open and closed string excitations (cf. mass-shell condition \cite{gsw}). Again, in the full partition function, $t$ has to be integrated over.

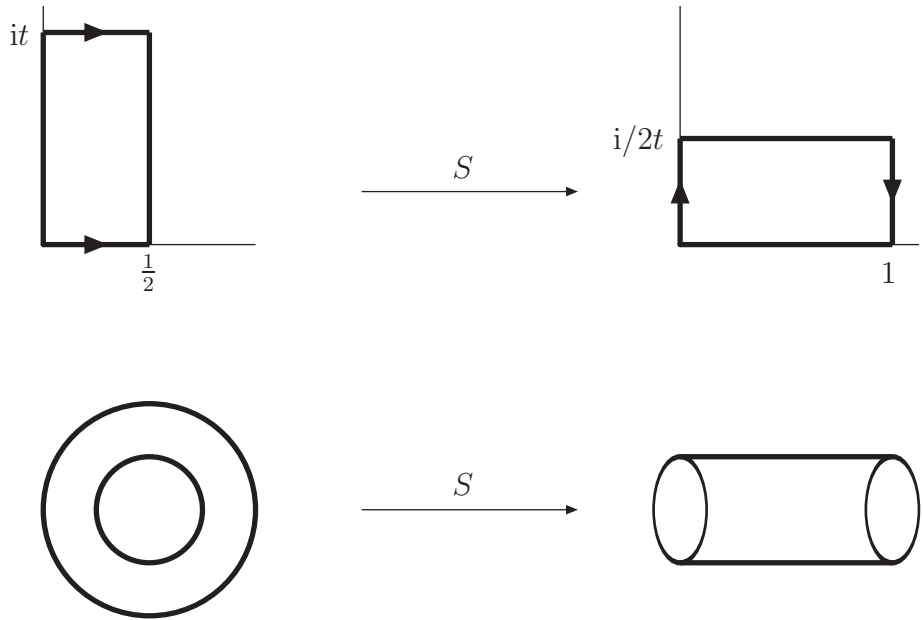
\begin{figure}
\SetScale{1}
\begin{center}
\begin{picture}(400,250)(0,50)
\LongArrow(160,80)(240,80)
\Text(195,90)[l]{$S$}
\SetWidth{2}
\BCirc(80,80){40}
\BCirc(80,80){20}
\Line(280,60)(360,60)
\Line(280,100)(360,100)
\Oval(280,80)(20,10)(0)
\Oval(360,80)(20,10)(0)
\SetWidth{.5}
\Line(40,180)(120,180)
\Line(40,180)(40,270)
\SetWidth{2}
\ArrowLine(40,180)(80,180)
\Line(40,180)(40,260)
\ArrowLine(40,260)(80,260)
\Line(80,260)(80,180)
\SetWidth{.5}
\Text(35,260)[r]{$\I t$}
\Text(80,170)[c]{$\um$}
\Vertex(40,180){1}
\LongArrow(160,200)(240,200)
\Text(195,210)[l]{$S$}
\Line(280,180)(370,180)
\Line(280,220)(280,270)
\SetWidth{2}
\ArrowLine(280,180)(280,220)
\Line(280,180)(360,180)
\Line(280,220)(360,220)
\ArrowLine(360,220)(360,180)
\SetWidth{.5}
\Text(275,220)[r]{${\I /2t}$}
\Text(360,170)[c]{$1$}
\Vertex(280,180){1}
\end{picture}
\end{center}
\caption{Annulus channel transformation}
\label{annulusS}
\end{figure}

The channel transformation of the annulus is again $S$. Scaling the annulus picture by $\um$, the transformation leads to the picture \ref{annulusS}.
In the direct channel we see an open string running in a loop, whereas in the transverse channel there is a {\em closed} string propagating between two states, which we call boundary states. Here the reason why there cannot be an open string theory without closed strings: the channel transformation entangles the two types of strings.

\subsection{Moebius strip}
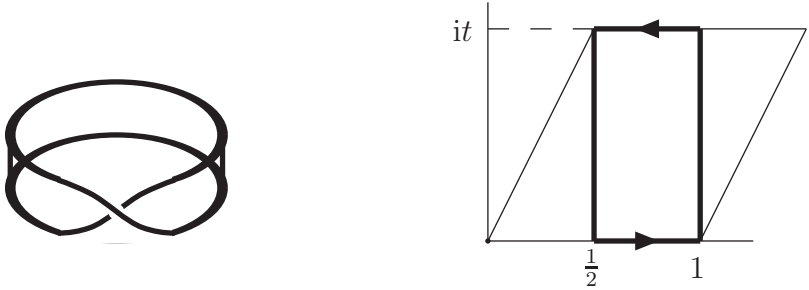
\begin{figure}
\SetScale{1}
\begin{center}
\begin{picture}(400,120)(0,40)
\SetWidth{2}
\Oval(80,60)(20,40)(0)
\Oval(80,80)(20,40)(00)
\Line(40,60)(40,80)
\Line(120,60)(120,80)
\CBox(60,40)(100,70){White}{White}
\Curve{(58,43)(70,45)(90,58)(102,63)}
\CCirc(80,51){2}{White}{White}
\Curve{(58,63)(70,58)(90,45)(102,43)}
\SetWidth{.5}
\Line(220,40)(260,40)
\Line(300,120)(340,120)
\DashLine(220,120)(260,120){10}
\Line(220,40)(260,120)
\Line(300,40)(340,120)
\SetWidth{2}
\ArrowLine(260,40)(300,40)
\ArrowLine(300,120)(260,120)
\Line(260,40)(260,120)
\Line(300,40)(300,120)
\SetWidth{.5}
\Line(280,40)(320,40)
\Line(220,40)(220,130)
\Text(215,120)[r]{$\I t$}
\Text(260,30)[c]{$\um$}
\Text(300,30)[c]{$1$}
\Vertex(220,40){1}
\end{picture}
\end{center}
\caption{Moebius strip}
\label{moebius1}
\end{figure}

Like the Klein bottle, the Moebius strip should symmetrize (or anti-symmetrize) the open string partition function. The anti-involution is the same as the annulus but this time $\tau$ is shifted to $\I t+\um$, which leads to fig. \ref{moebius1}.
For the bosonic string the partition function that accomplishes the $\Omega$-projection on the annulus partition function is
\be
M(t) = -\frac{1}{(2t)^{12}} \eta( {\sst \um}(\I t+1) )^{-24}. 
\label{bosonicmoebius}
\ee
When an internal theory is added, the overall sign of the Moebius strip is {\em a priori} undetermined. It can usually be fixed solving the tadpole cancellation equations (see section \ref{tadcancel}). Such equations give a minus sign for the pure bosonic string and this is what (\ref{bosonicmoebius}) shows, but can lead to plus signs in other cases.

\begin{figure}
\SetScale{1}
\begin{center}
\begin{picture}(400,250)(0,50)
\LongArrow(140,80)(220,80)
\Text(175,90)[l]{$P$}
\SetWidth{2}
\Oval(80,60)(20,40)(0)
\Oval(80,80)(20,40)(00)
\Line(40,60)(40,80)
\Line(120,60)(120,80)
\CBox(60,40)(100,70){White}{White}
\Curve{(58,43)(70,45)(90,58)(102,63)}
\CCirc(80,51){2}{White}{White}
\Curve{(58,63)(70,58)(90,45)(102,43)}
\SetWidth{.5}
\SetWidth{2}
\Line(260,60)(340,60)
\Line(260,100)(340,100)
\Oval(260,80)(20,10)(0)
\Oval(340,80)(20,10)(0)
\SetWidth{.5}
\Line(255,65)(265,95)
\Line(255,95)(265,65)
\Line(20,180)(60,180)
\Line(80,180)(120,180)
\Line(20,180)(20,270)
\Text(15,260)[r]{$\I t$}
\Text(60,170)[c]{$\um$}
\Text(100,170)[c]{$1$}
\Vertex(20,180){1}
\Line(100,260)(140,260)
\DashLine(20,260)(60,260){3}
\Line(20,180)(60,260)
\Line(100,180)(140,260)
\SetWidth{2}
\ArrowLine(60,180)(100,180)
\ArrowLine(100,260)(60,260)
\Line(60,180)(60,260)
\Line(100,180)(100,260)
\SetWidth{.5}
\LongArrow(140,210)(220,210)
\Text(175,220)[l]{$P$}
\Line(240,180)(320,180)
\Line(280,260)(360,260)
\Line(240,180)(240,270)
\Text(235,260)[r]{$\I /4t$}
\Text(280,170)[c]{$\um$}
\Line(280,178)(280,182)
\Text(320,170)[c]{$1$}
\Vertex(240,180){1}
\Line(260,220)(240,180)
\Line(340,220)(360,260)
\DashLine(240,260)(280,260){3}
\SetWidth{2}
\ArrowLine(320,180)(340,220)
\Line(260,220)(320,180)
\Line(280,260)(340,220)
\ArrowLine(280,260)(260,220)
\SetWidth{.5}
\end{picture}
\end{center}
\caption{Moebius channel transformation}
\label{moebiusP}
\end{figure}
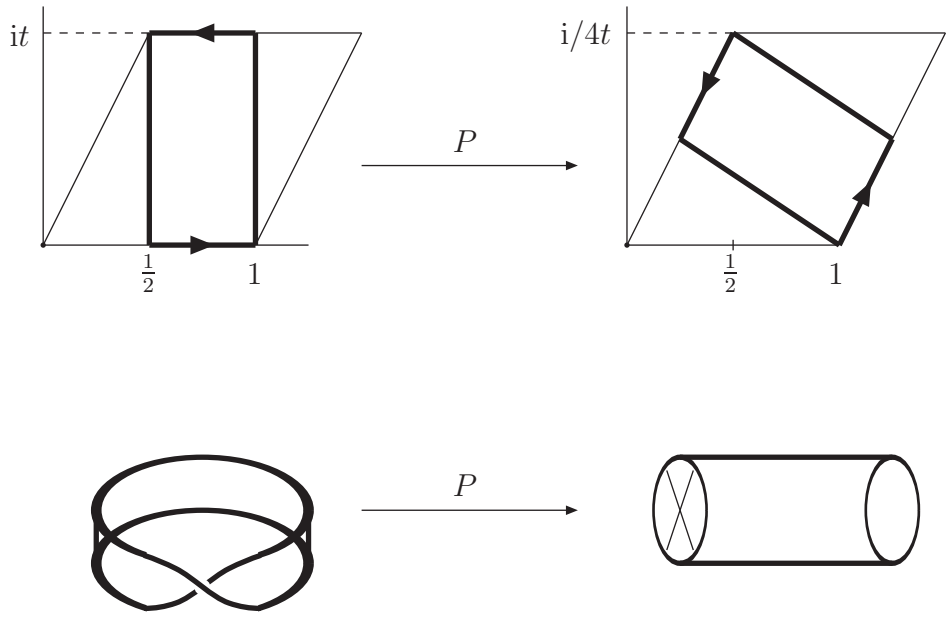

The channel transformation for the Moebius is slightly different. It is now $P=TS^2TS$, which leads to picture \ref{moebiusP}.
In the direct channel we have an open string running in a loop and coming back to itself with opposite orientation. In the transverse channel a closed string propagates between a boundary and a crosscap state, or vice-versa. In the end the open string partition function is the average $\um(A+M)$. The full partition functions $K,A,M$ have the $t$-dependence integrated with measure the same measure as the torus: $dt/t^2$.

Having set the stage to the study of open string theories, we now apply the machinery to go beyond the bosonic string.

\chapter{Open descendants}\label{opendes}
If a string theory is built upon an internal conformal field theory, the internal theory will manifest itself in the string perturbative series. At the tree level, for instance, vertex operators for transition amplitudes will decompose into a space-time and an internal part. A correlator will then split up into the product of a space-time contribution and an internal part. At the level of one loop and higher, the principle is the same, but the loop diagrams will put strict integrality conditions on the internal theory.

In this chapter we construct open descendants, which are open string theories built out of closed string ones. We will work in a general setting, the world-sheet theory being a tensor product of free bosons and an internal sector. We will see the restrictions imposed upon the internal theory appearing explicitly and interpret them. This would seem to leave out strings theories with world-sheet supersymmetry, but, by means of a procedure called `bosonic string map' \cite{bsm}, one can map the partition function of a superstring theory into that of a bosonic one. In this way, all the superstring partition function calculations can be carried out more easily by studying its faithful bosonic map. The results of this chapter can therefore be straightforwardly extended to the superstring case.

\section{Construction of the unoriented theory}
In a string model, one of the first things one would like to know is its spectrum and the simplest transition amplitudes. After that, one can ask whether the theory is consistent at higher amplitudes. (In this thesis we take a top-to-bottom approach. The spectrum is derived by looking at the one-loop partition functions and the simple transition amplitudes are studied when checking for consistency at higher orders.) Our construction of open string theories will revolve around the one-loop vacuum diagrams, which will introduce a series of concepts and a few very simple consistency requirements which are nevertheless restrictive enough to help classifying the internal theories themselves.

\subsection{General torus}
The torus for a general string theory, involving a space-time part and internal sectors, can be evaluated straight away, using considerations from the previous chapter. We get
\be
T = \int_{\cal F} \frac{d\tau d\bar{\tau}}{t^2} (2t)^{-d/2} \, 
    \eta(\tau)^{-d} \eta(\bar{\tau})^{-d} \; T(\tau,\bar{\tau}), \;\;\; 
T(\tau,\bar{\tau}) = \sum_{ij} \chi_i(\tau)\, Z_{ij}\, 
                     \overline{\chi}_{j} (\bar{\tau}).
\label{torus}
\ee
The $\tau$-integration is done here over the real and imaginary parts of $\tau=s+\I t$, restricted to a fundamental region ${\cal F}$ to take modular invariance into account. The contribution from the internal sector is summarized in $T(\tau,\bar{\tau})$. The torus (\ref{torus}) has the explicit form of the tensor product of a bosonic partition function times an internal theory partition function
\be
T \;\;\sim\;\;
\underbrace{(2t)^{-d/2} \eta(\tau)^{-d}\eta(\bar{\tau})^{-d}}_{\text{space-time}}\;\; \underbrace{T(\tau,\bar{\tau})}_{\text{internal}}
\ee
One can check that (\ref{torus}) is explicitly modular invariant if the invariant matrix from the internal sector satisfies $[S,Z]=[T,Z]=0$.

The internal partition function $T(\tau,\bar{\tau})$ can be evaluated once we know the world-sheet spectrum of the internal theory (which is {\em not} the space-time one). The space-time contribution for the Klein bottle, Annulus and Moebius strip is then simply the one evaluated in the previous chapter. The internal contribution is however not directly calculable. Indeed, for instance, finding a Klein bottle projection that leaves out unoriented states of $T(\tau,\bar{\tau})$ and the associated annulus and Moebius strip is the problem to be solved.

\subsection{Transverse channels for Klein bottle, Annulus and Moebius strip}
Deriving general expressions for the internal sectors' Klein bottle, annulus and Moebius strip is more easily solved in the transverse channel description. In this channel, the $\chi=0$ surfaces are seen as closed string transition amplitudes between boundary and crosscap states
\be
K(t) = \langle C| e^{-tH_{\text{\scriptsize closed}}} |C\rangle, \;\;\;\;
A(t) = \langle B| e^{-tH_{\text{\scriptsize closed}}} |B\rangle, \;\;\;\; 
M(t) = \langle B| e^{-tH_{\text{\scriptsize closed}}} |C\rangle+\langle C\leftrightarrow B\rangle , \nn
\ee
with $|B\rangle$ and $|C\rangle$ a boundary and crosscap state respectively. So, in terms of pictures, a closed string state emerges from a boundary or crosscap state, propagates for a time $t$ and disappears into a boundary or crosscap again. This is a tree-level amplitude, so it is a correlator between two special states. The first thing to do is then to find out what are these states.

\subsubsection{Gluing conditions}
If one looks at the transverse channel annulus picture, where a closed string is propagating between two boundaries, and changes the world-sheet space and time direction, one gets an open string running in a loop with end points attached to a hyper-surface - a D-brane. From the point of view of the open string, in the Neumann directions momentum should not flow from the open string to the hyper-surface. Now, in the same way a closed string can be mapped to the complex plane, the open string with boundary conditions $a$ and $b$ can be mapped via $\omega = e^z$ to a complex half-plane, with the boundary condition $a$ imposed along the real line $\re(z)<0$ and $b$ along $\re(z)>0$ (see chapter \ref{sewchap}). When this mapping is done, the no-flow condition implies \cite{cardy}
\be
T(z)=\overline{T}(\bar{z}), \label{TatB}
\ee
at the end points of the open string. The change in boundary conditions is mediated by a boundary field at $z=0$. It is only the Virasoro algebra which must be preserved at the boundary. If the chiral algebra is extended, the extra symmetries need not be preserved there. It can, of course, still happen that they are preserved, but it can also happen that they are preserved only up to an automorphism, or that they are not be preserved at all. The two latter cases are often referred to as `symmetry-breaking boundaries'. In the case that the extra symmetries are preserved up to an automorphism we have
\be
J(z)=\omega\left( \overline{J}(\bar{z}) \right), \label{JatB}
\ee
at the end points of the open string, with $\omega$ an automorphism of the chiral algebra. This was the case in \cite{fuchssbb} \cite{foe}. Some cases where the extended symmetries are not preserved at all were studied in \cite{alphainduction} and \cite{nimrep}.

\begin{figure}
\SetScale{1}
\begin{center}
\begin{picture}(400,100)(0,20)
\LongArrow(40,20)(40,120)
\LongArrow(20,40)(120,40)
\Line(100,100)(100,110)
\Line(100,100)(110,100)
\Text(35,120)[r]{$\I s$}
\Text(35,100)[c]{$\pi$}
\Text(105,105)[c]{$z$}
\Text(80,30)[c]{$\beta$}
\Text(120,30)[c]{$t$}
\LongArrow(160,60)(240,60)
\SetWidth{2}
\Line(40,40)(80,40)
\Line(40,100)(80,100)
\ArrowLine(80,40)(80,100)
\ArrowLine(40,40)(40,100)
\SetWidth{.5}
\SetWidth{2}
\BCirc(320,70){40}
\BCirc(320,70){20}
\SetWidth{.5}
\LongArrow(320,20)(320,120)
\LongArrow(270,70)(370,70)
\Line(380,100)(380,110)
\Line(380,100)(390,100)
\Text(385,106)[c]{$\xi$}
\Text(345,60)[c]{$1$}
\Text(375,62)[c]{$e^{2\pi/\beta}$}
\Vertex(340,70){3}
\Vertex(360,70){3}
\end{picture}
\end{center}
\caption{World-sheet duality transformation}
\label{wstrafo}
\end{figure}
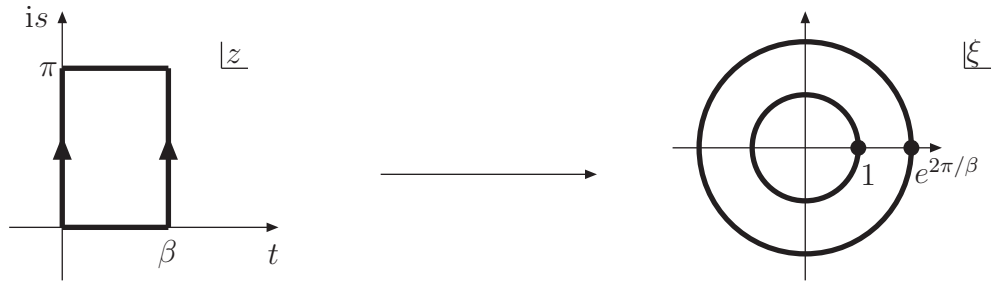

Considering (\ref{TatB}) and (\ref{JatB}), imposing the no-flow conditions on the plane and interchanging the interpretation of world-sheet time and space again brings us back to the closed string picture. This can be done in a series of moves. Define the cylinder in the half-plane by $z=t+\I s,\; s\in[0,\pi],t\in[0,\beta]$. Now swap $s\leftrightarrow t$, rescale by $2\pi/\beta$ and finally use the conformal map $z \rightarrow \xi = e^{\frac{2\pi \I }{\beta}z}$. This is summarized in fig. \ref{wstrafo}.
The cylinder is thus brought to an annulus on the full complex $\xi$-plane. Time in the full plane runs radially, so we see a closed string emerging from a boundary state at radius $|\xi|=1$, propagating and disappearing into another boundary or crosscap at $|\xi|=e^{\frac{2\pi^2}{\beta}}$. In the case of a crosscap, the calculation involves the crosscap identification $\xi \sim -1/{\xi^*}$, and is in the rest similar. The quantized no-flow conditions in the complex plane become operator equations on the boundary and crosscap states
\be
\left( L_n -  \overline{L}_{-n} \right) |B\rangle = 0, &&
\left( J_n - (-1)^{h_J} \; \omega \left( \overline{J}_{-n} \right) \right) |B\rangle = 0, \nn \\
\left( L_n - (-1)^{n} \overline{L}_{-n} \right) |C\rangle = 0, &&
\left( J_n - (-1)^{h_J+n} \; \omega \left( \overline{J}_{-n} \right) \right) |C\rangle = 0. \label{gluing}
\ee
In the following, we write the conditions for $J_n$'s only, since the $L_n$'s are just a subset thereof. The space-time free bosons $X^\mu$ have a symmetry current $\partial X$, with real modes $\alpha_n$ (see chapter \ref{freebosons}), and the automorphism $\omega$ has two solutions: $\omega(\alpha_n) = \pm \alpha_n$, corresponding to Neumann and Dirichlet boundary conditions respectively \cite{recknagel}. In this thesis we will consider only the trivial automorphism for the internal conformal field theory.

\subsubsection{Ishibashi states}
The closed strings states that solve the gluing conditions are the so-called Ishibashi states \cite{ishi}. For trivial automorphism, the boundary and crosscap Ishibashi states are
\be
|i\rangle\!\rangle_{\!B} = \sum_{\text{\scriptsize rep}\;i} |i\rangle \otimes U |\bar{i}\rangle, \;\;\;
|i\rangle\!\rangle_{\!C} = \sum_{\text{\scriptsize rep}\;i} |i\rangle \otimes U (-1)^{\overline{L}_0-h_i} |\bar{i}\rangle. \label{ishi}
\ee
The sums are over the whole chiral algebra representation $|i\rangle$. The operator $U$ is an anti-unitary operator satisfying the commutation relation $\overline{J}_n U = (-1)^{h_J} U \overline{J}_n$. Using this commutator, the hermiticity condition $J_n^\dagger = J_{-n}$ and the fact that the holomorphic and anti-holomorphic matrix elements $(J_n)_{ij}=(\overline{J}_n)_{ij}$ are equal, we can show that (\ref{ishi}) is indeed a solution to the gluing conditions (\ref{gluing}). From the definition, we see that there exists an Ishibashi state whenever a representation is allowed to couple to its charge conjugate in the anti-holomorphic sector. Whether or not this happens can be seen from the torus modular matrix $Z_{ij}$, so have thus $\#$Ishibashi states = $\Tr(Z_{ij^c})$.

Ishibashi states actually do not belong to the physical Hilbert space because the sum over the whole representation gives them an infinite norm. (For the bosonic string they are coherent states.) But this is not a problem since their explicit form is not needed. What will be needed instead are the inner products
\be
&& _{B\!}\langle\!\langle i | e^{-tH_{\text{\scriptsize closed}}} 
     | j \rangle\!\rangle_{\!B} = 
         \delta_{ij} \chi_j(\I t/\pi), \nn \\
&& _{C\!}\langle\!\langle i | e^{-tH_{\text{\scriptsize closed}}} 
     | j \rangle\!\rangle_{\!C} = 
         \delta_{ij} \chi_j(\I t/\pi), \\
&& _{B\!}\langle\!\langle i | e^{-tH_{\text{\scriptsize closed}}} 
     | j \rangle\!\rangle_{\!C} = \; 
   _{C\!}\langle\!\langle i | e^{-tH_{\text{\scriptsize closed}}} 
     | j \rangle\!\rangle_{\!B} = \delta_{ij} 
         \hat{\chi}_j({\sst \um} + \I t/\pi), \nn
\ee
where we recall $\hat{\chi}_j = \sqrt{T_{jj}} \; \chi_j \equiv e^{-\I \pi (h_i - \frac{c}{24})} \chi_j$. For each Ishibashi state a {\em transverse field} propagates in the closed string channel. From now on we label transverse fields by $m,n$ and chiral fields by $i,j$. Later on, boundary conditions will be labeled $a,b$.

\subsection{Boundary and crosscap states}
The general solution for the gluing conditions (\ref{gluing}) is a linear combination of Ishibashi states. A boundary and crosscap state would then be
\be
|B_a\rangle = \sum_m B_{ma}     |m\rangle\!\rangle_{\!B}, \;\;\;
|C  \rangle = \sum_m \Gamma_{m} |m\rangle\!\rangle_{\!C},
\ee
where $B_{ma}$ are the boundary coefficients and $\Gamma_m$ the crosscap coefficients. These coefficients measure the strength of the interaction of the conformal field $\phi_{m\bar{m}}$ with the boundary or crosscap state respectively. The dual boundary and crosscap states are defined by the bras $
\langle B_a|= {_{B\!}\langle\!\langle m|}  \sum_m B_{ma}^* \,
\langle C  |= {_{C\!}\langle\!\langle m|}  \sum_m \Gamma_m^*$.
The boundary conditions $a,b$ can be interpreted geometrically as D-brane positions in the compact space.

With the boundary and crosscap states defined, we can finally evaluate the internal transverse channel amplitudes for the Klein bottle, annulus and Moebius strip. Before we do that, we introduce Chan-Paton factors $n_a$ that count the number of times a D-brane of type $a$ appears in the theory. A boundary state $|B_a\rangle$ then contributes $n_a$ times to the annulus or Moebius strip. One could in principle define a similar multiplicity for the crosscap, but in practice it seems only one type of crosscap is allowed in a string theory. Such a crosscap multiplicity would also spoil closed string integrality and for these two reasons we will only consider one type of crosscap state. With Chan-Paton factors the transverse channel amplitudes are
\be
\tilde{K} &=& \int_0^{\infty} dt \,\langle C | 
              e^{-tH_{\text{\scriptsize closed}}} | C \rangle = \int dt \, 
              \sum_m \Gamma_m^* \Gamma_m \, \chi_m (\I t/\pi), \nn \\ 
\tilde{A} &=& \int_0^{\infty} dt \,\langle B_a | 
              e^{-tH_{\text{\scriptsize closed}}} | B_b \rangle = \int dt\, 
              \sum_{m} B_{ma}^* B_{mb} n_a n_b \, \chi_m(\I t/\pi), \\
\tilde{M} &=& \int_0^{\infty} dt \,\langle C | 
              e^{-tH_{\text{\scriptsize closed}}} | B_a \rangle + 
              \langle C \leftrightarrow B \rangle  = \int dt \, 
              \sum_m (\Gamma_m^* B_{ma} + B_{ma}^* \Gamma_m )  n_a \, 
              \hat{\chi}_m ({\sst \um} + \I t/\pi). \nn
\ee
We have put tildes in $K,A,M$ to denote they were computed in the transverse channel.

\subsection{General Klein bottle, annulus and Moebius strip}
We can now apply the channel transformations to get the final form of the Klein bottle, annulus and Moebius strip amplitudes. We have seen in chapter \ref{cftchap} that the channel transformation is $S$ for the Klein bottle and annulus and $TST^2S$ for the Moebius strip. Let us take the Klein bottle first. We rescale the integration variable as $t\rightarrow \pi t$ and do an $S$-transformation on $\tilde{K}$. This gives
\be
\tilde{K} = 
   \int_0^{\infty} dt\, \sum_m \Gamma_m^* \Gamma_m \, \chi_m(\I t) = 
   \int_{0}^{\infty} dt \,\sum_{im} \Gamma_m^* \Gamma_m S_{im} \, 
   \chi_i(\I/t) = K.
\ee
Now we do another change of variables $t\rightarrow 1/2t$, so that we get the same character argument as the space-time $\eta$-function. We get, in the new variable,
\be
K = \um \int_0^{\infty} \frac{dt}{t^2} 
    \sum_{im} \Gamma_m^* \Gamma_m S_{im} \, \chi_i (2\I t).
\ee
And this is the form of the Klein bottle in the direct channel. A similar calculation for the annulus gives
\be
A = 2 \int_0^{\infty} \frac{dt}{t^2} 
    \sum_{im} S_{im} B_{ma}^* B_{mb} n_a n_b \, \chi_i(\I t/2).
\ee
This time we used $t\rightarrow 2/t$ to get the same internal character argument as the space-time part. For the Moebius strip one has first to extract the phase from the hatted character, then perform the channel transformation $TST^2S$, and finally reinsert the phase into the hatted character. In the end we get
\be
M= \um \int_0^{\infty} \frac{dt}{t^2} \sum_{im} P_{im} 
   (\Gamma_m^* B_{ma} + B_{ma}^* \Gamma_m) n_a \, 
   \hat{\chi}_i({\sst \um}+\I t/2),
\ee
with $P=\sqrt{T}ST^2S\sqrt{T}$. Note that the direct channel internal quantities have the same integration measure as the space-time part. This is required for the integrations over the modular parameters to have a sensible interpretation.

\section{The spectrum}
Having transformed the internal sector transverse channel amplitudes to the direct channel, we can now check the spectrum of the full theory. The correct space-time contribution \cite{theguilty} must be tensored with the internal sector. When all this is done and normalizations are fixed, the direct channels of the full theory are
\be
T &=& \int_{\cal F} \frac{d\tau d\bar{\tau}}{t^2} 
   (2t)^{-d/2} \eta(\tau)^{-d} \eta(\bar{\tau})^{-d} 
   \sum_{ij} \chi_i(\tau) \, Z_{ij} \, \overline{\chi}_j(\bar{\tau}), \nn \\
K &=& \int_0^\infty \frac{dt}{t^2} 
   (2t)^{-d/2} \eta(2\I t)^{-d} \sum_i K_i \, \chi_i(2\I t), \nn \\
A &=& \int_0^\infty \frac{dt}{t^2} 
   (2t)^{-d/2} \eta(\I t/2)^{-d} \sum_i A_{ia}^{\phantom{ia}b} n_a n_b \, 
   \chi_i(\I t/2), \\
M &=& \int_0^\infty \frac{dt}{t^2} 
   (2t)^{-d/2} \eta({\sst \um}+\I t/2)^{-d} \sum_i M_{ia} n_a \, 
   \hat{\chi}_i({\sst \um}+\I t/2), \nn
\ee
where we defined the Klein bottle, annulus and Moebius strip coefficients
\be
K_i =               
   \sum_m S_{im} \Gamma_m^* \Gamma_m, \;\;\; 
A^{\ph{ia}b}_{ia} = 
   \sum_m S_{im} B_{ma} B_{mb}^*, \;\;\; 
M_{ia} =            
   \sum_m P_{im} 
   \Big( \um ({\Gamma_m B_{ma}^* + \Gamma_m^* B_{ma}}) \Big). 
\label{KAMcoeff}
\ee
These quantities can, as we will now see, be interpreted as spectrum degeneracies and should therefore obey certain positivity and integrality conditions.

Recall the quantities. The space-time dimension, or number of free bosons $X^\mu$, is $D=d+2$ with $d$ the number of light-cone dimensions. The torus integration is over a fundamental region ${\cal F}$ to account for modular invariance. The symmetric matrix $Z_{ij}$ has integer entries, with $i,j$ running through the set of irreducible representations of the chiral algebra of the internal theory. The coefficients $K_i, A_{ia}^{\ph{ia}b}, M_{ia}$ are the Klein bottle, annulus and Moebius strip coefficients of the internal sector.

\subsection{Closed spectrum}
The projection ${\um}(T+K)$ gives us the spectrum of the closed sector. For this purpose it will be useful to merge the space-time and internal characters' trace expansions, so we define `full' characters as
\be
\psi_i(\tau) = \eta(\tau)^{-d} \chi_i(\tau) = \sum_{n\geq 0} d_n^i q^{h_i+n-1},
\label{fullchar}
\ee
with $d_n^i$ meaning the degeneracy of the full character at excitation level $n$. Already here we find something interesting. The exponent of $q$ is the mass-squared of a particular state. We thus see that for internal conformal weights between 0 and 1, extra tachyonic states appear. Since these have no space-time excitations, they are scalars. If the internal weights are above one, we have massive space-time scalars. In space-time supersymmetric theories, the array of extra tachyonic states is projected out because they are not space-time supersymmetric states.

Having written (\ref{fullchar}), the expressions for the torus and direct channel Klein bottle can now be added. We get
\be
&&\um(T+K) =\int_{s=0,\,t=0}^{1,\,\infty}  \frac{ds\,dt}{t^2} (2t)^{-d/2} \times \nn\\ && \times \;\;\um \left( \sum_{\stackrel{\sst i,j}{m,n \geq 0}} d_m^i Z_{ij} d_n^j \, q^{h_i+m-1}\bar{q}^{h_j+n-1} + \sum_{\stackrel{\sst i}{m \geq 0}} d_m^i K_i\, q^{h_i+m-1} \right). \label{T+K}
\ee
Now we perform the integration over $s$. Take the diagonal part $Z_{ii}$ first. When $i=j$ the $s$-integration forces $m=n$ and we get
\be
\um(T+K) = \int_0^{\infty} \frac{dt}{t^2} (2t)^{-d/2}
           \sum_{\stackrel{\sst i}{m\geq 0}} 
           \left( \frac{(d_m^i)^2 Z_{ii} + d_m^i K_i}{2} 
           \right)q^{h_i+m-1}.
\label{T+Kii}
\ee
Since the result is an unoriented partition function, the quantity between brackets represents the total unoriented degeneracy at level $m$, so it must be an integer number. This can be achieved if the torus and Klein bottle coefficients obey
\be
\um(Z_{ii}+K_i) \in \mathbb{N}_0. \label{KBint}
\ee
This is our first integrality constraint. It may not the only way to obtain an integral spectrum; a given internal theory may well have $d_n^i$'s such that integrality is respected even if (\ref{KBint}) does not hold. However, (\ref{KBint}) is the only way to get integers at all excitation levels, that also leads to a sensible interpretation of the multiplicities in terms of chiral algebra representations.

Note that we have used all the Klein bottle coefficients to define (\ref{T+Kii}), so for the case $i\neq j$ there will be only contributions from the torus. For the special case of the identity character, $i=j=0$, at excitation level $m=n=1$ we have the degeneracies for the graviton and dilaton. In this case $\um(Z_{00}+K_0)=1$, otherwise we would have multiple gravitons and dilatons. This means $Z_{00}=K_0=1$. From the purely conformal field theory point of view, $Z_{00}=1$ is the condition for a unique vacuum, which we naturally want. Then $K_0=+1$ otherwise we have no vacuum.

In the case $i\neq j$ the $s$-integration enforces $h_i-h_j+m-n=0$ and we get
\be
\um(T+K) = \int_0^{\infty} \frac{dt}{t^2} (2t)^{-d/2}\sum_{\stackrel{\sst i>j}{m\geq 0}} \left( \frac{d_{m+\Delta_{ij}}^i \left(Z_{ij}+Z_{ji}\right)  d_m^j }{2} 
\right)q^{h_i+m-1}, \label{T+Kij}
\ee
with $\Delta_{ij}=|h_i\!-\!h_j|$. Again the quantity between brackets should be an integer, and it is so when $Z_{ij}$ is symmetric. Non-symmetric $Z_{ij}$ appear in heterotic theories, where the left and right chiral algebras are different, but we do not consider these theories. Also, since the world-sheet parity operator $\Omega$ interchanges the left and right Hilbert spaces, the assumption that a conformal field theory should allow for crosscaps only makes sense if the modular invariant is symmetric.

\subsubsection{Open spectrum}
The annulus and Moebius strip add up to
\be
&&\um(A+M) = \int_{0}^\infty \frac{dt}{t^2} (2t)^{-d/2} \times \nn \\ 
&&\times \;\; \um\left(
\sum_{\stackrel{\sst i,a,b}{m\geq 0}} d_m^i A_{ia}^{\ph{ia}b} n_a n_b \, q^{\um (h_i+m-1)} +
\sum_{\stackrel{\sst i,a}{m\geq 0}} d_m^i M_{ia} n_a (-1)^m\, q^{\um (h_i+m-1)} \right). \label{M+A}
\ee
This requires that
\be
\um \left( \sum_{ab} A_{ia}^{\ph{ia}b} n_a n_b + 
           \sum_a M_{ia} n_a \right) \in \mathbb{N}_0.
\label{M+A2}
\ee
The correct interpretation of this result requires however further discussion.

\section{Orientation matters}
\label{nimish}
In this section we construct and analyze the unoriented annulus. This will enable us to write down the unoriented open string partition function and to present a set of requirements that supplement positivity and integrality of the partition functions as a guideline for constructing unoriented theories. 

\subsection{Oriented versus unoriented annuli}
The annulus amplitude $\langle B_a| e^{-tH} |B_b\rangle$ is an oriented amplitude, since it is a scattering from brane $b$ to brane $a$. In the unoriented theory the $\Omega$-projection swaps end-points, so the unoriented set of annulus coefficients should be symmetric in $a,b$.

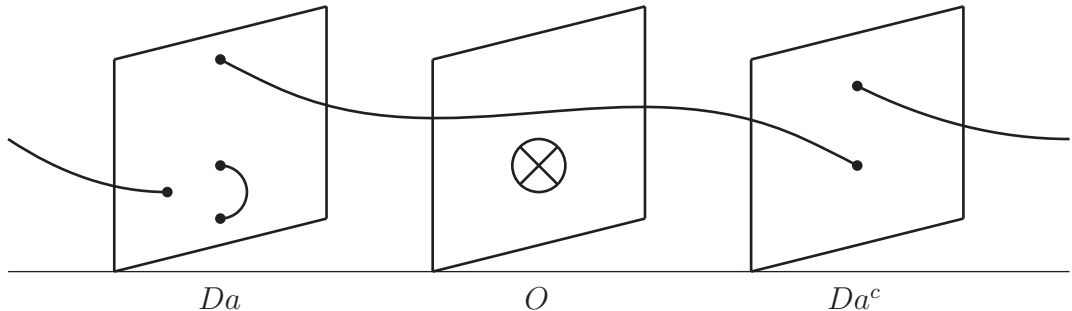
\begin{figure}
\SetScale{1}
\begin{center}
\begin{picture}(400,150)(0,0)
\Line(0,20)(400,20)
\SetWidth{1}
\Line(40,20)(120,40)
\Line(40,100)(120,120)
\Line(40,20)(40,100)
\Line(120,40)(120,120)
\Text(80,10)[c]{$Da$}
\BCirc(200,60){10}
\Line(193,53)(207,67)
\Line(193,67)(207,53)
\Line(160,20)(240,40)
\Line(160,100)(240,120)
\Line(160,20)(160,100)
\Line(240,40)(240,120)
\Text(200,10)[c]{$O$}
\Line(280,20)(360,40)
\Line(280,100)(360,120)
\Line(280,20)(280,100)
\Line(360,40)(360,120)
\Text(320,10)[c]{$D{a}^c$}
\Curve{(80,100)(100,90)(300,70)(320,60)}
\Vertex(320,60){2}
\Vertex(80,100){2}
\Curve{(320,90)(360,75)(400,70)}
\Vertex(320,90){2}
\Curve{(0,70)(30,55)(60,50)}
\Vertex(60,50){2}
\CArc(80,50)(10,270,90)
\Vertex(80,40){2}
\Vertex(80,60){2}
\SetWidth{.5}
\end{picture}
\end{center}
\caption{D-branes and O-planes}
\label{DandO}
\end{figure}

From the conformal field theory point of view, the different boundary labels $a$ mean different conformally invariant boundary conditions. From the geometrical point of view, the different labels correspond to D-branes at different locations inside the compactification manifold. In an unoriented theory, orientifold O-planes are added to the picture. These planes introduce mirror (image) branes, with labels ${a}^c$, as can be seen from fig. \ref{DandO}.
Depending on the specific model at hand, the mirror branes may or may not coincide with the branes themselves (see for instance the O0 and O2 orientifolds of \cite{wzwbert}). When the mirror image is the brane itself we have $a={a}^c$, and when the image is a different brane we have $a \neq {a}^c$. Without O-planes, a stack of $n$ D-branes gives rise to a $U(n)$ Chan-Paton gauge group \cite{tasi}. When an O-plane is added to the picture, branes that are mapped into themselves give rise to $SO(n)$ and $Sp(n)$ gauge groups, whereas strings that stretch between a brane $Da$ and its mirror image $Da^c$ give rise to $U(n)$ groups.

This suggests that the partition function $\um (A+M)$ should describe the scattering of strings from a brane to the orientifold reflection of another brane. That requires us to write down an annulus for this case, symmetric in the boundary indices, which we designate by $A^\Omega_{iab}$. This annulus coefficient is the natural coefficient for the unoriented string.

We can construct $A^\Omega_{iab}$ in the following way. The world-sheet parity operator $\Omega$ interchanges the left and right Hilbert spaces and may further have some involutive action on the representations themselves, hence the superscript on $A^\Omega_{iab}$, referring to the possible different actions of $\Omega$ on these representations. The action of $\Omega$ on boundary Ishibashi states is\footnote{In general, the Ishibashi states can be degenerate, in which case the $\Omega$ become matrices \cite{nimrep}. Here that does not happen because we will be considering automorphism invariants only.}
\be
\Omega |m\rangle\!\rangle = \Omega^m |m^c\rangle\!\rangle, \;\;\; \Omega^m = (\Omega^{m^c})^*,
\ee
where $\Omega^m$ is an eigenvalue and the hermiticity condition comes from the fact that $\Omega$ is a unitary involution. The transverse channel computation of $\tilde{A}=\langle B_a | e^{-tH} \Omega |B_b \rangle$ and transformation to the direct channel yields simply
\be
A_{iab}^\Omega = \sum_m S_{im} B_{ma}^* \Omega^m B_{m^c b}.
\ee
One can easily check that this expression is symmetric in $a$ and $b$, although not explicitly. To reach explicit symmetry we need a relation like
\be
B_{m^c b}^*=C^m B_{mb}, \label{CPT}
\ee
with $C^m (C^{m^c})^*=1$, which relates the emission of a closed string state to the absorption of its charge conjugate state. Such relation should hold due to CPT invariance (invariance under combined conjugation, parity and time-reversal). This leads to
\be
A_{iab}^\Omega = \sum_m S_{im} B_{ma} B_{mb} \, g^m, \;\;\; g^m = \Omega^m C^m,
\label{uA}
\ee
an annulus coefficient which is now manifestly symmetric in the boundary indices $a,b$ and can, as we will see, be related to the oriented one. Before we relate the two annuli, let us explore some of the properties of $A_{ia}^{\ph{ia}b}$

\subsection{NIMreps and completeness conditions}
Positivity and integrality of the one-loop partition functions and modular invariance are simple consistency checks of a conformal field theory. It was argued in \cite{gannon} that completeness of boundaries could be taken as an additional consistency requirement. Completeness means there should exist a boundary projection operator such that $\sum_a |B_a\rangle\langle B_a|=1$. Following \cite{completeness} this requirement can be put into the form
\be
\sum_b A_{ia}^{\ph{ia}b} A_{jb}^{\ph{jb}c} = 
\sum_k N_{ij}^{\ph{ij}k} A_{ka}^{\ph{ka}c}. \label{AA=NA}
\ee
This equation defines a non-negative integer matrix representation of the fusion algebra or `NIMrep'. If one knows the internal theory's $S$-matrix, one also knows the fusion rules and consequently the oriented annulus coefficients can be determined by solving (\ref{AA=NA}), which can be done either analytically or numerically. From the oriented annulus one can then determine the boundary coefficients up to a phase. Solving (\ref{AA=NA}) is therefore a possible starting point for the open descendants construction.

Let us now define `reflection coefficient' as $R_{ma}=\sqrt{S_{m0}} B_{ma}$. In order for the oriented annulus to give rise to $U(n)$ Chan-Paton gauge groups we need $A_{0a}^{\ph{0a}b}=\delta_{ab}$. With this and (\ref{AA=NA}) we can derive \cite{stanevnotes} \cite{trace}
\be
\sum_m R_{ma} R_{mb}^* = \delta_{ab}, \;\;\;
\sum_a R_{ma} R_{na}^* = \delta_{mn}. \label{RR*=1}
\ee
Equation (\ref{RR*=1}) is actually equivalent to (\ref{AA=NA}), as was proved in \cite{zuber}. The point here is that completeness implies that $R_{ma}$ is invertible, which means that the number of different conformally invariant boundary conditions is equal to the number of Ishibashi states. This claim has been proven in \cite{frs}.

When searching for NIMreps, we can find zero, one or more than one solutions. If no solution exists, the conformal field theory at hand cannot be formulated in surfaces with boundaries, which is deemed as a sign of inconsistency. Modular invariants of automorphism type (see section \ref{modinvs}) are expected to yield only one NIMrep, and this will be indeed the case in the example of chapter \ref{freebosons}. Multiple NIMreps are typically present in cases with $Z_{ii^c}>1$. In the following we assume the NIMrep to be unique. After having found a NIMrep, we can construct $A_{iab}^\Omega$ as follows.

\subsection{S-NIMreps and boundary conjugation}
The unoriented annulus coefficient $A_{iab}^\Omega$ can be related to the oriented one $A_{ia}^{\ph{ia}b}$ by means of $A_{0ab}^\Omega$. This particular unoriented annulus coefficient can be used to raise and lower boundary indices
\be
A_{ia}^{\ph{ia}b} = \sum_c A_{iac}^\Omega A_{0bc}^\Omega. \label{uA->A}
\ee
It is simple to see that the right-hand-side does indeed satisfy (\ref{AA=NA}) if completeness holds. Remember that the set of coefficients $A_{iab}^\Omega$ is in principle not unique - there may exist various sets of $A_{iab}^\Omega$ that lead to the same $A_{ia}^{\ph{ia}b}$. Each set of different $A_{iab}^\Omega$ is a therefore a different symmetrization of $A_{ia}^{\ph{ia}b}$ and we refer to it as an S-NIMrep. Using (\ref{KAMcoeff}) and (\ref{uA->A}) one can also show \cite{nimrep} that a set of $A_{iab}^\Omega$ can always be cast into the form (\ref{uA}). 

To get sensible gauge groups like $SO(n)$, $Sp(n)$ or $U(n)$, the coefficient $A_{0ab}^\Omega$ must be an involution, { i.e.}
\be
A_{0ab}^\Omega=C^B_{ab} = \delta_{ab^c}. \label{C^B}
\ee
The matrix $C_{ab}^B$ acts then as a boundary conjugation matrix.
We now have two ways to arrive at an S-NIMrep. We can either:
\bi
\item{Compute $\langle B_b | e^{-tH} \Omega | B_a \rangle$ and transform to the open string channel, or}
\item{Compute $\langle B_b | e^{-tH} | B_a \rangle$ and symmetrize the result using $C^B_{ab}$.}
\ei
Comparing both results we see that the effect of $\Omega$ on a boundary state is
\be
\Omega|B_a\rangle = C^B_{ab} |B_b\rangle.
\ee
As expected, it maps a boundary state into its conjugate, in accordance with the geometrical picture of fig. \ref{DandO}.

\subsection{Crosscaps}
The argument above requires us to evaluate Klein bottle and Moebius strip amplitudes with an insertion of $\Omega$ too: $\tilde{K}=\langle C | e^{-tH} \Omega | C \rangle$, $\tilde{M}= \langle B_b | e^{-tH} \Omega | C \rangle+\langle B\leftrightarrow C \rangle$. Since the crosscap corresponds to an orientifold plane, we expect the crosscap state to be an eigenstate of $\Omega$ and a fixed point, {i.e.}
\be
\Omega | C \rangle = \pm | C \rangle. \label{omegaC=C}
\ee
To determine which sign is the correct one, we note that $\Omega$ always acts on the boundary states with a plus sign, so we expect the same to happen for the crosscap state. Furthermore, the minus sign would imply $M_{ia} \hat{\chi}_i= -M_{i{a}^c}\hat{\chi}_i$, which would violate spectrum integrality for $M_{0a}$ if the boundary is self-conjugate (see below). For these reasons we take the plus sign to be the correct one.

If $\Omega | C \rangle = | C \rangle$, we can derive $\Gamma_{m^c}=\Omega^m \Gamma_m$. This, together with the assumption that the crosscap Ishibashi state also satisfies the CPT relation (\ref{CPT}), when applied to the transverse channel expressions with $\Omega$, leads to\footnote{The Klein bottle and Moebius coefficients should also have an $\Omega$ label attached: $K^\Omega_i$, $M_{ia}^\Omega$, but we drop it to simplify the notation.}
\be
K_i = \sum_m S_{im} \Gamma_m^2 \, g^m, \;\;\; M_{ia} = \sum_m P_{im} \Gamma_m B_{ma} \, g^m. \label{KMg^i}
\ee

An additional condition that can be used to determine whether or not a given symmetrization matrix $C^B_{ab}$ has an underlying orientifold symmetry is the following. Using the crosscap CPT formula and some algebra we can derive $M_{ia}=M_{i^c a^c}$, which is interpreted as follows. An arbitrary symmetrization matrix $C^B_{ab}$ always lead to an annulus of the form $A^\Omega_{iab}=\sum_m S_{im} B_{ma} V^m B_{mb}$. If we now write $V^m=C^m \Omega^m$ we come to the correct expression for the unoriented annulus, but nothing guarantees us that the $\Omega^m$ obtained this way is a symmetry of the theory. In order to be a symmetry, the condition $M_{ia}=M_{i^c a^c}$ must be satisfied. If it's not satisfied, then $\Omega^m$ cannot be a symmetry. This issue has been studied in detail in \cite{tft2}.

One final consistency check is to verify that the channel transformations give vanishing boundary and crosscap coefficients on non-Ishibashi states. Remember that Ishibashi states only exist for chiral fields that couple to their charge conjugate in the torus, so if $Z_{ii^c}=0$, we should have $B_{ia}=\Gamma_i=0$.
For boundary coefficients the completeness hypothesis that $B_{ia}$ is invertible implies vanishing of $B_{ia}$ for $i$ not Ishibashi. For the crosscap coefficients, this can be done writing $M_{ia}=\sum_m P_{im} X_{ma}$. Then, using (\ref{pe1}) we can derive $X_{ia} = B_{ia} D_{i}$, with $D_i$ still undetermined. Substituting this into (\ref{pe2}) shows that the Klein bottle must be of the form
\be
K_i = \sum_m S_{im} D_m^2 (g^m)^{-1}. \label{KasD}
\ee
Defining now $D_m = g^m \Gamma_m$ gives us back $K_i$ and $M_{ia}$. Now note that due to $B_{ia}$ vanishing on non-Ishibashi labels, $D_i$ can be chosen to vanish on those labels too without any loss of generality. This proves the sum over $m$ on (\ref{KasD}) can be carried out through Ishibashi labels only.

\subsection{Open spectrum}
We can now analyze the open string spectrum. Assuming $n_a=n_{a^c}$ (which is reasonable if we look at picture \ref{DandO}), equation (\ref{M+A2}) should then be integer whatever the set of Chan-Paton factors $n_a$ is. Then we can replace the summation in $b$ for a summation in $b^c$ and use $A_{ia}^{\ph{ia}b^c}=A_{iab}^{\Omega}$ to write (\ref{M+A2}) as
\be
\um \left( 
    \sum_{ab} A_{iab}^\Omega n_a n_b + \sum_a M_{ia} n_a 
    \right) \in \mathbb{N}_0.
\ee
For self-conjugate boundaries $a=a^c$ we get
\be
\um(A+M) = \int_{0}^\infty \frac{dt}{t^2} (2t)^{-d/2} 
\sum_{\stackrel{\sst i}{m\geq 0}} 
   \left( \frac{ A_{iaa}^{\Omega} n_a^2 + (-1)^m M_{ia} n_a }{2} 
   \right) d_m^i q^{\um (h_i+m-1)}. \label{M+Aaa}
\ee
The number inside brackets times $d_m^i$ can be interpreted as the total open string state degeneracy at excitation level $m$ and should therefore be an integer. That will be so if we require the integrality condition
\be
\um(A_{iaa}^{\Omega} + M_{ia}) \in \mathbb{N}_0.
\ee
From the identity character $i=0$ we get the gauge group. At the first excitation level $m=1$, the state is massless and the degeneracy coming from the space-time transverse excitations endows it with a space-time index $\mu$, making it a vector boson. Then the internal sector generates $\um(A_{0aa}^{\Omega} n_a^2 - M_{0a} n_a)$ copies of this boson. Since $A_{0aa}^\Omega=\pm M_{0a}=1$, the total number of vector bosons coming from self-conjugate boundaries is $\um n_a(n_a \mp 1)$, hinting at a gauge group $SO(n_a)$ or $Sp(n_a)$, depending on the overall sign of the Moebius strip. For $i\neq 0$, in the simple case where the annulus and Moebius coefficients are $1$, we get scalars in the adjoint of $Sp(n_a)$. For more general annulus and Moebius coefficients there can be ambiguities in interpreting the counting of states, but we expect in general only one interpretation to give sensible representations. From the above it is also explicitly clear that the minus sign in (\ref{omegaC=C}) would lead to non-integer multiplicities.

For non-self-conjugate boundaries $a\neq a^c$ we have
\be
&&\um(A+M) = \int_{0}^\infty \frac{dt}{t^2} (2t)^{-d/2}
             \sum_{\stackrel{\sst i}{m\geq 0}} d_m^i q^{\um (h_i+m-1)} 
             \times \\
&& \left(\frac
         { A_{iaa}^{\Omega} n_a^2 + A_{ia^c a^c}^{\Omega} n^2_{a^c} 
         + A_{iaa^c}^{\Omega} n_a n_{a^c} + A_{ia^c a}^{\Omega} n_{a^c} n_a 
         + (-1)^m M_{ia} n_a + (-1)^m M_{ia^c} n_{a^c} }{2} 
   \right) . \nn
\ee
We get again positive integer degeneracies if
\be
\um(A_{iaa}^{\Omega} + M_{ia}) \in \mathbb{N}_0.
\ee
For $i=0$ we get, since $A_{0ab}^\Omega=\delta_{ab^c}$, $n_a n_{a^c}$ vector bosons, which hints at $U(n_a)$ gauge group. For $i\neq 0$ and annuli and Moebius coefficients equal to 1, the degeneracy at level zero is $\um(n_a^2 + n_a) + \um(n_{a^c}^2 + n_{a^c}) + n_a n_{a^c}$, which are space-time scalars respectively in the symmetric tensor, in the conjugate symmetric tensor and in the adjoint representations of $U(n_a)$. Again, for more general annulus and Moebius coefficients, we expect only one sensible interpretation to the degeneracy.

For the non-diagonal annulus case we get
\be
\um(A+M) = \int_{0}^\infty \frac{dt}{t^2} (2t)^{-d/2} 
\sum_{\stackrel{\sst i,a>b}{m\geq 0}} 
   \left( 
      \frac{      A_{iab}^{\Omega} n_a n_b +
                  A_{iba}^{\Omega} n_b n_a}{2}
   \right) d_m^i q^{\um (h_i+m-1)}. \label{M+Aab}
\ee
Positivity and integrality is guaranteed from (\ref{AA=NA}) and from the symmetry of $A^\Omega_{iab}$ in the boundary indices.

\subsection{U-NIMreps}
We have seen that the condition $\um(A_{iaa}^\Omega+M_{ia})\in \mathbb{N}_0$ is the correct integrality condition for the unoriented open string. With this we can define a U-NIMrep as an S-NIMrep which further has Klein bottle and Moebius strips coefficients satisfying $\um(Z_{ii} + K_i) \in \mathbb{N}_0$ and $\um(A_{iaa}^\Omega+M_{ia})\in \mathbb{N}_0$ and the polynomial equations of \cite{trace} \cite{tft2}
\be
\sum_b A_{ia}^{\ph{ia}b} M_{jb} &=& \sum_k Y_{ij}^{\ph{ij}k} M_{ka},
\label{pe1}\\
\sum_{ab} C^{B}_{ab} M_{ia}M_{jb} &=& \sum_{k} Y_{\ph{k}ij}^k K_k, \label{pe2}
\ee
with $Y_{ij}^{\ph{ij}k}$ an integer, defined by \cite{planar}
\be
Y_{ij}^{\ph{ij}k} = \sum_{m} \frac{S_{im}P_{jm}P^*_{km}}{S_{0m}}.
\ee
These polynomial equations can be derived from (\ref{RR*=1}) and can be used to determine the Klein bottle and Moebius coefficients numerically. U-NIMreps can only be defined if the invariant matrix $Z_{ij}$ is symmetric.

\subsection{Summary of consistency conditions}\label{posint}
We can now enumerate the set of consistency conditions the open descendant construction should have. The unoriented string theory descending from a closed oriented string theory based on an internal conformal field theory that has a symmetric modular invariant torus partition function should also be consistent on surfaces with boundaries and crosscaps. For that we require:
\begin{enumerate}
\item{Existence of a NIMrep, { i.e.} a set of non-negative integers 
$A_{ia}^{\ph{ia}b}=A_{i^c b}^{\ph{i^c b}a}$ satisfying (\ref{AA=NA});}
\item{Existence of an S-NIMrep, { i.e.} at least one boundary conjugation matrix $C^B_{ab}$ such that $A_{iab}^\Omega=A_{ia}^{\ph{ia}c}\, C^B_{cb}$ is 
symmetric in the boundary indices;}
\item{Existence of a U-NIMrep, { i.e.} a set of Klein bottle and Moebius strip coefficients $K_i, M_{ia}$, such that $\um(A_{iaa}^\Omega+M_{ia})$ and $\um(Z_{ii}+K_i)$ are non-negative integers with $K_0=+1$ and $M_{ia}=M_{i^c {a}^c}$, and that (\ref{pe1}-\ref{pe2}) hold.}
\end{enumerate}
This is a set of world-sheet consistency requirements based on positivity and integrality requirements of the one-loop partition functions. In section \ref{sewchap} we discuss more general conditions that can be used to prove consistency to all orders in perturbation theory, not only at the one-loop level.

Note that if one cannot find an S- or U-NIMrep for a given NIMrep, this would mean that even though the conformal field theory can in principle\footnote{A NIMrep is a necessary but not sufficient condition for consistency on surfaces with boundaries.} be consistently defined on all oriented surfaces, it is inconsistent in the presence of crosscaps. In the same way one argues that a conformal field theory is inconsistent if it cannot be defined on surfaces with boundaries, one can conjecture that a conformal field theory is again inconsistent if cannot be defined on surfaces with crosscaps. Again, the conjecture that existence of a U-NIMrep is necessary for a conformal field theory to be consistent only makes sense if $Z_{ij}$ is symmetric.

\subsection{Boundaries and crosscaps from integral data}
Solving (\ref{AA=NA}) yields the oriented annulus coefficients. This gives us the boundary coefficients $B_{ma}$ up to a phase. Next, by finding a symmetrization matrix $C^B_{ab}$ one reaches the unoriented annulus coefficients, which determine $B_{ma}$ up to a sign, provided we know $\Omega^m$. These eigenvalues are in general difficult to find, but if we look at the unoriented annulus coefficients
\be
A_{iab}^\Omega = \sum_m S_{im} B_{ma} B_{mb} \, g^m, \label{Aiab}
\ee
we see that we can absorb a factor $\sqrt{g^m}$ into $B_{ma}$ and with that define a new boundary coefficient $B_{ma}^\prime = \sqrt{g^m} B_{ma}$, which can now be determined from (\ref{Aiab}) up to an overall $m$-dependent sign. Getting this new boundary coefficient $B^\prime$ is much easier than trying to find explicit solutions to $\Omega^m$ that are consistent. (In \cite{foe}, \cite{kbpaper} and \cite{2ndpaper} this was implicitly done.) At the level of the one-loop partition functions, determining one boundary coefficient or the other is, of course, the same.

The crosscap coefficient can also be recast into the form $\Gamma_m^\prime = \sqrt{g^m} \Gamma_m$, so that we remove $g^m$ from the problem at the level of partition functions. From (\ref{KMg^i}) we can get $\Gamma^\prime$, again up to an overall sign. If we consider $B^\prime$ and $\Gamma^\prime$, the positivity and integrality problem is invariant under $\Gamma_m^\prime \rightarrow \varepsilon_m \Gamma_m^\prime, \;B_{ma}^\prime \rightarrow \varepsilon_m B_{ma}^\prime$.

\subsubsection{Classifying algebra}
Telling what $B_{ma}$ is exactly can in principle be done because the boundary coefficients are expected to form a representation of the `classifying algebra' \cite{class}
\be
B_{ma} B_{na} = \sum_p X_{mn}^{\ph{mn}p} \,S_{0p} B_{pa},
\ee
which is sensitive to the phase/sign changes and should thus determine $B_{ma}$ exactly.  Knowing the structure constants $X_{mn}^{\ph{mn}p}$ would enable us to find also $\Omega^m$ and $C^m$, but unfortunately these constants are expressed in terms of model-dependent data like operator product coefficients and fusing matrices.

So, at this point we cannot do better than considering the boundary and crosscap coefficients $B_{ma}^\prime$ and $\Gamma_m^\prime$ when trying to find solutions for the positivity and integrality constraints. Accordingly, in the explicit solutions presented below, the boundary and crosscap coefficients are $B^\prime$ and $\Gamma^\prime$. However, in order to keep the notation simple, we will drop the primes from $B_{ma}^\prime$ and $\Gamma_m^\prime$.

\subsubsection{Methods for constructing open descendants}
Starting from a modular invariant torus one can construct its open descendants by finding its S- and U-NIMreps. This can be done either from first principles or by letting a computer program solve the positivity and integrality constraints numerically. One approach is, for instance, to postulate boundary and crosscap coefficients and verify they lead to Klein bottle, annulus and Moebius coefficients satisfying the positivity and integrality requirements of section \ref{posint}. This is a good approach if the case at hand has some underlying symmetry which allows an educated guess at the boundary and crosscap coefficients.

An alternative way is to search for solutions to the NIMrep equation (either analytically or numerically), finding a boundary conjugation matrix and then solve the polynomial equations to find a Klein bottle and Moebius strip consistent with the positivity and integrality requirements. This method can be used in cases where there are no clues as to what the boundary and crosscap coefficients look like.

\subsection{Other consistency requirements}\label{otherreqs}

The set of consistency requirements of the previous section are essential to have a sensible unoriented string theory. In this section we discuss other conjectured consistency conditions (also at the one-loop level) that are at the moment on less solid grounds than the ones above.

Amongst these are reality of the crosscap coefficients $\Gamma_m$ \cite{gammareal} and the (yet unproved) trace formula of \cite{trace}
\be
\sum_a A_{\ph{i}aa}^i = \frac{S_{ij}}{S_{0j}} Y^j_{\ph{j}00} K_j.
\ee
There is not much to say here, except that all consistent models known so far obey these two conditions, whereas many models that are inconsistent at other levels also fail to fulfill these two conditions.

We turn now to a condition not yet well understood, but which seems to play an important role nonetheless.

\subsubsection{Klein bottle constraint}
Since $\Omega$ is a symmetry of the theory, it must respect the bulk interactions and operator products \cite{tasi}. Therefore its eigenvalues should be conserved in fusion: $\Omega^m \Omega^n \Omega^p \geq 0$ if $N_{nm}^{\ph{nm}p}\neq 0$. In the spectrum of closed strings the projection is implemented by the Klein bottle, so this suggests that \cite{kbpaper} \cite{kb2}
\be
K_i K_j K_k \geq 0, \;\;\; \text{if $N_{ij}^{\ph{ij}k}\neq 0$}.
\ee
The Klein bottle projection should thus be preserved in fusion. But this is known to be violated in some otherwise consistent cases \cite{2ndpaper}. However, violations only occur in cases where $N_{ij}^{\ph{ij}k} \geq 2$ and even. Since two (or any even number of) anti-symmetrized representations can combine into a symmetrized one, it is not clear if these violations should be regarded as an inconsistency. What seems to be clear is that violations when $N_{ij}^{\ph{ij}k}$ is an odd number do lead to inconsistencies at various levels, so we will take that as a clear sign of trouble. It has been argued in \cite{even} that cases involving simple currents like those of \cite{2ndpaper} the fusion rules which lead to violations of the Klein bottle constraint are always even.

In \cite{nimrep} further examples were explored that violated the Klein bottle constraint. Some of these, like for instance automorphism invariants of $A_{1,9}$, also violated other requirements and could be discarded, but in the case of extension invariants of Wess-Zumino-Witten (WZW) models \cite{olive}, the extended theory fulfilled the Klein bottle constraint, whereas one of the U-NIMreps of the unextended theory violated it.

It is thus clear that Klein bottle constraint is still not well understood and a more rigorous formulation of it is necessary.

\subsubsection{Tadpole cancellation}\label{tadcancel}
This consistency condition is well understood, but it is more of a space-time requirement. In particular, it allows us to determine the Chan-Paton factors $n_a$.

The tadpole cancellation conditions see that the whole theory is divergence-free at one-loop. Recall that the integrations over the modular parameters for the Klein bottle, annulus and Moebius strip go all the way to infinity, thus potentially leading to divergences. The sum $\tilde{K}+\tilde{M}+\tilde{A}$ leads, in the transverse channel, to
\be
\tilde{K}+\tilde{M}+\tilde{A} = \int_0^\infty dt \sum_{\stackrel{\sst m}{n\geq 0}} 
        d_n^m e^{-2\pi t (h_m+n-1)}
        \left( \sum_a 2^{-D/2} B_{ma} n_a + 
               (-1)^n \varepsilon_m \Gamma_m \right)^2,
\label{tadpole}
\ee
where the factor $2^{-D/2}$ comes from the space-time contribution to the transverse channel and $\varepsilon_m$ a relative sign between $B$ and $\Gamma$, which fixes the Moebius projection. The tachyonic terms $n=0,\,h_m<1$ are highly divergent, but can be regularized by analytic continuation using $\int_0^\infty \! ds\, e^{as}\!=\!-1/a$ and are anyway expected to disappear in supersymmetric theories. The divergences coming from the mode excitations $n=1,\,h_m=0$ and internal fields $n=0,\,h_m=1$ cannot be regularized since the analytic continuation gives $-1/0$, therefore one must require that for field with $h_m=0,1$
\be
\sum_a  2^{-D/2} B_{ma} n_a + (-1)^{\delta_{m0}} \varepsilon_m \Gamma_m = 0,
\label{tadpolekill}
\ee
which can be solved for integer Chan-Paton factors.

The boundary and crosscap coefficients are in general irrational numbers, so one would normally expect no solutions to the Diophantine equation (\ref{tadpolekill}). However, in practice solutions seem to be embarrassingly abundant. A thorough study remains to be done to answer why this is so. Some steps in that direction have been taken in \cite{marijn}.

It should be noticed that some of the massless tadpole divergences are not necessarily fatal \cite{polcai}. Tadpoles of physical fields can be dealt with by shifting the vacuum into a stable background. Tadpoles of unphysical fields cannot couple to a background and therefore must be eliminated. Unphysical fields are those that have been projected out by the Klein bottle or Moebius strip. In space-time supersymmetric theories there are tadpoles coming from the NS-NS sector (Neveu-Schwarz) and tadpoles coming from R-R sector (Ramond). Depending on the Klein bottle projection, one of them is physical and the other unphysical. However, in these theories, both tadpoles come from the same supersymmetry multiplet, so, if we cancel one of the tadpoles, the other is automatically canceled as well.

Tadpole cancellation is, in a way, the analog of modular invariance for open string theory. It is an important tool that not only determines the final space-time spectrum but also guarantees it will be free of gauge and gravitational anomalies in the low-energy effective action \cite{polcai} \cite{tadpole}, just like modular invariance does for closed strings \cite{tadbert}.

When a string theory has D-branes that are space-time filling (i.e. that fill out all the non-compact dimensions), there will be tadpoles, whose cancellation requires introduction of O-planes. If the theory has D-branes that are not space-time filling, like the type IIA superstring, it can be tadpole free because the solitonic D-brane solutions are localized solutions to the equations of motion and these do not generate tadpoles. One can think of this in terms of R-R flux cancellation: the flux coming from non space-time filling branes can escape to infinity via the non-compact directions orthogonal to the D-brane. But if the brane is space-time filling, all these directions are taken. Then the flux has no where to go and must therefore annihilate on a source of negative R-R charge, the O-plane.

\section{Solutions to the integrality constraints}\label{modinvs}
The first step in the construction of open descendants is a left-right symmetric torus modular invariant partition function. The torus provides the operator spectrum, which tells us how the representations of the chiral algebra combine into a full conformal field theory. Furthermore, it is the object where the unoriented Klein bottle projection acts. A short list of modular invariants types is therefore in order.
\subsection{Modular invariant types} \label{mipfs}
\subsubsection{Automorphism invariants}
These invariants are characterized by a torus of the form $Z_{ij}=\delta_{i,\pi(j)}$:
\be
T = \sum_{ij} \chi_i \, \delta_{i,\pi(j)} \, \overline{\chi}_{{\pi(j)}}. \label{auto}
\ee
To satisfy $T$-invariance, the permutation of the chiral labels must be such that $h_i = h_{\pi(i)}\; \text{mod} \; N$. Any conformal field theory with a maximally extended symmetry algebra will necessarily be of automorphism kind \cite{naturality}.

An important kind of automorphism invariant is the C-diagonal, or Cardy invariant, which has $Z_{ij}= C_{ij} = \delta_{ij^c}$. This was the first invariant for which a general solution for the boundary coefficients was found \cite{cardy}, that was later generalized to unoriented surfaces in \cite{sagnottiopen}. Diagonal invariants have $Z_{ij}=\delta_{ij}$ but their apparent simplicity can be misleading as they are sometimes inconsistent due to lack of NIMreps \cite{nimrep}. Another way to obtain automorphism invariants is by means of simple currents \cite{simple}. Simple current invariants form a very large and important subset of modular invariants.

\subsubsection{Extension invariants}
In this case the chiral algebra has unused extended symmetry. There are various types of extensions, but the most common ones are simple current based extensions, for which case the torus modular invariant is typically of the form
\be
T=\sum_{i} |\chi_{i_1}+\cdots+\chi_{i_n}|^2 + \sum_f n_f |\chi_f|^2, \label{extension}
\ee
with $f$ labeling the simple current's fixed-points and $n_f$ an integer. This form shows explicitly the characters of the unextended algebra combining into a larger character. Extension invariants have links to symmetry-breaking boundary conditions \cite{fuchssbb} \cite{foe}, but we will not pursue that subject here.

There exist also modular invariants that are combinations of automorphisms and extensions. A torus of type
\be
T=\sum_i \left(\chi_{i_1}+\cdots+\chi_{i_n}\right)\overline{\left(\chi_{\pi(i_1)}+\cdots+\chi_{\pi(i_n)}\right)} + \sum_f n_f \chi_f \overline{\chi}_{\pi(f)} \label{autoext}
\ee
would be the general form of such an invariant.

\subsubsection{Exceptional invariants}
If the modular invariant is neither C-diagonal nor based in some simple current, the invariant is said to be {\em exceptional}. Solutions for the exceptional cases are hard to find since there are no symmetry principles underlying them. One however can still try a case-specific analysis of exceptional invariants.

An exceptional invariant can take many forms when written in terms of characters, so there is no typical example thereof. The list of known exceptional is somewhat scattered. For some of the results check \cite{galois} and section \ref{exceptable}.

\subsection{The Cardy-Rome example}
In this section we introduce a solution originally proposed by Cardy and later extended by Sagnotti {et al.} We also introduce a slight modification of it, which will show up in the solutions of the orbifold exceptional invariants of next chapter.

The Cardy-Rome solution takes the C-diagonal torus partition function. For the C-diagonal invariant all chiral labels are Ishibashi labels and every chiral field is a transverse field. There is thus a one-to-one correspondence between chiral labels and boundary conditions. The boundary and crosscap coefficients found by Cardy and Sagnotti {et al.} are
\be
   B_{ma}=\frac{S_{ma}}{\sqrt{S_{0m}}}, \;\;\; 
   \Gamma_m = \frac{P_{0m}}{\sqrt{S_{0m}}}.
\ee
These lead to the following Klein bottle, annulus and Moebius coefficients
\be
K_i = Y_{i00}=\nu_i,\;\;\; A_{ia}^{\ph{ia}b} = N_{ia}^{\ph{ia}{b}}, \;\;\; 
M_{ia} = Y_{{i}a0}. 
\ee
The $\nu_i$ is also known as Frobenius-Schur indicator \cite{bantay}. This is an index for a chiral label $i$, such that
\be
\nu_i = \left\{ \ba{cl} 
           +1 & \text{if $i$ is real} \\
           -1 & \text{if $i$ is pseudo-real} \\
            0 & \text{if $i$ is complex}
        \ea\right. 
\label{fsindicator}
\ee
By means of the following property of $Y_{ij}^{\ph{k}}$ found by Bantay \cite{bantay},
\be
\left| Y_{i0}^{\ph{i0}k} \right| \leq N_{\ph{k}ii}^k, \;\;\;
Y_{i0}^{\ph{i0}k} = N_{\ph{k}ii}^k \; \text{mod 2},
\ee
we can prove that the Klein bottle and Moebius integrality conditions are satisfied. The Klein bottle constraint is trivially satisfied because the Frobenius-Schur indicator is conserved in fusion. All other consistency requirements are also met.

\subsubsection{Non-trivial Klein bottles}
We can do a little twist of the above by using any simple current the conformal field theory may eventually have. Take again the C-diagonal invariant and a simple current $J$. Postulate new boundary and crosscap coefficients \cite{kbpaper}
\be
B_{ma}^{[J]}=\frac{S_{(Jm),a}}{\sqrt{S_{mJ}}}, \;\;\; \Gamma_m^{[J]} = \frac{P_{Jm}}{\sqrt{S_{Jm}}}.
\ee
The new Klein bottle, annulus and Moebius coefficients are
\be
K_i^{[J]} = Y_{iJJ^c},\;\;\;     A_{ia}^{[J]b} = N_{(Ji)a}^{\ph{(Ji0a}b},\;\;\; M_{a}^{[J]i} = Y^{\ph{(J^c a)J}i}_{(J^c a)J}.  \label{NTKB}
\ee
Again, one can check that these coefficients satisfy the integrality conditions. The modified Klein bottle coefficient can be interpreted as different choices for the symmetrization of the $\Omega^m$ projection. Klein bottle coefficients of the form (\ref{NTKB}) go by the nickname `non-trivial Klein bottle'. Proving integrality can be done using \cite{borisov}
\be
\left| Y_{iJ}^{\ph{iJ}k} \right| \leq N_{\ph{J^c k}ii}^{J^c k}, \;\;\;
Y_{iJ}^{\ph{iJ}k} = N_{\ph{J^c k}ii}^{J^c k} \; \text{mod 2},
\ee
together with some properties of the $S$- and $P$-matrices and a bit of algebra. The Klein bottle constraint can be proved using some properties of $Y_{ij}^{\ph{ij}k}$ and conservation of a quantity called `simple current charge' in fusion. See \cite{kbpaper} for details. Again, all other consistency requirements are met.

\section{U-NIMreps for other invariants}
In this section we review some of the current results regarding U-NIMreps for invariants other than the Cardy one.

\subsection{Simple current invariants}
As we have seen in section \ref{mipfs}, simple currents can be used to construct modular invariant partition functions. The question is then what are the open descendants that go with these tori.

The first steps in this direction were taken in \cite{earlyzuber}, which wrote down NIMreps for $SU(2)$ and some $SU(3)$ WZW models. In \cite{usecross} U-NIMreps for $SU(2)$ WZW models were written. Then in \cite{class} boundary coefficients for automorphism invariants induced by $\mathbb{Z}_2$ simple currents were given.

Later on, in \cite{fuchssbb} a generalization of the boundary and annulus coefficients to extension invariants was presented. There it was also noticed that from the point of view of the unextended theory some of the boundary conditions would break part the symmetry of the extended theory. This corresponded to having $\omega$'s in (\ref{gluing}) other than the identity. 

In \cite{2ndpaper}, boundary and crosscap coefficients for automorphism invariants that preserve all of the extended symmetries were given. This concluded the construction of open descendants for $\mathbb{Z}_2$ simple current extensions and also included non-trivial Klein bottles. Crosscaps for $\mathbb{Z}_2$ extension invariants were written in \cite{twistklein}.

The subject was finally wrapped up in \cite{foe} and \cite{lenthesis}, where universal formulas for a general simple current invariant were written down that summarize today's knowledge of the open descendants construction on these models. It is important to stress that these formulas work for {\em any} simple current invariant, be it automorphism, extension or a combination thereof and can be shown to lead to Klein bottle, annulus and Moebius coefficients that satisfy positivity and integrality. The universal formulas provide furthermore a lot of physical insight. For instance, the issues of symmetry-breaking boundaries and their relation to unextended and extended chiral algebras is now well understood.

\subsection{Diagonal invariants}
Pure diagonal invariants $Z_{ij}=\delta_{ij}$ were tackled in \cite{biftex} for the case of WZW models. The procedure there was to look at an extended chiral algebra ${\cal A}^{\text{E}}$ and consider an orbifold sub-group thereof, with chiral algebra ${\cal A}$, that preserved only part of the bulk symmetry. The symmetry-breaking boundaries of the orbifold theory were related to a charge conjugation automorphism $\omega=C$ on (\ref{JatB}) and these corresponded to the symmetry-preserving boundary conditions of the diagonal invariant of the ${\cal A}^{\text{E}}$-theory. 

However, in spite of the simplicity of their torus, diagonal invariants do not necessarily correspond to sensible conformal field theories. The point is that some models do not admit NIMreps for the diagonal invariant. In \cite{nimrep} some examples were presented where this happens. Presumably the orbifold theory does not exist for those examples.

That the diagonal invariant might be unphysical is the surprising result mentioned in the introduction. It is certainly an interesting and important problem to be addressed in future research.

\subsection{Exceptional invariants}\label{exceptable}
Exceptional invariants, as the name suggests, typically have no universal mechanism that can be used to derive S- and U-NIMreps from first principles. Their analysis is thus usually done case by case.

The first results for these type of invariants were the NIMreps of the $E$-type invariants of $SU(2)$ WZW models \cite{earlyzuber}, whose `$E_7$' invariant was later supplemented with U-NIMreps by \cite{usecross}. In \cite{alphainduction}, boundary coefficients for the ${G}_{2,3}$ WZW exceptional extension invariant were written. Chapter \ref{freebosons} presents S- and U-NIMreps for yet another case, the exceptional invariants of extended free boson orbifolds \cite{excep}. The free boson orbifolds are an explicit example of how one can use the integral data from the formalism developed in the present chapter to induce U-NIMreps for a class of exceptional modular invariants.

New results for exceptional invariants were presented in \cite{nimrep}. These are the exceptional automorphisms and extensions of WZW models (see \cite{olive} for notation). WZW automorphisms were classified in \cite{autogannon}. WZW extensions can be related to conformal embeddings ($X \subset Y$, in the table below and classified in \cite{embed}), higher spin extensions (HSE, some of which were classified in \cite{mero}) or simple current extensions (SC) admitting exceptional U-NIMreps in addition to those covered by \cite{foe}. In table \ref{extab} we present the outcome. These results are based on a complete computer search and are presented from the point of view of the unextended theory. Note also that, since all conformal embeddings are at level 1, the level index in the embedding CFT was dropped.

The data presentation is best explained with an example. Take for instance $A_{2,3}$. This model is first extended by a simple current and then embedded in $SO(8)$. It has 2 NIMreps, the first of which gives rise to 2 S-NIMreps and the other to 3 S-NIMreps. Then each of these S-NIMreps gives rise respectively to (0+1) and (0+1+1) U-NIMreps. 

When the conformal field theory is complex, one can extend either the C-diagonal or the diagonal invariant. The latter is marked with an asterisk in HSE$^*$. In the extension invariants, following the discussion of section \ref{otherreqs}, we have allowed for violations of the Klein bottle constraint (but no other condition). These are marked with an asterisk in the U-NIMrep column. For more details the reader is referred to \cite{nimrep}.

It is possible to get boundary and crosscap coefficients the automorphism invariant models. As an example we take $E_{4,8}$. The boundaries are $R_{ma}=\pm {\frac{2}{\sqrt{17}}}\sin{(2\pi l/17)}, \; l=1,\ldots,8$. For the extension models it is not possible to extract boundaries and crosscaps from the integral data because of the degeneracy labels of Ishibashi states (cf. \cite{nimrep}).

\bigskip\bigskip\bigskip

\begin{table}[h]
\bc
\begin{tabular}{lccc}
Modular invariant                 &   NIMreps   &   S-NIMreps   &    U-NIMreps        \\ \hline
$A_{1,10} \subset SO(5)$          &      1      &       2       &       1+1           \\
$A_{1,16}$ `$E_7$' invariant      &      1      &       1       &        1            \\
$A_{1,28} \subset G_2$            &      1      &       1       &        1            \\
$A_{2,3}$ SC/ $\subset SO(8)$     &      2      &      2+3      &   (0+1)+(0+1+$1^*$) \\
$A_{2,5} \subset SU(6)$           &      1      &       2       &       1+$1^*$       \\
$A_{2,6}$ SC                      &      2      &      2+1      &     (0+1)+1         \\
$A_{2,9} \subset E_6$             &      3      &     2+2+2     &  (1+0)+(1+0)+(0+0)  \\
$A_{3,2}$ SC2/ $\subset SU(6)$    &      1      &       4       &     1+1+1+1         \\
$A_{3,4} \subset SO(15)$          &      1      &       4       &     0+0+1+1         \\
$A_{9,2}$ HSE$^*$                 &      1      &       2       &       1+1           \\
$B_{2,3} \subset SO(10)$          &      2      &      4+4      & (0+0+1+$1^*$)+(0+0+1+1) \\
$B_{2,7} \subset SO(14)$          &      2      &      4+4      & (0+0+1+$1^*$)+(0+0+1+1) \\
$B_{2,12} \subset E_8$            &      1      &       4       &        1            \\
$B_{12,2}$ HSE                    &      1      &       2       &        1            \\
$C_{3,2} \subset SO(14)$          &      2      &      4+4      & (0+0+1+$1^*$)+(0+0+1+1) \\
$C_{3,4} \subset SO(21)$          &      1      &      16       &   $1+1+14\times 0$  \\
$C_{4,3} \subset SO(27)$          &      1      &      16       &   $1+1+14\times 0$  \\
$C_{7,2}$ HSE                     &      2      &      4+4      & (0+0+1+$1^*$)+(0+0+1+1) \\
$C_{10,1}$ HSE                    &      1      &       2       &       1+1           \\
$D_{7,3}$ HSE$^*$                 &      2      &      2+2      &   (0+0)+(0+2)       \\
$D_{7,3}$ HSE                     &      2      &      2+2      &   (0+0)+(1+1)       \\
$D_{9,2}$ HSE                     &      2      &      2+5      &  $(1+1)+5\times 0$  \\
$D_{9,2}$ HSE$^*$                 &      3      &     4+4+4     & $(1+1+0+0)+(4\times 0)+(4\times0)$ \\
$E_{6,4}$ HSE$^*$                 &      2      &      2+2      &   (0+0)+(1+$1^*$)       \\
$E_{7,3}$ HSE                     &      2      &      4+4      & (0+0+1+1)+(0+0+1+1) \\
$E_{8,4}$ automorphism            &      1      &       1       &        1            \\
$F_{4,3}$ automorphism            &      1      &       4       &        1            \\
$F_{4,3} \subset SO(26)$          &      2      &      4+4      & (0+0+1+$1^*$)+(0+0+1+$1^*$) \\
$G_{2,3} \subset E_6$             &      2      &      2+2      &   (0+1)+(0+1)       \\
$G_{2,4}$ automorphism            &      1      &       4       &        1            \\
$G_{2,4} \subset SO(14)$          &      2      &      4+4      & (0+0+1+$1^*$)+(0+0+1+$1^*$) \\
\end{tabular}
\caption{NIMreps for exceptional invariants of WZW models}
\label{extab}
\ec
\end{table}

\chapter{Free bosons} \label{freebosons}
In this chapter we study internal conformal field theories with central charge $c=1$. From the geometrical point of view, these theories correspond to compactifying one dimension of a String Theory. For the bosonic case this means going to $D=25$, and for superstring theories to $D=9$. If one takes tensor products of various of these theories, one goes further down in dimension, so, restricting oneself to the $c=1$ case is only limiting as far as it means selecting one of the possible branches in the huge tree of compactification possibilities.

We concentrate on $c=1$ conformal field theories that are not tensor product theories nor the limit $n \rightarrow \infty$ of Virasoro minimal models \cite{watts}. This restricts us to a few cases only. For these cases the chiral algebra can have extended symmetries, which we will take into account in order to lower the number of primary fields so that a rational conformal field theory is reached. With the models identified, the integrality conditions will be used to write down boundary and crosscap coefficients.

\section{Classification of $c=1$ theories}
The space of $c=1$ theories is summarized by the following picture
\begin{figure}[h]
\SetScale{1}
\begin{center}
\begin{picture}(400,200)(0,0)
\Vertex(100,50){2}
\DashLine(100,50)(150,50){3}
\Vertex(150,50){2}
\LongArrow(150,50)(300,50)
\Vertex(200,50){2}
\LongArrow(200,50)(200,200)
\DashLine(200,0)(200,36){3}
\Line(200,50)(200,47)
\Vertex(200,0){1}
\Vertex(150,174){2}
\Vertex(150,185){2}
\Vertex(150,196){2}
\Vertex(200,50){2}
\Text(98,40)[l]{$0$}
\Text(140,40)[l]{$\sqrt{2}$}
\Text(120,25)[l]{\scriptsize (Self-dual point)}
\Text(185,40)[l]{$2\sqrt{2}$}
\Text(290,40)[l]{$R$}
\Text(140,85)[l]{$R_{\text{orb}}\!=\!\sqrt{2}$}
\Text(171,200)[l]{$R_{\text{orb}}$}
\LongArrow(187,75)(197,55)
\Text(191,2)[l]{$0$}
\Text(139,198)[c]{T}
\Text(139,186)[c]{O}
\Text(139,174)[c]{I}
\end{picture}
\end{center}
\label{c1}
\caption{The space of $c=1$ theories (with $\alpha^{\prime}=2$)}
\end{figure}
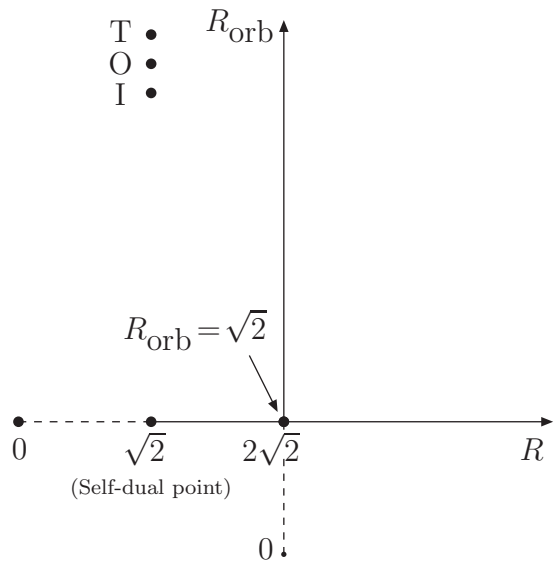
\noindent
We see there are three types of $c=1$ theories: free bosons (horizontal line), orbifolds thereof (vertical line) and three special theories, nicknamed TOI (three dots). The lines are the moduli space of a continuous parameter. For the free boson we call it $R$ and for the orbifold $R_{\text{orb}}$. There is no formal proof that this picture is complete, and there are indeed indications that it is not \cite{c=1inc}.

\subsubsection{Free bosons} These are simply conformal field theories of a compact field $X$, with the same classical action of the bosonic string $X(\st)$, subject to periodic boundary conditions 
\be 
X(\sigma,\tau) \simeq X (\sigma+2\pi,\tau)+2\pi \omega R,
\ee
with $\omega$ the winding number. This relaxed periodicity condition, as compared to the usual closed string $X(\sigma,\tau)=X(\sigma+2\pi,\tau)$, reflects the fact that the space-time coordinate $X$ is periodic, with $\omega$ the number of times the string winds around the compact dimension. The geometrical interpretation corresponding to this conformal field theory is that of a coordinate compactified in a circle of radius $R$. At special values of the radius $R$ the chiral algebra extends, as we will see below.

The free boson theory at radius $R$ is actually equivalent to the same theory at $2/R$. This is $T$-duality (see below). Because of it, it is enough to consider only $R \in [\sqrt{2},+\infty]$ to get all inequivalent free boson theories. We could have chosen $[0,\sqrt{2}]$ instead (the dashed line in fig. \ref{c1}), but the choice above is more convenient for our purposes.

\subsubsection{Free boson orbifolds} This is the same compact field $X$, but this time with the identifications 
\be 
X(\sigma,\tau) \simeq 
X(\sigma+2\pi,\tau)+2\pi \omega R_{\text{\scriptsize{orb}}}, 
\;\;\; X \simeq -X.
\ee
Geometrically, the orbifold theory can be interpreted as compactification in a line segment. Of all the $c=1$ theories, orbifolds have the richest structure.

The line that represents the moduli space of the orbifold theory actually touches that of the free boson: the free boson theory at $R=2\sqrt{2}$ is the same theory as the orbifold of the free boson at self-dual radius $R_{\text{\scriptsize{orb}}}=\sqrt{2}$.

\subsubsection{TOI theories} 
These were discovered by Ginsparg in \cite{ginsparg} by taking special orbifolds of the $R=\sqrt{2}$ circle theory. Boundary and crosscap states were investigated in \cite{cap}, using the simple current methods of \cite{foe} together with the formalism of \cite{fuchssbb}. Higher-loop consistency conditions of TOI models will not be checked, so these will be skipped from now on.

\section{Extended free bosons and orbifolds} 
Here we present some details regarding the free boson and orbifold conformal field theories with which we will work throughout the rest of this book.

\subsection{Free bosons}
The compactified free boson conformal field theory is defined by the action 
\be 
S=\frac{1}{4\pi \alpha^\prime}\int\! d^2\sigma \; \partial_a  X \partial^a  X, 
\label{WSaction} 
\ee 
with the free boson field subject to the periodicity $X(\sigma,\tau)=X(\sigma+2\pi,\tau)+2\pi R$. We use the conventions of \cite{nickbert}, except for the closed string periodicity, on which we use a conformal dilation $(\sigma,\tau)\rightarrow (2\sigma,2\tau)$ to arrive at the same periodicity conventions of (\ref{bcs}).

After canonical quantization, the general solution for the compact free boson quantum field is
\be 
 X(\sigma,\tau) = \hat{q} + \alpha^\prime \hat{p} \tau + \hat{w} R \sigma + 
    \frac{\I}{2}\sqrt{2\alpha^\prime} \sum_{n\neq 0} \frac{1}{n}\alpha_n 
    e^{-\I n(\tau+\sigma)} + \frac{1}{n}\bar{\alpha}_n e^{-\I n(\tau-\sigma)}.
\ee 
with $\hat{q},\hat{p},\alpha_n$ the usual position, momentum and creation/annihilation operators. The operator $\hat{w}$ is the winding operator. Winding modes are a typical stringy phenomena; they cannot exist for point-particle field theories because particles cannot wrap. Single-valuedness of the string wave-function around the compact dimension requires 
\be 
e^{\I pX(\sigma+2\pi,\tau)} = e^{\I pX(\sigma,\tau)}, 
\ee 
which restricts the eigenvalues of the string linear momentum to the Kaluza-Klein values $p=n/R,\; n\in Z$. We now split $X$ into left- and right-components $X(\sigma,\tau)=X_L(\tau+\sigma) + X_R(\tau-\sigma)$. Defining left- and right-momenta we get 
\be 
X(\sigma,\tau) &=& \hat{q}_L + \hat{q}_R + 
   \um\alpha^\prime \hat{p}_L (\tau+\sigma) + 
   \um\alpha^\prime \hat{p}_R (\tau-\sigma) \nn\\ &&+ 
      \frac{\I}{2}\sqrt{2\alpha^\prime} 
   \sum_{n\neq 0} \frac{1}{n}\alpha_n e^{-\I n(\tau+\sigma)} + 
                  \frac{1}{n}\bar{\alpha}_n e^{-\I n(\tau-\sigma)}, 
\ee 
with $\hat{p}_{L},\hat{p}_R$ the left and right momentum operators, which have eigenvalues respectively 
\be 
p_L=\frac{n}{R}+\frac{mR}{\alpha^\prime},\;\;\; 
p_R=\frac{n}{R}-\frac{mR}{\alpha^\prime}. 
\ee 
The integers $n,m$ are usually referred to as momentum and winding numbers. (The winding number is the eigenvalue of the winding operator $\hat{w}$.) In the following, we take the convention of fig. \ref{c1} and set $\alpha^\prime=2$. Note that interchanging simultaneously $n \rightarrow m$ and $R \rightarrow 2/R$ leaves the spectrum invariant - this is the statement of $T$-duality.

We now go from the cylinder coordinates $(\sigma,\tau)$ to the usual coordinates on the complex plane $(z,\bar{z})$ via the conformal transformation 
\be 
z=e^{\I (\tau+\sigma)}, \;\;\; \bar{z}=e^{\I (\tau-\sigma)}. 
\ee 
In terms of the coordinates on the plane, 
the free boson quantum field becomes 
\be 
X(z,\bar{z}) &=& \hat{q}_L + \hat{q}_R -\I \left(\hat{p}_L \log{z} + 
                 \hat{p}_R \log{\bar{z}}\right) + 
                 \I \sum_{n\neq 0} \frac{1}{n} \left( \alpha_n z^{-n} + 
                 \bar{\alpha}_n \bar{z}^{-n}\right) \nn \\ &=& 
   X_L(z) + \overline{X}_R(\bar{z}). 
\ee 

The free boson chiral algebra contains the Virasoro algebra, generated by $1$ and $T(z)$, and the operator $\partial X(z)$\footnote{We abbreviate 
$\partial_z X_L(z)$ by $\partial X(z)$ and 
$\partial_{\bar{z}} \overline{X}_R(\bar{z})$ 
by $\overline{\partial} \overline{X}(\bar{z})$.}. Note that due to the logarithmic branch cut when $z\rightarrow e^{2\pi i}z$, the field $X(z,\bar{z})$ itself is {\em not} a conformal field. The conformal fields are instead combinations of highest-weight states of the free boson algebra that are created by acting on the vacuum state with the so-called vertex operators 
\be 
V_{p_L,p_R}(z,\bar{z}) = \,:e^{\I p_L X(z)+\I p_R \overline{X}(\bar{z})}: \;\;\; |p_L,p_R\rangle = V_{p_L,p_R}(0,\bar{0})|0\rangle, 
\ee 
where $p_L,p_R$ are left and right momentum eigenvalues and 
the dots $::$ denote the usual normal ordering of the oscillator modes $\alpha_n$. In order to keep the notation uncluttered, we drop the dots from now on, with the understanding that all vertex operators are normally ordered.

If one considers the Virasoro algebra only, all the $V_{p_L,p_R}(z,\bar{z})$ are primary fields, but we can use extra symmetries to extend the chiral algebra further. Consider integer spin currents of the form 
\be 
V_\lambda = e^{\I \lambda X(z)}, 
\;\;\; \lambda^2 \in 2\mathbb{Z}, \;\;\;  h_{V_\lambda}=\um \lambda^2.
\label{vext}
\ee 
These are simple currents and can be used to extend the chiral algebra \cite{bertnotes}. As in any extension, the representations of the chiral algebra enlarge and fields that were primary before the extension can become descendants afterwards. For the free boson what happens is that the infinity of states created from $V_{p_L,p_R}(z,\bar{z})$ can be reorganized into a finite set of primary fields $\phi_{k\bar{k}}\zz$ and the extended conformal field theory becomes rational. By considering the operator product expansion of the vertex operators with the chiral extension currents 
\be
V_\lambda (z) \; V_{p_L,p_R}(w,\bar{w}) \sim 
   (z-w)^{p_L \lambda} \; 
   e^{\I \left( (p_L + \lambda)X(w)+p_R\overline{X}(\bar{w}) \right)}+\cdots  
\ee
we see that if operator locality (no branch cuts) is to be preserved, the extension is only possible when $R^2=2N$, with $N$ positive integer. For more details of this construction see for instance \cite{bertnotes} or \cite{nickbert}.

The extended free boson conformal field theory contains the following set of chiral fields

\begin{table}[h]
\begin{center}
\begin{tabular}{cccc}
   {Field} & {weight} & {description} & {diagonal vertex operator} \\ \hline   
      0    &     0    &   {identity}  &     1     \\ 
      $\Phi_k,\; k=1,\ldots,2N-1$  &  $k^2/4N$  & {charged fields} &  
      $e^{\I \frac{k}{R}X(z,\bar{z})}$
\end{tabular}
\caption{Free boson chiral labels}
\end{center}
\end{table}
\noindent
We added a column with the form of the vertex operator for the diagonal theory $Z_{ij}=\delta_{ij}$, with whom we can calculate the correlators which will be needed to test perturbative consistency at higher loops. We use the term `chiral field' or `chiral label' to designate chiral halves of full conformal fields. The labels $\Phi_k,\overline{\Phi}_{\bar{k}}$ lead then to $\phi_{k\bar{k}}\zz$. Again, full conformal fields are objects in ${\cal A} \otimes \overline{\cal A}$, whose precise form is determined by the modular invariant. Of the charged chiral fields, $k=N$ is real ($\Phi_k=\Phi_k^c$), the others complex ($\Phi_{k}\neq \Phi_k^c$).

The operator product expansion of the full conformal field in the diagonal theory can be evaluated explicitly by manually redoing the normal ordering. The result is 
\be 
\phi_{i}(z,\bar{z}) \; \phi_{j}(w,\bar{w})\sim |z-w|^{\frac{2ij}{R^2}} \; \phi_{i+j}(w,\bar{w})+\cdots
\ee 
Since in the diagonal theory the left-right coupling is $\bar{i}=i$, we abbreviate $C_{i\bar{i},j\bar{j}}^{\ph{i\bar{i},j\bar{j}} k\bar{k}}$ to $C_{ij}^{\ph{ij}k}$ (and likewise for the primary fields $\phi_{i\bar{i}}\rightarrow \phi_{i}$). So we see the operator product coefficients in this field normalization are simply $C_{\phi_i \phi_j}^{\ph{\phi_j \phi_i}\phi_{i+j}}=1$. Note also that a primary field $\phi_{a+b}$ will not exist if $a+b>2N$. In this case the operator product is into $\phi_{2N-a-b}$, at a {\em descendant} level. This is because from the unextended point of view $\phi_{a+b}$ for $a+b>N$ was a primary field, but in the extended theory it becomes a descendant. Another important operator product is 
\be
-\partial{X(z)}\overline{\partial}{\overline{X}(\bar{z})} \, \phi_k(w,\bar{w}) \sim \frac{\frac{k^2}{R^2}}{|z-w|}\, \phi_{k}(w,\bar{w})+\cdots 
\ee 
from which we get the orbifold theory operator product $C_{J\phi_k}^{\ph{\phi_{k}J}\phi_k}=\frac{k^2}{R^2}$ (see below).

The extended free boson chiral algebra is generated by $1,\; T(z), \;\partial X(z), \;e^{\I nRX(z)}$. This is the $U(1)$ affine algebra at level $2N$, or in short $U(1)_{2N}$, which is why the compactification radius of the free boson is sometimes referred in the literature as $R_{U(1)}$.

\subsubsection{Free bosons open descendants}

The construction of open descendants for the case of free bosons is very simple because all possible modular invariants are just simple current automorphism and extension invariants. For these, the boundary and crosscap coefficients can be calculated using the formalism of \cite{foe}.

\subsection{Free bosons orbifolds}

We can take an orbifold of the extended free boson theory by dividing out the $\mathbb{Z}_2$ symmetry $X(z,\bar{z}) \rightarrow -X(z,\bar{z})$ and keeping the states invariant under this symmetry. This projects out part of the spectrum, so the chiral algebra will not contain contain the operators $\partial X(z)$ and $e^{\I nR X(z)}$, but we can still form invariant combinations such as $\cos{(nRX(z))}$. From now on we will be working with the orbifold theory exclusively, so we drop the subscript from $R_{\text{\scriptsize orb}}$.

Since the unprojected field content was complete, removing some of the fields renders the theory non-local, e.g. branch cuts will appear in some operator products. To restore consistency one has to add new fields, the twisted sector fields. The chiral field content of the extended orbifold and diagonal vertexes turns out to be

\begin{table}[h]
\begin{center}
\begin{tabular}{cccc}
{Field} & {weight}& {description} & {diagonal vertex operator} \\ \hline 
      0      &       0     &    {identity}           &    1    \\ 
      J      &       1     &    {simple current}     &  
          $-\partial X(z)\overline{\partial}\overline{X}(\bar{z})$      \\ 
      $\Phi^{1}$   &    ${N}/{4}$  &    {splitting of $\Phi_N$}  & 
         $2\sin{\left( \frac{N}{R}X(z) \right)} 
           \sin{\left( \frac{N}{R}\overline{X}(\bar{z})\right)}$         \\ 
      $\Phi^{2}$   &    {N}/{4}  &    {splitting of $\Phi_N$}  & 
         $2\cos{\left(\frac{N}{R}X(z)\right)} 
           \cos{\left( \frac{N}{R}\overline{X}(\bar{z})\right)}$      \\ 
      $\Phi_k,\; {k=1,\ldots,N-1}$  & ${k^2}/{4N}$ & {fixed points of $J$}  & 
            ${\sqrt{2}}\cos{\left( \frac{k}{R}(X(z)+\overline{X}(\bar{z})) 
              \right)}$ \\ 
      $\sigma_{1,2}$  & ${1}/{16}$  & {twisted sector} & {(see below)} \\            
      $\tau_{1,2}$    & ${9}/{16}$  & {twisted sector} & {(see below)}
\end{tabular}
\caption{Orbifold chiral labels}
\label{orbifoldvertexoperators}
\end{center}
\end{table} 
\noindent
(See \cite{c=1orientifolds} for more information on the $\Phi^i$ diagonal vertex operators.) We have inserted some overall normalizations to ensure that the two-point functions are proportional to unity. We did not include the explicit form of the twisted sector vertex operators, since these will not play a role\footnote{The explicit form can be found for example on \cite{twistver}.}. The simple current $J$ can be used to extend the orbifold theory back into the parent free boson theory at $R^2={2N}$. The diagonal vertex operators coming from $\Phi^{i},\; i={1,2}$ form the conformal field $\phi^i\zz$ and likewise the diagonal vertex $\Phi_k$ gives rise to $\phi_k \zz$. The form of the vertex operators for the diagonal conformal fields $\phi^{1,2}$ and $\phi_k$ is slightly different, even though they come from similar free boson fields, but this but a straightforward generalization of the vertex operators one gets by studying the equivalence between orbifold $N=2$ and the tensor product of two Ising models (see for instance the textbook \cite{cftbook} or \cite{robthesis}).

\subsubsection{Field complexification for $N$ odd}
The conformal fields $\phi^{1,2}$ behave differently depending on whether $N$ is odd or even. They are real fields for $N$ even and complex fields for $N$ odd. The twist fields also show this behavior. Since the orbifold vertex operators (\ref{orbifoldvertexoperators}) will always give rise to real fields, for the $N$ odd case one must complexify the $\phi^i$ fields by means of the following complex linear combinations 
\be 
\hat{\phi  }^{1}   & \!\!\! = \displaystyle{\frac{1}{\sqrt{2}}}
                       (\phi^1   + \I \phi^2),   \;\;\;  
\hat{\sigma}_{1}   & \!\!\! = \frac{1}{\sqrt{2}}(\sigma_1 + \I \sigma_2), 
                     \;\;\;  
\hat{\tau  }_{1}     = \frac{1}{\sqrt{2}}(\tau_1   + \I \tau_2),   \nn \\ 
\hat{\phi  }^{2}   & \!\!\! = \displaystyle{\frac{1}{\sqrt{2}}}
                       (\phi^2   - \I \phi^2),   \;\;\; 
\hat{\sigma}_{2}   & \!\!\! = \frac{1}{\sqrt{2}}(\sigma_1 - \I \sigma_2), 
                     \;\;\;  
\hat{\tau  }_{2}     = \frac{1}{\sqrt{2}}(\tau_1 - \I \tau_2).
\label{Noddvertexes}
\ee 
These are the true conformal fields for $N$ odd. 

\subsubsection{Orbifold fusion rules}
The fusion rules of the orbifold theory are needed to find the NIMreps. Since the orbifold $S$-matrix depends on the compactification parameter $N$ \cite{dvvv}, the orbifold fusion rules change accordingly. In the following we define  
\be 
[k] = \left\{ \ba{cl} |k| & \mbox{if $-N<k<N$, $k\neq 0$} \\  
                     2N-k & \mbox{if $k>N$} \\
                     2N+k & \mbox{if $k<-N$} 
      \ea \right.
\label{orblabelmap}
\ee 
This is to make sure the labels of $\Phi_k$ fall into the allowed range $k=1,\ldots,N-1$. For $N$ even we have fusion rules (with $\Phi$ standing for any field and taking $i\neq j$), 
\be 
0 \times \Phi = \Phi & J \times J = 0 & J \times \Phi^i=\Phi^j \nn \\  
J \times \Phi_k=\Phi_k & J \times \sigma_i=\tau_i & J\times \tau_i=\sigma_i \nn
\ee 
\be 
\Phi^i \times \Phi^i= 0 & \Phi^i \times \Phi^j = J & \Phi^i \times \Phi_k = \Phi_{N-k} \nn  \\  
\Phi^i \times \sigma_i=\sigma_j & \Phi^i \times \sigma_j=\tau_i & ({\sst N/2} \; \mbox{odd}) \nn \\ 
\Phi^i \times \sigma_i=\tau_j & \Phi^i \times \sigma_j=\sigma_i & ({\sst N/2} \; \mbox{even}) 
\ee 
\be 
&&\Phi_k \times \Phi_k = 0+J+\Phi_{[2k]}  \nn\\ 
&&\Phi_k \times \Phi_{N-k} = \Phi^1+\Phi^2+\Phi_{[N-2k]}  \nn \\ 
&&\Phi_k \times \Phi_l = \Phi_{[k+l]} + \Phi_{[k-l]} \nn \\ 
&&\Phi_{2k} \times \tau_i=\sigma_j+\tau_j \nn \\
&&\Phi_{2k-1} \times \sigma_i=\sigma_i+\tau_i \nn 
\ee 
Cases not covered by this table can be derived from it using the totally-symmetric nature of $N_{ijk}$ (all indices lowered). For $N$ odd the fusion rules are mostly the same, but there are a few changes 
\be 
\Phi^i \times \Phi^i= J & \Phi^i \times \Phi^j = 0 \nn 
\ee 
\be 
\Phi^i \times \sigma_i=\sigma_j & \Phi^i \times \sigma_j=
   \tau_i & (  {\sst   (N-1)/2}  \; \mbox{odd}) \\ 
\Phi^i \times \sigma_i=\tau_j & \Phi^i \times \sigma_j=
   \sigma_i & ( {\sst  (N-1)/2}  \; \mbox{even}) \nn
\ee 
The reader interested in more details about the orbifold construction is referred to \cite{bertnotes} and \cite{dvvv}. 
 
\section{Open descendants for orbifolds} 
 Most of the modular invariant partition functions of free boson orbifolds are just simple current and extension invariants. The boundary and crosscap coefficients for these cases are straightforward to evaluate by means of \cite{foe}. However, the orbifold possesses a class of modular invariants that is exceptional. In some cases, it is possible to find some underlying symmetry principle and from it derive boundary coefficients \cite{alphainduction}, but in general this is not possible and the only way to solve the problem is empirically. This is the case for the orbifold: its exceptional invariants are related to automorphisms of the fusion rules that are not related to any simple current. From the geometrical point of view, exceptional invariants correspond to compactifications on a circle of fractional radius. The results of this section appeared in \cite{excep}.

\subsubsection{The exceptional torus}
The exceptional invariant of the orbifold theory can be built using an automorphism $\omega$ whose simultaneous action on all labels of a fusion coefficient leaves the coefficient invariant: $N_{ij}^{\ph{ij}k}=N_{\omega(i) 
\omega(j)}^{\ph{\omega(i) \omega(j)}\omega(k)}$. This automorphism acts non-trivially on the chiral labels $\Phi_k$ only, as described in appendix \ref{EA}. The torus modular invariant partition function is then 
\be 
T=\sum_{ij} \chi_i(\tau) \; \delta_{i,\omega(j)} \; \overline{\chi}_j(\bar{\tau}). \label{orbexceptorus} 
\ee 
This torus will be modular invariant for $N$ odd and such that its prime number decomposition contains at least two different prime factors \cite{galois} \cite{autogannon}. The first $N$ that comply to these requirements are 15, 21, 33, 35, 45, etc. In this thesis we will focus on the case $N=p_1 \times p_2$, with $p_1 < p_2$ and both prime. The automorphism acts in such a way that the fields\footnote{We denote by $\dot{p}$ multiples of $p$, with zero included, when possible.} $\Phi_{k}, \; k=\dot{p_1},\dot{p_2}$ couple with themselves on the torus (self-couple), whereas the remaining $\Phi_k$ couple amongst themselves crosswise. Since the $\Phi_k$ are real, the transverse fields (or Ishibashi labels) are $0$, $J$ and $\Phi_{k,} \; k=\dot{p_1},\dot{p_2}$, in total $p_1+p_2$ of them.
 
We can construct another exceptional invariant replacing $\delta_{i,\omega(j)}$ with the charge conjugation matrix $C_{i,\omega(j)}$ in (\ref{orbexceptorus}). We will call the two invariants as ``diagonal + automorphism" (D+A) and  ``Cardy + automorphism" (C+A) respectively. In C+A $\Phi^i, \sigma_i, \tau_i$ become transverse fields, raising the total number of these to $p_1+p_2+6$.

Geometrically, these two tori correspond to free boson compactification on the fractional circle of radius $R^2=2p_1/p_2$ and its $T$-dual.

\subsubsection{Klein bottle projections}
Having the torus, we can look for the Klein bottle projection. As usual, the Klein bottle has to be such that non-transverse fields do not propagate in the transverse channel; in other words, that their crosscap coefficient $\Gamma$ vanishes. The simplest way to get $\Gamma$ is to invert (\ref{KAMcoeff}), from which we conclude that the signs $\varepsilon_j=K_j/Z_{jj}$ of the direct channel Klein bottle have to be such that
\be 
\Gamma^2_i  = \sum_j S_{ij} \varepsilon_j Z_{jj} = 0, \;\;\; \forall_i: Z_{ii^c} = 0. \label{sumrule} 
\ee 
In the D+A case this condition is satisfied for the 
``trivial" Klein bottle projection, $K_i=1$ for all the fields coupling diagonally on the torus. By the mechanism explained in chapter \ref{opendes}, the simple current $J$ generates a second Klein bottle, this one with $K_i = -1$ on the four twist fields $\sigma_i$ and $\tau_i$ and $K_i=1$ for the other diagonal fields. There are several other Klein bottle choices that satisfy  the sum rule (\ref{sumrule}), but if in addition we impose the Klein bottle constraint  these are the only two that are allowed.

In the C+A case we observe first of all that surprisingly the trivial choice 
$K_i=1$ for all the fields coupling diagonally on the torus violates the sum rule (\ref{sumrule}). There are however several 
Klein bottle choices satisfying the sum rule, and two of them also
satisfy the Klein bottle constraint.
One has $K_{\Phi_k}=-1$ when $k$ is an odd multiple of $p_1$ and $K_i=1$ for the remaining fields coupling diagonally on the torus, and the other has $K_{\Phi_k}=-1$ when $k$ is an odd multiple of $p_2$ and $K_i=1$ for the remaining fields coupling diagonally on the torus. These two Klein bottle choices are again related by the action of a simple current, this time $\Phi^i$.

\subsubsection{Annulus and Moebius strip}
 For the two aforementioned Klein bottles of D+A and C+A it is possible to numerically solve and symmetrize the NIMrep equation (\ref{AA=NA}) and derive unoriented annulus coefficients for low $N$. By inspection a general formula for an annulus for arbitrary $N=\dot{p}_1 \times \dot{p}_2$ can then be postulated. Finally, using (\ref{KAMcoeff}) we can solve for boundary and crosscap coefficients to get a complete set of boundary and crosscap coefficients.

Working backwards, starting from the boundary and crosscap coefficients discovered, one can check positivity and integrality of the postulated Klein bottle, annulus and Moebius strip. The proof, albeit straightforward, is rather lengthy and will therefore not be written here. One needs simple algebra plus the Gauss summation formula \cite{gauss}
\be
    \sum_{k=1}^N e^{2\pi \I \frac{{k(k+g)}}{2N}} = 
    \sqrt{N}\,e^{\frac{\pi \I}{4} \left( 1-\frac{g^2}{N} \right)} 
    \;\;\; \text{for $N$ odd},
\ee
to simplify the expressions arising.

\subsection{D+A invariant} 
The boundary and crosscap coefficients of the D+A torus are as follows. We define U-crosscap coefficient as $U_m = \sqrt{S_{0m}} \Gamma_m$. This is the analog of the reflection coefficient $R_{ma} = \sqrt{S_{0m}} B_{ma}$. Recall also that the reflection coefficient includes the orientifold factor $\sqrt{g^m}$.

The D+A invariant has one NIMrep, which can be symmetrized in four ways. Of these four S-NIMreps, three do not admit U-NIMreps and are thus viewed as unphysical. The fourth S-NIMrep leads to two U-NIMreps.

\subsubsection{D+A crosscaps} 

For the D+A case and trivial Klein bottle, the U-crosscap coefficients are 
\be 
\begin{array}{c|cccccc} 
 \; &  0 & J & \Phi^i & \Phi_k & \sigma_i & \tau_i \\ \hline 
U_m 
   & \displaystyle{{\um
        \left( \frac{1}{\sqrt{p_1}}+\frac{1}{\sqrt{p_2}} \right)}}
   & \displaystyle{{\um
        \left( \frac{1}{\sqrt{p_1}}-\frac{1}{\sqrt{p_2}} \right)}}   
   & 0 & \ldots, \underbrace{\frac{1}{\sqrt{p_2}}}_{k = 2\dot{p_1}}, \ldots, 
     \underbrace{\frac{1}{\sqrt{p_1}}}_{k = 2\dot{p_2}},\ldots  
   & 0 & 0 \end{array} 
\label{U} 
\ee
The dots stand for zero entries and the under-brackets show exactly for which $k$ is the $\Phi_k$ U-crosscap coefficient non-vanishing. We used the sign freedom $U_m \> \varepsilon_m U_m, \; R_{ma} \> \varepsilon_{m} R_{ma}$ to define the coefficients such that for this Klein bottle they are all positive.

The second Klein bottle, with $K_i=-1$ for the twist fields, has $U_0$ and $U_J$ interchanged and also a minus sign for $\Phi_{k,}\;k=2\dot{p_1}$. Note that for both cases all non-transverse fields have vanishing U-crosscap coefficient, as expected.

\subsubsection{D+A boundaries}
Before displaying the reflection coefficients, we first classify the possible boundary conditions that are found from inverting the annulus. It turns out that three types of boundary conditions are possible: 
\begin{itemize}

\item{Type $b$. This type contains two boundary conditions, $b_1$ and $b_2$, regardless of $N$.}  
 
\item{Type $a_1$. These split further into two subsets, $a_{1f}$ and $a_{1f}^\prime$, with $f$ an odd integer ranging from $1$ to $p_1\!-\!2$. Each of these subsets contains thus $(p_1\!-\!1)/2$ boundary conditions, amounting to $p_1\!-\!1$ boundary conditions coming from this type of boundary.} 
 
\item{Type $a_2$. Similar to type $a_1$. It splits into subsets $a_{2f}$ and $a^\prime_{2f}$ (with odd $f$ ranging from $1$ to $p_2\!-\!2$ this time). It contains $p_2\!-\!1$ boundary conditions.} 
 
\end{itemize} 
There are $p_1 + p_2$ boundary conditions in total, as many as the transverse fields.

\clearpage 

The reflection coefficients are the same for both Klein bottle projections, as there is only one physical S-NIMrep. These are
\be 
\begin{array}{c|cccccc} 
 \;&  0 & J & \Phi^l & \Phi_k & \sigma_l & \tau_l \\ \hline 
   R_{m,b_1}  & \displaystyle{ {\frac{1}{\sqrt{2p_1}}} } 
              & \displaystyle{ {\frac{-1}{\sqrt{2p_1}}}} 
              & 0  & \ldots, 
                  \underbrace{\frac{2(-1)^n}{\sqrt{2p_1}}}_{k=2n{p_2}}, 
                     \ldots &  0 & 0   \\ \hline 
   R_{m,b_2}  & \displaystyle{  {\frac{1}{\sqrt{2p_2}}} } 
              & \displaystyle{ {\frac{1}{\sqrt{2p_2}}} } & 0 & \ldots, 
                  \underbrace{\frac{2(-1)^n}{\sqrt{2p_2}}}_{k=2n{p_1}}, 
                     \ldots & 0 & 0 \\ \hline 
   R_{m,a_{1f}} &  \displaystyle{ {\frac{1}{\sqrt{2p_1}}} }
                &  \displaystyle{ {\frac{-1}{\sqrt{2p_1}}}} & 0 & \ldots, 
     \underbrace{{\frac{2 \cos{(\frac{\pi n f}{2p_1}})}{\sqrt{2p_1}}}}_{k=n                     {p_2}}, \ldots & 0 & 0   \\ \hline 
   R_{m,a^\prime_{1f}} &  \displaystyle{ {\frac{1}{\sqrt{2p_1}}} }
                       &  \displaystyle{ {\frac{-1}{\sqrt{2p_1}}} }& 0 & 
     \ldots, \underbrace{{\frac{2(-1)^n  \cos{(\frac{\pi n f}{2p_1}})}{\sqrt{2p_1}}}}_{k=n {p_2}}, \ldots & 0 & 0   \\ \hline 
   R_{m,a_{2f}} &  \displaystyle{ {\frac{1}{\sqrt{2p_2}}}} 
                &  \displaystyle{ {\frac{1}{\sqrt{2p_2}}}} & 0 & \ldots, 
     \underbrace{{\frac{2\cos{(\frac{\pi n f}{2p_2}})}{\sqrt{2p_2}}}}_{k=n {p_1}}, \ldots & 0 & 0   \\ \hline 
   R_{m,a^\prime_{2f}} &  \displaystyle{ {\frac{1}{\sqrt{2p_2}}}} 
                       &  \displaystyle{ {\frac{1}{\sqrt{2p_2}}}} & 0 & 
              \ldots, \underbrace{{\frac{2(-1)^n  \cos{(\frac{\pi n f}{2p_2}})}{\sqrt{2p_2}}}}_{k=n {p_1}}, \ldots & 0 & 0 
\end{array} \label{R} 
\ee 
As expected, the non-transverse fields have vanishing reflection coefficients. Suppressing the zero columns in (\ref{R}), we come to an orthogonal, invertible square matrix, as required by the completeness conditions. Note that for the limiting case $f=p_l$ the boundaries $a_{lf}$ and $a'_{lf}$ are both equal to $b_l$. It is however convenient to treat the $b$-boundaries separately.

The boundary conjugation matrix is trivial. All boundaries are self-conjugate. With the knowledge of the U-crosscap and reflection coefficients we can now calculate direct channel annulus and Moebius strip coefficients.

\subsubsection{D+A annulus}
For briefness we present only the diagonal annulus $A^\Omega_{{i}aa}$, which contains the most important information concerning integrality. The remaining off-diagonal annuli $A_{iab}, a\neq b$ can easily be derived too. Using (\ref{KAMcoeff}) we get 
\be
\begin{array}{c|ccccccc} 
\; &  0 & J & \Phi^i & \Phi_k & \sigma_i & \tau_i & \;\\ \hline 
A_{i,b_1,b_1} & 1 & 1 & 1 & \ldots, \underbrace{2}_{k=\dot{p_1}},\ldots & 0 & 0 & \\ 
A_{i,b_2,b_2} & 1 & 1 & 1 & \ldots, \underbrace{2}_{k=\dot{p_2}},\ldots & 0 & 0 & \\ 
 
A_{i,a_{1f},a_{1f}} = A_{i,a_{1f}^{\prime},a_{1f}^{\prime}} & 1 & 1 & 0 & \ldots, \underbrace{1}_{k\pm f= \text{\scriptsize even} \;\dot{p_1}}, \ldots, \underbrace{2}_{k= \text{\scriptsize even} \;\dot{p_1}},\ldots & 0 & 0 &\\ 
A_{i,a_{2f},a_{2f}} = A_{i,a_{2f}^{\prime},a_{2f}^{\prime}} & 1 & 1 & 0 & \ldots, \underbrace{1}_{k\pm f= \text{\scriptsize even} \;\dot{p_2}}, \ldots, \underbrace{2}_{k=  \text{\scriptsize even} \;\dot{p_2}},\ldots & 0 & 0 &\\ 
\end{array} \label{A} 
\ee 
We dropped the superscript $\Omega$ because there is only one S-NIMrep.

\subsubsection{D+A Moebius strip}
Evaluating (\ref{KAMcoeff}) for the trivial Klein bottle projection gives us the Moebius strip of the first U-NIMrep
\be
\begin{array}{c|cccccccc} 
\; & \;&  0 & J & \Phi^i & \Phi_k & \sigma_i & \tau_i & \;\\ \hline 
M_{i,b_1} & & 1 & 1 & 1 & \ldots, \underbrace{2}_{k= \text{\scriptsize odd}\;\dot{p_1}},\ldots & 0 & 0 & \\ 
M_{i,b_2} & & 1 & 1 & 1 & \ldots, \underbrace{2}_{k= \text{\scriptsize odd}\;\dot{p_2}},\ldots & 0 & 0 & \\ 
M_{i,a_{1f}} = M_{i,a_{1f}^\prime} & & 1 & 1 & 0 & \ldots, \underbrace{1}_{k \pm f = \text{\scriptsize even}\;\dot{p_1}},\ldots & 0 & 0 & \\ 
M_{i,a_{2f}} = M_{i,a_{2f}^\prime} & & 1 & 1 & 0 & \ldots, \underbrace{1}_{k \pm f = \text{\scriptsize even}\;\dot{p_2}},\ldots & 0 & 0 & 
\end{array} \label{Mt} 
\ee 
For the second U-NIMrep the Klein bottle has $K_i=-1$ on the twist fields and the Moebius is 
\be 
\begin{array}{c|cccccccc} 
\; & \;&  0 & J & \Phi^i & \Phi_k & \sigma_i & \tau_i & \;\\ \hline 
M_{i,b_1} & & -1 & -1 & 1 & \ldots, \underbrace{2}_{k= \text{\scriptsize odd}\;\dot{p_1}},\ldots & 0 & 0 & \\ 
M_{i,b_2} & & 1 & 1 & -1 & \ldots, \underbrace{-2}_{k=\text{\scriptsize odd}\;\dot{p_2}},\ldots & 0 & 0 & \\ 
M_{i,a_{1f}} = M_{i,a_{1f}^{\prime}} & & -1 & -1 & 0 & \ldots, \underbrace{1}_{k\pm f = \text{\scriptsize even} \;\dot{p_1}},\ldots & 0 & 0 & \\ 
M_{i,a_{2f}} = M_{i,a_{2f}^{\prime}} & & 1 & 1 & 0 & \ldots, \underbrace{-1}_{k\pm f = \text{\scriptsize even}\;\dot{p_2}},\ldots & 0 & 0 & 
\end{array} \label{M} 
\ee
It is straightforward to show that (\ref{A}-\ref{M}) respect the positivity and integrality conditions for the open sector and that (\ref{R}) leads to a positive integer off-diagonal annulus $A_{iab}$.

Note also that both Klein bottle projections come from the same S-NIMrep. The difference lies in the signs of the Moebius, so for the two cases some branes are identified with different signs.

\subsection{C+A invariant}
 Most quantities are similar to the D+A case, but there are some differences nevertheless. There is again one NIMrep which leads to two S-NIMreps, each of which in turn produces one U-NIMrep.

We present the results for the Klein bottle with $K^{\Phi_k}=-1$ when $k$ is an odd multiple of $p_1$ and $K^i=1$ for remaining the fields that couple diagonally on the torus. Since in the C+A case no quantity is proportional to the difference $p_1-p_2$, the results for the second Klein bottle are obtained by simply interchanging $p_1 \leftrightarrow p_2$ on the formulas below. 

\subsubsection{C+A crosscaps}
The U-crosscap coefficients are 
\be 
\begin{array}{c|cccccc} 
\; &  0 & J & \Phi^i & \Phi_k & \sigma_i & \tau_i \\ \hline 
U_m &  
 \displaystyle{ {\frac{1}{2\sqrt{p_2}}} } &  
 \displaystyle{ {\frac{1}{2\sqrt{p_2}}} } & 
 \displaystyle{ {\frac{1}{2\sqrt{p_1}}} } & 
   \ldots,
\underbrace{\frac{1}{\sqrt{p_1}}}_{k=\text{\scriptsize odd}\;\dot{p_2}},
   \ldots, 
\underbrace{\frac{1}{\sqrt{p_2}}}_{k=\text{\scriptsize even}\;\dot{p_1}},
   \ldots & 0 & 0 \\ 
\label{Uc+a} 
\end{array} 
\ee

\subsubsection{C+A boundaries}
Next we classify the possible boundaries conditions. As in the (D+A) case we get a certain number fixed boundaries along with an expanding part which depends on the odd parameter $f$. 
\begin{itemize} 
\item{Type $a, b, c, d$. There are two subtypes of $a$-type boundaries, $a$ and $a^\prime$ which are real (as in self-conjugate). The boundaries $b,c$ and $d$ are unique but complex, thus admitting conjugate boundaries $b^\prime,c^\prime$ and $d^\prime$ respectively. In total these four types generate eight boundary conditions.} 
\item{Type $E_{f}$. This type of boundary is real and splits into subtypes $E_{f}$ and $E_{f}^\prime$, with odd $f$ ranging from 1 to $p_1\!-\!2$, and contributes with a total of $p_1\!-\!1$ boundary conditions.} 
\item{Type $F_{f}$. This type is complex and thus splits into $F_{f}$ and its conjugate $F_{f}^\prime$, with off $f$ ranging from 1 to $p_2\!-\!2$, and contributes with $p_2\!-\!1$ boundary conditions.} 
\end{itemize} 
The total number of boundaries is thus $p_1+p_2+6$, as many as the transverse fields, as expected. 
 
The reflection coefficients are 
\be  
\begin{array}{c|cccccc} 
 \;&  0 & J & \Phi^j & \Phi_k & \sigma_j & \tau_j \\ \hline 
R_{m,a} &  
   {\frac{1}{\sqrt{8p_1}}} &  
   {\frac{1}{\sqrt{8p_1}}} &  
   {\frac{1}{\sqrt{8p_1}}} &  
      \ldots, \underbrace{\sst \frac{1}{\sqrt{2p_1}}}_{k=n{p_2}}, 
      \ldots &  \frac{1}{\sqrt{8}} & \frac{1}{\sqrt{8}} \\ \hline 
R_{m,a^\prime} &  
   \frac{1}{\sqrt{8p_1}} &  
   \frac{1}{\sqrt{8p_1}} &  
   \frac{1}{\sqrt{8p_1}} & 
      \ldots, \underbrace{\sst \frac{1}{\sqrt{2p_1}}}_{k=n{p_2}}, 
      \ldots &  \frac{-1}{\sqrt{8}} & \frac{-1}{\sqrt{8}} \\ \hline 
R_{m,b} &  
   \frac{1}{\sqrt{8p_1}} &  
   \frac{1}{\sqrt{8p_1}} &  
   \frac{-1}{\sqrt{8p_1}}  &  
      \ldots, \underbrace{\sst \frac{(-1)^n}{\sqrt{2p_1}}}_{k=n{p_2}}, 
      \ldots &  \frac{\I \sigma_{1j}}{\sqrt{8}} 
             &  \frac{\I \sigma_{1j}}{\sqrt{8}}  \\ \hline 
R_{m,c} &  
   \frac{1}{\sqrt{8p_2}} &  
   \frac{-1}{\sqrt{8p_2}}&  
   \frac{\I \epsilon\sigma_{1j}}{\sqrt{8p_2}} &  
      \ldots, \underbrace{\sst \frac{(-1)^n}{\sqrt{2p_2}}}_{k=2n{p_1}}, 
      \ldots, \underbrace{\sst \frac{\I}{\sqrt{2p_2}}}_{k=(2n-1){p_1}}, 
      \ldots & \frac{ e^{\I \pi\sigma_{1j}/4}}{\sqrt{8}} & 
               \frac{-e^{\I \pi\sigma_{1j}/4}}{\sqrt{8}}  \\ \hline 
R_{m,d} &  
   \frac{1}{\sqrt{8p_2}} &  
   \frac{-1}{\sqrt{8p_2}}&  
   \frac{\I \epsilon\sigma_{1j}}{\sqrt{8p_2}} & 
      \ldots, \underbrace{\sst \frac{(-1)^n}{\sqrt{2p_2}}}_{k=2n{p_1}}, 
      \ldots, \underbrace{\sst \frac{\I }{\sqrt{2p_2}}}_{k=(2n-1){p_1}}, 
      \ldots & \frac{-e^{\I \pi\sigma_{1j}/4}}{\sqrt{8}} & 
               \frac{ e^{\I \pi\sigma_{1j}/4}}{\sqrt{8}} \\ \hline 
R_{m,E_{f}} &  
   \frac{1}{\sqrt{2p_1}} &  
   \frac{1}{\sqrt{2p_1}} &  
   \frac{1}{\sqrt{2p_1}} & ., 
      \underbrace{\sst \frac{ 2(-1)^n \cos{( \frac{\pi 2nf}{2p_1} )} } 
         {\sqrt{2p_1}} }_{k=2n{p_2}}, ., 
      \underbrace{\sst  \frac{  2 \delta_{\!f} (-1)^n \sin{(\frac{\pi 
         (2n-1)f}{2p_1})}} {\sqrt{2p_1}  }}_{k=(2n-1){p_2}},.& 0 & 0  
      \\ \hline 
R_{m,E_{f}^\prime} &  
   \frac{1}{\sqrt{2p_1}} &  
   \frac{1}{\sqrt{2p_1}} &  
   \frac{-1}{\sqrt{2p_1}}& ., \underbrace{\sst  \frac{ 2(-1)^n 
         \cos{( \frac{\pi 2nf}{2p_1} )} } {\sqrt{2p_1}} }_{k=2n{p_2}}, ., 
         \underbrace{\sst  \frac{  -2 \delta_{\!f} (-1)^n  
         \sin{(\frac{\pi (2n-1)f}{2p_1})}} 
        {\sqrt{2p_1}  }}_{k=(2n-1){p_2}},.& 0 & 0 \\ \hline 
R_{m,F_{f}} &  
   \frac{1}{\sqrt{2p_2}} &  
   \frac{-1}{\sqrt{2p_2}} &  
   \frac{\I \sigma_{1j}}{\sqrt{2p_2}} & ., 
      \underbrace{\sst  \frac{ 2\cos{( \frac{\pi 2nf}{2p_2} )} } 
      {\sqrt{2p_2}} }_{k=2n{p_1}},., 
      \underbrace{\sst  \frac{  -2\I \delta_{\!f} (-1)^n  
      \sin{(\frac{\pi (2n-1)f}{2p_2})}} {\sqrt{2p_2}  }}_{k=(2n-1){p_1}},.
   & 0 & 0 \\  
\end{array} \label{Rc+a}
\ee \be 
\epsilon=(-1)^{(N+1)/2},\;\;\; \delta_f=(-1)^{(p_1+f)/2},\;\;\; \sigma_{1j}=2\delta_{1j}-1. \nn   
\ee 
Again, both the U-crosscap and reflection coefficients vanish on the non-transverse fields.

The reflection coefficients for boundary conditions $b^\prime,c^\prime,d^\prime$ and $F_f^\prime$ are simply the complex conjugates of their unprimed counterparts. The boundary conjugation matrix is off-diagonal for $b,c,d$ and $F_f$, with the primed boundaries as conjugates. 

Since we have two S-NIMreps we must use the label $\Omega$ in the annulus coefficients. Define $\Omega_1$ and $\Omega_2$, with the latter standing for the data with $p_1$ and $p_2$ interchanged.
The diagonal annulus and Moebius strip that go with the U-crosscap and reflection coefficients of (\ref{Uc+a}-\ref{Rc+a}) are

\subsubsection{C+A annulus}
\be
\begin{array}{c|cccccccc} 
\; &  O & J & \Phi^1 & \Phi^2 & \Phi_k & \sigma_i & \tau_i & \; \\ \hline 
   A^{\Omega_x}_{i,a,a} = 
   A^{\Omega_x}_{i,a^\prime,a^\prime} & 1 & 0 & 0 & 0 &
      \ldots, \underbrace{1}_{k=\text{\scriptsize even}\;\dot{p_1}},
      \ldots & 0 & 0 \\ 
   A^{\Omega_x}_{i,b,b} = 
   A^{\omega_x}_{i,b^\prime,b^\prime}& 0 & 1 & 0 & 0 & 
      \ldots, \underbrace{1}_{k=\text{\scriptsize even}\;\dot{p_1}},
      \ldots & 0 & 0 \\ 
   A^{\Omega_x}_{i,c,c} = 
   A^{\Omega_x}_{i,d,d}& 0 & 0 & 0 & 1 & 
      \ldots, \underbrace{1}_{k=\text{\scriptsize odd}\;\dot{p_2}},
      \ldots  & 0 & 0 \\ 
   A^{\Omega_x}_{i,c^\prime,c^\prime} = 
   A^{\Omega_x}_{i,d^\prime,d^\prime} & 0 & 0 & 1 & 0 &
      \ldots, \underbrace{1}_{k=\text{\scriptsize odd}\;\dot{p_2}},
      \ldots & 0 & 0\\ 
   A^{\Omega_x}_{i,E_{f},E_{f}} = 
   A^{\Omega_x}_{i,E_{f}^\prime,E_{f}^\prime}& 1 & 1 & 0 & 0 &  ., 
      \underbrace{1}_{k\pm f=\text{\scriptsize odd}\;\dot{p_1}},., 
      \underbrace{2}_{k=\text{\scriptsize even}\;\dot{p_1}},. & 0 & 0 \\ 
   A^{\Omega_x}_{i,F_{f},F_{f}} =
   A^{\Omega_x}_{i,F_{f}^\prime,F_{f}^\prime}& 0 & 0 & 1 & 1 &  ., 
      \underbrace{1}_{k\pm f=\text{\scriptsize even}\;\dot{p_2}},., 
      \underbrace{2}_{k=\text{\scriptsize odd}\;\dot{p_2}},. & 0 & 0 \\ 
\label{Ac+a} 
\end{array} 
\ee 

\subsubsection{C+A Moebius strip}
\be 
\begin{array}{c|cccccccc} 
\; &  O & J & \Phi^1 & \Phi^2 & \Phi_k & \sigma _i & \tau_i &  \;\\ \hline 
   M_{i,a} = 
   M_{i,a^{\prime}} & 1 & 0 & 0 & 0 & 
      \ldots, \underbrace{(-1)^{k/2}}_{k=\text{\scriptsize even}\;\dot{p_1}},
      \ldots & 0 & 0 \\ 
   M_{i,b} = 
   M_{i,b^{\prime}}& 0 & 1 & 0 & 0 & 
      \ldots, \underbrace{-(-1)^{k/2}}_{k=\text{\scriptsize even}\;\dot{p_1}},
      \ldots & 0 & 0 \\ 
   M_{i,c} = 
   M_{i,d} & 0 & 0 & 0 & 1 & 
      \ldots, \underbrace{1}_{k=\text{\scriptsize odd}\;\dot{p_2}},
      \ldots & 0 & 0 \\ 
   M_{i,c^{\prime}} = 
   M_{i,d^{\prime}} & 0 & 0 & 1 & 0 & 
      \ldots, \underbrace{1}_{k=\text{\scriptsize odd}\;\dot{p_2}},
      \ldots & 0 & 0 \\ 
   M_{i,E_{f}} & 1 & 1 & 0 & 0 & 
      \ldots, \underbrace{(-1)^{k/2}}_{k \pm 
         f=\text{\scriptsize odd}\;\dot{p_1}},\ldots & 0 & 0 \\ 
   M_{i,E_{f}^{\prime}} & 1 & 1 & 0 & 0 & 
      \ldots, \underbrace{-(-1)^{k/2}}_{k \pm 
         f=\text{\scriptsize odd}\;\dot{p_1}},\ldots & 0 & 0 \\ 
   M_{i,F_{f}} & 0 & 0 & -\epsilon & \epsilon & 
      \ldots, \underbrace{1}_{k\pm 
         f=\text{\scriptsize even}\;\dot{p_2}},\ldots & 0 & 0 \\ 
   M_{i,F_{f}^{\prime}} & 0 & 0 & \epsilon & -\epsilon & 
      \ldots, \underbrace{1}_{k\pm 
         f=\text{\scriptsize even}\;\dot{p_2}},\ldots & 0 & 0 \\ 
\label{Mc+a} 
\end{array} 
\ee 
Positivity and integrality of the diagonal annulus and Moebius strip is explicit. Again it can be shown that (\ref{Rc+a}) leads to positive integer off-diagonal annulus.

Since in $B_{ma} \Omega^m_x C^m B_{mb}$ only $\Omega^m$ depends on the orientifold projection, we can determine the ratio $\Omega_1^m/\Omega_2^m$ by expanding and analyzing the annulus coefficients for both U-NIMreps. We find that the orientifold ratio is $\pm \I$ for $m$ a twist field and $\pm 1$ on the other fields. The $\Omega^m$ turn out to be phases, which is not a problem provided $\Omega^m=(\Omega^{m^c})^*$.

We see that the exceptional invariants of the orbifold theory are conformal field theories with a very rich structure that displays many theoretical properties. A possibility for future work would be to generalize the results of this section to arbitrary $N=\prod_n p_n$. It would also be nice to determine more of the quantities involved, like for instance $\Omega^m$. As mentioned before, this requires knowledge of fusing matrices. These matrices are also needed to examine consistency at higher loops. This is the path we now turn to.

\chapter{Consistency of the perturbative expansion}
\label{sewchap}
In the last chapters we presented positivity and integrality requirements that can be used to build a consistent String Theory. When met, these one-loop requirements ensure a sensible spectrum. One might then wonder how can perturbative consistency be checked to higher orders in string perturbation theory, particularly since as the Euler number grows, direct evaluation of the possible diagrams gets nigh impossible. The solution is to use some sort of induction mechanism, from which consistency at order $N$ would imply consistency at order $N+1$.

Such an induction mechanism exists and it is called ``sewing constraints''. It works as follows. When one sews together two world-sheet surfaces that individually lead to consistent transition amplitudes, in the sense that the latter are unambiguous and factorize correctly, the resulting surface will have consistent transition amplitudes as well, if the sewing constraints are satisfied. These constraints have been developed step by step, starting with the closed string constraints of \cite{sonoda} \cite{ms}. Later on, in \cite{lewellen} constraints involving oriented open strings were formulated. Finally, in \cite{sagnotticross} constraints for unoriented strings appeared. The sewing constraints are definitely necessary for consistency, although it is not yet clear they are sufficient, because a complete analysis for surfaces with crosscaps has not yet been done.

The explicit form of the various sewing constraints equations all have one point in common: they need specific data from the conformal field theory at hand. This was not the case with positivity and integrality of the closed and open sectors; there the results were general and needed only very basic data like conformal weights, the $S$-matrix and the modular invariant. For checking the sewing constraints, in addition to the basic data we will need also information from the theory's correlators. To prove perturbative consistency of a String Theory one would in principle have to find or postulate this information and verify that it indeed is a solution to the sewing constraints.

The sewing constraints actually relate some of the model-specific quantities, so not all of them need to be independent. Following this line of reasoning, in \cite{frs} it is shown that the knowledge of chiral data and a special symmetric Frobenius algebra is sufficient to generate correlators of a full conformal field theory and to prove that they will automatically satisfy the sewing constraints on oriented surfaces. In particular, knowledge of only one conformally invariant boundary condition is enough to build the remaining ones and to derive the form of the torus and annulus partition function. The method of \cite{frs} provides a lot of data, but it uses properties of the case-specific fusing and braiding matrices, which would have to be computed for each case, should one be interested in precise figures of the chiral data. The fusing and braiding matrices are therefore a very crucial piece of data upon which much of the conformal field theory revolves. It is therefore the aim of the final chapters of this thesis to go as far as possible into the explicit evaluation of these quantities for $c=1$ theories.

Before doing this, we briefly review sewing constraints explicitly to show where they come from. For this purpose we review also boundary conformal field theory.

\section{Boundary conformal field theory}
In the same way closed strings can be mapped to a complex plane, open strings can be mapped into a complex half-plane with $\re(z)\geq 0$. Time flows again in circles from $z=0$ to infinite and the real line $\re(z)<0$ corresponds to the the open string end point at $\sigma=\pi$ and $\re(z)<0$ to the end point at $\sigma=0$. In chapter \ref{cftchap} the closed conformal field theory was studied, so now it is time to have a look at the open string conformal field theory, which is conformal field theory on the half-plane (instead of the full plane, which is the sphere).

In chapter \ref{opendes}, the boundaries of the annulus were analyzed in the transverse channel, where they were defined in terms of Ishibashi states, which are again closed string states. Because of closed-open string world-sheet duality, we would then like to know what happens at a true boundary of a surface where a conformal field theory lives. If the two ends of an open string have different boundary conditions, say $a$ and $b$, there must be some disturbance along the real boundary line $\re(x)=0$ such that the boundary condition changes from $a$ to $b$. That disturbance is called a {\em boundary field}. This can be seen from the picture
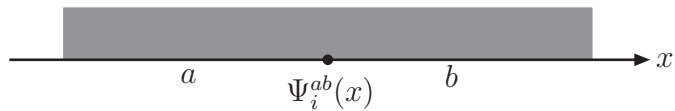
\begin{figure}[h]
\SetScale{1}
\begin{center}
\begin{picture}(400,50)(0,50)
\CBox(100,60)(300,80){White}{Gray}
\SetWidth{1}
\LongArrow(80,60)(320,60)
\SetWidth{.5}
\Vertex(200,60){2}
\Text(145,54)[l]{$a$}
\Text(245,54)[l]{$b$}
\Text(185,48)[l]{$\Psi_{i}^{ab}(x)$}
\Text(325,60)[l]{$x$}
\end{picture}
\end{center}
\label{bdfield}
\caption{Boundary field}
\end{figure}
\noindent
The boundary field $\Psi_i^{ab}(x)$ is then an insertion at point $x$ in the real line mediating the change from boundary conditions $a$ to $b$, which has a chiral label attached. An heuristic way to see why the chiral label appears is to notice that if we sew two half-planes through the real line, we get a full complex plane, where fields have the two holomorphic and anti-holomorphic labels. It is natural then for the boundary fields to have only one chiral label.

The number of boundary fields $\Psi^{ab}_i(x)$ depends on the torus modular invariant. The invariant determines how many Ishibashi states there are, which is equal to the number of independent conformally invariant boundary conditions. Then the number of irreducible representations of the extended chiral algebra gives the number of chiral labels. Acting on the boundary field with the creation operators of the chiral algebra generates boundary descendant fields.

\subsection{Boundary operator products}
Naturally, one can have more than one boundary field present at the real line of the complex half-plane. This corresponds to an array of open strings scattering off a disk diagram. Then two questions arise: what happens when two boundary fields come close to each other and what happens when a bulk field approaches a boundary? In these cases, one can define a short-distance operator product expansion, just like he does for bulk fields. These expansions take the form
\be
   \Psi_{i}^{ab}(x) \Psi_{j}^{bc}(y) &\sim& \sum_k (x-y)^{h_k-h_i-h_j} \; C_{ijk}^{abc} \; \Psi_{k}^{ac}(y)\; + 
       \; \cdots \;\;\; \text{as $x\rightarrow y$}, \nn \\
   \phi_{i\bar{i}}(z,\bar{z}) &\sim& \sum_k (2\,\im(z))^{h_k-h_i-h_{\bar{i}}} \; B^{a}_{(i\bar{i})k} \; \Psi_k^{aa}(\re(z)) \; + 
       \; \cdots\;\;\; \text{as $\im(z)\rightarrow 0$}.
\ee
The dots represent, as usual, descendants contributions. So, when two boundary fields come close, the change of boundary conditions can be expressed as if a linear combination of boundary fields was present, with strength given by the {\em boundary-boundary operator product coefficients}. Likewise, when a bulk field approaches a boundary of type $a$, it dissipates into a sum of boundary fields, with strength given by the {\em bulk-boundary operator product coefficients}. This last coefficient can be related to the boundary coefficients $B_{ma}$ (see for instance \cite{runkelad}).

\subsection{Conformal blocks}
We have seen in chapter \ref{cftchap} that correlation functions of bulk conformal fields $\phi_{i\bar{i}}\zz$ on the sphere are heavily constrained by conformal invariance. The four-point function is the first correlator whose form isn't totally constrained. This correlator can be written in a canonical form, like for instance (\ref{1234corr}), but can also be expanded as a bilinear sum of functions called `conformal blocks' or `chiral blocks' \cite{bpz}. This can be done in three ways
\be
&& \langle 0| \phi_{i\bar{i}}(z_1,\bar{z}_1) \phi_{j\bar{j}}(z_2,\bar{z}_2) 
              \phi_{k\bar{k}}(z_3,\bar{z}_3) \phi_{l\bar{l}}(z_4,\bar{z}_4) 
   |0 \rangle  \;\;\; \text{for}\; |z_1|\!>\!|z_2|\!>\!|z_3|\!>\!|z_4| 
              \mbox{\hspace{1cm}} \nn \\ && \mbox{\hspace{1cm}}= 
   \sum_{p\bar{p}}^{{}^{{}^{}}} 
      C_{i\bar{i},j\bar{i}}^{\ph{i\bar{i},j\bar{j}}p\bar{p}} \, 
      C_{k\bar{k},l\bar{l},p\bar{p}} \,
      {\cal F}_p^{ijkl}(z_1,z_2,z_3,z_4)\,
      \overline{\cal F}_{\bar{p}}^{\bar{i} \bar{j} 
          \bar{k} \bar{l}}(\bar{z}_1,\bar{z}_2,\bar{z}_3,\bar{z}_4), 
          \;\;\; \text{$S-$channel} \nn \\ &&\mbox{\hspace{1cm}}=
   \sum_{p\bar{p}} C_{i\bar{i},k\bar{k}}^{\ph{i\bar{i},k\bar{k}}p\bar{p}} \, 
      C_{j\bar{j},l\bar{l},p\bar{p}} \,{\cal F}_p^{ikjl}(z_1,z_3,z_2,z_4)\,
      \overline{\cal F}_{\bar{p}}^{\bar{i} \bar{k} 
          \bar{j} \bar{l}}(\bar{z}_1,\bar{z}_3,\bar{z}_2,\bar{z}_4),
          \;\;\; \text{$U-$channel} \label{4pt} \\ && \mbox{\hspace{1cm}}= 
   \sum_{p\bar{p}} C_{i\bar{i},l\bar{l}}^{\ph{i\bar{i},l\bar{l}}p\bar{p}} \, 
      C_{k\bar{k},j\bar{j},p\bar{p}} \,{\cal F}_p^{ljki}(z_1,z_4,z_3,z_2)\,
      \overline{\cal F}_{\bar{p}}^{\bar{l} \bar{j} 
          \bar{k} \bar{i}}(\bar{z}_1,\bar{z}_4,\bar{z}_3,\bar{z}_2).
          \;\;\; \text{$T-$channel} \nn
\ee
The functions ${\cal F}_{{p}}$ and $\overline{\cal F}_{\bar{p}}$ are the chiral (or holomorphic) and anti-chiral (or anti-holomorphic) conformal blocks for correlator $\langle\phi_i\phi_j\phi_k\phi_l\rangle$. The expansions (\ref{4pt}) have the following meaning. One can take operator products of the fields, effectively reducing the four-point function to an infinite combination of two-point functions. There are three ways of contracting the four fields to form a two-point function, respectively $(ij)(kl)$, $(ik)(jl)$ and $(il)(kj)$, which correspond to the three expressions above and are called the $S$-, $U$- and $T$-channel expansions.

Conformal invariance allows us to fix three of the four insertion points. The usual choice is $(z_1,z_2,z_3,z_4)\rightarrow(\infty,1,z,0)$ and will be henceforth referred to as `canonical insertion points'. In terms of this choice, equation (\ref{4pt}) becomes\footnote{In other conventions, like for instance \cite{fuchsbook}, the canonical insertions are different, which leads to some swapping of the chiral labels.}
\be
&& \lim_{\eta,\bar{\eta}\rightarrow \infty} 
   \eta^{-2h_i}\bar{\eta}^{-2{h}_{\bar{i}}}
      \langle 0| \phi_{i\bar{i}}(\eta,\bar{\eta})\phi_{j\bar{j}}(1,\bar{1})
                 \phi_{k\bar{k}}(z,\bar{z})\phi_{l\bar{l}}(0,\bar{0}) 
      |0\rangle \mbox{\hspace{2cm}} \nn \\ &&\mbox{\hspace{1cm}}= 
   \sum_{p\bar{p}}^{{}^{{}^{}}} 
      C_{i\bar{i},j\bar{j}}^{\ph{i\bar{i},j\bar{j}}p\bar{p}} \, 
      C_{k\bar{k},l\bar{l},p\bar{p}} \, {\cal F}_p^{ijkl}(z)\,
         \overline{{\cal F}}_{\bar{p}}^{\bar{i}\bar{j}\bar{k}\bar{l}}(\bar{z})
         \nn \\ &&\mbox{\hspace{1cm}}= 
   \sum_{p\bar{p}} C_{i\bar{i},l\bar{l}}^{\ph{i\bar{i},l\bar{l}}p\bar{p}} \, 
      C_{k\bar{k},j\bar{j},p\bar{p}} \, {\cal F}_p^{ijkl}(1-{z}) \, 
         \overline{\cal F}_{\bar{p}}^{\bar{i}\bar{j}\bar{k}\bar{l}}
         (1-\bar{z}), \label{can4pt} \\ && \mbox{\hspace{1cm}}= 
   \sum_{p\bar{p}}  C_{l\bar{l},j\bar{j}}^{\ph{l\bar{l},j\bar{j}}p\bar{p}} \, 
      C_{k\bar{k},i\bar{i},p\bar{p}} \, {\cal F}_p^{ljki}(1/z)\, 
         \overline{{\cal F}}_{\bar{p}}^{\bar{l}\bar{j}\bar{k}\bar{i}}
         (1/\bar{z}) \, z^{-2h_k}\bar{z}^{-2h_{\bar{k}}}. \nn
\ee
The factor $\eta^{-2h_i}\bar{\eta}^{-2h_{\bar{i}}}$ is put in to absorb the divergence coming from pulling $z_1$ to infinity. 

The conformal block functions ${\cal F}_p^{ijkl}(z)$ are in general multi-valued functions on the $z$-plane with branch points at $z=\infty,1,0$. To get an unambiguous definition for them, we take a branch cut running from $-\infty$ to 1 along the real line. While correlators are only defined when there is radial ordering of the insertion points, the conformal blocks are defined throughout the whole $z$-plane (except on the branch cut) by analytic continuation. Also, they obey certain differential equations and conformal Ward identities that can be used to determine their explicit form \cite{bpz}.

We can assign a conformal block a picture:
\begin{figure}[ht]
\SetScale{1}
\begin{center}
\begin{picture}(500,40)(-20,10)
\Text(0,20)[l]{$\langle \phi_{i\bar{i}} \phi_{j\bar{j}} 
\phi_{k\bar{k}} \phi_{l\bar{l}} \rangle \!=\! 
   C_{i\bar{i},j\bar{j}}^{\ph{i\bar{i},j\bar{j}}p\bar{p}} \; 
   C_{k\bar{k},l\bar{l},p\bar{p}}\; 
      \left| {\cal F}_p^{ijkl} \right|^2 \!=\! 
   C_{i\bar{i},j\bar{j}}^{\ph{i\bar{i},j\bar{j}}p\bar{p}} \; 
   C_{k\bar{k},l\bar{l},p\bar{p}} \; \cdot$}
\ArrowLine(290,20)(310,20)
\ArrowLine(310,20)(330,20)
\ArrowLine(350,20)(330,20)
\ArrowLine(310,50)(310,20)
\ArrowLine(330,50)(330,20)
\Text(290,13)[l]{$i$}
\Text(318,13)[l]{$p$}
\Text(348,13)[l]{$l$}
\Text(308,56)[l]{$j$}
\Text(328,56)[l]{$k$}
\Text(355,20)[l]{$\cdot$}
\ArrowLine(360,20)(380,20)
\ArrowLine(380,20)(400,20)
\ArrowLine(420,20)(400,20)
\ArrowLine(380,50)(380,20)
\ArrowLine(400,50)(400,20)
\Text(360,13)[l]{$\bar{i}$}
\Text(388,13)[l]{$\bar{p}$}
\Text(418,13)[l]{$\bar{l}$}
\Text(398,56)[l]{$\bar{k}$}
\Text(378,56)[l]{$\bar{j}$}
\end{picture}
\end{center}
\caption{Graphical representation of a conformal block}
\label{cblocks}
\end{figure}
\noindent
This is not the most natural convention for the arrows, like the one used in \cite{ms} \cite{frs}, but it is the most convenient one for our purposes since the known expressions for correlators of free boson and orbifold theories have all ingoing lines. From the channel expansions, we see that the operator product coefficients and the conformal blocks can be normalized together. We take as normalization
\be
{\cal F}_p^{ijkl}(z)\sim 1 \times  z^{h_p-h_k-h_l}(1+\cdots)
\label{cblocknorm}
\ee
As we will see in next section, in the presence of extended chiral algebras this definition deserves some comments.

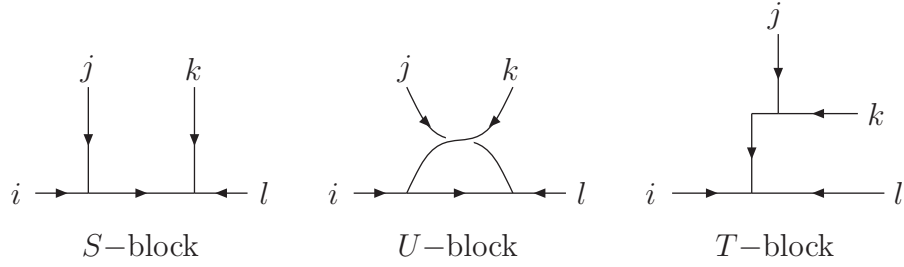
\begin{figure}[ht]
\SetScale{1}
\begin{center}
\begin{picture}(400,90)(0,0)
\ArrowLine(40,20)(60,20)
\ArrowLine(60,20)(100,20)
\ArrowLine(120,20)(100,20)
\ArrowLine(60,60)(60,20)
\ArrowLine(100,60)(100,20)
\Text(80,0)[c]{$S-$block}
\Text(35,20)[r]{$i$}
\Text(60,67)[c]{$j$}
\Text(100,67)[c]{$k$}
\Text(125,20)[l]{$l$}
\ArrowLine(160,20)(180,20)
\ArrowLine(180,20)(220,20)
\ArrowLine(240,20)(220,20)
\Curve{(180,60)(185,50)(215,30)(220,20)}
\CBox(195,35)(205,45){White}{White}
\Curve{(180,20)(185,30)(215,50)(220,60)}
\ArrowLine(187,47)(188,46)
\ArrowLine(213,47)(212,46)
\Text(200,0)[c]{$U-$block}
\Text(155,20)[r]{$i$}
\Text(180,67)[c]{$j$}
\Text(220,67)[c]{$k$}
\Text(245,20)[l]{$l$}
\ArrowLine(280,20)(310,20)
\ArrowLine(360,20)(310,20)
\ArrowLine(310,50)(310,20)
\ArrowLine(320,80)(320,50)
\ArrowLine(350,50)(320,50)
\Line(320,50)(310,50)
\Text(320,0)[c]{$T-$block}
\Text(275,20)[r]{$i$}
\Text(320,87)[c]{$j$}
\Text(355,50)[l]{$k$}
\Text(365,20)[l]{$l$}
\end{picture}
\end{center}
\caption{Conformal blocks in three channels}
\label{cblock2}
\end{figure}

The three channels for expanding the bulk four-point function lead to three different sets of conformal blocks, with pictures as in fig. \ref{cblock2}.
Since the three expansions represent the same thing, the three different basis of conformal blocks must be related by some linear, invertible relation. Therefore there must exist two sets of matrices that interpolate between these pictures. These are the duality matrices, or {\em braiding} and {\em fusing} matrices. We will define these matrices in the next chapter. They will enable us to execute moves and changes in the conformal blocks pictures, so that one can use graphical calculus with pictures to derive identities instead of having do deal with the explicit infinite sums.

The requirement that the three channels produce equal four-point functions will force a relation between operator product coefficients and fusing matrices. This will be our first sewing constraint.

\subsection{Boundary correlation functions}
In the same way bulk correlators can be decomposed into {bilinear} combinations of holomorphic and anti-holomorphic conformal blocks, it turns out that boundary correlators can be decomposed into {\em linear} combinations of the same blocks.

Conformal invariance fixes the form of the boundary one-, two- and three-point functions. By suitably normalizing the bulk and boundary fields, these three amplitudes on the disk with boundary condition $a$ are\footnote{When a disk with boundary condition $a$ is mapped into the complex half-plane, the boundary condition at $x=-\infty$ is the same as the one at $x=+\infty$. Therefore the cyclicity in the boundary indices on (\ref{boundary123}).}
\be
\langle \Psi_{i}^{aa}(x) \rangle_a &=& \delta_{i0}\; \langle \Psi_0^{aa} \rangle_a = \delta_{i0} \; \alpha^a, \nn \\
\langle \Psi_i^{ab}(x) \Psi_j^{ba}(y) \rangle_a &=& \frac{\delta_{ij^c}\; C^{aba}_{ij^c 0}\; \alpha^a}{(x-y)^{-2h_i}}, \label{boundary123}\\
\langle \Psi_i^{ab}(x) \Psi_j^{bc}(y) \Psi_k^{ca}(t) \rangle_a &=&
\frac{\delta_{N_{ijk^c},1} \; C^{abc}_{ijk^c} \;C^{aca}_{k^c k 0} \;\alpha^a}
{(x-y)^{-h_k+h_i+h_j} (y-t)^{-h_i+h_j+h_k} (x-t)^{-h_j+h_i+h_k}}. 
 \nn
\ee
It is possible to write the sewing constraints in such a way that the normalization factor $\alpha^a$ drops out from all expressions. The formulas of next section are written in this form.

\begin{figure}[ht]
\SetScale{1}
\begin{center}
\begin{picture}(200,40)(55,25)
\Text(110,20)[r]{$\phi_{i\bar{i}}\zz$}
\SetWidth{.5}
\LongArrow(120,20)(162,20)
\SetWidth{.5}
\CBox(185,10)(235,20){White}{Gray}
\Line(180,20)(240,20)
\ArrowLine(200,40)(200,20)
\ArrowLine(210,60)(210,40)
\ArrowLine(230,40)(210,40)
\Line(210,40)(200,40)
\Text(185,25)[c]{$a$}
\Text(235,25)[c]{$a$}
\Text(235,40)[l]{$\bar{i}$}
\Text(210,70)[c]{$i$}
\end{picture}
\end{center}
\caption{Bulk field decomposition}
\label{bulkblo}
\end{figure}
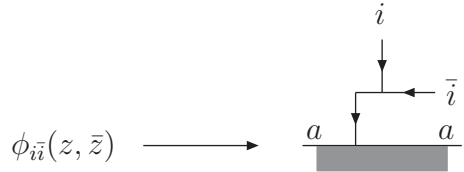

 A general correlation function on a surface with boundaries will contain various insertions of boundary fields and bulk fields. The usual way to deal with the bulk fields is to use a bulk-boundary operator product expansion to rewrite them as combinations of boundary fields and then evaluate the resulting correlator, which will consist solely of boundary fields and must therefore be a conformal block. This is sometimes referred to as the `doubling trick', and corresponds pictorially to fig. \ref{bulkblo} \cite{zuber}.

A general correlator on a surface with one boundary, having $M$ boundary field insertions and $N$ bulk field insertions would then be
\be
\langle 
   \Psi_{i_1}^{a_1 a_2}(x_1) \cdots \Psi_{i_M}^{a_M a_1}(x_M)  \; 
   \phi_{j_1\bar{j}_1}(z_1,\bar{z}_1) \cdots  
   \phi_{j_N\bar{j}_N}(z_N,\bar{z}_N) \rangle,
\ee
which after the doubling trick becomes a chiral $(M\!+\!2N)$-point conformal block. For a surface with two or more boundaries there will be traces over boundary field insertions for each boundary.

\section{Sewing constraints}
The sewing constraints are a set of equations derived from the various ways one has to write correlators in the bulk, boundary and in the presence of crosscaps, as sums of conformal blocks. We discuss them now for a maximally extended chiral algebra and arbitrary automorphism invariant. The requirement of maximallity constrains the fusion rules to $N_{ij}^{\ph{ij}k} \leq 1$ and therefore the degeneracy labels $\alpha,\beta,\gamma,\delta$ of \cite{ms} drop out. For this case, the model-specific pieces of data needed are
\bi
\item{$C_{i\bar{i},j\bar{j}}^{{\ph{i\bar{i},j\bar{j}}}k\bar{k}}$: bulk-bulk operator product coefficients;}

\item{$C_{ijk}^{abc}$: boundary-boundary operator product coefficients;}

\item{$B^a_{(i\bar{i})k}$: bulk-boundary operator product coefficients;}

\item{$\epsilon_{i\bar{i}^c}$: a sign, determining the symmetry of a conformal field as it approaches a crosscap;}

\item{$F_{pq}[ijkl]$: fusing matrix. A square matrix relating different basis of conformal blocks;}

\item{$\xi_{ij}^k$: a sign coming from interchanges of vector spaces in conformal blocks.}
\ei
The last two are purely chiral data. They only depend on the holomorphic chiral algebra at hand and not on how it couples to its anti-holomorphic partner. The sign $\epsilon_{i\bar{i}^c}$ should be related to the orientifold projector $\Omega^i$, although the relation is not yet clear. If we multiply $\xi_{ij}^k$ with a factor $e^{\pm \I \pi(h_k-h_i-h_j)}$ we get a quantity that is referred to by \cite{frs} as $R_{\pm}^{(ij)k}$. Now we can define the braiding matrix as $B^\pm=R^\pm F R^\mp$. (This relation can also be derived from pictures.)

\subsection{Bulk constraints}
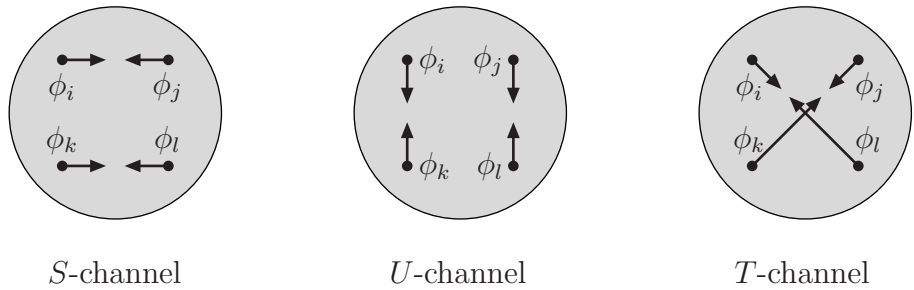
\begin{figure}[ht]
\SetScale{1}
\begin{center}
\begin{picture}(400,100)(0,10)
\GCirc(60,60){40}{0.85}
\Vertex(40,40){2}
\Vertex(80,40){2}
\Vertex(40,80){2}
\Vertex(80,80){2}
\Text(40,70)[c]{$\phi_{i}$}
\Text(80,70)[c]{$\phi_{j}$}
\Text(40,50)[c]{$\phi_{k}$}
\Text(80,50)[c]{$\phi_{l}$}
\Text(60,0)[c]{$S$-channel}
\SetWidth{1}
\LongArrow(40,40)(55,40)
\LongArrow(80,40)(65,40)
\LongArrow(40,80)(55,80)
\LongArrow(80,80)(65,80)
\SetWidth{.5}
\GCirc(190,60){40}{0.85}
\Vertex(170,40){2}
\Vertex(210,40){2}
\Vertex(170,80){2}
\Vertex(210,80){2}
\Text(175,80)[l]{$\phi_{i}$}
\Text(207,80)[r]{$\phi_{j}$}
\Text(175,40)[l]{$\phi_{k}$}
\Text(207,40)[r]{$\phi_{l}$}
\Text(190,0)[c]{$U$-channel}
\SetWidth{1}
\LongArrow(170,40)(170,55)
\LongArrow(210,40)(210,55)
\LongArrow(170,80)(170,65)
\LongArrow(210,80)(210,65)
\SetWidth{.5}
\GCirc(320,60){40}{0.85}
\Vertex(300,40){2}
\Vertex(340,40){2}
\Vertex(300,80){2}
\Vertex(340,80){2}
\Text(300,70)[c]{$\phi_{i}$}
\Text(340,70)[l]{$\phi_{j}$}
\Text(300,50)[c]{$\phi_{k}$}
\Text(340,50)[l]{$\phi_{l}$}
\Text(320,0)[c]{$T$-channel}
\SetWidth{1}
\LongArrow(300,40)(325,65)
\LongArrow(340,40)(315,65)
\LongArrow(300,80)(310,70)
\LongArrow(340,80)(330,70)
\SetWidth{.5}
\end{picture}
\end{center}
\caption{Bulk sewing constraint}
\label{sewbulk}
\end{figure}

The sewing constraints involving bulk fields only were originally written by \cite{sonoda} \cite{bpz} as a duality equation for four-point functions on the sphere. In \cite{ms} they were reinterpreted as consistency constraints linking the various ways of sewing together three-point conformal blocks. In addition, the authors of \cite{ms} derived a constraint for the one-point function on the torus. The constraints for the four-point function on the sphere and the one-point function on the torus are schematically depicted in fig. \ref{sewbulk}.
As stated before, duality of the four-point function on the sphere is just the statement that the three channels should yield the same result. Using the graphical representation for conformal blocks we can derive
\be
C_{i\bar{i},k\bar{k}}^{\ph{i\bar{i},k\bar{k}}q\bar{q}}
C_{j\bar{j},l\bar{l},q\bar{q}} 
&=& \sum_{p\bar{p}} C_{i\bar{i},j\bar{j}}^{\ph{i\bar{i},j\bar{j}}p\bar{p}}
C_{k\bar{k},l\bar{l},p\bar{p}} \; \xi_{lk}^{p^c} \xi_{kq}^{i^c} 
\xi_{\bar{l}\bar{k}}^{\bar{p}^c} \xi_{\bar{k}\bar{q}}^{\bar{i}^c}\times \nn \\
&& e^{\I \pi(h_i-h_{\bar{i}} + h_l-h_{\bar{l}} - h_p+h_{\bar{p}} - h_q + h_{\bar{q}})} \; F_{p^c q}\! \left[ \ba{cc} j & l \\ i^c & k \ea \right] \! F_{\bar{p}^c\bar{q}}\! \left[ \ba{cc} \bar{j} & \bar{l} \\ \bar{i}^c & \bar{k} \ea \right], \label{CC=FFCC} \\
C_{i\bar{i},l\bar{l}}^{\ph{i\bar{i},l\bar{l}}q\bar{q}}
C_{k\bar{k},j\bar{j},q\bar{q}} 
&=& \sum_{p\bar{p}} C_{i\bar{i},j\bar{j}}^{\ph{i\bar{i},j\bar{j}}p\bar{p}}
C_{k\bar{k},l\bar{l},p\bar{p}} \; F_{p q}\! \left[ \ba{cc} j & l \\ i & k \ea \right] \! F_{\bar{p}\bar{q}}\! \left[ \ba{cc} \bar{j} & \bar{l} \\ \bar{i} & \bar{k} \ea \right], \label{CC=FFCC2}
\ee
The first equation is the statement of $S-U$ duality and the second $S-T$ duality. Duality of the one-point function on the torus leads, in the simple case $\xi_{ij}^k=1$, to
\be
S_{ij}(p)&=&S_{00}(0)\;e^{-\I \pi h_p} \sum_r {\frac{ F_{i^c 0}\! \left[ \ba{cc} i & i \\ p^c & p \ea \right] \;
 B^-_{p^c r^c}\! \left[ \ba{cc} i & j \\ i^c & j \ea \right] \! 
 B^-_{r^c 0}\! \left[ \ba{cc} j & i \\ i^c & j \ea \right]
}{ F_{00}\! \left[ \ba{cc} p & p \\ p^c & p \ea \right] \! F_{p^c 0}\! \left[ \ba{cc} j & j \\ j^c & j \ea \right] \! F_{p^c 0}\! \left[ \ba{cc} i & i \\ i^c & i \ea \right] }}. \label{sas=b}
\ee
Actually, equation (\ref{sas=b}) is a subset of the more general equation $SaS^{-1}=b$, which relates the two ways to sew a torus one-point function (via the $a$-cycle or the $b$-cycle). This subset contains all the non-redundant information though \cite{ms}. Later on this equation will be checked on the orbifold theory for the simple case of $p=0$.

The completeness theorem of \cite{ms} together with the analysis of \cite{sonoda}, claims that if a conformal field theory satisfies this set of equations, then that theory will be consistent on any surface constructed by sewing together closed, oriented sufaces. Actually, this claim has been questioned and repaired in \cite{MSrigorous}.

\subsection{Boundary constraints}
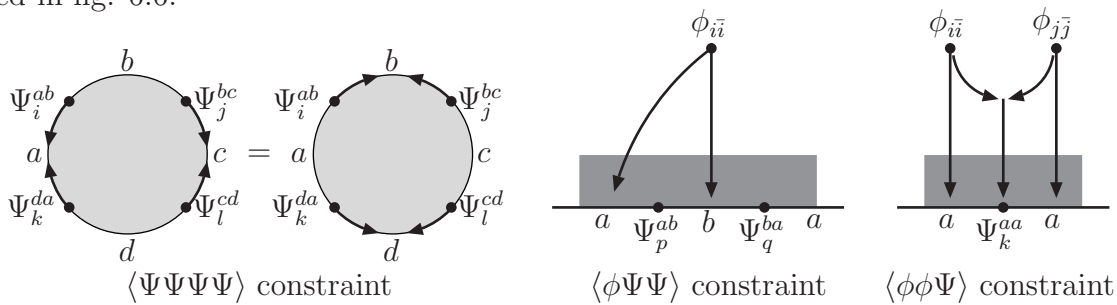
\begin{figure}[ht]
\SetScale{1}
\begin{center}
\begin{picture}(500,90)(-30,0)
\GCirc(50,50){30}{0.85}
\Vertex(28,30){2}
\Vertex(28,70){2}
\Vertex(72,30){2}
\Vertex(72,70){2}
\Text(23,70)[r]{$\Psi_{i}^{ab}$}
\Text(75,70)[l]{$\Psi_{j}^{bc}$}
\Text(23,30)[r]{$\Psi_{k}^{da}$}
\Text(75,30)[l]{$\Psi_{l}^{c d}$}
\Text(100,0)[c]{$\langle \Psi\Psi\Psi\Psi \rangle$ constraint}
\SetWidth{1}
\LongArrowArc(50,50)(30,135,175)
\LongArrowArcn(50,50)(30,225,185)
\LongArrowArcn(50,50)(30,45,5)
\LongArrowArc(50,50)(30,315,355)
\SetWidth{0.5}
\Text(50,86)[c]{$b$}
\Text(15,50)[c]{$a$}
\Text(50,14)[c]{$d$}
\Text(85,50)[c]{$c$}
\Text(100,50)[c]{$=$}
\GCirc(150,50){30}{0.85}
\Vertex(128,30){2}
\Vertex(128,70){2}
\Vertex(172,30){2}
\Vertex(172,70){2}
\Text(123,70)[r]{$\Psi_{i}^{ab}$}
\Text(175,70)[l]{$\Psi_{j}^{bc}$}
\Text(123,30)[r]{$\Psi_{k}^{da}$}
\Text(175,30)[l]{$\Psi_{l}^{c d}$}
\SetWidth{1}
\LongArrowArc(150,50)(30,45,80)
\LongArrowArcn(150,50)(30,135,100)
\LongArrowArcn(150,50)(30,315,280)
\LongArrowArc(150,50)(30,225,260)
\SetWidth{0.5}
\Text(150,86)[c]{$b$}
\Text(115,50)[c]{$a$}
\Text(150,14)[c]{$d$}
\Text(185,50)[c]{$c$}
\CBox(220,30)(310,50){White}{Gray}
\Vertex(250,30){2}
\Vertex(290,30){2}
\Vertex(270,90){2}
\SetWidth{1}
\LongArrow(270,90)(270,35)
\LongArrowArc(330,10)(100,127,165)
\Line(210,30)(320,30)
\SetWidth{.5}
\Text(270,100)[c]{$\phi_{i\bar{i}}$}
\Text(250,20)[c]{$\Psi_{p}^{ab}$}
\Text(290,20)[c]{$\Psi_{q}^{ba}$}
\Text(230,25)[c]{$a$}
\Text(270,25)[c]{$b$}
\Text(310,25)[c]{$a$}
\Text(270,0)[c]{$\langle \phi\Psi\Psi \rangle$ constraint}
\CBox(350,30)(410,50){White}{Gray}
\Vertex(360,90){2}
\Vertex(400,90){2}
\Vertex(380,30){2}
\SetWidth{1}
\LongArrow(360,90)(360,35)
\LongArrow(400,90)(400,35)
\LongArrowArc(380,90)(19,188,265)
\LongArrowArcn(380,90)(19,352,275)
\LongArrow(380,71)(380,35)
\Line(340,30)(420,30)
\SetWidth{.5}
\Text(360,100)[c]{$\phi_{i\bar{i}}$}
\Text(400,100)[c]{$\phi_{j\bar{j}}$}
\Text(380,20)[c]{$\Psi_{k}^{aa}$}
\Text(360,25)[c]{$a$}
\Text(400,25)[c]{$a$}
\Text(380,0)[c]{$\langle \phi\phi\Psi \rangle$ constraint}
\end{picture}
\end{center}
\caption{Boundary sewing constraints}
\label{sewbd}
\end{figure}

The set of sewing constraints was enlarged to conformal field theories on surfaces with boundaries by \cite{lewellen}. Three new constraint equations are found: crossing symmetry of the four-point function of boundary fields, duality of the bulk-boundary-boundary four-point function and duality of the bulk-bulk-boundary five-point function. The pictures that go with this are represented in fig. \ref{sewbd}.

Looking at the picture, we see that the boundary four-point function $\langle \Psi\Psi\Psi\Psi \rangle$ can be expressed in two ways, depending on whether we take contractions $(ij)(kl)$ or $(ik)(jl)$. Note that the contraction $(il)(kj)$ is impossible because the boundary fields are constrained to the real line.

The bulk-boundary-boundary $\langle \phi \Psi\Psi \rangle$ correlator can be expanded in two ways, depending on which particular boundary the field faces ($a$ or $b$).

For the bulk-bulk-boundary $\langle \phi\phi\Psi \rangle$ we can either contract the bulk fields before pulling them to the boundary or not. The equations that go with these three constraints involve, as the ones above, fusing matrices. These matrices depend, as we will see later, on a few so-called `gauge' choices. The gauge that we will take later simplifies some fusing matrices elements (check chapter \ref{orbfus}). In this gauge the boundary sewing constraints coming from the picture above have the form respectively
\be
&& C_{jkq^c}^{bcd} \; C_{ilq}^{dab} \; C_{q^c q0}^{bdb} = \sum_q
   C_{ijp^c}^{abc} \; C_{klq}^{cda} \; C_{p^c p0}^{aca}
      F_{pq}\! \left[ \ba{cc} j & k \\ i & l \ea  \right], \label{cc=Fcc} \\
&& C_{plq}^{abb} \;B^a_{(i\bar{i})l}\; C^{aba}_{qq^c 0} =
   \sum_{km} C_{pqk^c}^{aba}\; B^a_{(i\bar{i})k} \;
   C_{kk^c0}^{aaa} \; \xi_{m^cp}^i \;
      \xi_{p^cp}^0 \;\xi_{qp^c}^{k^c} \;\xi^i_{p^c m^c}\; 
      \hat{\nu}_{k\bar{i}mlq} \times \nn \\
&& \text{\hspace{.5cm}} e^{\I \pi(\um(h_k+h_l)+2h_m-2h_i-h_p-h_q)}
      F_{km}\!   \left[ \ba{cc} \bar{i} & q \\ i & p \ea  \right] 
      F_{m^cl}\! \left[ \ba{cc} i & \bar{i} \\ p & q \ea  \right], 
         \label{ccB=FFccB} \\
&& B^a_{(i\bar{i})p}\; B^b_{(j\bar{j})q}\; 
   C^{aaa}_{pqk^c} \; C^{aaa}_{kk^c0} = \sum_{rm\bar{m}}
   C^{\ph{i\bar{i},j\bar{j}}m\bar{m}}_{i\bar{i},j\bar{j}}\; 
   B^a_{(m\bar{m})k} \; C^{aaa}_{kk^c0} \; \hat{\nu}_{pij\bar{m}k}\; 
   \times \nn \\
&& \text{\hspace{.5cm}}  
   e^{\I \frac{\pi}{2}(h_p-h_q-h_i+h_j+h_{\bar{i}}+h_{\bar{j}}+h_m-
      h_{\bar{m}}+h_k-2h_r)}
   F_{q^cr}\!   \left[ \ba{cc} k & \bar{j} \\ p & j \ea  \right] 
      F_{pm}\!   \left[ \ba{cc} \bar{i} & j \\ i & r \ea  \right] 
      F_{r^c\bar{m}}\!   
         \left[ \ba{cc} \bar{i} & \bar{j} \\ {m} & k \ea  \right]\!\!. 
\label{BBcc=CBcFF}
\ee
The quantity $\hat{\nu}_i$ is defined as the Frobenius-Schur indicator for self-conjugate $i$ and $+1$ for complex $i$. Then $\hat{\nu}_{i_1\ldots i_n}= \hat{\nu}_{i_1}\cdots\hat{\nu}_{i_n}$. These constraints must be satisfied for the conformal field theory to be consistent on surfaces with boundaries of any order in string perturbation theory.

\subsection{Crosscap constraints}
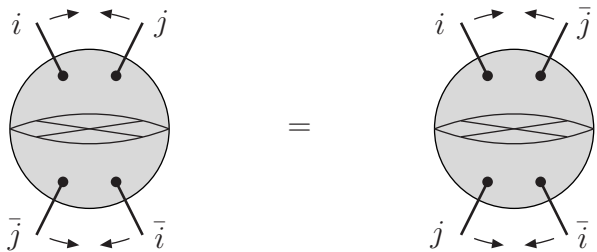
\begin{figure}[ht]
\SetScale{1}
\begin{center}
\begin{picture}(400,100)(0,0)
\GCirc(120,40){30}{0.85}
\LongArrowArc(120,40)(44,70,85)
\LongArrowArcn(120,40)(44,110,95)
\LongArrowArc(120,40)(44,250,265)
\LongArrowArcn(120,40)(44,290,275)
\Curve{(90,40)(110,35)(150,40)}
\Curve{(90,40)(110,45)(150,40)}
\Line(100,37)(140,43)
\Line(100,43)(140,37)
\Vertex(110,20){2}
\Vertex(130,20){2}
\Vertex(110,60){2}
\Vertex(130,60){2}
\SetWidth{1}
\Line(110,20)(100,0)
\Line(130,20)(140,0)
\Line(110,60)(100,80)
\Line(130,60)(140,80)
\SetWidth{.5}
\Text(95,80)[r]{$i$}
\Text(145,80)[l]{$j$}
\Text(95,0)[r]{$\bar{j}$}
\Text(145,0)[l]{$\bar{i}$}
\Text(200,40)[c]{$=$}
\GCirc(280,40){30}{0.85}
\LongArrowArc(280,40)(44,70,85)
\LongArrowArcn(280,40)(44,110,95)
\LongArrowArc(280,40)(44,250,265)
\LongArrowArcn(280,40)(44,290,275)
\Curve{(250,40)(270,35)(310,40)}
\Curve{(250,40)(270,45)(310,40)}
\Line(260,37)(300,43)
\Line(260,43)(300,37)
\Vertex(270,20){2}
\Vertex(290,20){2}
\Vertex(270,60){2}
\Vertex(290,60){2}
\SetWidth{1}
\Line(270,20)(260,0)
\Line(290,20)(300,0)
\Line(270,60)(260,80)
\Line(290,60)(300,80)
\SetWidth{.5}
\Text(255,80)[r]{$i$}
\Text(305,80)[l]{$\bar{j}$}
\Text(255,0)[r]{$j$}
\Text(305,0)[l]{$\bar{i}$}
\end{picture}
\end{center}
\caption{Crosscap constraint}
\label{sewcross}
\end{figure}

Finally, we check the constraints on unoriented surfaces. Not much is known about these. The only constraint of this kind was originally derived in \cite{sagnotticross} and later applied to $SU(2)$ WZW models \cite{usecross}, where it was used to derive crosscap coefficients. From the picture below, we see it relates two ways of doing the short-distance expansion of a bulk two-point function in the presence of a crosscap. The crosscap identification makes it impossible to distinguish the field $\phi_{i\bar{i}}\zz$ from $\phi_{\bar{i}i}(\bar{z},z)$, thus there are two ways of writing the two-point correlator: either contracting $i$ with $j$ or with $\bar{j}$, as fig. \ref{sewcross} shows.

Upon lifting the two expressions to a chiral correlator, an equality arises. In the same gauge for the fusing matrices as the boundary sewing constraints, it reads

\be
\epsilon_{i\bar{i}^c}\;\Gamma_k \;C_{i\bar{i},j\bar{j}}^{\ph{i\bar{i},j\bar{j}}k\bar{k}^c} = 
\sum_l \Gamma_l \;C_{\bar{i}i,j\bar{j}}^{\ph{i\bar{i},j\bar{j}}l\bar{l}^c} \; 
\xi_{kj}^{i^c} \xi_{\bar{j}\bar{i}}^k \; \hat{\nu}_{i\bar{i}kl} \;e^{\I \pi(h_i-h_{\bar{i}}+h_j-h_{\bar{j}})} \;
F_{k^cl}\!   \left[ \ba{cc} i & \bar{j} \\ j & \bar{i} \ea  \right]. \label{eGC=GCF}
\ee
Note that one cannot contract $i$ with $\bar{i}$ because the anti-involution defined by crosscap keeps the two labels diametrically apart. The factor $\epsilon_{k\bar{k}^c}$ is a sign that determines the symmetry of an Ishibashi primary field in the presence of a crosscap.

As opposed to the bulk and boundary sewing constraints, the crosscap sewing constraint probably incomplete in the sense that there may be more constraints on unoriented surfaces that one needs to verify. As it is known from topology, a torus with one crosscap is topologically equivalent to a sphere with three crosscaps. This is expected to give an equation of type $\Gamma\Gamma\Gamma=\Gamma$, which would fix the normalization of the crosscap coefficients \cite{gamma3}. However, the precise form of constraint has not yet been investigated.

\subsection{The role of duality matrices}
As we have seen above, the sewing constraints involve the fusing and braiding matrices. With their knowledge one can solve some of the sewing constraints for non-trivial quantities like operator product coefficients. One can also work the other way around and derive them by looking for operator product coefficients and with this solving the sewing constraints for the fusing matrices. So we see the duality matrices are a vital tool which is decisive to solve a conformal field theory. Even more because these matrices can, as stated in the beginning of this chapter, lead to direct building of most conformal field theory quantities via the method of \cite{frs}. This motivation enough to search for these quantities, whose importance recently leaped to a major role.

In the case of the free boson and orbifolds thereof, we are in a good position to solve the sewing constraints because some of the chiral and non-chiral quantities are easy to get from direct evaluation of the correlators. The bulk operator products, for instance, can be calculated because the free boson field mode expansion is known and tractable. Once we know the operator product coefficients and the $N$-point correlators \cite{kn}, we can, for instance, extract the conformal blocks and evaluate duality matrices, which can in turn be used in the various sewing constraints equations. In most cases however, operator product coefficients are not so easy to get, which makes the explicit determination of duality matrices an important thing to do, since finding these may be the simplest way to solve the theory.

The free boson and its orbifold is one of the simplest non-trivial examples for which there is independent knowledge of model-dependent quantities like operator products, and this can make the whole process solvable (at least to a certain extent). Finding the free boson and orbifold duality matrices complements the information from operator products and can therefore be a way to verify some of the sewing constraints and with this the validity of the procedure of \cite{frs}.

\chapter{Orbifold fusing matrices}
\label{orbfus}
In this chapter we explicitly evaluate orbifold duality matrices. We will take the modular invariant to be the diagonal invariant $Z_{ij}=\delta_{ij}$. Since the duality matrices are chiral data, they are independent of the modular invariant we start from, so we might as well choose the invariant that is easier to deal with and has the simplest form for the vertex operators of primary fields. By inserting the vertex operators of table \ref{orbifoldvertexoperators} into correlators and using some known results, we will be able to derive explicit expressions for four-point functions containing untwisted sector fields only. We can then decompose the correlators into bilinear sums of conformal blocks and extract the latter, with which we can look for the duality matrices that interpolate between the various conformal blocks. For the case of four-point correlators involving twisted fields, the results of \cite{vafa} \cite{dixon} can be used to get the conformal blocks needed to derive duality matrices.

Duality matrices obey the so-called polynomial equations of \cite{ms}. These are three basic identities, one called the pentagon and two the hexagons, derived from graphical calculus that are fundamental at tree-level. Not only that, they can actually reach much further. As was noted in \cite{runkelad} \cite{sew=pent}, for a C-diagonal modular invariant there is a one-to-one correspondence between chiral labels and boundary labels and as such boundary operator product coefficients turn out simply to {\em coincide} with the fusing matrices. Then the sewing constraint involving $\langle \Psi\Psi\Psi\Psi \rangle$ reduces to the pentagon equation. Thus, checking the pentagon equation for orbifold fusing matrices is already equivalent to solving one of the orbifold sewing constraints.

It will not be possible in this work to solve all sewing constraints for the various operator product coefficients, as the fusing matrices do not always over-determine the problem, but it will be possible to get and discuss many of the fusing matrices and verify they obey the key equations.

\section{Defining duality matrices}
First of all we need to define the duality matrices and setup conventions. When we take canonical insertion points, there is some index reshuffling and the braiding and fusing matrices that relate the bulk four-point function $U$- and $T$-channels to the $S$-channel become, for the simple case where the fusion rules are 0 or 1,
\be
&&{\cal F}_p^{ijkl}(z) = B^\pm_{pq}\!\left[ \ba{cc} j&k \\ i&l\ea \right] 
  {\cal F}_p^{ikjl}(1/z), \;\;\; \text{for $\im(z) >0 , <0$} \label {defB} \\
&&{\cal F}_p^{ijkl}(z) = F_{pq}\!\left[ \ba{cc} j&k \\ i&l\ea \right] 
  {\cal F}_p^{ilkj}(1-z).\label{defBF}
\ee
The corresponding picture is \ref{bflink}.
\begin{figure}[ht]
\SetScale{1}
\begin{center}
\begin{picture}(500,90)(0,0)
\ArrowLine(20,20)(40,20)
\ArrowLine(40,20)(60,20)
\ArrowLine(80,20)(60,20)
\ArrowLine(40,50)(40,20)
\ArrowLine(60,50)(60,20)
\Text(15,20)[r]{$i$}
\Text(50,10)[c]{$p$}
\Text(85,20)[l]{$l$}
\Text(40,57)[c]{$j$}
\Text(60,57)[c]{$k$}
\ArrowLine(200,20)(220,20)
\ArrowLine(220,20)(240,20)
\ArrowLine(260,20)(240,20)
\ArrowLine(220,50)(220,20)
\ArrowLine(240,50)(240,20)
\Text(195,20)[r]{$i$}
\Text(230,10)[c]{$q$}
\Text(265,20)[l]{$l$}
\Text(220,57)[c]{$k$}
\Text(240,57)[c]{$j$}
\ArrowLine(380,20)(400,20)
\ArrowLine(440,20)(400,20)
\ArrowLine(400,40)(400,20)
\ArrowLine(430,40)(410,40)
\Line(410,40)(400,40)
\ArrowLine(410,60)(410,40)
\Text(375,20)[r]{$i$}
\Text(395,30)[r]{$q$}
\Text(445,20)[l]{$l$}
\Text(410,67)[c]{$j$}
\Text(435,40)[l]{$k$}
\Text(140,20)[c]{$= \;B_{pq}^\pm 
\left[ \ba{cc} j & k \\ i & l \ea \right] \; \cdot$}
\Text(320,20)[c]{$= \;F_{pq} \left[ \ba{cc} j & k \\ i & l \ea \right]\;\cdot$}
\end{picture}
\end{center}
\caption{Braiding and fusing of conformal blocks}
\label{bflink}
\end{figure}
The operation of braiding/debraiding and fusing/defusing can be used at the level of the pictures. Equations (\ref{defBF}) will be our means to determine free boson orbifold fusing matrices. Note that the conformal block ${\cal F}_p^{ijkl}$ has its cut at $z=1$ running to $-\infty$ parallel to the real axis, while the cut for ${\cal F}_p^{ikjl}$ also starts at $z=1$ but runs to $+\infty$, so that the whole real line splits the complex $z$-plane into two halves. Since there are two regions, there are also two braiding matrices: $B^\pm$. For the fusing matrix, the block on the left-hand-side has again the cut at $z=1$ running to $-\infty$ parallel to the real axis, but now the cut for ${\cal F}_p^{ilkj}$ runs from $2$ to $+\infty$, leaving the region between $z=1$ and 2 free of cuts. Since the operator product that defines the $T$-channel is valid precisely in this region, the region where $F$ is defined is free of cuts and there is thus only one fusing matrix.

\subsection{Chiral vertex operators and gauge choices}
An alternative way to understand what a conformal block is, is to refer to the concept of chiral vertex operators \cite{cvo}. A chiral vertex operator $\Phi_{ij}^k(z)$ is an intertwiner from representations $i,j$ into representation $k$ at position $z$. In the rest of the discussion we assume the fusion rules to be 0 or 1 (see also the comments on section \ref{Fdegen}). A conformal block is nothing but the chiral correlator of four chiral vertex operators. The link is
\be
{\cal F}_p^{ijkl}(z) = \langle i| \Phi_{jp}^{i}(1) \Phi_{kl}^{p}(z) |l\rangle,
\ee
with the ingoing and outgoing states defined by $|l\rangle = \Phi_{l0}^l (0)|0\rangle$ and $\langle i^c |= \lim_{\eta\rightarrow \infty} \eta^{2h_i} \langle 0| \Phi_{ii^c}^0 (\eta)$. The normalization of chiral vertex operators can be set by looking at the chiral three-point function
\be
\langle i| \Phi_{jk}^i(z) |k\rangle = ||\Phi_{jk}^i||\; z^{h_i-h_j-h_k}.
\ee
The normalization factor $||\Phi_{jk}^i||$ can in principle be any complex number we want. We will call a change in the normalization factor of chiral vertex operators a ``gauge transformation''. Such action makes $||\Phi_{jk}^i|| \rightarrow \lambda_{jk}^i ||\Phi_{jk}^i||$, which changes the normalization of the conformal block. The fusing matrices change accordingly as
\be
F_{pq}\left[\ba{cc} j & k \\ i & l \ea \right] \rightarrow 
   \frac{\lambda_{jp^c}^{i^c} \lambda_{kl}^{p^c}}{\lambda_{ql}^{i^c} \lambda_{jk}^q}
F_{pq}\left[\ba{cc} j & k \\ i & l \ea \right]
= F^{\text{\scriptsize new}}_{pq}\left[\ba{cc} j & k \\ i & l \ea \right]. \label{gaugefreedom}
\ee
A fusing matrix is therefore defined only up to gauge transformation. In other words, two sets of fusing matrices represent different physical conformal field theories only if they cannot be related via a gauge transformation of type (\ref{gaugefreedom}). Fixing conformal block normalization as (\ref{cblocknorm}) still allows for $\lambda_{ij}^k$ that are signs (or phases).

\subsection{Conformal block normalization}
The asymptotic normalization of a conformal block (\ref{cblocknorm}) is only valid when one considers the Virasoro algebra alone. In an extended chiral algebra, the concept of primary and descendant is redefined and conformal blocks must be renormalized accordingly. Later in this chapter we will see that a conformal block ${\cal F}_p^{ijkl}(z)$ can only be consistently normalized to unit when $i,j,k,l,p$ are primaries. For the extended orbifold, we take the normalization
\be
 {\cal F}_p^{ijkl}(z) \sim 1\times z^{h_p-h_k-h_l}(1+\cdots) \;\;\; \text{as}\;z\rightarrow 0, \;\;\;\text{when $i,j,k,l,p$ are primaries.} \label{Fasymp}
\ee
If one of the labels propagates at a descendant level, the conformal block cannot be asymptotically normalized to unity without the duality matrices starting to depend on the descendant level; a fact that is not consistent with their definition \cite{ms}. In this case we have to define its normalization by referring to the unextended algebra. We will come back to this point later.

\subsection{Duality matrices from the three-point function}
Some fusing and braiding matrices can be readily evaluated by looking at the three-point function. For the braiding matrix we get
\be
B^{\pm}_{p q}\!\left[\ba{cc} j & k \\ i & 0 \ea\right] = \xi_{jk}^{i^c} \; \delta_{p^c k}\delta_{q^c j} \; e^{\pm \I \pi (h_i-h_j-h_k)}.
\ee
The factor $\xi_{jk}^{i}$ is the eigenvalue of the mapping of the vector space of couplings $V_{jk}^i$ to $V_{kj}^i$. Since doing the interchange twice brings us back to $V_{jk}^{i}$, the eigenvalue is at most a sign and is symmetric in the lower indexes.

The three-point fusing matrix is naively 1, but in fact depends on the normalization of chiral vertex operators. In \cite{frs} it is argued that it is always possible to find a normalization such that
\be
F_{k^c i^c}\!\left[\ba{cc} j & k \\ i & 0 \ea\right] =  
F_{i j}\!\left[\ba{cc} 0 & j \\ i & k \ea\right] = 
F_{k^c j}\!\left[\ba{cc} j & 0 \\ i & k \ea\right] = 1,\;\;\;
F_{i k^c}\!\left[\ba{cc} i & j \\ 0 & k \ea\right] = \hat{\nu}_i \hat{\nu}_j
\hat{\nu}_k. \label{FRSgauge}
\ee
(See below (\ref{BBcc=CBcFF}) for a definition of $\hat{\nu}$.) This natural choice is particularly simple for the orbifold theory, where all the Frobenius-Schur indicators are +1. This choice is taken below throughout orbifold computations.

\subsection{Pentagon and hexagon identities}
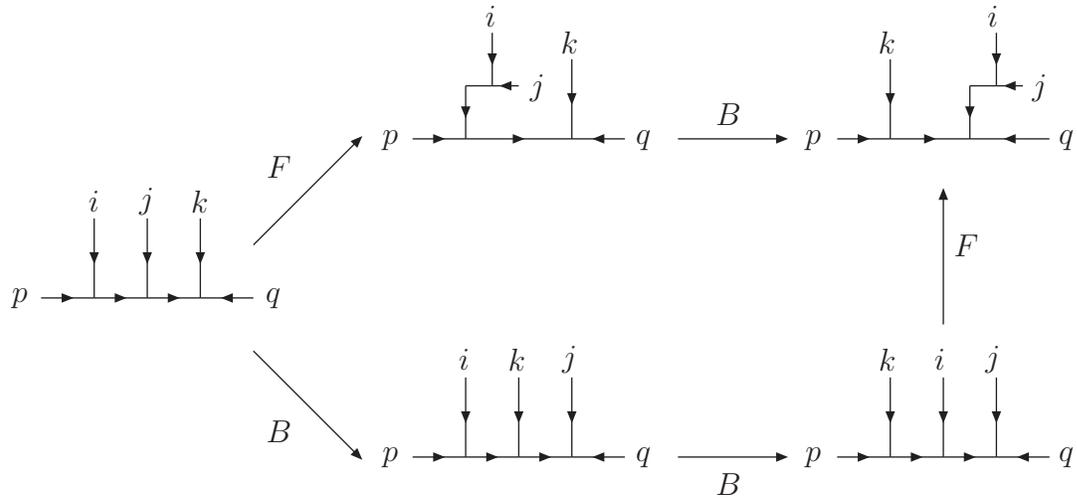
\begin{figure}
\SetScale{1}
\begin{center}
\begin{picture}(400,200)(0,0)
\ArrowLine(0,100)(20,100)
\ArrowLine(20,100)(40,100)
\ArrowLine(40,100)(60,100)
\ArrowLine(80,100)(60,100)
\ArrowLine(20,130)(20,100)
\ArrowLine(40,130)(40,100)
\ArrowLine(60,130)(60,100)
\Text(-5,100)[r]{$p$}
\Text(85,100)[l]{$q$}
\Text(20,137)[c]{$i$}
\Text(40,137)[c]{$j$}
\Text(60,137)[c]{$k$}
\ArrowLine(140,40)(160,40)
\ArrowLine(160,40)(180,40)
\ArrowLine(180,40)(200,40)
\ArrowLine(220,40)(200,40)
\ArrowLine(160,70)(160,40)
\ArrowLine(180,70)(180,40)
\ArrowLine(200,70)(200,40)
\Text(135,40)[r]{$p$}
\Text(225,40)[l]{$q$}
\Text(160,77)[c]{$i$}
\Text(180,77)[c]{$k$}
\Text(200,77)[c]{$j$}
\ArrowLine(300,40)(320,40)
\ArrowLine(320,40)(340,40)
\ArrowLine(340,40)(360,40)
\ArrowLine(380,40)(360,40)
\ArrowLine(320,70)(320,40)
\ArrowLine(340,70)(340,40)
\ArrowLine(360,70)(360,40)
\Text(295,40)[r]{$p$}
\Text(385,40)[l]{$q$}
\Text(320,77)[c]{$k$}
\Text(340,77)[c]{$i$}
\Text(360,77)[c]{$j$}
\ArrowLine(140,160)(160,160)
\ArrowLine(160,160)(200,160)
\ArrowLine(220,160)(200,160)
\ArrowLine(160,180)(160,160)
\Line(160,180)(170,180)
\ArrowLine(170,200)(170,180)
\ArrowLine(180,180)(170,180)
\ArrowLine(200,190)(200,160)
\Text(135,160)[r]{$p$}
\Text(225,160)[l]{$q$}
\Text(170,207)[c]{$i$}
\Text(185,180)[l]{$j$}
\Text(200,197)[c]{$k$}
\ArrowLine(300,160)(320,160)
\ArrowLine(320,160)(350,160)
\ArrowLine(380,160)(350,160)
\ArrowLine(320,190)(320,160)
\ArrowLine(350,180)(350,160)
\Line(360,180)(350,180)
\ArrowLine(360,200)(360,180)
\ArrowLine(370,180)(360,180)
\Text(295,160)[r]{$p$}
\Text(385,160)[l]{$q$}
\Text(360,207)[c]{$i$}
\Text(375,180)[l]{$j$}
\Text(320,197)[c]{$k$}
\LongArrow(80,80)(120,40)
\LongArrow(80,120)(120,160)
\LongArrow(240,40)(280,40)
\LongArrow(240,160)(280,160)
\LongArrow(340,90)(340,140)
\Text(90,50)[c]{$B$}
\Text(90,150)[c]{$F$}
\Text(260,170)[c]{$B$}
\Text(260,30)[c]{$B$}
\Text(350,120)[c]{$F$}
\end{picture}
\end{center}
\caption{Pentagon identity}
\label{pentpic}
\end{figure}

Playing around with the pictorial representation of conformal blocks one can derive several identities \cite{ms}. In fig. \ref{pentpic} 
have one such basic identity. Writing the whole picture above in terms of only fusing matrices we arrive to the pentagon identity
\be
\sum_s 
      F_{q^c s}
          \left[ \begin{array}{cc} j & k \\ p & b \end{array} \right]
      F_{pl}
          \left[ \begin{array}{cc} i & s \\ a & b \end{array} \right]
      F_{s^c r}
          \left[ \begin{array}{cc} i & j \\ l^c & k \end{array} \right]=
      F_{pr}
          \left[ \begin{array}{cc} i & j \\ a & q \end{array} \right]
      F_{q^c l}
          \left[ \begin{array}{cc} r & k \\ a & b \end{array} \right].
\ee
The signs $\xi_{ij}^k$ all cancel precisely. 
Setting $b=0$ on fig. \ref{pentpic} we get
\be
F_{pq}\left[ \begin{array}{cc} j & k \\ i & l
\end{array}
\right]=\xi_{kl}^q \xi_{pl}^i \; e^{\mp \I \pi(h_i+h_k-h_p-h_q)}B^{\pm}_{pq}\left[ \begin{array}{cc} j & l \\ i & k \end{array}
\right],\label{b->f}
\ee
from which we conclude that fusing and braiding matrices are related by phases, so there is really only one duality matrix. Nevertheless, it is often convenient to consider the two matrices separately.

\begin{figure}
\SetScale{1}
\begin{center}
\begin{picture}(400,200)(0,0)
\ArrowLine(0,100)(20,100)
\ArrowLine(20,100)(40,100)
\ArrowLine(60,100)(40,100)
\ArrowLine(20,130)(20,100)
\ArrowLine(40,130)(40,100)
\Text(-5,100)[r]{$i$}
\Text(65,100)[l]{$l$}
\Text(20,137)[c]{$j$}
\Text(40,137)[c]{$k$}
\ArrowLine(100,40)(120,40)
\ArrowLine(120,40)(140,40)
\ArrowLine(160,40)(140,40)
\ArrowLine(120,70)(120,40)
\ArrowLine(140,70)(140,40)
\Text(95,40)[r]{$i$}
\Text(165,40)[l]{$k$}
\Text(120,77)[c]{$j$}
\Text(140,77)[c]{$l$}
\ArrowLine(240,40)(260,40)
\ArrowLine(300,40)(260,40)
\ArrowLine(260,60)(260,40)
\ArrowLine(270,80)(270,60)
\ArrowLine(280,60)(270,60)
\Line(270,60)(260,60)
\Text(235,40)[r]{$i$}
\Text(305,40)[l]{$j$}
\Text(270,87)[c]{$l$}
\Text(285,60)[l]{$k$}
\ArrowLine(100,160)(120,160)
\ArrowLine(160,160)(120,160)
\ArrowLine(140,180)(130,180)
\ArrowLine(130,200)(130,180)
\ArrowLine(120,180)(120,160)
\Line(130,180)(120,180)
\Text(95,160)[r]{$i$}
\Text(165,160)[l]{$l$}
\Text(130,207)[c]{$j$}
\Text(145,180)[l]{$k$}
\ArrowLine(240,160)(260,160)
\ArrowLine(260,160)(280,160)
\ArrowLine(300,160)(280,160)
\ArrowLine(260,190)(260,160)
\ArrowLine(280,190)(280,160)
\Text(235,160)[r]{$i$}
\Text(305,160)[l]{$k$}
\Text(260,197)[c]{$l$}
\Text(280,197)[c]{$j$}
\ArrowLine(340,100)(360,100)
\ArrowLine(360,100)(380,100)
\ArrowLine(400,100)(380,100)
\ArrowLine(360,130)(360,100)
\ArrowLine(380,130)(380,100)
\Text(335,100)[r]{$i$}
\Text(405,100)[l]{$j$}
\Text(360,137)[c]{$l$}
\Text(380,137)[k]{$k$}
\LongArrow(60,120)(80,140)
\LongArrow(180,160)(220,160)
\LongArrow(320,140)(340,120)
\LongArrow(340,80)(320,60)
\LongArrow(220,40)(180,40)
\LongArrow(80,60)(60,80)
\Text(60,60)[c]{$B$}
\Text(60,140)[c]{$F$}
\Text(200,170)[c]{$B$}
\Text(200,30)[c]{$B$}
\Text(340,140)[c]{$B$}
\Text(340,60)[c]{$F$}
\end{picture}
\end{center}
\caption{Hexagon identity 1}
\label{hex1pic}
\end{figure}
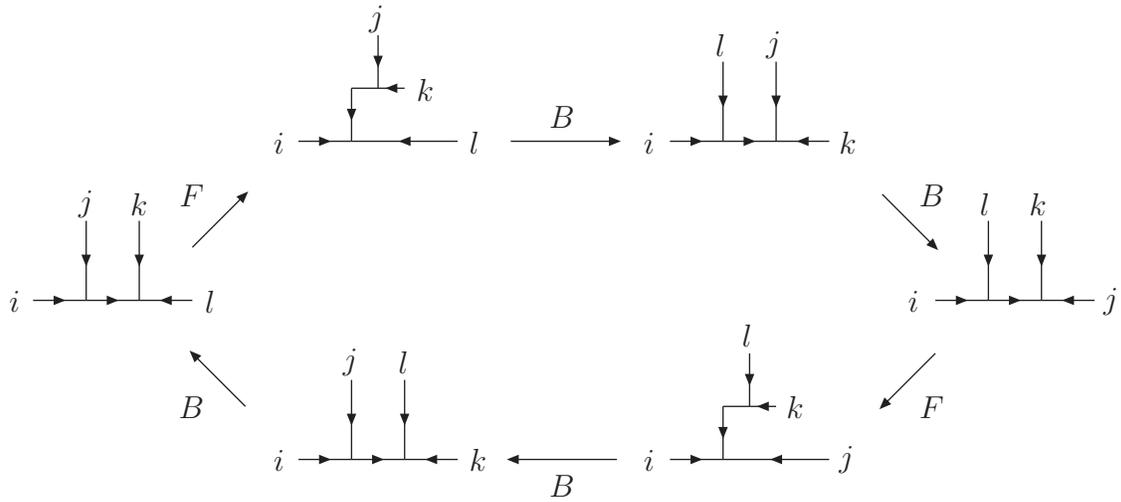

\begin{figure}
\SetScale{1}
\begin{center}
\begin{picture}(400,200)(0,20)
\ArrowLine(0,100)(20,100)
\ArrowLine(20,100)(40,100)
\ArrowLine(60,100)(40,100)
\ArrowLine(20,130)(20,100)
\ArrowLine(40,130)(40,100)
\Text(-5,100)[r]{$i$}
\Text(65,100)[l]{$l$}
\Text(20,137)[c]{$j$}
\Text(40,137)[c]{$k$}
\ArrowLine(100,40)(120,40)
\ArrowLine(120,40)(140,40)
\ArrowLine(160,40)(140,40)
\ArrowLine(120,70)(120,40)
\ArrowLine(140,70)(140,40)
\Text(95,40)[r]{$i$}
\Text(165,40)[l]{$k$}
\Text(120,77)[c]{$j$}
\Text(140,77)[c]{$l$}
\ArrowLine(240,40)(260,40)
\ArrowLine(300,40)(260,40)
\ArrowLine(260,60)(260,40)
\ArrowLine(270,80)(270,60)
\ArrowLine(280,60)(270,60)
\Line(270,60)(260,60)
\Text(235,40)[r]{$i$}
\Text(305,40)[l]{$k$}
\Text(270,87)[c]{$j$}
\Text(285,60)[l]{$l$}
\ArrowLine(100,160)(120,160)
\ArrowLine(160,160)(120,160)
\ArrowLine(140,180)(130,180)
\ArrowLine(130,200)(130,180)
\ArrowLine(120,180)(120,160)
\Line(130,180)(120,180)
\Text(95,160)[r]{$i$}
\Text(165,160)[l]{$l$}
\Text(130,207)[c]{$j$}
\Text(145,180)[l]{$k$}
\ArrowLine(240,160)(260,160)
\ArrowLine(260,160)(280,160)
\ArrowLine(300,160)(280,160)
\ArrowLine(260,190)(260,160)
\ArrowLine(280,190)(280,160)
\Text(235,160)[r]{$i$}
\Text(305,160)[l]{$k$}
\Text(260,197)[c]{$l$}
\Text(280,197)[c]{$j$}
\ArrowLine(340,100)(360,100)
\ArrowLine(400,100)(360,100)
\ArrowLine(360,120)(360,100)
\ArrowLine(370,140)(370,120)
\ArrowLine(380,120)(370,120)
\Line(370,120)(360,120)
\Text(335,100)[r]{$i$}
\Text(405,100)[l]{$k$}
\Text(370,147)[c]{$l$}
\Text(385,120)[l]{$j$}
\LongArrow(60,120)(80,140)
\LongArrow(180,160)(220,160)
\LongArrow(320,140)(340,120)
\LongArrow(320,60)(340,80)
\LongArrow(180,40)(220,40)
\LongArrow(60,80)(80,60)
\Text(60,60)[c]{$B$}
\Text(60,140)[c]{$F$}
\Text(200,170)[c]{$B$}
\Text(200,30)[c]{$F$}
\Text(340,140)[c]{$F$}
\Text(340,60)[c]{$B$}
\end{picture}
\end{center}
\caption{Hexagon identity 2}
\label{hex2pic}
\end{figure}
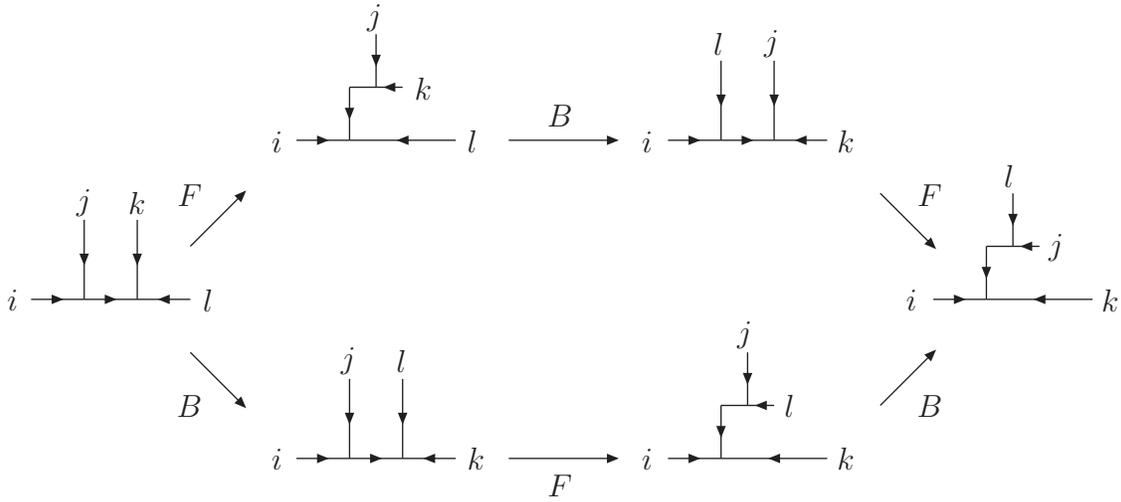

Another important identity comes from picture \ref{hex1pic}. 
It leads to the hexagon identity
\be
\sum_s
F_{p^c s}\!\left[\begin{array}{cc} j & k \\ i & l \end{array}\right]
F_{s^c q}\!\left[\begin{array}{cc} l & k \\ i & j \end{array}\right] \;
\xi_{sl}^{i^c} \xi_{jk}^s \xi_{qj}^{i^c} \xi_{lk}^q = \delta_{pq}. 
\label{hexfalse}
\ee
So we see the hexagon gives us more information on the fusing matrices plus something about the $\xi_{jk}^i$ eigenvalues. There is another hexagon equation, which is derived from the picture \ref{hex2pic}
and leads to
\be
\sum_q 
F_{pq}\!\left[\begin{array}{cc} j & k \\ i & l \end{array}\right]
F_{q^c r}\!\left[\begin{array}{cc} l & j \\ i & k \end{array}\right] 
\xi_{lq}^{i^c} \; e^{\mp \I \pi h_q}
= F_{pr}\!\left[\begin{array}{cc} j & l \\ i & k \end{array}\right]
\xi_{jl}^r \xi_{lk}^{p^c}\; e^{\pm \I \pi(h_p+h_r-h_i-h_j-h_k-h_l)}, \label{hextrue}
\ee
with $\pm$ depending on the sense of the braidings, so there are actually three hexagons. Actually, one can also show the hexagons are not all independent. We can for instance use (\ref{hextrue}) with the plus sign on the the same equation with the minus sign and checked that this yields the first hexagon (\ref{hexfalse}).

The pentagon and the two hexagons are the fundamental polynomial equations at tree-level. Any other equation relating fusing and braiding matrices can be shown to be a combination of the pentagon and/or hexagon by the completeness theorem of \cite{ms}.

\subsection{Free boson fusing matrices}
For completeness, we present the fusing matrices for the free boson. In the unextended free boson case, the fusing matrices are just $F=1$ for all valid indices. In the extended case it is not so simple. Some general results for this case were presented in \cite{ms}. In \cite{frs} results for the gauge (\ref{FRSgauge}) were presented. They are
\be
F_{\{j+k\}^c,\{i+j\} }\!\left[\begin{array}{cc} 
      i & j \\ 
      {\{i+j+k\}}^c & k 
   \end{array}\right]&=& (-1)^{ (i+j+1) \big( j\sigma(i+j+k) + 
      (j+k)(\sigma(i+j)+\sigma(j+k))  \big) }, \nn \\
      \xi_{ij}^{\{i+j\}}&=&(-1)^{(i+j)\sigma(i+j)},
\ee
with $\{i+j\}=i+j+2Nn$ with $n$ the unique integer such that $\{i+j\}\in [0,2N[$, and $\sigma(i+j)$ the $n$ for which $\{i+j\}\in [0,2N[$. So we see clearly the non-trivial sign structure of $\xi_{ij}^k$ arising.

The extended free boson theory is perhaps the simplest example of non-trivial fusing matrices. There is a for a small number of cases where fusing matrices have been studied. These include the Virasoro minimal models \cite{virafus} and $SU(2)_k$ WZW models \cite{su2fus}. See also \cite{morefusion} for some results on coset models.

\section{Fusing matrices for the untwisted sector}
\label{Fdegen}
To determine the orbifold fusing matrices we need the explicit form of orbifold four-point correlators, and we start off by examining the correlators involving four fields from the untwisted sector.

Since the untwisted vertex operators of the orbifold theory are combinations of vertex operators of the free boson theory, the untwisted orbifold correlators are simply combinations free boson correlators. The later are given by the Koba-Nielsen formula \cite{kn}, which is a formula for the correlator of an arbitrary number of exponentials of free boson fields on the sphere 
\be
\langle \prod_{i=1}^q e^{\I 
   \left( p_{Li} X(z_i)+ p_{Ri}\overline{X}(\bar{z}_i)\right)} 
\rangle = \prod_{i<j} 
   z_{ij}^{p_{Li} p_{Lj}} \bar{z}_{ij}^{p_{Ri} p_{Rj}} \times 
   \delta_{\sum_{i=1}^q p_{Li},0}\;\delta_{\sum_{i=1}^q p_{Ri},0}. 
\label{koba-nielsen} 
\ee 
(Remember the exponentials are implicitly normally ordered.) The delta functions ensure space-time momentum conservation; in other words, they impose the fusion rules. In the extended free boson theory with diagonal invariant, primary fields have $p_L=p_R$ and we have a simplification 
\be
\langle \prod_{i=1}^q e^{\I p_i X(z_i,\bar{z}_i)} \rangle = 
   \prod_{i<j} |z_{ij}|^{2p_i p_j}\times \delta_{\sum_{i=1}^q p_i,0}. 
\label{koba-nielsen2} 
\ee 
In the orbifold theory, along with primary fields built out of exponentials of $X(z,\bar{z})$, we also have the primary field $J=-\partial X \overline{\partial X}$. In the free boson theory this was a descendant of the identity, but becomes a primary in the orbifold. Orbifold correlators with $J$ present can also be derived from (\ref{koba-nielsen2}) by means of appropriate partial differentiations. We will use (\ref{koba-nielsen2}) and derivatives thereof as needed.

As a reminder for the notation, we distinguish once again chiral labels (or chiral fields) and conformal fields
\bi
\item{Chiral labels: $0,J,\Phi^i,\Phi_k,\sigma_i,\tau_i$. Highest-weight states of the chiral algebra upon which a representation module of the algebra can be built. A general chiral label is designated by italic indices $i,j,k,l,p,q$. Duality matrices are chiral objects and depend thus on the chiral labels.}
\item{Conformal fields: $1,J=\partial X \overline{\partial X},\phi^{i\bar{i}},\phi_{k\bar{k}},\sigma_{i\bar{i}},\tau_{i\bar{i}},\; i=1,2$ these are functions of $\zz$ and form combinations of chiral labels. Since we take the torus diagonal invariant $\bar{i}=i$, so we abbreviate the conformal fields by $0,J,\phi^i,\phi_k,\sigma_i,\tau_i$ (and likewise for operator product coefficients). Their vertex operators are given by table \ref{orbifoldvertexoperators}. The canonical correlators of four conformal fields are single-valued in the $z$-plane and divide into conformal blocks.}
\ei
In the $N$ odd case remember that the conformal fields $\phi^i,\sigma_i,\tau_i$ go to the complex combinations $\hat{\phi}^i,\hat{\sigma}_i,\hat{\tau}_i$. For the twisted fields we use the same notation for chiral fields and conformal fields for simplicity, but the difference should be clear from the context.

\subsubsection{Examination of the fusion rules}
Before we start evaluating orbifold correlators, we look at the orbifold fusion rules, which are always a helpful tool to probe the problem. A nice feature of the orbifold theory is that the fusion rules are always zero or one: $N_{ij}^{\phantom{ij}k} \in \{0,1\}$. Not only does this removes the fusing matrix degeneracy labels $\alpha,\beta,\gamma,\delta$ of \cite{ms}, which would appear in (\ref{defB}) and (\ref{defBF}), it also allows us to draw the following conclusion. The fusing matrix relates bases on the space of conformal blocks of dimensions. In our conventions, this dimensional relation is $\sum_p N_{ab}^{\ph{ab}p} N_{pcd} = \sum_q N_{bc}^{\ph{bc}q} N_{qad}$, and for the untwisted sector we see that three things can happen 
\be 
\sum_p N_{ab}^{\ph{ab}p} N_{pcd} = \sum_q N_{bc}^{\ph{bc}q} N_{qad} = \left\{ 
\begin{array}{cl} 
 0 & \rightarrow \;
F_{pq}\!\left[\begin{array}{cc} b & c \\ a & d \end{array}\right] = 0 \\ 
     1 & \rightarrow \;\mbox{Fusing matrix is 1x1} \\ 
   > 1 & \rightarrow \;\mbox{Fusing matrix is 2x2 or bigger} \\ 
\end{array} 
\right. \label{inspectingfusing} 
\ee 
Note that any fusing matrix where one of the entries is a simple current will automatically be 1x1. The only case of the untwisted sector with 2x2 or bigger matrices is then when all the four entries are fields of type $\Phi_k$ (fixed points of $J$). 
 
There are many matrices to be evaluated. We start off with 1x1 matrices and then proceed to the bigger ones. In the end an appendix is made with the results for easy reading.

\subsection{Using the polynomial equations to solve for unknowns}
 Computing the fusing matrices by extracting conformal blocks from the untwisted four-point functions and relating them via (\ref{defBF}) will not provide us with exact results. The calculations will show ambiguities related to the chiral algebra extension and these will have to be dealt with.

To solve the ambiguities we resort to comparison with fusing matrices of some known models to which the orbifold theory is equivalent, at certain radii. From this comparison we postulate a general form for the fusing matrices. After sorting out the ambiguities, and assuming the \cite{frs} gauge for $F[0ijk]$, we get a naive result, which will satisfy the polynomial equations up to signs.

To get a concrete result we allow some sign freedom in the fusing matrices elements naively calculated and $\xi_{ij}^k$ to vary, with the help of a computer, force the outcome to satisfy the pentagon and hexagon equations. In the end it turns out that it is possible to find sign changes in the naive result and a set of $\xi_{ij}^k$ such that all the pentagon and hexagon equations are satisfied. The result happens not to be unique because the gauge choice is not completely fixed by $F[0ijk]$. Then, again by means of a computer, we search for gauge transformations $\lambda_{ij}^k$ in (\ref{gaugefreedom}), that can interpolate between the various solutions. The result is that all solutions fall into the same gauge class, and therefore what we have is essentially a unique result for the fusing and braiding matrices. All these results have been checked up to $N=21$. {\em Every pentagon and hexagon equation was tested up to this} $N$. Given the mod 4 periodicity of the orbifold structures involved, there is thus reason to believe these behavior will go on indefinitely so that our solution is valid for all $N$.

As said above, it will be necessary to introduce corrections to the results derived solely by relating conformal blocks via (\ref{defBF}). It turns out that these corrections show up only when there are fields propagating at descendant level in the correlators. This hints at the chiral algebra extension being responsible for the changes, since the extension transforms some primaries into descendants.

 \subsection{Explicit calculation for the untwisted sector}
The result for untwisted orbifold fusing matrices will depend on whether $N$ is even or odd. Given the various types of chiral fields, we split the the calculations below into several cases. In each case we deal with $N$ even first and $N$ odd afterwards. Generally the $N$ odd case is very similar to $N$ even, but there are occasional complications. If a fusing matrix $F[ijkl]$ does not show up in the list below for certain values of the chiral labels, then it is not allowed by the fusion rules. We will also take canonical insertion points for the correlators, so that we don't need to write down the $z_i$ dependence explicitly. Also, all free boson correlators that are used to determine the orbifold ones have been checked for the conformal Ward identities of \cite{bpz}.

\subsubsection{Case $F[ijk0]$} 
To define these fusing matrices we take the gauge of \cite{frs}, which is the one for which (\ref{FRSgauge}) holds. Such gauge is a convenient choice for our case because the Frobenius-Schur indicator is always +1 in free boson orbifolds. The result for $F[ijk0]$ is thus simply 
\be 
 F\!\left[ \begin{array}{cc} 
              j & k \\ i & 0 \end{array}
           \right] = 1,\;\; \mbox{if fusion possible.}
\label{abc0} 
\ee 
This holds regardless of where the identity field is, e.g. $F[ij0k]=F[i0jk]=F[0ijk]=1$, and for both $N$ odd and even. This result can also be derived from the three-point functions.

\subsubsection{Case $F[JJJJ]$}
 For this case we need the correlator $\langle JJJJ \rangle$. It can be derived from (\ref{koba-nielsen2}) by taking the following series of actions. Take $q=4$, split the correlator into chiral and anti-chiral parts, write the chiral result in the form $\prod_{i<j} e^{p_i p_j \log{z_{ij}}}$, take derivatives with respect to $p_i$, set $p_i=0$, take derivatives with respect to $z_i$ and finally take  canonical insertions points. After doing the above, we come to 
\be 
\langle JJJJ \rangle = 
\left| 1+\frac{1}{(1-z)^2} + \frac{1}{z^2}\right|^2. 
\ee 
There is then only one conformal block. It is 
\be 
{\cal F}_0^{JJJJ}(z) = 1+\frac{1}{(1-z)^2} + \frac{1}{z^2}. 
\ee 
This block goes like $z^{-2}$, which from (\ref{Fasymp}) confirms propagation of the identity in the $S$-channel. Having the block, we use the fusing equation (repeated for convenience) 
\be 
{\cal F}^{ijkl}_p(z) = \sum_{q} F_{pq}\!\left[\ba{cc} j&k \\ i&l \ea \right] {\cal F}^{ilkj}_q(1-z), \label{fe} 
\ee 
to derive the fusing matrix, which is simply 
\be 
 F\!\left[\begin{array}{cc} J & J \\ J & J \end{array}\right] = 1.
\ee 
Again valid for all $N$. 
 
\subsubsection{Case $F[JJ\Phi^i\Phi^i]$} 
 Next we have matrices of type $F[JJ\Phi^i\Phi^i]$ and permutations thereof. The correlators relevant for this case involve free boson correlators of type $\langle \partial X \overline{\partial X} \partial X \overline{\partial X} e^{\I aX\zz} e^{\I bX\zz} \rangle$. These can be evaluated by the same method we used to determine $F[JJJJ]$: differentiate the Koba-Nielsen chiral formula with respect to $p_1,p_2$, set $p_1,p_2$ to zero, differentiate with respect to $z_1,z_2$, push the insertions to the canonical points and finally impose momentum conservation (this last step forces $b=-a$). There are six relevant correlators. Written as $S$-channel expansions at canonical insertion points they are 
\be 
\langle 
  \partial X \overline{\partial X}
  \partial X \overline{\partial X}
  e^{\I\frac{a}{R}X\zz} e^{-\I\frac{a}{R}X\zz} 
\rangle &=& 
  1 \left| 
    \frac{1\!-\!z+(\frac{a}{R})^2 z^2}{z^{(\frac{a}{R})^2}(1\!-\!z)}  
    \right|^2, \nn \\
\langle 
  \partial X \overline{\partial X} e^{ \I\frac{a}{R}X\zz} 
  \partial X \overline{\partial X} e^{-\I\frac{a}{R}X\zz} 
\rangle &=&
    \left( (\frac{a}{R})^2 \right)^2 
    \left| 
    \frac{1-\frac{1}{(\frac{a}{R})^{2}}z(1\!-\!z)}{z(1\!-\!z)} 
    \right|^2 , \nn \\   
\langle 
  \partial X \overline{\partial X}
  e^{\I\frac{a}{R}X\zz} e^{-\I\frac{a}{R}X\zz} 
  \partial X \overline{\partial X}
\rangle &=& 
    \left( (\frac{a}{R})^2 \right)^2    
    \left| 
    \frac{(1\!-\!z)^2+\frac{z}{(\frac{a}{R})^2}}{z(1\!-\!z)^{(\frac{a}{R})^2}} 
    \right|^2, \nn \\ 
\langle 
  e^{ \I\frac{a}{R}X\zz} 
  \partial X \overline{\partial X}
  \partial X \overline{\partial X}
  e^{-\I\frac{a}{R}X\zz}
\rangle &=&
    \left((\frac{a}{R})^2\right)^2
    \left| 
    \frac{(1\!-\!z)^2+\frac{z}{(\frac{a}{R})^2}}{z(1\!-\!z)^2}
    \right|^2, \label{ddee} \\  
\langle 
  e^{ \I\frac{a}{R}X\zz} 
  \partial X \overline{\partial X}
  e^{-\I\frac{a}{R}X\zz}
  \partial X \overline{\partial X}
\rangle &=&
    \left( (\frac{a}{R})^2\right)^2
    \left| 
    \frac{1-\frac{1}{(\frac{a}{R})^2}(1\!-\!z)z}{z(1\!-\!z)} 
    \right|^2, \nn \\
\langle
  e^{\I\frac{a}{R}X\zz} e^{-\I\frac{a}{R}X\zz} 
  \partial X \overline{\partial X}
  \partial X \overline{\partial X} 
\rangle &=& 
  1 \left| 
    \frac{1\!-\!z+(\frac{a}{R})^2 z^2}{z^{2}(1\!-\!z)} 
    \right|^2. \nn
\ee
Outside the absolute square brackets we have the operator product coefficients. (We have also abbreviated the chiral components of the free boson field to $X=X(z),\,\overline{X}(\bar{z})=\overline{X}.$) The conformal blocks behave asymptotically as expected, since every line in the $S$-channel propagates at the primary level. This result is used whenever there are correlators with two $J$ fields. 

The vertex operators for the orbifold $\phi^i$ conformal fields can, as usual, be expanded into combinations of free boson vertex operators. We will need orbifold correlators of type $\langle JJ\phi^i\phi^j\rangle$ and permutations, so we do the splitting and get 
\be 
\langle JJ\phi^i\phi^j \rangle = 
   \langle \partial X \partial X
   \left(
       e^{\I\frac{N}{R}X} + (-1)^i e^{\I\frac{N}{R}X} 
   \right) 
   \left( 
       e^{\I\frac{N}{R}X} + (-1)^j e^{\I\frac{N}{R}X}  
   \right) 
   \times C.C. \rangle \label{JJi,j} 
\ee 
(Note that labels $i,j \in \{1,2\}$ and $\I$ is the imaginary unit.) Here $C.C.$ stands for complex conjugate, e.g. a symmetric anti-holomorphic part for the correlator. In principle we should also multiply this correlator by an overall factor coming from the normalizations of the four fields, but since we just want to compare the conformal blocks for this correlator with the blocks for the correlator $\langle J\phi^j\phi^i J \rangle$, this overall factor drops out and can be ignored. We will neglect such overall factors systematically in order to keep the notation as clean as possible. As we see from expanding (\ref{JJi,j}), the orbifold correlator $\langle JJ\phi^i\phi^j\rangle$ indeed reduces to a sum of free boson correlators (in this case, four of them).

\subsubsection{Cocycle factors} 
The product of the two chiral halves in (\ref{JJi,j}) looks to be nothing but a simple factorization. However there is a slight complication. The unextended free boson vertex operators $V_{P_L,P_R}$ are bosonic and should thus commute. However, if we evaluate the operator product of $V_{1}V_{2}$ and compare it with the operator product of $V_{2}V_{1}$, we see they differ by an overall factor $(-1)^{P_{1L}P_{2L}-P_{1R}P_{2R}}$. This factor can be removed by inserting cocycle factors into the definition of the vertex operators \cite{cocycle} \cite{jungnickel}. Such cocycle factors are operators acting on a space of their own and produce the minus signs needed to restore commutation of bosonic vertex operators. There are many ways to represent them, but to cut a long story short, here we simply take a Pauli $\sigma$-matrix representation for them \cite{cocyclesigma}. So we redefine our free boson vertex operators as 
\be 
   V_{P_L,P_R}(z,\bar{z}) = 
     e^{-\I\frac{\pi}{2}(nm)} \sigma_3^m \sigma_1^n \; 
   : e^{\I p_L X(z)+\I p_R \overline{X}(\bar{z})} : \; , 
\ee 
The numbers $m,n$ are the integers coming from $p_{L,R}=\frac{n}{R}\pm \um mR$ and $\sigma_{1,3}$ are the usual Pauli matrices (not to be confused with the twisted sector $\sigma_{1,2}$ fields). Note that in the extended free boson theory we reorganized the momenta such that the momenta of the primary fields is $p_L=p_R$, which makes $m$, the winding number, zero. Thus in the orbifold we only need to worry about cocycles when we deal with fields $\phi^i$, since these are the only ones with different left and right momenta. (Eventually chiral descendants with $p_L\neq p_R$ would also need cocycle factors, but we will not need those fields.) So, whenever we encounter non-symmetrical chiral halves in correlators, we should insert the above phases and Pauli matrices. Then we need to anti-commute the $\sigma_1$ matrices to the right at both sides of (\ref{defBF}), so as to ensure we are comparing equal things in the cocycle space. The net effect is to change sums of factors into differences in some correlators. We will not write down the cocycle factors explicitly in our formulas, but will refer to them when needed. 

On the first case at hand, $F[JJ\Phi^i\Phi^j]$, there is no chiral asymmetry and, after applying momentum conservation to (\ref{JJi,j}), we get the $N$ even result 
\be 
\langle JJ\phi^i \phi^j \rangle &=& 
\langle 
   \partial X \partial X  
   \left(
      (-1)^i e^{ \I\frac{N}{R}X} e^{-\I\frac{N}{R}X} +  
      (-1)^j e^{-\I\frac{N}{R}X} e^{ \I\frac{N}{R}X} 
   \right) 
\times C.C. \rangle \nn \\ 
&=& 
\langle \partial X \overline{\partial X} \partial X \overline{\partial X} 
   \left( 
      e^{ \I\frac{N}{R}X} e^{ \I\frac{N}{R}\overline{X}} 
      e^{-\I\frac{N}{R}X} e^{-\I\frac{N}{R}\overline{X}} + 
      e^{-\I\frac{N}{R}X} e^{ \I\frac{N}{R}\overline{X}}
      e^{ \I\frac{N}{R}X} e^{-\I\frac{N}{R}\overline{X}} 
   \right. \nn \\ 
   &&+ \left. 
      e^{ \I\frac{N}{R}X} e^{-\I\frac{N}{R}\overline{X}}
      e^{-\I\frac{N}{R}X} e^{ \I\frac{N}{R}\overline{X}} +  
      e^{-\I\frac{N}{R}X} e^{-\I\frac{N}{R}\overline{X}}
      e^{ \I\frac{N}{R}X} e^{ \I\frac{N}{R}\overline{X}} 
   \right)
\rangle. \label{JJij} 
\ee 
We notice that $i=j$, otherwise the result vanishes. This is consistent with the fusion rules for $N$ even. In the second line, the winding number is $m=1$ for the second and third terms of the sum and the cocycle factors for all the four terms are respectively $\sigma_3^N \sigma_3^N, \sigma_1 \sigma_1, \sigma_1 \sigma_1, \sigma_3^N \sigma_3^N$. All these products of Pauli matrices square to one, so even though in principle we could have had cocycle factors, they turn out to be +1 here. This goes through for all other correlators with two $J$ and two $\phi^i$. 
Since all cocycle factors are +1, the four terms of (\ref{JJij}) are equal and the conformal block is 
\be
{\cal F}_{0}^{JJ\Phi^i\Phi^i}(z) = z^{-N/2}\frac{1-z+\frac{N}{2}z^2}{1-z}. 
\ee 
This goes like $z^{-N/2}$, signaling propagation of the identity in the $S$-channel, as expected from the fusion rules for $N$ even. To extract a fusing matrix we need the block for $\langle J\phi^i\phi^i J \rangle$. With a calculation similar to (\ref{JJij}), it turns out to be 
\be 
{\cal F}_{\Phi_j}^{J\Phi^i\Phi^iJ}(z) = \frac{1}{z(1-z)^{N/2}}\left((1-z^2) + \frac{2z}{N}\right). 
\ee 
Asymptotics go like $z^{-1}$, signaling propagation of $J$ in the $S$-channel, again as expected from $N$ even fusion rules. From (\ref{fe}) we can now determine the fusing matrix $F[JJ\Phi^i\Phi^i]$. The remaining conformal blocks needed for the case $F[JJ\Phi^i\Phi^i]$ and permutations are 
\be 
  {\cal F}_{\Phi_j}^{\Phi^i JJ \Phi^i}(z) = 
     \frac{1}{z} + \frac{2}{N(1-z)^2} \;\;\;
  {\cal F}_{0}^{\Phi^i \Phi^i JJ}(z) = 
     \frac{1}{z^2} + \frac{N}{2(1-z)},\nn \\ 
  {\cal F}_{\Phi_j}^{\Phi^i J \Phi^i J}(z) = 
     \frac{1}{z(1-z)} + \frac{2}{N} \;\;\;
  {\cal F}_{\Phi_j}^{J \Phi^i J \Phi^i}(z) = 
     \frac{1}{z} \left( \frac{1}{1-z} - \frac{2z}{N} \right). 
\ee
Using (\ref{fe}) we come to
\be 
&&   F\!\left[ \begin{array}{cc}  
        J & \Phi^{i} \\ 
        J & \Phi^{i} 
        \end{array}\right]= \frac{N}{2}\;\;\;
     F\!\left[ \begin{array}{cc}  
        \Phi^{i} & \Phi^{i} \\ 
           J     &    J 
        \end{array}\right]= \frac{2}{N}\;\;\;
     F\!\left[ \begin{array}{cc}  
        \Phi^{i} &    J \\ 
           J     & \Phi^{i} 
        \end{array}\right]=1
\nn \\
&&   F\!\left[ \begin{array}{cc}  
           J     &    J \\ 
        \Phi^{i} & \Phi^{i} 
        \end{array}\right]= \frac{2}{N}\;\;\;
     F\!\left[ \begin{array}{cc}  
        \Phi^{i} & J \\ 
        \Phi^{i} & J 
        \end{array}\right]= \frac{N}{2} \;\;\;
     F\!\left[ \begin{array}{cc}  
           J     & \Phi^{i} \\ 
        \Phi^{i} &    J 
        \end{array}\right]= 1.
\ee 
Remember this result is for $N$ even. For deriving the $N$ odd result one must replace $\phi^{1,2} \rightarrow \hat{\phi}^{1,2} = \frac{1}{\sqrt{2}} (\phi^1 \pm \I \phi^2)$. There are no further subtleties and after reevaluating the correlators and extracting the new conformal blocks the result for $N$ odd is ($i \neq j$) 
\be 
&&   F\!\left[ \begin{array}{cc}  
        J & {\Phi}^{i} \\ 
        J & {\Phi}^{j} 
        \end{array}\right]= \frac{N}{2}\;\;\;
     F\!\left[ \begin{array}{cc}  
        {\Phi}^{i} & {\Phi}^{j} \\ 
            J      &     J 
        \end{array}\right]= \frac{2}{N} \nn,\;\;\; 
     F\!\left[ \begin{array}{cc}  
        {\Phi}^{i} &     J \\ 
            J      & {\Phi}^{j} 
        \end{array}\right]= 1 \nn \\ 
&&   F\!\left[ \begin{array}{cc}  
            J      &     J \\ 
        {\Phi}^{i} & {\Phi}^{j} 
        \end{array}\right]= \frac{2}{N}\;\;\; 
     F\!\left[ \begin{array}{cc}  
        {\Phi}^{i} & J \\ 
        {\Phi}^{j} & J 
        \end{array}\right]= \frac{N}{2}\;\;\; 
     F\!\left[ \begin{array}{cc}  
            J      & {\Phi}^{j} \\ 
        {\Phi}^{i} &     J 
        \end{array}\right]= 1. 
\ee
This is consistent with the fusion rules of $N$ odd. 
 
\subsubsection{Case $F[JJ\Phi_k \Phi_k]$} 
 Next up we have type $F[JJ\Phi_k \Phi_k]$. Again we split the orbifold vertex operators for $\Phi_k$ into free bosons ones and using (\ref{ddee}) we see that the combinations are such that we get same conformal blocks as for the $\langle JJ\phi^i \phi^i \rangle$ case, but with $N/2$ replaced by $k^2/2N$. The final result is 
\be 
&&   F\!\left[ \begin{array}{cc}  
        J & \Phi_k \\ 
        J & \Phi_k 
        \end{array}\right]= \frac{k^2}{2N}\;\;\;
     F\!\left[ \begin{array}{cc}  
        \Phi_k & \Phi_k \\ 
        J & J 
        \end{array}\right]= \frac{2N}{k^2}\;\;\;
     F\!\left[ \begin{array}{cc}  
        \Phi_k & J \\ 
        J & \Phi_k 
        \end{array}\right]= 1 \nn \\ 
&&   F\!\left[ \begin{array}{cc}  
        J & J \\ 
        \Phi_k & \Phi_k 
        \end{array}\right]= \frac{2N}{k^2}\;\;\;
     F\!\left[ \begin{array}{cc}  
        \Phi_k & J \\ 
        \Phi_k & J 
        \end{array}\right]=\frac{k^2}{2N}\;\;\;
     F\!\left[ \begin{array}{cc}  
        J & \Phi_k \\ 
        \Phi_k & J
        \end{array}\right]= 1. 
\ee 
Since there are no $\Phi^i$ fields here, this result is also valid for the $N$ odd case. 
 
\subsubsection{Case $F[J \Phi_k \Phi_a \Phi_b]$} 
 In this case several interesting features arise. We require $a+b\neq N$, since $a+b=N$ leads to a different case. First we need the free boson correlators $\langle \partial X \overline{\partial X} e^{\I\frac{a}{R}X\zz} e^{\I\frac{b}{R}X\zz} e^{\I\frac{c}{R}X\zz}\rangle$, which again we can get by taking derivatives of the Koba-Nielsen formula 
\be 
\langle 
   \partial X \overline{\partial X} 
    e^{\I\frac{(-c-d)}{R}X\zz} e^{\I\frac{c}{R}X\zz} e^{\I\frac{d}{R}X\zz}
\rangle &=& 
   \left(\frac{c+d}{R}\right)^2 
   \left| 
    z^{\frac{cd}{R^2}} \frac{1-\frac{cz}{c+d}}{(1-z)^{\frac{c(c+d)}{R^2}}}
   \right|^2, \nn\\ 
\langle 
   e^{\I\frac{(-c-d)}{R}X\zz}
   \partial X \overline{\partial X} 
   e^{\I\frac{c}{R}X\zz} e^{\I\frac{d}{R}X\zz}
\rangle &=& 
   \left(\frac{c+d}{R}\right)^2
   \left| 
   z^{\frac{cd}{R^2}} \frac{1-\frac{dz}{c+d}}{1-z}
   \right|^2, \nn\\ 
\langle 
   e^{\I\frac{(-b-d)}{R}X\zz} e^{\I\frac{b}{R}X\zz} 
   \partial X \overline{\partial X}
   e^{\I\frac{d}{R}X\zz}
\rangle &=& 
   \left(\frac{d}{R}\right)^2
   \left|
   \frac{1-\frac{(b+d)z}{d}}{z(1-z)}
   \right|^2, \nn\\ 
\langle
   e^{\I\frac{(-b-c)}{R}X\zz} e^{\I\frac{b}{R}X\zz} e^{\I\frac{c}{R}X\zz} 
   \partial X \overline{\partial X}
\rangle &=& 
   \left(\frac{c}{R}\right)^2
   \left|
   \frac{(1-z)^{\frac{bc}{R^2}}\left( 1+\frac{bz}{c} \right)}{z}
   \right|^2. 
\label{deee} 
\ee 
Let us look at the correlators we want to use. The orbifold correlator $\langle J \phi_k \phi_a \phi_b \rangle$ can be expanded in the $S$-channel. 
Now the following can happen. The fusion rules are $\Phi_a \times \Phi_b = \Phi_{[a+b]} + \Phi_{[a-b]}$, so whenever $a+b > N$, then $k=2N-a-b$ instead of $a+b$. However, from (\ref{deee}) we see the conformal block doesn't care about this. Its asymptotics are always that of propagation of a chiral field $\Phi_{a+b}$, whether or not it in fact exists. If it doesn't, the field that is truly propagating is $\Phi_{2N-a-b}$, but at a {\em descendant} level such that the asymptotics of the block it generates match the result of the Koba-Nielsen formula. In general, the asymptotics behavior of a conformal block allows for an integer $M$ as follows
\be 
{\cal F}_{p}^{ijkl}(z) \sim z^{h_{p} - h_k - h_l + M}(1+\cdots), \;\;\mbox{as}\;z\rightarrow 0. 
\ee 
When dealing with pure Virasoro algebra $M$ is always zero, but in extensions or whenever $N_{ij}^{\phantom{ij}k} > 1$, it can happen that the integer $M$ is larger than zero. Actually, since in an extended theory some fields that were primaries prior to the extension eventually become descendants afterwards, this is  bound to happen. 
 
Which gives rise to the following question. What happens to the normalization of the conformal blocks if the intermediate state propagates at the descendant level? If propagation is at the primary level then the blocks can be normalized to 1, but if it is at descendant level that may not be so. That is because we can only normalize conformal blocks once. After that normalization is fixed, the normalizations for the contributions of descendants are fixed uniquely and cannot be set back to 1 {\em ad libitum}. In the case of descendants propagation one needs pull the operator product coefficients of the primary fields outside the conformal block. What remains is then the true conformal block of the descendant. Explicitly, what we mean is that if $(p,M)$ is a descendant of $p$ at level $M$, the normalization of a descendant block is \cite{fuchsbook}
\be 
{\cal F}_{(p,M)}^{ijkl}(z) \sim \beta_{ij}^{(p,M)} \beta^{(p,M)}_{kl} \times z^{h_p-h_k-h_l+M}(1+\cdots), \;\; C_{ij}^{\ph{ij}(p,M)} =  \beta_{ij}^{(p,M)} \beta^{(p,M)}_{kl} C_{ij}^{\ph{ij}p}. \label{blockdesc}
\ee 
The $\beta$ coefficients can be determined from conformal invariance in the general case \cite{bpz}, but in the simple case of free bosons and orbifolds thereof, they can be obtained explicitly from the operator products. This remark is the key to the accurate derivation of the fusing matrices when $a+b>N$. If we naively normalize the descendants blocks to 1, the fusing matrix starts depending on the descendant we take. This is wrong since the fusing matrix is the same for all descendants \cite{ms}.

Having settled block normalization, we can now use the information to compute the $F[Jkab]$ fusing matrices. First we expand the orbifold correlator $\langle J \phi_k \phi_a \phi_b \rangle$ into free boson factors 
\be 
\langle J \phi_k \phi_a \phi_b \rangle &=&  
   \langle \left(
      \partial X 
      e^{i\frac{k}{R}X} e^{i\frac{a}{R}X} e^{i\frac{b}{R}X}   + 
      \partial X 
      e^{i\frac{k}{R}X} e^{i\frac{-a}{R}X} e^{i\frac{b}{R}X}  \right.
\nn \\ &&+\; \left.
      \partial X 
      e^{i\frac{k}{R}X} e^{i\frac{a}{R}X} e^{i\frac{-b}{R}X}  + 
      \partial X 
      e^{i\frac{k}{R}X} e^{i\frac{-a}{R}X} e^{i\frac{-b}{R}X} \right)
   \times C.C. \rangle
\label{Jkab}
\ee 
Again there will be overall factors, but they won't matter for evaluating the fusing matrix. Momentum conservation demands that the sum of the momenta of the various terms of (\ref{Jkab}) vanishes. The first term does not contribute due to this. As for the other three, if one of them conserves momentum, the remaining two will not. This allows us to split the calculation into four sub-cases. First suppose $a+b<N$. We have three possibilities: $k=a+b, \; k=a-b, \; k=b-a$. For $k=a\!+\!b$ only the last term of (\ref{Jkab}) contributes and using (\ref{deee}) the result is 
\be
\langle J \phi_k \phi_a \phi_b \rangle_{k=a+b} = 
   \frac{(a+b)^2}{2N} 
   \left| 
      z^{\frac{ab}{2N}}\frac{1-\frac{az}{a+b}}{(1-z)^{\frac{a(a+b)}{2N}}}
   \right|^2 = 
      C_{\phi_J \phi_k}^{\ph{\phi_J \phi_k}\phi_k} 
      C_{\phi_a \phi_b \phi_k} 
   \left| 
      {\cal F}_{\Phi_{a+b}} (z)
   \right|^2.
\label{Jkab<N} 
\ee 
The conformal block is 
\be 
{\cal F}_{\Phi_{a+b}}^{J \Phi_k \Phi_a \Phi_b}(z) = 
   z^{\frac{ab}{2N}}\frac{1-\frac{az}{a+b}}{(1-z)^{\frac{a(a+b)}{2N}}}, 
\ee 
and has the correct asymptotics. To derive the fusing matrix we must compare this with the block coming from the correlator $\langle J \phi_b \phi_a \phi_k \rangle$. After doing this we come to $F[Jkab]=b/(b+a)$. There are also no subtleties with $k=\pm(a-b)$, but when $k=2N-a-b$ the matter is however different, due to propagations at descendant level. In this case the relevant operator product coefficient is $C_{J\phi_k}^{\ph{J\phi_k}\phi_k} = (2N-a-b)^2/R^2$. However, (\ref{deee}) still gives the same result as (\ref{Jkab<N}). Therefore, as stated above, we must pull this factor outside the conformal block. So far so good, but now we see when $k=2N-a-b$ there is no way to achieve momentum conservation in (\ref{Jkab}) at the primary level, so we must take a descendant field in one of the external legs of the correlator in order to be able to extract a fusing matrix. The easiest thing to do is to take a descendant of the extension current. The first descendant with respect to the free boson extension currents (\ref{vext}) has vertex operator $\cos{\big(\frac{(-a-b)X(z)}{R}\big)}$. The free boson decomposition for this descendant simply yields factors $e^{-\I(a+b)X(z)/R}$. Inserting these vertex operators and their anti-holomorphic partners into (\ref{deee}) and pulling out the primary operator product coefficient we come to 
\be 
\langle J \phi_k \phi_a \phi_b \rangle_{k=2N-a-b} &=& 
      \frac{(2N-a-b)^2}{2N} 
   \left| 
      \frac{a+b}{2N-a-b} \; z^{\frac{ab}{2N}} 
      \frac{1-\frac{az}{a+b}}{(1-z)^{\frac{a(a+b)}{2N}}}
\right|^2  \nn \\ &=& 
      C_{\phi_J \phi_k}^{\ph{\phi_J \phi_k}\phi_k} 
      C_{\phi_a \phi_b \phi_k}
   \left| 
      {\cal F}_{\Phi_{2N-a-b}} (z) 
   \right|^2. 
\label{Jkab>N} 
\ee 
Inside the absolute square brackets we have the conformal block, which as expected does not normalize to one: 
\be
{\cal F}_{\Phi_{2N-a-b}}^{J \Phi_k \Phi_a \Phi_b}(z) = 
   \frac{a+b}{2N-a-b} \; z^{\frac{ab}{2N}} 
   \frac{1-\frac{az}{a+b}}{(1-z)^{\frac{a(a+b)}{2N}}}. 
\ee 
Extracting the fusing matrix is now simple. We get $F[Jkab]=b/(2N-a-b)$. Following this method we can derive all the $F[Jabc]$ matrices 
\be
F\!\left[ \begin{array}{cc}  
      \Phi_k & \Phi_a \\ 
      J & \Phi_b 
   \end{array}\right]= 
      \left\{ \begin{array}{l} 
      \frac{b}{b-a},\; k=\pm(a-b)  \\ 
      \frac{b}{b+a},\; k=a+b       \\ 
      \frac{-b}{2N-a-b}, \; k=2N-a-b \end{array} 
      \right. &
F\!\left[ \begin{array}{cc}  
      \Phi_b & J \\ 
      \Phi_a & \Phi_k 
   \end{array}\right]= 
      \left\{ \begin{array}{l} 
      \frac{b}{b-a},\; k=\pm(a-b)  \\ 
      \frac{b}{b+a},\; k=a+b       \\ 
      \frac{-b}{2N-a-b},\; k=2N-a-b \end{array} 
   \right. \nn \\
F\!\left[ \begin{array}{cc}  
      J & \Phi_a \\ 
      \Phi_k & \Phi_b 
   \end{array}\right]= 
      \left\{ \begin{array}{l} 
      \frac{a}{a-b},\; k=\pm(a-b)  \\ 
      \frac{a}{a+b},\; k=a+b       \\ 
      \frac{-a}{2N-a-b}, k=2N-a-b \end{array} 
   \right. & 
F\!\left[ \begin{array}{cc}  
      \Phi_b & \Phi_k \\ 
      \Phi_a & J 
   \end{array}\right]=
      \left\{ \begin{array}{l} 
      \frac{a}{a-b},\; k=\pm(a-b)  \\ 
      \frac{a}{a+b},\; k=a+b       \\ 
      \frac{-a}{2N-a-b},\;k=2N-a-b \end{array} 
   \right. \nn \\
\ee 
This result valid for all $N$. Note however that the above expressions have an extra minus sign in the case where $a+b>N$ as compared to what one would get from the naive calculation. These signs do not follow from anything we said so far; they are appear because the pentagon and hexagon identities cannot be solved without them.

\subsubsection{Case $F[J\Phi_k \Phi_{N-k} \Phi^i]$}
Again we expand the orbifold correlators into free boson ones and use (\ref{deee}). Let us take the ordering $F[J\Phi_k \Phi_{N-k}\Phi^{i}]$. The expansion into free boson factors is 
\be
&& \langle J\phi_k \phi_{N-k} \phi^{1,2} \rangle =  
   \langle \partial X \overline{\partial X} 
   \left( 
      e^{\I\frac{ k}{R}X} e^{\I\frac{ k}{R}\overline{X}} +  
      e^{\I\frac{-k}{R}X} e^{\I\frac{-k}{R}\overline{X}} 
   \right) 
\times \nn \\ 
&& \left( 
      e^{\I\frac{N-k}{R}X}    e^{\I\frac{N-k}{R}\overline{X}} +  
      e^{\I\frac{-(N-k)}{R}X} e^{\I\frac{-(N-k)}{R}\overline{X}} 
   \right) 
   \left( 
      e^{\I\frac{N}{R}X} + (-1)^i e^{\I\frac{-N}{R}{X}} 
   \right) 
   \left( 
      e^{\I\frac{N}{R}\overline{X}} + (-1)^i e^{\I\frac{-N}{R}\overline{X}} 
   \right) 
   \rangle \nn \\ 
&&=\langle \partial X
      e^{\I\frac{k}{R}} e^{\I\frac{N-k}{R}X} e^{\I\frac{-N}{R}X} \times C.C. 
   \rangle = 
      \frac{k^2}{2N} 
   \left| 
      z^{\frac{k-N}{2}} \frac{1-\frac{(N-k)z}{k}}{(1-z)^{\frac{k(k-N)}{2N}}} 
   \right|^2. 
\ee 
Notice that in going from the first to the second line all other possible combinations vanish due to momentum conservation. This includes the terms with different left and right chiral halves, which may have led to cocycle factors. The conformal block is 
\be
{\cal F}_{\Phi_k}^{J\Phi_k\Phi_{N-z}\Phi^i}(z) = z^{\frac{k-N}{2}} \frac{1-\frac{(N-k)z}{k}}{(1-z)^{\frac{k(k-N)}{2N}}}. 
\ee 
The other relevant blocks can be determined similarly. Then using (\ref{fe}) we come to 
\be
\ba{lll}
F\!\left[ \begin{array}{cc}  
      \Phi_k & \Phi_{N-k} \\ 
      J & \Phi^{i} 
   \end{array}\right]_{{}_{{}_{{}_{{}_{}}}}}
                    = \; {\displaystyle \frac{N}{k}}       &
F\!\left[ \begin{array}{cc} 
      \Phi^{i} & \Phi_{N-k} \\ 
      J & \Phi_k 
   \end{array}\right]= \; {\displaystyle \frac{k}{N}}      &
F\!\left[ \begin{array}{cc}  
      \Phi_{N-k} & \Phi^{i} \\ 
      J & \Phi_k 
   \end{array}\right]= \; {\displaystyle \frac{k}{k-N}}    \\ 
F\!\left[ \begin{array}{cc}  
      J & \Phi_{N-k} \\ 
      \Phi_k & \Phi^{i} 
   \end{array}\right]_{{}_{{}_{{}_{{}_{}}}}}
                     = \; {\displaystyle \frac{k-N}{k}}     &
F\!\left[ \begin{array}{cc} 
      J & \Phi_{k} \\ 
      \Phi^{i} & \Phi_{N-k} 
   \end{array}\right]= \; {\displaystyle \frac{k}{N}}       &
F\!\left[ \begin{array}{cc}  
      J & \Phi^{i} \\ 
      \Phi_k & \Phi_{N-k} 
   \end{array}\right]= \; {\displaystyle \frac{N}{k}}      \\
F\!\left[ \begin{array}{cc}  
      \Phi_k & J \\ 
      \Phi^{i} & \Phi_{N-k} 
   \end{array}\right]_{{}_{{}_{{}_{{}_{}}}}}
                     = \; {\displaystyle \frac{k}{k-N}}     &
F\!\left[ \begin{array}{cc} 
      \Phi_{N-k} & J \\ 
      \Phi_{k} & \Phi^{i} 
   \end{array}\right]= \; {\displaystyle \frac{N-k}{N}}     &
F\!\left[ \begin{array}{cc}  
      \Phi^{i} & J \\ 
      \Phi_k & \Phi_{N-k} 
   \end{array}\right]= \; {\displaystyle \frac{N}{N-k}}    \\ 
F\!\left[ \begin{array}{cc}  
      \Phi^{i} & \Phi_{N-k} \\ 
      \Phi_k & J 
   \end{array}\right]= \; {\displaystyle \frac{k}{k-N}}     &
F\!\left[ \begin{array}{cc} 
      \Phi_{N-k} & \Phi_{k} \\ 
      \Phi^{i} & J 
   \end{array}\right]= \; {\displaystyle \frac{N}{k}}       &
F\!\left[ \begin{array}{cc}  
      \Phi_{N-k} & \Phi^{i} \\ 
      \Phi_k & J 
   \end{array}\right]= \; {\displaystyle \frac{k}{N}}
\ea
\label{Jk(N-k)i}
\ee 
The result for $N$ odd actually turns out to be the same. After doing the substitutions $\phi^i \rightarrow \hat{\phi}^i$, expanding the correlators and suppressing the vanishing terms we get 
\be
\langle J \phi_k \phi_{N-k} \hat{\phi}^{i} \rangle = 
   \langle (-1-\I (-1)^i) \partial X
   e^{\I\frac{k}{R}X} e^{\I\frac{N-k}{R}X} e^{\I\frac{-N}{R}X} 
   \times C.C.\rangle 
\ee 
The factor $(-1-\I (-1)^i)$ is the operator product coefficient for $\hat{\phi}^i \; {\phi}_k \sim (-1-\I (-1)^i) {\phi}_{N-k}+\cdots$, as one can check by decomposing the orbifold fields into free boson factors. Therefore the conformal block normalization is the same of $N$ even and so is the result for the fusing matrices. The pentagon and hexagon do not require any sign changes here.

\subsubsection{Case $F[\Phi_a \Phi_a \Phi^i \Phi^{j}]$}

Here cocycles do play a role. To see this consider our first sub-case, the $\langle \phi_a \phi^i \phi_a \phi^{j} \rangle$ correlator 
\be 
&& \langle \phi_a \phi^i \phi_a \phi^{j} \rangle =  
   \langle \left(  
      e^{\I\frac{ a}{R}X} e^{\I\frac{ a}{R}\overline{X}} +  
      e^{\I\frac{-a}{R}X} e^{\I\frac{-a}{R}\overline{X}}  
   \right) 
   \left( (e^{\I\frac{N}{R}X} + (-1)^i e^{\I\frac{-N}{R}{X}}) 
          (e^{\I\frac{N}{R}\overline{X}} + (-1)^i 
           e^{\I\frac{-N}{R}\overline{X}} ) 
   \right) \times \nn \\ 
&&{}\;\;\; \left(  
      e^{\I\frac{ a}{R}X} e^{\I\frac{ a}{R}\overline{X}} + 
      e^{\I\frac{-a}{R}X} e^{\I\frac{-a}{R}\overline{X}}  
   \right)
   \left( (e^{\I\frac{N}{R}X} + (-1)^j e^{\I\frac{-N}{R}{X}}) 
          (e^{\I\frac{N}{R}\overline{X}} + (-1)^j 
           e^{\I\frac{-N}{R}\overline{X}} ) 
   \right) \rangle \nn \\ 
&&=\langle 
      e^{\I\frac{ a}{R}X} e^{\I\frac{ N}{R}X} 
      e^{\I\frac{-a}{R}X} e^{\I\frac{-N}{R}X} \times C.C.  
   + (-1)^{i+j} e^{\I\frac{a}{R}{X}} e^{\I\frac{N}{R}{X}}
      e^{\I\frac{-a}{R}{X}} e^{\I\frac{-N}{R}{X}} 
      e^{\I\frac{ a}{R}\overline{X}} e^{\I\frac{-N}{R}\overline{X}}
      e^{\I\frac{-a}{R}\overline{X}} e^{\I\frac{ N}{R}\overline{X}} \nn \\ 
&&{}\;\;\;+ 
      e^{\I\frac{a}{R}X} e^{\I\frac{-N}{R}X} e^{\I\frac{-a}{R}X} 
      e^{\I\frac{N}{R}X} \times C.C.  
   + (-1)^{i+j} e^{\I\frac{a}{R}{X}} e^{\I\frac{-N}{R}{X}}
      e^{\I\frac{-a}{R}{X}}e^{\I\frac{N}{R}{X}} 
      e^{\I\frac{ a}{R}\overline{X}} e^{\I\frac{ N}{R}\overline{X}}
      e^{\I\frac{-a}{R}\overline{X}} e^{\I\frac{-N}{R}\overline{X}}  
\rangle \nn \\
&&=\left| 
      \frac{z^{a/2}}{(1-z)^{a/2}} + (-1)^{i+j+a} \frac{(1-z)^{a/2}}{z^{a/2}} 
   \right|^2. 
\label{aaii,j} 
\ee 
In going from the first to the second line we applied momentum conservation. Since the expression in the second line has terms which are not left-right symmetric, one has to check for cocycles. The second and fourth terms of the sum on the second line have both factors $\sigma_3^a \sigma_1 \sigma_3^a \sigma_1$. When we push $\sigma_1$ through $\sigma_3^a$ we get an extra sign $(-1)^a$, and that the reason this factor appears in the last line. The conformal block ${\cal F}_{\Phi_{N-a}}$ is just the quantity inside the absolute square and has the correct asymptotics. Comparing it with the $T$-channel blocks we get a fusing matrix $F=(-1)^{i+j+a}$.  
 
We can compare the result (\ref{aaii,j}) with the tensor product of two Ising models, to which extended orbifold theory is known to be equivalent at $N=2$. The double Ising correlator corresponding to (\ref{aaii,j}) can be evaluated straightforwardly using standard methods and blocks extracted. For the same gauge choice, the fusing matrix is $-1$, so the cocycle factor $(-1)^a$ gives precisely the minus sign which necessary to make the results of the two conformal field theories match.

The permutations of the $\langle \phi_a \phi^i \phi_a \phi^j \rangle$ correlator can also be calculated similarly. For instance 
\be 
&& \langle \phi_a \phi_a \phi^i \phi^{j} \rangle =  
   \langle \left(  
      e^{\I\frac{ a}{R}X} e^{\I\frac{ a}{R}\overline{X}} +  
      e^{\I\frac{-a}{R}X} e^{\I\frac{-a}{R}\overline{X}}  
   \right) \left(  
      e^{\I\frac{ a}{R}X} e^{\I\frac{ a}{R}\overline{X}} + 
      e^{\I\frac{-a}{R}X} e^{\I\frac{-a}{R}\overline{X}}  
   \right) \times \nn \\ 
&&{}\; \left( ( 
      e^{\I\frac{N}{R}X} + (-1)^i e^{\I\frac{-N}{R}{X}}) (  
      e^{\I\frac{N}{R}\overline{X}} + (-1)^i e^{\I\frac{-N}{R}\overline{X}} ) 
   \right) \left( ( 
      e^{\I\frac{N}{R}X} + (-1)^j e^{\I\frac{-N}{R}{X}}) (  
      e^{\I\frac{N}{R}\overline{X}} + (-1)^j e^{\I\frac{-N}{R}\overline{X}} ) 
   \right) \rangle \nn \\ 
&&= \langle 
      e^{\I\frac{a}{R}X} e^{\I\frac{-a}{R}X} e^{\I\frac{N}{R}X}
      e^{\I\frac{-N}{R}X}\times C.C.  + (-1)^{i+j} e^{\I\frac{a}{R}{X}}
      e^{\I\frac{-a}{R}{X}} e^{\I\frac{N}{R}{X}} e^{\I\frac{-N}{R}{X}} 
      e^{\I\frac{a}{R}\overline{X}} e^{\I\frac{-a}{R}\overline{X}}
      e^{\I\frac{-N}{R}\overline{X}} e^{\I\frac{N}{R}\overline{X}}  
   \nn \\ &&{} \;\;\;+ 
      e^{\I\frac{a}{R}X} e^{\I\frac{-a}{R}X} e^{\I\frac{-N}{R}X}
      e^{\I\frac{N}{R}X}\times C.C. + (-1)^{i+j} e^{\I\frac{a}{R}{X}}
      e^{\I\frac{-a}{R}{X}} e^{\I\frac{-N}{R}{X}} e^{\I\frac{N}{R}{X}} 
      e^{\I\frac{a}{R}\overline{X}} e^{\I\frac{-a}{R}\overline{X}}
      e^{\I\frac{N}{R}\overline{X}}e^{\I\frac{-N}{R}\overline{X}}  
   \rangle \nn \\ 
&&= \frac{1}{|z|^{N}} 
   \left|  {(1-z)^{a/2}} + (-1)^{i+j} {(1-z)^{-a/2}} \right|^2.
\label{aiaj}
\ee
The cocycles matrices are $\sigma_3^a\sigma_3^a\sigma_1\sigma_1$, so there are no sign changes in this case. The conformal block is the quantity inside the absolute square (divided by 2 when $\eta=1$ or by $a$ otherwise) and is to be compared to the $T$-channel result 
\be 
\langle \phi_a \phi^{j} \phi^i \phi_a \rangle = \frac{1}{|1-z|^{N}}\left| z^{a/2}+\frac{(-1)^{i+j}}{z^{a/2}}\right|^2.  
\ee 
This yields $F[aa11]=1/2,\; F[aa12]=1/a$. Repeating the process for all cases we get for $N$ even 
\be
\ba{ll}
F\!\left[ \begin{array}{cc}  
      \Phi^i & \Phi_a \\ 
      \Phi_a & \Phi^{j} 
   \end{array}\right]= {\displaystyle (-1)^{i+j+a}}
& F\!\left[ \begin{array}{cc}  
      \Phi_a & \Phi^{j} \\ 
      \Phi^i & \Phi_a 
   \end{array}\right]_{{}_{{}_{{}_{{}_{}}}}}= {\displaystyle (-1)^{i+j+a}} 
\\ 
F\!\left[ \begin{array}{cc} 
      \Phi_a & \Phi^{i} \\ 
      \Phi_a & \Phi^{i} 
   \end{array}\right]= {\displaystyle \frac{1}{2}} \;\;\;
F\!\left[ \begin{array}{cc} 
      \Phi_a & \Phi^{i} \\ 
      \Phi_a & \Phi^{j} 
   \end{array}\right]= {\displaystyle \frac{1}{a}}  
& F\!\left[ \begin{array}{cc} 
      \Phi^i & \Phi^{i} \\ 
      \Phi_a & \Phi_{a} 
   \end{array}\right]= {\displaystyle 2} \;\;\;
F\!\left[ \begin{array}{cc} 
      \Phi^i & \Phi^{j} \\ 
      \Phi_a & \Phi_a 
   \end{array}\right]_{{}_{{}_{{}_{{}_{}}}}}= {\displaystyle a}
\\ 
F\!\left[ \begin{array}{cc} 
      \Phi^{i} & \Phi_a \\ 
      \Phi^{i} & \Phi_a 
   \end{array}\right]= {\displaystyle\frac{1}{2}} \;\;\;
F\!\left[ \begin{array}{cc} 
      \Phi^j & \Phi_a \\ 
      \Phi^i & \Phi_a 
   \end{array}\right]= {\displaystyle\frac{1}{a}} 
& F\!\left[ \begin{array}{cc} 
      \Phi_a & \Phi_a \\ 
      \Phi^{i} & \Phi^{i} 
   \end{array}\right]= {\displaystyle 2} \;\;\;
F\!\left[ \begin{array}{cc}
      \Phi_a & \Phi_a \\ 
      \Phi^i & \Phi^j 
   \end{array}\right]= {\displaystyle a}
\ea
\ee 
In the last two lines, $i\neq j$. 
The $N$ odd result differs. Upon substitution $\phi^i \rightarrow \hat{\phi}^i$ we get 
\be 
\langle \phi_a \hat{\phi}^i \phi_a \hat{\phi}^{j} \rangle_{\text{$N$odd}} &=&  
   (1-(-1)^{i+j}) \langle \phi_a {\phi}^1 \phi_a {\phi}^{1} 
   \rangle_{\text{$N$ even} \rightarrow \text{$N$ odd}} \nn\\  && + \; 
   \I((-1)^i+(-1)^j) \langle \phi_a {\phi}^1 \phi_a {\phi}^{2} 
   \rangle_{\text{$N$ even} \rightarrow \text{$N$ odd}}.
\ee 
Since the factors $(1-(-1)^{i+j}),\, \I((-1)^i+(-1)^j)$ turn out to be odd $N$ operator product coefficients, the fusing matrices' normalizations remain intact. Ultimately, in the $N$ odd case the net effect of the substitution $\phi^i \rightarrow \hat{\phi}^i$ is to interchange fusing matrices with $i=j$ for those with $i\neq j$ and vice-versa. We could have also guessed this just by looking at the $N$ odd fusion rules. The fusing matrices are then
\be
\ba{ll}
F\!\left[ \begin{array}{cc}  
      \Phi^i & \Phi_a \\ 
      \Phi_a & \Phi^{j} 
   \end{array}\right]= {\displaystyle -(-1)^{i+j+a}}
& F\!\left[ \begin{array}{cc}  
      \Phi_a & \Phi^{j} \\ 
      \Phi^i & \Phi_a 
   \end{array}\right]_{{}_{{}_{{}_{{}_{}}}}}= {\displaystyle -(-1)^{i+j+a}} 
\\ 
F\!\left[ \begin{array}{cc} 
      \Phi_a & \Phi^{i} \\ 
      \Phi_a & \Phi^{j} 
   \end{array}\right]= {\displaystyle \frac{1}{2}} \;\;\;
F\!\left[ \begin{array}{cc} 
      \Phi_a & \Phi^{i} \\ 
      \Phi_a & \Phi^{i} 
   \end{array}\right]= {\displaystyle \frac{1}{a}}  
& F\!\left[ \begin{array}{cc} 
      \Phi^i & \Phi^{j} \\ 
      \Phi_a & \Phi_{a} 
   \end{array}\right]= {\displaystyle 2} \;\;\;
F\!\left[ \begin{array}{cc} 
      \Phi^i & \Phi^{i} \\ 
      \Phi_a & \Phi_a 
   \end{array}\right]_{{}_{{}_{{}_{{}_{}}}}}= {\displaystyle a}
\\ 
F\!\left[ \begin{array}{cc} 
      \Phi^{j} & \Phi_a \\ 
      \Phi^{i} & \Phi_a 
   \end{array}\right]= {\displaystyle \frac{1}{2}} \;\;\;
F\!\left[ \begin{array}{cc} 
      \Phi^i & \Phi_a \\ 
      \Phi^i & \Phi_a 
   \end{array}\right]= {\displaystyle \frac{1}{a}} 
& F\!\left[ \begin{array}{cc} 
      \Phi_a & \Phi_a \\ 
      \Phi^{i} & \Phi^{j} 
   \end{array}\right]= {\displaystyle 2} \;\;\;
F\!\left[ \begin{array}{cc}
      \Phi_a & \Phi_a \\ 
      \Phi^i & \Phi^i 
   \end{array}\right]= {\displaystyle a}
\ea
\ee
Again, in the second and third line $i \neq j$. Solving the pentagon and hexagon does not bring any sign flips.

\subsubsection{Case $F[\Phi^i \Phi_a \Phi_b \Phi_c]$} 
 
Here we can have again propagations at descendant levels. Cocycle factors are trivial here because when we have only one field $\phi^i$ in a correlator, the left and right chiral halves of left-right asymmetric terms can never conserve momentum simultaneously. Expanding the correlator $\langle \phi^i \phi_a \phi_b \phi_c \rangle$ we get 
\be 
&& \langle \phi^i \phi_k \phi_a \phi_b\rangle = \langle  
   \left( ( e^{\I\frac{N}{R}X} + (-1)^i e^{\I\frac{-N}{R}X} )
          ( e^{\I\frac{N}{R}\overline{X}} + (-1)^i 
            e^{\I\frac{-N}{R}\overline{X}}) \right) \nn \\ 
&&{} \;\;\; \times
   \left( e^{\I\frac{ a}{R}X} e^{\I\frac{ a}{R}\overline{X}} +
          e^{\I\frac{-a}{R}X} e^{\I\frac{-a}{R}\overline{X}} \right) 
   \left( e^{\I\frac{ b}{R}X} e^{\I\frac{ b}{R}\overline{X}} + 
          e^{\I\frac{-b}{R}X} e^{\I\frac{-b}{R}\overline{X}} \right) 
   \left( e^{\I\frac{ c}{R}X} e^{\I\frac{ c}{R}\overline{X}} + 
          e^{\I\frac{-c}{R}X} e^{\I \frac{-c}{R}\overline{X}} \right) 
   \rangle \nn \\ 
&&= \mbox{8 terms, of which only one is non-vanishing}. 
\ee 
Suppose for instance that $k=a+b-N$. Then $\langle \phi^i \phi_k \phi_a \phi_b\rangle = |1-z|^{-ab/N}|z|^{b(a+b-N)/N}$ and nothing more contributes. Comparing with the $T$-channel gives $F=1$. This is same every other case and the naive fusing matrix is simply 1 whenever fusion is possible. But in this case the pentagon and hexagon equations require some sign flips when there is propagation at descendant levels, which in this case happens whenever $a+b>N$ or $b+c>N$. The correct result will depend on whether $N$ is even or odd. For $N$ even we get
\be 
F\!\left[ \begin{array}{cc}
\Phi_a & \Phi_b \\ 
\Phi^i & \Phi_c 
\end{array}\right]= \left\{ \ba{cl} -(-1)^{i+b+\frac{N}{2}} & \mbox{if $a=b+c-N$} \\ -(-1)^{i+b+\frac{N}{2}} & \mbox{if $c=a+b-N$} \\ 1 & \mbox{else}\ea \right. 
\ee 
And for $N$ odd 
\be 
F\!\left[ \begin{array}{cc} 
\Phi_a & \Phi_b \\ 
\Phi^i & \Phi_c 
\end{array}\right]= \left\{ \ba{cl} -(-1)^{i+b+\frac{N-1}{2}} & \mbox{if $a=b+c-N$} \\ (-1)^{i+b+\frac{N-1}{2}} & \mbox{if $c=a+b-N$} 
\\  1 & \mbox{else}\ea \right. 
\ee 
The permutations $F[c\Phi^i ab]$, $F[bc\Phi^i a]$ and $F[abc\Phi^i]$ are similar, the field taking the role of $\Phi_b$ now being the one opposite $\Phi^i$.
\subsubsection{Case $F[\Phi^i\Phi^j\Phi^k\Phi^l]$} 
After expanding the vertex operators we get 
\be
\langle \phi^i \phi^j \phi^k \phi^l \rangle &=& 
   \langle \left( e^{\I\frac{N}{R}X} + (-1)^i 
      e^{\I\frac{-N}{R}X} \right) \ldots 
   \left( e^{\I\frac{N}{R}X} + (-1)^l e^{\I\frac{-N}{R}X} \right) \times C.C. 
   \rangle \nn \\ 
&=&\langle ((-1)^{i+j} e^{\I\frac{ N}{R}X} e^{\I\frac{ N}{R}X} 
           e^{\I\frac{-N}{R}X} e^{\I\frac{-N}{R}X} + 
           (-1)^{i+j} e^{\I\frac{-N}{R}X} e^{\I\frac{-N}{R}X} 
           e^{\I\frac{N}{R}X}e^{\I\frac{N}{R}X}\nn \\ 
&&         (-1)^{i+k} e^{\I\frac{-N}{R}X} e^{\I\frac{N}{R}X} 
           e^{\I\frac{-N}{R}X} e^{\I\frac{ N}{R}X} +(-1)^{i+k} 
           e^{\I\frac{ N}{R}X} e^{\I\frac{-N}{R}X}
           e^{\I\frac{ N}{R}X} e^{\I\frac{-N}{R}X} \label{ijklNeven} \\ 
&&         +(-1)^{i+l} e^{\I\frac{N}{R}X} e^{\I\frac{-N}{R}X}
           e^{\I\frac{-N}{R}X} e^{\I\frac{N}{R}X} +(-1)^{i+l} 
           e^{\I\frac{-N}{R}X} e^{\I\frac{N}{R}X} e^{\I\frac{N}{R}X}
           e^{\I\frac{-N}{R}X} ) \times C.C.\rangle \nn\\ 
&=&\left| (-1)^{i+j} (1-z)^{-N/2} z^{N/2} + (-1)^{i+k} (1-z)^{-N/2} z^{N/2} + 
          (-1)^{i+l} (1-z)^{N/2} z^{-N/2} \right|^2. \nn 
\ee
All other terms vanish due to momentum conservation. Note also that $(-1)^{i+j+k+l}=1$ otherwise the correlator vanishes (which we can also see from the fusion rules). There are no cocycles here because the left-right symmetric terms contribute with cocycle matrix $\sigma_3^N$ and, since $N$ is even, the $\sigma_1$'s coming from left-right anti-symmetric terms commute. For $N$ odd there will be some cocycles though. The $N$ even conformal blocks are 
\be 
&& {\cal_F}_{0}^{\Phi^i\Phi^i\Phi^i\Phi^i}(z) 
   =\frac{1+z^N+(1-z)^N}{2z^{N/2}(1-z)^{N/2}} \;\;\; 
{\cal_F}_{0}^{\Phi^i\Phi^i\Phi^j\Phi^j}(z) 
   =\frac{1-z^N+(1-z)^N}{2z^{N/2}(1-z)^{N/2}}\nn\\ 
&& {\cal_F}_{J}^{\Phi^i\Phi^j\Phi^j\Phi^i}(z) 
   =\frac{1+z^N-(1-z)^N}{Nz^{N/2}(1-z)^{N/2}} \;\;\;
{\cal_F}_{J}^{\Phi^i\Phi^j\Phi^i\Phi^j}(z) 
   =\frac{1-z^{N}-(1-z)^{N}}{Nz^{N/2}(1-z)^{N/2}}, 
\ee 
leading to $N$ even fusing matrices 
\be 
F\!\left[ \begin{array}{cc} 
      \Phi^i & \Phi^i \\ 
      \Phi^i & \Phi^i 
   \end{array}\right]= 1 \;\;\; 
F\!\left[ \begin{array}{cc} 
      \Phi^j & \Phi^i \\ 
      \Phi^i & \Phi^j 
   \end{array}\right]= 1 \;\;\;
F\!\left[ \begin{array}{cc} 
      \Phi^j & \Phi^j \\ 
      \Phi^i & \Phi^i 
   \end{array}\right]= \frac{2}{N} \;\;\; 
F\!\left[ \begin{array}{cc} 
      \Phi^i & \Phi^j \\ 
      \Phi^i & \Phi^j 
   \end{array}\right]= \frac{N}{2}. 
\ee 
For $N$ odd we substitute $\phi^i \rightarrow \hat{\phi}^i$. This leads to 
\be 
\langle \hat{\phi}^1\hat{\phi}^1\hat{\phi}^1\hat{\phi}^1 
\rangle_{\text{$N$ odd}}&=& 
   \langle {\phi}^1{\phi}^1{\phi}^1{\phi}^1 - 
           {\phi}^1{\phi}^1{\phi}^2{\phi}^2 - 
           {\phi}^1{\phi}^2{\phi}^1{\phi}^2 - 
           {\phi}^1{\phi}^2{\phi}^2{\phi}^1 
   \rangle_{\text{$N$ even}\rightarrow \text{$N$ odd}}\nn \\ 
\langle \hat{\phi}^1\hat{\phi}^1\hat{\phi}^2\hat{\phi}^2 
\rangle_{\text{$N$ odd}}&=& 
   \langle {\phi}^1{\phi}^1{\phi}^1{\phi}^1 - 
           {\phi}^1{\phi}^1{\phi}^2{\phi}^2 +
           {\phi}^1{\phi}^2{\phi}^1{\phi}^2 + 
           {\phi}^1{\phi}^2{\phi}^2{\phi}^1
   \rangle_{\text{$N$ even}\rightarrow \text{$N$ odd}}\nn \\ 
\langle \hat{\phi}^1\hat{\phi}^2\hat{\phi}^2\hat{\phi}^1 
\rangle_{\text{$N$ odd}}&=& 
   \langle {\phi}^1{\phi}^1{\phi}^1{\phi}^1 + 
           {\phi}^1{\phi}^1{\phi}^2{\phi}^2 +
           {\phi}^1{\phi}^2{\phi}^1{\phi}^2 - 
           {\phi}^1{\phi}^2{\phi}^2{\phi}^1
   \rangle_{\text{$N$ even}\rightarrow \text{$N$ odd}} \label{ijklNodd} \\ 
\langle \hat{\phi}^1\hat{\phi}^2\hat{\phi}^1\hat{\phi}^2 
\rangle_{\text{$N$ odd}}&=& 
   \langle {\phi}^1{\phi}^1{\phi}^1{\phi}^1 + 
           {\phi}^1{\phi}^1{\phi}^2{\phi}^2 -
           {\phi}^1{\phi}^2{\phi}^1{\phi}^2 + 
           {\phi}^1{\phi}^2{\phi}^2{\phi}^1
   \rangle_{\text{$N$ even}\rightarrow \text{$N$ odd}}. \nn
\ee 
So we must see now how the cocycles change (\ref{ijklNeven}) for $N$ odd. In this case we must pay attention when chirally asymmetric terms appear, e.g. terms of type $G\overline{H}$. Such terms can produce extra signs. We get 
\be 
\langle {\phi}^i{\phi}^j{\phi}^k{\phi}^l 
\rangle_{\text{$N$ even} \rightarrow \text{$N$ odd}}&=&  
   \langle |A|^2 + (-1)^{i+l} A\overline{B} -(-1)^{i+k}A\overline{C} + 
           |B|^2 + (-1)^{i+l} B\overline{A} \nn \\ && 
                 + (-1)^{i+j} B\overline{C} +|C|^2 
                 + (-1)^{i+j} C\overline{B} -(-1)^{i+k} C\overline{A}
\rangle,
\ee
\vspace{-.7cm} 
\be
A = e^{\I\frac{ N}{R}X} e^{\I\frac{ N}{R}X}
    e^{\I\frac{-N}{R}X} e^{\I\frac{-N}{R}X},\;\; 
B = e^{\I\frac{ N}{R}X} e^{\I\frac{-N}{R}X}
    e^{\I\frac{ N}{R}X} e^{\I\frac{-N}{R}X},\;\;
C = e^{\I\frac{ N}{R}X} e^{\I\frac{-N}{R}X}
    e^{\I\frac{-N}{R}X} e^{\I\frac{ N}{R}X}. \nn
\ee 
Indeed we see minus signs appearing for $A\overline{C}$ and $C\overline{A}$ due to the cocycle factors. Doing the sums to obtain the hatted fields of (\ref{ijklNodd}) we get 
\be
\langle \hat{\phi}^1\hat{\phi}^1\hat{\phi}^1\hat{\phi}^1 
\rangle_{\text{$N$ odd}}&=& 
     -\left| (1-z)^{-N/2}z^{N/2} - (1-z)^{-N/2}z^{-N/2} + (1-z)^{N/2} z^{-N/2} 
      \right|^2, \nn \\ 
\langle \hat{\phi}^1\hat{\phi}^1\hat{\phi}^2\hat{\phi}^2 
\rangle_{\text{$N$ odd}}&=& 
      \left| (1-z)^{-N/2} z^{N/2} + (1-z)^{-N/2}z^{-N/2} - (1-z)^{N/2}z^{-N/2}
      \right|^2, \nn \\ 
\langle \hat{\phi}^2\hat{\phi}^2\hat{\phi}^1\hat{\phi}^1 
\rangle_{\text{$N$ odd}}&=& 
      \left| -(1-z)^{-N/2} z^{N/2} + (1-z)^{-N/2}z^{-N/2} + (1-z)^{N/2}z^{-N/2}
      \right|^2, \\ 
\langle \hat{\phi}^1\hat{\phi}^2\hat{\phi}^1\hat{\phi}^2 
\rangle_{\text{$N$ odd}}&=& 
      \left| (1-z)^{-N/2} z^{N/2} + (1-z)^{-N/2}z^{-N/2} + (1-z)^{N/2}z^{-N/2}
      \right|^2. \nn
\ee
The blocks coming from these correlators give rise to $N$ odd matrices 
\be 
F\!\left[ \begin{array}{cc} 
      \Phi^i & \Phi^i \\ 
      \Phi^i & \Phi^i 
   \end{array}\right]= 1 \;\;\;
F\!\left[ \begin{array}{cc}
      \Phi^j & \Phi^i \\ 
      \Phi^i & \Phi^j 
   \end{array}\right]= 1 \;\;\; 
F\!\left[ \begin{array}{cc} 
      \Phi^i & \Phi^j \\ 
      \Phi^i & \Phi^j 
   \end{array}\right]= \frac{2}{N} \;\;\; 
F\!\left[ \begin{array}{cc} 
      \Phi^j & \Phi^j \\ 
      \Phi^i & \Phi^i 
   \end{array}\right]= \frac{N}{2}.
\ee 
And this is again left untouched by the polynomial equations.
 
\subsubsection{Cases $F[\Phi_a\Phi_b\Phi_c\Phi_d]$} 
So far the fusing matrices were 1x1. For the $F[\Phi_a\Phi_b\Phi_c\Phi_d]$ case they can and will in general be larger than 1x1. The naive results will be valid for odd and even $N$, but there will be special cases involving chiral fields $\Phi_{N/2}$, which obviously only exist in the $N$ even case. Furthermore there are no cocycle factors, since we are dealing with left-right symmetric vertex operators. Still, new features arise. 
 
Expanding the orbifold correlator $\langle \Phi_a \Phi_b \Phi_c \Phi_d \rangle$ into free boson factors we get 
\be 
\langle \Phi_a \Phi_b \Phi_c \Phi_d \rangle &=& \frac{1}{4} 
   \langle ( e^{\I\frac{ a}{R}X} e^{\I\frac{ a}{R}\overline{X}}  +
             e^{\I\frac{-a}{R}X} e^{\I\frac{-a}{R}\overline{X}} ) 
           ( e^{\I\frac{ b}{R}X} e^{\I\frac{ b}{R}\overline{X}}  +
             e^{\I\frac{-b}{R}X} e^{\I\frac{-b}{R}\overline{X}} ) \nn \\ 
&& \times  ( e^{\I\frac{ c}{R}X} e^{\I\frac{ c}{R}\overline{X}}  + 
             e^{\I\frac{-c}{R}X} e^{\I\frac{-c}{R}\overline{X}} ) 
           ( e^{\I\frac{ d}{R}X} e^{\I\frac{ d}{R}\overline{X}}  +
             e^{\I\frac{-d}{R}X} e^{\I\frac{-d}{R}\overline{X}} ) 
\rangle \nn \\ &=& \um 
   \left( |1-z|^{\frac{2bc}{2N}} |z|^{\frac{2cd}{2N}} \delta_{a+b+c+d,0} + 
          |1-z|^{\frac{2bc}{2N}} |z|^{\frac{2cd}{2N}} \delta_{-a+b+c+d,0} 
\right. \nn \\ && + 
          |1-z|^{\frac{-2bc}{2N}} |z|^{\frac{ 2cd}{2N}} \delta_{a-b+c+d,0} + 
          |1-z|^{\frac{-2bc}{2N}} |z|^{\frac{-2cd}{2N}} \delta_{a+b-c+d,0}  
\label{abcd} \\ && + 
          |1-z|^{\frac{ 2bc}{2N}} |z|^{\frac{-2cd}{2N}} \delta_{a+b+c-d,0} + 
          |1-z|^{\frac{-2bc}{2N}} |z|^{\frac{ 2cd}{2N}} \delta_{a+b-c-d,0}  
\nn \\ && + \left. 
          |1-z|^{\frac{ 2bc}{2N}} |z|^{\frac{-2cd}{2N}} \delta_{a-b+c-d,0} + 
          |1-z|^{\frac{-2bc}{2N}} |z|^{\frac{-2cd}{2N}} \delta_{a-b-c+d,0} 
\right). \nn
\ee 
So we have eight terms that may contribute. As we shall see, some correlators have to be evaluated at descendant level and for this reason the first term can sometimes contribute when one of the external legs is a descendant field. In the following, we abbreviate $\Phi_a,\Phi_b,\Phi_c,\Phi_d,\Phi_k$ by the charge indices $a,b,c,d,k$ in order to keep the notation clean.

\subsubsection{Case $F[ \Phi_a \Phi_a \Phi_a \Phi_a ]$} 
 In this case we have to distinguish two possibilities: $a=N/2$ and $a\neq N/2$. First take $a\neq N/2$. After setting $a,b,c,d=a$ in (\ref{abcd}) we get 
\be 
\langle aaaa \rangle = \um
   \left( |1-z|^{-2x}|z|^{2x} + |1-z|^{-2x}|z|^{-2x} +|1-z|^{2x}|z|^{-2x}
   \right),\;\;\; x=\frac{a^2}{2N}. \label{Faaaa}
\ee
For $2a<N$ the fusion rule $\Phi_a\times\Phi_a =  0+J+ \Phi_{2a}$ tells us what to expect in the $S$-channel. Getting the correct conformal block asymptotics requires a rewriting of the correlator, so we use 
\be 
   |1-z|^A + \frac{1}{|1-z|^A} = \frac{1}{2|1-z|^A}{
   \left( |1+(1-z)^A|^2 + |1-(1-z)^A|^2 \right).} 
\label{blockreshuffling} 
\ee 
With this we can reorganize the correlator and extract the conformal blocks, which turn out to be 
\be 
{\cal F}_{0}^{aaaa}(z) &=& \um z^{-x} (1-z)^{-x} (1+(1-z)^{2x}), \nn \\ 
{\cal F}_{J}^{aaaa}(z) &=& 
   \frac{1}{2x} z^{-x} (1-z)^{-x} (1-(1-z)^{2x}), \nn \\ 
{\cal F}_{2a}^{aaaa}(z) &=& z^x (1-z)^{-x}. 
\ee 
When $2a>N$ the fusion is $\Phi_a\times\Phi_a = 0+J+\Phi_{2(N-a)}$, so we would expect a conformal block behaving as 
\be 
{\cal F}_{2(N-a)}^{aaaa} \sim z^{h_{2(N-a)} - 2h_a}, 
\ee 
but the block extracted from the correlator always behaves as $z^{h_{2a} - 2h_a}$ regardless. This is again what we encountered before. If $2a>N$ the fusion, which should be into (the non-existent) $\Phi_{2a}$, is actually into a descendant of $\Phi_{2(N-a)}$, at level $2a-N$. So, while the fusion rule $N_{aa}^{\ph{aa}2(N-a)}=1$, the respective operator product coefficients only start to be non-vanishing at descendant level $M=2a-N$. As we said above, in principle one would need to absorb a descendant $\beta$ coefficient into the descendant block, but in this case from the explicit operator product expansion we see that $\beta=1$. When writing the fusing matrix, we take care of both $2a>N$ and $2a<N$ by writing $[2a]$, as in (\ref{orblabelmap}). The fusing equation gives us a fusing matrix 
\be
F\! \left[ \begin{array}{cc} a & a \\ a & a \end{array} \right] = 
   \begin{array}{c|ccc} & 0 & J & \Phi_{[2a]} \\ \hline 
                        0 & \um & \frac{x}{2} & \um \\ 
                        J & \frac{1}{2x} & \um & -\frac{1}{2x} \\ 
                        \Phi_{[2a]} & 1 & -x & 0  
   \end{array}\;\;\; x=\frac{a^2}{2N}.
\ee
This is valid for $a \neq N/2$ and all $N$. (Rows and columns correspond to the labels $p,q$ of $F_{pq}[ijkl]$.)

When $N$ is even, there is the extra case $a=N/2$, for which the fusion rules change. They are $\Phi_a \times \Phi_a = 0 + J + \Phi^1 + \Phi^2$, so there there should be an extra conformal block. The $0$ and $J$ blocks remain the same but the block for $\Phi_{2a}$ is expected to split into two. However from the explicit expression (\ref{Faaaa}) the correlator remains the same, so we conclude that the block 
\be 
   {\cal F}_{\Phi^{i}}^{\frac{N}{2}\frac{N}{2}\frac{N}{2}\frac{N}{2}} = 
   z^x (1-z)^{-x}, \;\;\; x=N/8,
\ee 
is in fact degenerate. Thus both $\Phi^{1}$ and $\Phi^2$ share the same conformal block function. Due to this degeneracy, the fusing equation cannot give us the exact form of the fusing matrix. For instance, we cannot get more than $F_{0\Phi^1}+F_{0\Phi^2}=1/2$ and $F_{J\Phi^1}+F_{J\Phi^2}=-1/2x$ from it. We see no reason to treat matrix elements involving $\Phi^{1}$ and $\Phi^2$ in a different, non-symmetric way, like for instance $F_{0\Phi^1}=1/8, F_{0\Phi^2}=3/8$, so we choose $F_{0\Phi^1}=F_{0\Phi^2}=1/4$. At present we don't know of any way to lift this degeneracy, which appears due to the chiral algebra extension. A comparison with fusing matrices of the double Ising model also hints at a symmetric splitting, and this also is why we choose this way to split the result. The pentagon and hexagon will in the end tell us whether this way of splitting is adequate or not. Also, at this stage we do not know anything about matrix elements $F_{\Phi^i\Phi^j}$ except that some of them must add up to zero. By looking at the descendant correlator $\langle \frac{-3N}{2}\frac{N}{2}\frac{N}{2} \frac{N}{2} \rangle$, we can improve on this and get $|F_{\Phi^i\Phi^j}|=1/2$, although again with the assumption of symmetric splitting. (Again, this matches the Ising squared result.) Putting together all this information, we get to a candidate fusing matrix of 
\be
F\! \left[ \begin{array}{cc}
       \frac{N}{2} & \frac{N}{2} \\ 
       \frac{N}{2} & \frac{N}{2} 
    \end{array} \right] 
  = \begin{array}{c|cccc} 
       & 0 & J & \Phi^{1} & \Phi^2 \\ \hline 
       0 & \um & \frac{x}{2} & \frac{1}{4} & \frac{1}{4} \\ 
       J & \frac{1}{2x} & \um & -\frac{1}{4x} & -\frac{1}{4x} \\ 
       \Phi^{1} & 1 & -x & \um (-1)^{\frac{N}{2}} & 
          -\um (-1)^{\frac{N}{2}} \\ 
       \Phi^2 & 1 & -x & -\um (-1)^{\frac{N}{2}} & \um (-1)^{\frac{N}{2}}  
    \end{array} \;\;\; x=\frac{N}{8}. 
\ee
And in the end this result does turn out to be the one that satisfies the pentagon and hexagon identities. Also, these identities could not be solved without the symmetric splitting assumption.

\subsubsection{Case $F[ \Phi_a \Phi_a \Phi_a \Phi_b ]$ and permutations} 
We require $b \neq a$. Starting with $F[aaab]$, we set the indices in (\ref{abcd}) and see we get something non-vanishing if $3a-b=0$. This signals propagation of field $\Phi_{[a-b]}$ in the $S$-channel. The correlator is
\be 
   \langle aaab \rangle = \um |1-z|^{2x}|z|^{\pm 2x'}, 
   \;\;\; x=\frac{a^2}{2N}, \;\;\; x'=\frac{ab}{2N}. 
\ee 
The propagating field $\Phi_{[a-b]}$ has conformal block 
\be 
{\cal F}_{[a-b]}^{aaab}(z) = (1-z)^x z^{-x'}. 
\ee 
To get fusing matrices we compare with 
\be 
   \langle abaa \rangle = \um |1-z|^{-2x'}|z|^{2x}. 
\ee 
Blocks for the propagating field $\Phi_{[a-b]}$ are 
\be 
{\cal F}_{[a-b]}^{abaa}(z) = (1-z)^{-x'} z^{x}. 
\ee 
The fusing matrix is then simply 1. This is however not the whole story; there are three ways more to get a non-vanishing fusing matrix. First, if $3a-b=2N$ there can be $S$-channel propagation of $\Phi_{2(N-a)}$ at the descendant level. Second and third, if $3a+b=2N$ there can be $S$-channel propagation at descendant level of either $\Phi_{[a+b]}$ (when $a+b>N$) or $\Phi_{2(N-a)}$ (when $2a>N$). For all these three cases we can take the first extension descendant for $\phi_b$, evaluate blocks for $\langle aaa(b-2N)\rangle$ and compare with blocks $\langle a(b-2N)aa \rangle$. We get the same $F=1$ in the end. Precisely whenever descendants are propagating, the pentagon and hexagon equations impose some sign changes here. In the end we get 
\be
F\!\left[ \begin{array}{cc} a & a \\ a & b \end{array} \right]=
   \left\{ \ba{cl} 1 & \mbox{if $b=3a$} \\ (-1)^a & \mbox{else} \ea 
   \right. 
\ee 
This is also valid for any permutations of $F[aaab]$ and for any $N$. Except for some special cases, in the matrix $F[abcd]$ is in general 1x1. Later on we will see that the result above is a part of the more general result 
(\ref{genabcd}).

\subsubsection{Case $F[ \Phi_a \Phi_a \Phi_b \Phi_b ]$ and permutations} 
Here the possibility $a+b=N$ leads again to different fusion and is treated separately. Start with $\langle aabb \rangle$. Fusion is $\Phi_a \times \Phi_a = 0+J+\Phi_{[2a]}$. We expect thus three blocks, but in fact we have only two since propagation of $\Phi_{[2a]}$ requires $a=b$ (already done) or $a=N-b$ (treated separately). Setting indices in (\ref{abcd}) we have 
\be 
\langle aabb \rangle = \um 
   \left( |1-z|^{-2x'}|z|^{-2x}+|1-z|^{2x'}|z|^{-2x} 
   \right), \;\;\; x=\frac{b^2}{2N},\;\;\; x'=\frac{ab}{2N}. 
\ee 
With reorganization of the absolute values we come to conformal blocks 
\be 
{\cal F}_{0}^{aabb}(z) &=& \frac{1}{ 2(1-z)^{x'}z^{x}}(1+(1-z)^{2x'}), \nn \\ 
{\cal F}_{J}^{aabb}(z) &=& \frac{1}{2x(1-z)^{x'}z^{x}}(1-(1-z)^{2x'}). 
\ee
The fusion equation requires now $\langle abba \rangle$ for comparison. That turns out to be 
\be 
\langle abba \rangle = \um 
   \left( |1-z|^{-2x}|z|^{-2x'}+|1-z|^{-2x}|z|^{2x'} 
   \right). 
\ee 
From the fusion rules we see that the fields propagating are $\Phi_a\times \Phi_b = \Phi_{[a+b]} + \Phi_{[a-b]}$, with blocks 
\be 
{\cal F}_{[a+b]}^{abba}(z) = (1-z)^{-x} z^{x'},  \;\;\;
{\cal F}_{[a-b]}^{abba}(z) = (1-z)^{-x} z^{x'}.
\ee 
The result stands for both $a+b<N$ and $a+b>N$ since the descendant operator product coefficients are 1. The remaining case is $\langle abab \rangle$, which gives 
\be
\langle abab \rangle = \um 
   \left( |1-z|^{2x}|z|^{-2x} + |1-z|^{-2x}|z|^{2x}
   \right), \;\;\; x=\frac{ab}{2N},
\ee
with conformal blocks
\be 
{\cal F}_{\Phi_{[a+b]}}^{abab}(z) = (1-z)^{-x} z^{x}, \;\;\;
{\cal F}_{\Phi_{[a-b]}}^{abab}(z) = (1-z)^{x} z^{-x}. 
\ee
Extracting the fusing matrices for these three cases is straightforward. We get
\be 
&& F\!\left[ \begin{array}{cc}  
      b & a \\ 
      a & b 
      \end{array}\right]= 
   \begin{array}{c|cc}  
      & \Phi_{[a+b]} & \Phi_{[a-b]} \\ \hline 
      \Phi_{[a+b]} & 0 & \epsilon \\ 
      \Phi_{[a-b]} & \epsilon & 0 
   \end{array} \;\;\; \epsilon=
      \left\{ \ba{cl} (-1)^{a+b} & \mbox{if $a+b>N$} \\ 1 & \mbox{else} \ea 
      \right. \nn \\ 
&& F\!\left[ \begin{array}{cc}  
      a & c \\ 
      a & c 
      \end{array}\right]= 
   \begin{array}{c|cc}  
      & \Phi_{[a+c]} & \Phi_{[a-c]} \\ \hline 
      0 &     \um    &    \um \\ 
      J & \frac{-1}{2x} & \frac{1}{2x} 
   \end{array} \;\;\; 
F\!\left[ \begin{array}{cc}  
      c & c \\ 
      a & a 
   \end{array}\right]= 
   \begin{array}{c|cc}  
      &      0     &    J    \\ \hline 
      \Phi_{[a+c]}  &      1     &   -x   \\ 
      \Phi_{[a-c]}  &      1     &    x 
   \end{array} \;\;\; x=\frac{ac}{2N}
\ee
with sign $\epsilon$ coming from imposing the pentagon and hexagons identities.
 
For $b=N-a$ we have degeneracy of conformal blocks due to propagation of both $\Phi^i$ fields. In this case another interesting fact appears. Take 
\be 
\langle aa(N-a)(N-a) \rangle = \um
   \left( |1-z|^{-2x'}|z|^{-2x} + |1-z|^{2x'}|z|^{-2x} 
   \right). 
\ee 
From the asymptotics we see that the blocks for $0$ and $J$ are present but the block for $\Phi_{[2a]}$ turns out to be zero. This is unexpected, but can perhaps be a consequence of level mismatches in the towers of descendants that propagate coming from the left and right vertices of the $S$-channel. The existent blocks are 
\be 
{\cal F}_{0}^{aa(N-a)(N-a)}(z) &=& 
   \frac{1}{2(1-z)^{x'} z^{x}}(1+(1-z)^{2x'}), \nn \\ 
{\cal F}_{J}^{aa(N-a)(N-a)}(z) &=& 
   \frac{1}{2x (1-z)^{x'} z^{x}}(1-(1-z)^{2x'}). 
\ee 
For comparison with the $T$-channel we need $\langle a(N-a)(N-a)a\rangle$, which is 
\be 
\langle a(N-a)(N-a)a \rangle = \um 
   \left( |1-z|^{-2x}|z|^{-2x'} + |1-z|^{-2x}|z|^{2x'} 
   \right). 
\ee 
The last term here is a degenerate conformal block for both $\Phi^{1,2}$. Fusion is $\Phi_a \times \Phi_{N-a} = \Phi^1 +\Phi^2 +\Phi_{[N-2a]}$ and blocks 
\be 
{\cal F}_{\Phi^{1,2}}^{a(N-a)(N-a)a}(z) = (1-z)^{-x} z^{x'}, \;\;\;  
{\cal F}_{\Phi_{[N-2a]}}^{a(N-a)(N-a)a}(z) = (1-z)^{x} z^{-x'}. 
\ee
For the last case $\langle a(N-a)a(N-a)\rangle$ we have 
\be
\langle a(N-a)a(N-a) \rangle = \um 
   \left( |1-z|^{2x}|z|^{-2x} + |1-z|^{-2x}|z|^{2x} 
   \right), 
\ee 
and blocks 
\be 
{\cal F}_{\Phi^{1,2}}^{a(N-a)a(N-a)}(z) = (1-z)^{-x} z^{x}, \;\;\;
{\cal F}_{\Phi_{[N-2a]}}^{a(N-a)a(N-a)}(z) = (1-z)^{x} z^{-x}. 
\ee
By considering descendants correlators and assuming again a symmetric splitting for $\Phi^{1,2}$ we can set the overall normalization of the fusing matrix elements and let the pentagon and hexagon fix eventual signs. After doing all this we get for $N$ even
\be 
F\!\left[ \begin{array}{cc}  
      (N-a) &   a \\ 
      a  & (N-a) 
   \end{array}\right]&=& 
   \begin{array}{c|ccc}  
                &   \Phi^1   &   \Phi^2 & \Phi_{[N-2a]} \\ \hline 
      \Phi^1    &  \um (-1)^a    &  -\um (-1)^a    &      1 \\ 
      \Phi^2    & -\um (-1)^a    &   \um (-1)^a    &      1 \\ 
      \Phi_{[N-2a]} &    \um     &    \um   &      0 
   \end{array} \nn \\
F\!\left[ \begin{array}{cc}  
      (N-a) & (N-a) \\ 
      a & a 
   \end{array}\right]&=&
   \begin{array}{c|ccc}  
                    &     0     &     J    & \Phi_{[2a]} \\ \hline  
      \Phi^1        &     1     &    -x   &     (-1)^{\frac{N}{2}}     \\ 
      \Phi^2        &     1     &    -x   &    -(-1)^{\frac{N}{2}}     \\ 
      \Phi_{[N-2a]} &     1     &     x   &     0 
   \end{array} \;\;\; x=\frac{a(N-a)}{2N} \\ 
F\!\left[ \begin{array}{cc}  
      a & (N-a) \\ 
      a & (N-a) 
   \end{array}\right]&=& 
   \begin{array}{c|ccc}  
      &    \Phi^1  &   \Phi^2   &  \Phi_{[N-2a]}    \\ \hline 
      0  &      \frac{1}{4}     &   \frac{1}{4}       &  \um    \\ 
      J  &-\frac{1}{4x}         &  -\frac{1}{4x}      &  \frac{1}{2x} \\ 
      \Phi_{[2a]} & \um(-1)^{\frac{N}{2}} & -\um(-1)^{\frac{N}{2}} & 0 
\end{array} \nn
\ee
The $N$ odd result is similar, but has a few changes. Matrix $F[a(N-a)a(N-a)]$ remains the same, whereas the other two change to 
(again $x={a(N-a)}/{2N}$) 
\be 
F\!\left[ \begin{array}{cc}  
      (N-a) & (N-a) \\ 
      a & a 
   \end{array}\right]&=& 
   \begin{array}{c|ccc}  
                    &   0   &     J   & \Phi_{[2a]} \\ \hline  
      \Phi^1        &   1   &    -x   & -\epsilon(-1)^{\frac{N-1}{2}}   \\ 
      \Phi^2        &   1   &    -x   &  \epsilon(-1)^{\frac{N-1}{2}}   \\ 
      \Phi_{[N-2a]} &   1   &     x   &     0 
   \end{array} \;\;\; \epsilon=-1\;\mbox{if $2a>N$}, \nn \\
F\!\left[ \begin{array}{cc}  
      a & (N-a) \\ 
      a & (N-a) 
   \end{array}\right]&=&
   \begin{array}{c|ccc}  
         &    \Phi^1  &   \Phi^2   &  \Phi_{[N-2a]}    \\ \hline 
      0  &  \frac{1}{4}       &   \frac{1}{4}     &  \um    \\ 
      J  & -\frac{1}{4x}      &  -\frac{1}{4x}    &  \frac{1}{2x} \\ 
      \Phi_{[2a]} & -\um\epsilon(-1)^{\frac{N-1}{2}} & 
                     \um\epsilon(-1)^{\frac{N-1}{2}} & 0 
   \end{array} 
\ee 
The sign $\epsilon$ comes again from solving the polynomial equations.
 
\subsubsection{Case $F[ \Phi_a \Phi_a \Phi_b \Phi_c ]$ and permutations.} 
Most of the time only one field propagates in the $S$-channel and the fusing matrix is 1x1. For $N$ even we can have the special case of $a=N/2, \, c=N-b$, which gives a 2x2 matrix. If the matrix is 1x1, the result is part of the more general result (\ref{genabcd}) for 1x1 matrices $F[abcd]$. For the $N$ even special case $a=N/2, c=N-b$ the matrix is 2x2 and we get 
\be 
\langle \frac{N}{2} \frac{N}{2} b(N-b) \rangle = \um |1-z|^{-2x} |z|^{2x'}, 
   \;\;\; x=\frac{b}{4}, \;\;\; x'=\frac{b(N-b)}{2N}. 
\ee 
The conformal blocks are again degenerate 
\be 
{\cal F}_{\Phi^{1,2}}^{\frac{N}{2} \frac{N}{2} b(N-b)}(z)=(1-z)^{-x} z^{x'}. 
\ee
For comparing with the $T$-channel blocks we need $\langle \frac{N}{2} (N-b) b\frac{N}{2} \rangle$ 
\be 
\langle \frac{N}{2} (N-b) b \frac{N}{2} \rangle = \um |1-z|^{2x'} |z|^{-2x}. 
\ee 
From which the conformal block is 
\be
   {\cal F}_{\Phi_{[\frac{N}{2}-b]}}^{\frac{N}{2} (N-b) b \frac{N}{2}}(z) = 
   (1-z)^{x'} z^{-x}, 
\ee 
so from the asymptotics we see that the field $\Phi_{[\frac{N}{2}-b]}$ propagates at the descendant level. Using (\ref{fe}) and once again considering descendants correlators and symmetric $\Phi^{1,2}$ splitting, we get 
\be 
F\!\left[ \begin{array}{cc}
      \frac{N}{2} & b \\ 
      \frac{N}{2} & (N-b) 
   \end{array}\right]&=& 
   \begin{array}{c|cc}  
               & \Phi_{[\frac{N}{2}+b]} & \Phi_{[\frac{N}{2}-b]} \\ \hline 
       \Phi^1  &    -(-1)^b    &   1   \\ 
       \Phi^2  &     (-1)^b    &   1 
   \end{array}. 
\ee 
Following the same procedure for all permutations we get 
\be 
F\!\left[ \begin{array}{cc}  
      \frac{N}{2} & b \\
      \frac{N}{2} & (N-b)
   \end{array}\right]&=&\begin{array}{c|cc}  
             & \Phi_{[\frac{N}{2}+b]} & \Phi_{[\frac{N}{2}-b]} \\ \hline
      \Phi^1 &   -(-1)^b   &    1   \\ 
      \Phi^2 &    (-1)^b   &    1 
   \end{array} \nn \\
F\!\left[ \begin{array}{cc}  
      (N-b) & b \\ 
      \frac{N}{2} & \frac{N}{2} 
   \end{array}\right]&=&\begin{array}{c|cc}  
           & \Phi^1 & \Phi^2 \\ \hline 
      \Phi_{[\frac{N}{2}+b]} &   -\um(-1)^b   &    \um(-1)^b   \\ 
      \Phi_{[\frac{N}{2}-b]} &    \um   &    \um
   \end{array} \nn \\
F\!\left[ \begin{array}{cc}  
      \frac{N}{2} & \frac{N}{2} \\ 
      (N-b) & b 
   \end{array}\right]&=&\begin{array}{c|cc}  
           & \Phi^1 & \Phi^2 \\ \hline 
      \Phi_{[\frac{N}{2}+b]} &   -\um(-1)^b   &    \um(-1)^b   \\ 
      \Phi_{[\frac{N}{2}-b]} &    \um   &    \um 
   \end{array} \nn \\
F\!\left[ \begin{array}{cc}  
      b & \frac{N}{2} \\ 
      (N-b) & \frac{N}{2} 
   \end{array}\right]&=&\begin{array}{c|cc}  
             & \Phi_{[\frac{N}{2}+b]} & \Phi_{[\frac{N}{2}-b]} \\ \hline 
      \Phi^1 &   -(-1)^b   &    1   \\ 
      \Phi^2 &    (-1)^b   &    1 
\end{array} \label{type6} \\ 
F\!\left[ \begin{array}{cc}  
      (N-b) & \frac{N}{2}\\ 
      \frac{N}{2} & b 
   \end{array}\right]&=&\begin{array}{c|cc}  
           & \Phi_{[\frac{N}{2}+b]} & \Phi_{[\frac{N}{2}-b]} \\ \hline 
      \Phi_{[\frac{N}{2}+b]} & (-1)^b   &    0   \\ 
      \Phi_{[\frac{N}{2}-b]} &    0     &    1 
   \end{array} \nn \\
F\!\left[ \begin{array}{cc}  
      \frac{N}{2} & (N-b) \\ 
      b & \frac{N}{2}  
   \end{array}\right]&=&\begin{array}{c|cc}  
           & \Phi_{[\frac{N}{2}+b]} & \Phi_{[\frac{N}{2}-b]} \\ \hline 
      \Phi_{[\frac{N}{2}+b]} &  (-1)^b   &    0   \\ 
      \Phi_{[\frac{N}{2}-b]} &    0      &    1 
\end{array} \nn
\ee
Where again some of these signs are set by the pentagon and hexagon.
 
\subsubsection{Case $F[ \Phi_a \Phi_b \Phi_c \Phi_d ]$}
\label{sectionDDPP}
In most cases where $F[abcd]$ is allowed by the fusion rules, only one term survives in (\ref{abcd}) and the corresponding fusing matrix is naively 1. Imposing the pentagon and hexagon leads to sign flips. To express these the following is needed. In the $S$-channel conformal blocks the couplings $N_{ab}^{\ph{ab}p}$ and $N_{pcd}$ are at stake. For the $T$-channel we have at stake $N_{adq}$ and $N_{bc}^{\ph{bc}q}$. Let us define an order for the couplings; call $N_{ab}^{\ph{ab}p}$, $N_{pcd}$, $N_{adq}$ and $N_{bc}^{\ph{bc}q}$ the first, second, third and fourth coupling respectively. Now, if the fusion rules allow a 1x1 fusing matrix $F[abcd]$, then the labels $p,q$ of the fields $\Phi_p,\Phi_q$ propagating in the $S$- and $T$-channels are defined uniquely by $a,b,c,d$. We call a coupling $N_{\Phi_i \Phi_j}^{\ph{\Phi_i \Phi_j}\Phi_k}$ `primary' (type P) if $k=i+j$ or $k=[i-j]$, and `descendant' if $k=2N-i-j$ (type D). With couplings ordered as above, for a 1x1 matrix $F[abcd]$ the extra signs then behave as
\be
F\!\left[ \begin{array}{cc}  
      b & c \\ 
      a & d 
   \end{array}\right]=\left\{ \ba{cl}  
      (-1)^{q}   &  \mbox{if couplings are of type DDPP} \\ 
      (-1)^{p}   &  \mbox{if couplings are of type PPDD} \\ 
      (-1)^{c}   &  \mbox{if couplings are of type DPDP} \\ 
      (-1)^{a}   &  \mbox{if couplings are of type PDPD} \\ 
      (-1)^{b}   &  \mbox{if couplings are of type PDDP} \\ 
      (-1)^{d}   &  \mbox{if couplings are of type DPPD} \\ 
      1 & \mbox{else} \ea 
   \right. \;\;\;a \neq N-d, \;\;\; c\neq N-b.
\label{genabcd}
\ee
{(We could have also written $(-1)^{b+c}$ for DDPP and $(-1)^{a+b}$ for PPDD, which is an equivalent statement.)} Given any six labels $p,q,a,b,c,d$ leading to a 1x1 matrix $F[abcd]$, the four couplings will always fall into one and only one of the cases above. The result (\ref{genabcd}) is valid for any $N$ and for any labels $a,b,c,d$ that generate a 1x1 matrix. (So for instance the cases $F[aaad]$ and $F[aacd]$ and permutations thereof are automatically included in (\ref{genabcd}).) 

When $a,b,c,d$ take special values, the fusing matrices enlarge. This happens for $F[a(N-a)c(N-c)]$, $F[a(N-c)c(N-a)]$ and $F[ac(N-a)(N-c)]$. The correlator for the first case, $b=N-a$ and $d=N-c$, is 
\be
   \langle a(N-a)c(N-c) \rangle = \um |1-z|^{-2x}|z|^{2x'}, 
   \;\;\;x=\frac{(N-a)c}{2N},\;\;\;x'=\frac{(N-c)c}{2N}.
\ee
Only the $\Phi^i$ are propagating because $a\neq c$ forbids $\Phi_{2a}$ from propagating. The degenerate blocks are 
\be
{\cal F}_{\Phi^{1,2}}^{a(N-a)c(N-c)}(z) = (1-z)^{-x} z^{x'}. 
\ee 
This is to be compared the $T$-channel blocks coming from
\be 
\langle a(N-c)c(N-a) \rangle = \um |1-z|^{2x}|z|^{-2x'}. 
\ee 
Here the propagating fields should be $\Phi_{a\pm (N-a)}$, but like in the $\langle \frac{N}{2}\frac{N}{2}b(N-b) \rangle$ case, one of them does not propagate when all the external legs are at the primary level. The for the propagating block is 
\be
{\cal F}_{\Phi_{[a-(N-c)]}}^{a(N-c)c(N-a)}(z) = (1-z)^{x'} z^{-x}. 
\ee 
By considering descendants correlators, we would come to the impossible fusing matrix of 
\be 
F\!\left[ \begin{array}{cc}  
      (N-a) & c\\ 
      a & (N-c) 
   \end{array}\right]=\begin{array}{c|cc}  
             & \Phi_{a+(N-c)} & \Phi_{a-(N-c)} \\ \hline 
      \Phi^1 &    1   &    1   \\ 
      \Phi^2 &    1   &    1 
   \end{array}. 
\ee 
This clearly cannot be the correct answer since it is not invertible. The solution is to insert a minus sign in one of the matrix elements. The same happens for $F[a(N-c)c(N-a)]$, which would appear to have all elements equal to $1/2$. As usual, we let the pentagon and hexagon set the signs that are necessary. This time the extra signs will not only solve the pentagon and hexagon equations, they will also make the fusing matrices invertible. Solving for $N$ even we get
\be
F\!\left[ \begin{array}{cc}  
      (N-a) & c\\ 
      a & (N-c) 
   \end{array}\right]&=&\begin{array}{c|cc}  
             & \Phi_{[a+(N-c)]} & \Phi_{[a-(N-c)]} \\ \hline 
      \Phi^1 &   (-1)^{q+\frac{N}{2}}   &    1   \\ 
      \Phi^2 &  -(-1)^{q+\frac{N}{2}}   &    1 
   \end{array} \nn \\
F\!\left[ \begin{array}{cc}  
      (N-c) & c\\ 
      a & (N-a) 
   \end{array}\right]&=&\begin{array}{c|cc}  
              & \Phi^1 & \Phi^2 \\ \hline 
      \Phi_{[a+(N-c)]} &  \um(-1)^{p+\frac{N}{2}} 
                       & -\um(-1)^{p+\frac{N}{2}}   \\ 
      \Phi_{[a-(N-c)]} & \um  & \um
   \end{array} \nn \\
F\!\left[ \begin{array}{cc}  
      c & (N-a)\\ 
      a & (N-c) 
   \end{array}\right]&=&\begin{array}{c|cc}  
             & \Phi_{[c+(N-a)]} & \Phi_{[c-(N-a)]} \\ \hline 
      \Phi_{[a+c]} &    (-1)^{\frac{p+q+N}{2}}   &    0   \\ 
      \Phi_{[a-c]} &    0   &    (-1)^{\frac{p+q+N}{2}} 
\end{array} 
\ee
For $N$ odd we have 
\be 
F\!\left[ \begin{array}{cc}  
      (N-a) & c\\ 
      a & (N-c) 
   \end{array}\right]&=&\begin{array}{c|cc}  
             & \Phi_{[a+(N-c)]} & \Phi_{[a-(N-c)]} \\ \hline 
      \Phi^1 &   \eta(-1)^{q+\frac{N-1}{2}}   &    1   \\ 
      \Phi^2 &  -\eta(-1)^{q+\frac{N-1}{2}}   &    1 
   \end{array} \;\;\; \eta=\left\{ \ba{cl} -1  
      & \mbox{if $a<c$}  \\  1 
      & \mbox{if $a>c$} \ea \right.\nn \\
F\!\left[ \begin{array}{cc}  
      (N-c) & c\\ 
      a & (N-a) 
   \end{array}\right]&=&\begin{array}{c|cc}  
             & \Phi^1 & \Phi^2 \\ \hline 
      \Phi_{[a+(N-c)]} &    \um\eta(-1)^{p+\frac{N-1}{2}}  
                       &   -\um\eta(-1)^{p+\frac{N-1}{2}}   \\ 
      \Phi_{[a-(N-c)]} &    \um   &    \um 
   \end{array} \nn \\
F\!\left[ \begin{array}{cc}  
      c & (N-a)\\ 
      a & (N-c) 
   \end{array}\right]&=&\begin{array}{c|cc}  
             & \Phi_{[c+(N-a)]} & \Phi_{[c-(N-a)]} \\ \hline 
      \Phi_{[a+c]} &    -(-1)^{\frac{p+q+N}{2}}   &    0   \\ 
      \Phi_{[a-c]} &    0   &    -(-1)^{\frac{p+q+N}{2}} 
   \end{array} \label{ac(N-a)(N-c)} 
\ee
again with $p,q$ the fields of type $\Phi_k$ propagating in the $S-$ and $T-$channels.

The calculation of the fusing matrices for the untwisted sector is complete. The set of fusing matrices (\ref{abc0}-\ref{ac(N-a)(N-c)}) is the solution.

\subsubsection{Braiding eigenvalues}
The set of signs $\xi_{ij}^k$ are the remaining piece of data necessary for determining the braiding matrices. Define `free boson charge' $Q(i)$ as $Q(0)=Q(J)=0,\; Q(\Phi_k)=k,\; Q(\Phi^i)=N$. For the fusing matrices above, the pentagon and hexagon equations are solved for
\be
\xi_{ij}^k = \left\{ \ba{cl}
   \text{for $N$ even:} &  \xi_{ij}^k = 1 \;\;\; 
                           \forall \; i,j,k  \\ {} & {} \\
   \text{for $N$ odd:}  &  \left\{ \ba{ll} 
                \xi_{ij}^k =  -1 & \text{if $Q(i)+Q(j)+Q(k)=2N$} \\
                \xi_{ij}^k =   1 & \text{else} \ea \right.
         \ea \right.
\ee
With this, all fusing and braiding matrices for the untwisted sector are determined.

\subsection{Discussion of results}
As we have seen, it is not possible to arrive at the correct form of the duality matrices solely by extracting orbifold conformal blocks from its correlators and comparing them in the various expansion channels. To get the complete result we resorted to solving the pentagon and hexagon equations, with the gauge freedom of $F[0ijk]$ fixed. As can be seen from the fusing matrices and associated fusion rules, the sign flips that turned out to be necessary are all related to descendant propagation. This hints at the extension being responsible for the subtleties that force these sign changes, which cannot be sorted out at the level of conformal blocks. When all indices of $F_{pq}[ijkl]$ come from primary fields of the extended theory, the naive results require no changes on the fusing matrix and can therefore be trusted. In looking for more solutions to the polynomial equations, we did not allow for changes in these signs in the numerical calculations.

An analysis of the problem of how exactly the extension works at the chiral algebra level and at conformal block level might give answers to questions like where do the descendant signs come from and how can one calculate them, or how can one resolve the problem of conformal block degeneracy when two conjugate fields propagate in the correlator channel expansions. Some steps towards this end have been taken in \cite{frs}, although it is yet too soon to advance solutions.
The complete result for $F_{pq}[ijkl]$ and $\xi^{k}_{ij}$ is essentially unique, since all the possible sign changes coming from the pentagon and hexagon are gauge-related. The fact that every one of the equations was verified up $N=21$ is strong evidence for the correctness of our results. We have also verified the one-loop constraint (\ref{sas=b}) held for $p=0$.

While the results for the duality matrices are not enough to verify all the sewing constraints, a subset can immediately be checked for the Cardy modular invariant of the orbifold. In the Cardy case $Z_{ij}=C_{ij}$, it was noticed by \cite{zuber} and \cite{runkelad} that if one writes
\be
C_{ijk}^{abc} = F_{b^c k^c}\!\left[ \ba{cc} i & j \\ a^c & c \ea \right],
\label{c=F}
\ee
then, using the symmetries of the boundary operator product coefficients, he can check that duality of the boundary four-point function becomes the pentagon identity \cite{sew=pent}. Having found a set of untwisted fusing matrices that satisfies the pentagon identity is then also a proof that the orbifold untwisted sector satisfies the boundary four-point sewing constraint for the Cardy case.

\section{Fusing matrices for the mixed sector}
We now calculate fusing matrices involving chiral fields from the twisted sector. These divide into two cases: mixed and (pure) twisted. In the mixed sector we have correlators with two untwisted fields and two twisted ones (correlators with an odd number of twist fields vanish due to the fusion rules) and in the twisted sector we have correlators with four twisted fields. In this section we study mixed correlators and in the next section we present results for the twisted sector. Results will not be complete because the correlators involving twisted $\tau_i$ fields are not readily available. Nevertheless it is useful to cover some ground by calculating a few correlators involving $\sigma_i$ fields and see what problems and subtleties arise. Mixed correlators were studied in \cite{vafa} \cite{jungnickel} and twisted correlators in \cite{dvvv} \cite{vafa} \cite{dixon}.

To derive mixed fusing and braiding matrices one needs the explicit conformal blocks for the mixed sector.  The canonical mixed correlator reads \cite{vafa} \cite{jungnickel}
\be
\langle \sigma_i (\infty,\bar{\infty}) \phi_a(1,\bar{1}) \phi_b\zz
        \sigma_j(0,\bar {0}) 
\rangle = |z|^{\frac{-b^2}{2N}} 
   \left( \left| \frac{1-\sqrt{z}}{1+\sqrt{z}} \right|^{ \frac{ab}{N}} + 
          \left| \frac{1-\sqrt{z}}{1+\sqrt{z}} \right|^{-\frac{ab}{N}} 
   \right), 
\label{orbTUUT}
\ee
This correlator has an extra factor $|z|^{-b^2/2N}$ as compared to \cite{vafa}. That factor is necessary to give the correct conformal weights to the intermediate states propagating and the correct asymptotics to the conformal blocks \cite{jungnickel}. Also, the difference in the exponents of (\ref{orbTUUT}) with respect to \cite{vafa} is related to the choice of $\alpha^\prime$, which affects the normalization of the twisted sector mode expansions. In \cite{vafa} $\alpha^\prime=1/2$ whereas here $\alpha^\prime=2$.

The fusion rules determine for which $\sigma_i$ and $\sigma_j$ the correlators are zero/non-zero. With some algebra we can also derive correlators for the remaining mixed cases with the twisted fields in other positions other than the first and fourth and for the other types of untwisted fields. For briefness we concentrate on the case $\langle \sigma ab \sigma\rangle$ only.

If we decompose (\ref{orbTUUT}) into conformal blocks in the $S$-channel we get, since $\Phi_k \times \sigma_i = \sigma_m + \tau_m$ (with $m=1$ or 2 depending on the fusion rules), two blocks
\be
&& \langle \sigma_i (\infty,\bar{\infty}) \phi_a(1,\bar 1) 
   \phi_b(z,\bar z) \sigma_j(0,\bar 0) \rangle = \nn \\  
&& = C_{\sigma_i \phi_a}^{ \ph{\sigma_i \phi_a} \sigma_m} \, 
     C_{\phi_b \sigma_2 \sigma_m} 
   \left| {\cal F}_{\sigma_m}^{\sigma_i \Phi_a \Phi_b \sigma_j}(z)\right|^2 + 
     C_{\sigma_1 \phi_a}^{\ph{\sigma_1 \phi_a} \tau_m}
     C_{\phi_b \sigma_2 \tau_m}
   \left| {\cal F}_{\tau_m  }^{\sigma_i \Phi_a \Phi_b \sigma_j}(z)\right|^2.
\ee
The blocks ${\cal F}_{\sigma_m}^{\sigma_i \Phi_a \Phi_b \sigma_j}(z)$ and ${\cal F}_{\tau_m}^{\sigma_i \Phi_a \Phi_b \sigma_j}(z)$ must have $z\rightarrow 0$ asymptotics $z^{-b^2/4N}(1+\cdots)$ and $z^{-b^2/4N + 1/2}(1+\cdots)$ respectively. In (\ref{orbTUUT}) we have two terms, but if we try to identify each of them with a conformal block, we get the wrong asymptotics. Some reshuffling is thus needed. Using
\be
\left|\frac{1 - z}{1 + z}\right|^k + \left|\frac{1 - z}{1 + z}\right|^{-k} = 
   \um (1 - z^2)^{ -k} 
   \left( |(1+z)^k + (1-z)^k|^2  +  |(1+z)^k - (1-z)^k|^2 \right),
\ee
we can rewrite (\ref{orbTUUT}) as
\be
&&{\cal F}_{\sigma_m}^{\sigma_i \Phi_a \Phi_b \sigma_j}(z) =
   \um z^{-\frac{b^2}{4N}} (1-z)^{-\frac{ab}{2N}}
   \left((1+\sqrt{z})^{-\frac{ab}{N}}+(1-\sqrt{z})^{-\frac{ab}{N}}
   \right),\nn\\
&&{\cal F}_{\tau_m}^{\sigma_i \Phi_a \Phi_b \sigma_j}(z) = 
   \frac{N}{ab} z^{-\frac{b^2}{4N}} (1-z)^{-\frac{ab}{2N}}
   \left((1+\sqrt{z})^{\frac{ab}{N}}-(1-\sqrt{z})^{\frac{ab}{N}}
   \right),
\label{orbblocks}
\ee
which have the correct $z \> 0$ asymptotics. The operator product coefficients are 
\be
&& C_{\sigma_i\Phi_a}^{\ph{\sigma_i\Phi_a}\sigma_m}
   C_{\Phi_b\sigma_i\sigma_m}=1 \nn\\
&& C_{\sigma_i\Phi_a}^{\ph{\sigma_i\Phi_a}\tau_m}
   C_{\Phi_b\sigma_i\tau_m}=({2ab}/{N})^2. 
\ee
The natural choice for the $C_{\sigma_i\Phi_a}^{\ph{\sigma_i\Phi_a}\tau_m}$ coefficient is then $C_{\sigma_i\Phi_a}^{\ph{\sigma_i \Phi_a}\tau_m}= 2a^2/N$.

\subsection{Braiding matrices}
We are now ready to pick (\ref{orbblocks}) and insert them in (\ref{defBF}). We can derive braiding matrices. The equations to solve are (assuming $B^+$)
\be
{\cal F}^{\sigma_i \Phi_a \Phi_b \sigma_j}_{\sigma_m}(z) &=&
   z^{-\frac{a^2+b^2}{4N}}
      B^+_{\sigma_m \sigma_m} 
         \left[ \begin{array}{cc} 
            \Phi_a & \Phi_b \\ 
            \sigma_i &\sigma_j 
         \end{array}\right]
   {\cal F}^{\sigma_i \Phi_b \Phi_a \sigma_j}_{\sigma_m}(1/z) \nn \\
&&+z^{-\frac{a^2+b^2}{4N}}
      B^+_{\sigma_m \tau_m}
         \left[ \begin{array}{cc} 
            \Phi_a & \Phi_b \\ 
            \sigma_i &\sigma_j 
         \end{array}\right] 
   {\cal F}_{\tau_m}^{\sigma_i \Phi_b \Phi_a \sigma_j}(1/z), \nn \\
{\cal F}^{\sigma_i \Phi_a \Phi_b \sigma_j}_{\tau_m}(z) &=& 
   z^{-\frac{a^2+b^2}{4N}}
      B^+_{\tau_m \sigma_m}
         \left[ \begin{array}{cc} 
            \Phi_a & \Phi_b \\ 
            \sigma_i &\sigma_j 
         \end{array}\right]
   {\cal F}^{\sigma_i \Phi_b \Phi_a \sigma_j}_{\sigma_m}(1/z) \nn \\
&&+z^{-\frac{a^2+b^2}{4N}}
      B^+_{\tau_m \tau_m}
         \left[ \begin{array}{cc} 
            \Phi_a & \Phi_b \\ 
            \sigma_i &\sigma_j 
         \end{array}\right] 
   {\cal F}_{\tau_m}^{\sigma_i \Phi_b \Phi_a \sigma_j}(1/z).
\ee
Inserting (\ref{orbblocks}) in the above equalities and taking principle values for the square roots we see that the powers of $z$ nicely cancel out, leaving us with
\be
(1-z)^{-\frac{ab}{2N}}&\times & 
   \left( (1+\sqrt{z})^{\frac{ab}{N}} + 
          (1-\sqrt{z})^{\frac{ab}{N}} \right) = 
\nn \\ && B^{+}_{\sigma_m\sigma_m} (1-1/z)^{-\frac{ab}{2N}} 
   \left( (1+\sqrt{1/z})^{\frac{ab}{N}} + 
          (1-\sqrt{1/z})^{\frac{ab}{N}} \right) 
\nn \\ &&+ B^{+}_{\sigma_m\tau_m} \frac{N}{ab} 
          (1-1/z)^{-\frac{ab}{2N}}
   \left( (1+\sqrt{1/z})^{\frac{ab}{N}} - 
          (1-\sqrt{1/z})^{\frac{ab}{N}} \right), 
\nn \\ (1-z)^{-\frac{ab}{2N}}& \times & \frac{N}{ab}
   \left( (1+\sqrt{z})^{\frac{ab}{N}} - 
          (1-\sqrt{z})^{\frac{ab}{N}} \right) = 
\nn \\ && B^{+}_{\tau_m\sigma_m} (1-1/z)^{-\frac{ab}{2N}} 
   \left( (1+\sqrt{1/z})^{\frac{ab}{N}} + 
          (1-\sqrt{1/z})^{\frac{ab}{N}} \right) 
\nn \\ &&+ B^{+}_{\tau_m\tau_m} \frac{N}{ab} 
          (1-1/z)^{-\frac{ab}{2N}} 
   \left( (1+\sqrt{1/z})^{\frac{ab}{N}} - 
          (1-\sqrt{1/z})^{\frac{ab}{N}} \right). \nn
\ee
Pushing the factor $(-1/z)^{-\frac{ab}{2N}}$ to the left-hand-side and abbreviating $(1+\sqrt{1/z})^{\frac{ab}{N}} = x,$ $(1-\sqrt{1/z})^{\frac{ab}{N}} = y$ we get for $B^+$
\be
   e^{\I \pi k/2} (x+y) =   B^+_{\sigma \sigma} (x+y) + 
               \frac{N}{ab} B^+_{\sigma \tau  } (x-y),  \nn\\
   e^{\I \pi k/2} (x-y) =  \frac{ab}{N} B^+_{\tau   \sigma} (x+y) + 
                            B^+_{\tau   \tau  } (x-y). 
\ee
Squaring the above equalities, we see that they are satisfied by a braiding matrix
\be
B^+\left[ \begin{array}{cc} 
      \Phi_a & \Phi_b \\ 
      \sigma_i & \sigma_j
   \end{array} \right] = e^{\I \pi \frac{ab}{2N}} \times 
   \begin{array}{c|cc}
      & \sigma_m & \tau_m \\ \hline
      \sigma_m & \cos{(\frac{\pi ab}{2N} )} & 
      {\I \frac{ab}{N}} \sin{(\frac{\pi ab}{2N})} \\ 
      \tau_m & \frac{\I N}{{ab}} \sin{(\frac{\pi ab}{2N})} & 
      \cos{(\frac{\pi ab}{2N})} 
   \end{array}. 
\label{TUUTB}
\ee
This solution reduces for $N=2$ to the result one would get from the tensor product of two Ising models \cite{ms}. For $N$ odd one must replace the twist fields with the appropriate complex combinations and re-calculate the braiding.

The simplest hexagon equations are inversion relations. Since the braiding above matrices above satisfy the first hexagon equation for eigenvalues $\xi_{UT}^T=1$ and $\xi_{TT}^U=1$, and since there are no extension problems in the twist-twist-untwist couplings, it is likely that all $\xi_{UT}^T$ and $\xi_{TT}^U=1$ are +1.

\section{Fusing matrices for the twisted sector}
Calculation of the orbifold four twist fields correlators was originally done in \cite{vafa} and \cite{dixon}. The technique used there was lifting the four-point function on the sphere with four insertions to its double cover, a torus with four insertions, so that the multi-valuedness behaviour of the free boson field around twist fields vertex operators insertions disappeared and the result became single-valued. The details of the calculation are however intricate and therefore we willl write down the results only. With some reshuffling to get to our conventions, the result of \cite{vafa} and \cite{dixon} is
\be
\langle \sigma_{1}(\infty,\bar{\infty}) \sigma_{r}(1,\bar{1}) 
   \sigma_{r+s}(z,\bar{z}) \sigma_{s}(0,\bar{0}) \rangle
=\left| \frac{\theta_3(0|\tau)}{\theta_1^\prime(0|\tau)}\right| 
   {R}\sum_{n,m} e^{im s} 
        q^{  \frac{1}{2}\left( \frac{2n+r}{R} - \frac{mR}{2}   \right)^2} 
   \bar{q}^{ \frac{1}{2}\left( \frac{2n+r}{R} + \frac{mR}{2}   \right)^2}
\label{TTTTHV}
\ee
where $\sigma_{r,s}$ denote $\sigma_{1,2}$ twist fields. Careful with the notation now: $r+s=t$ is defined mod 2, and $t=0$ corresponds to $\sigma_1$ and $t=1$ to $\sigma_2$. So for instance $r=s=1$ means $\langle \sigma_{1} \sigma_{2} \sigma_{1} \sigma_{2} \rangle$.

The parameter $q$ is redefined as $q=e^{\I \pi \tau}$, with $\tau$ the modulus of the double cover torus. This modulus can be rewritten as $\tau=\I \,\bigK(\sqrt{z})/\bigK(\sqrt{1-z})$, with $\bigK$ the complete elliptic functions \cite{bateman} and $\theta$ the generalized theta functions
\be 
\theta \left[\ba{c} a \\ b \ea\right](z|\tau)=
\sum_{n} q^{(n+a)^2 \tau + 2(n+a)(z+b)},
\ee
with the classical Jacobi $\theta$ functions $\theta_{1,2,3,4}$ defined as
\be
\theta \!\left[\ba{c} \um \\ \um \ea\right]\!(z|\tau)= \theta_1(z|\tau), \;
\theta \!\left[\ba{c} \um \\ 0 \ea\right]\!(z|\tau)= \theta_2(z|\tau), \;
\theta \!\left[\ba{c} 0 \\ 0 \ea\right]\!(z|\tau)= \theta_3(z|\tau), \;
\theta \!\left[\ba{c} 0 \\ \um \ea\right]\!(z|\tau)= \theta_4(z|\tau), \nn
\ee
and $\theta_1^\prime$ the $\theta_1$ function with zero-mode removed \cite{bateman}. 

Other combinations of twist fields yield vanishing correlators due to the fusion rules. Correlators with $\sigma_2$ as the first field are obtained by replacing $\sigma_1 \leftrightarrow \sigma_2$. (This dihedral symmetry was noticed in \cite{dixon}.) Being essentially a computation on the torus, it is not surprising to see a sum over lattice momenta arising. The theta function argument $z$ will play no role in the following, so we will skip it from now on.

The result (\ref{TTTTHV}) follows our conventions with $\alpha^\prime = 2$ and world-sheet spatial parameter running from 0 to $2\pi$. It was obtained from \cite{vafa} by redefining the Hamidi-Vafa compactification radius such that $R^{HV}\rightarrow 2\pi R$. The world-sheet dilation then brings $2R\rightarrow R$. The correlator turns out not to be invariant under the $T-$duality transformation $R\rightarrow 2/R$. The reason for this was explained in \cite{robthesis}. Because of it, one has to specify which range of $R$ he takes and here we pick again the non-equivalent radii $R\in[\sqrt{2},\infty]$. When we do this, we must take a final step of substituting $R\rightarrow 2/R$ in the expression of \cite{vafa} for the four-twist correlator. After all this is done we come to (\ref{TTTTHV}).

Note also that this four-twist correlator holds for an unextended orbifold theory. To get a result for the extended orbifold requires some manipulations so that the lattice momenta in the summations come to a finite sum.

\subsubsection{Equivalence to the Ising squared model}
An interesting cross-check is to see whether in the case of $R=2$, where the orbifold is equivalent to the tensor product of two Ising models, these complicated-looking expressions for four-twist correlators yield the same result as the straightforward Ising squared calculation. The Ising squared field $(0,\sigma)$ (in the usual Ising notation) corresponds to the orbifold $\sigma_1$ twist field. At canonical insertions its four-point correlator is simply
\be
\langle (0,\sigma)(0,\sigma)(0,\sigma)(0,\sigma) \rangle &=& 
\langle 0000 \rangle \times 
\langle \sigma\sigma\sigma\sigma \rangle \nn \\
&=& 1 \times \um |z(1-z)|^{-1/4} 
    \left( \left| 1+\sqrt{1-z} \right| + 
    \left| 1-\sqrt{1-z} \right| \right). 
\label{Isingsssscorr}
\ee
This should be the same result as the orbifold correlator with $r=s=0,\; R=2$ in (\ref{TTTTHV}). We get
\be
\langle \sigma_1 \sigma_1 \sigma_1 \sigma_1 \rangle 
 = \left| \frac{\theta_3(\tau)}{\theta_1^\prime(\tau)}\right| 
{2}\sum_{n,m} q^{\frac{1}{2}(n - m)^2} \bar{q}^{\frac{1}{2}(n + m)^2}. 
\label{ising2orb2nr1}
\ee
Now we use the identity $\theta_1^\prime (\tau) = 4 \theta_2 \theta_3 \theta_4 (\tau)$ and rewrite the sums over $n$ and $m$ to come to
\be
\langle \sigma_1 \sigma_1 \sigma_1 \sigma_1 \rangle 
 = \left| \frac{1}{2\theta_2 \theta_4}\right| 
\sum_{\stackrel{\sst  i=0,1}{n,m}}
        q^{  {2}(\frac{i}{2} + m)^2} 
   \bar{q}^{ {2}(\frac{i}{2} + n)^2}. 
\label{ising2orb2nr2}
\ee
Note that in this form the exponents of $q$ are those of an extension, so we are in fact dealing with an extended orbifold theory. The sums are now the theta functions $\theta_3(2\tau),\theta_2(2\tau)$, which we can relate to combinations of other theta functions. When we do this, we get
\be
\langle \sigma_1 \sigma_1 \sigma_1 \sigma_1 \rangle 
 = \left| \frac{1}{2\theta_2 \theta_4}\right| 
\left( \left|\sqrt{\theta_3^2+\theta_4^2}\right|^2 + \left|\sqrt{\theta_3^2-\theta_4^2}\right|^2
\right) =\left|\frac{\theta_3^2}{\theta_2\theta_4} \right|.\label{ising2orb2nr3}
\ee
This looks very different from (\ref{Isingsssscorr}), but when we make use of the so-called `Inversion Problem' relations \cite{bateman} $z=\theta_2^4/\theta_3^4,\; 1-z=\theta_4^4/\theta_3^4$ on (\ref{Isingsssscorr}), we see that
\be
\langle (0,\sigma)(0,\sigma)(0,\sigma)(0,\sigma) \rangle = \left|\frac{\theta_3^2}{\theta_2\theta_4} \right| = \langle \sigma_1 \sigma_1 \sigma_1 \sigma_1 \rangle, \label{Ising2orb2equiv}
\ee
thus confirming that the two theories are indeed the same.

\subsection{Twisted conformal blocks}
The next step is now to express the correlator in terms of conformal blocks
\be
&&\langle \sigma_1 \sigma_r \sigma_{r+s} \sigma_s \rangle = \nn\\
&&= |z(1-z)|^{-1/4}\frac{1}{|\theta_3|^2} \sum_{m,n} (-1)^{ms} q^{\um (\frac{mR}{2} + \frac{2n+r}{R})^2} \bar{q}^{\um (\frac{mR}{2} - \frac{2n+r}{R})^2} = \sum_p C^2_p |{\cal F}_p(z)|^2. \label{TTTTgen}
\ee
Being a rational conformal field theory, the orbifold four-twist conformal blocks must have a finite sum over $p$, so we need to reshuffle the theta-functions. The manipulations that lead to fusing matrices are different for even and odd $N$.
\subsection{Case $N$ even}
We start with the $R^2=2N,\; N$  even case. Do the following steps in (\ref{TTTTgen}): pull out a $2/R$ factor from the brackets on the exponential, redefine $n \rightarrow n - mN/2$, split the $n$ sum $\sum_n f(n) = \sum_{n,i=0}^{N-1} f(i+nN)$, redefine $n-m=-m$. After doing this we come to
\be
&&\langle \sigma_1 \sigma_r \sigma_{r+s} \sigma_s \rangle =  \nn \\
&&= |z(1-z)|^{-1/4}\frac{1}{|\theta_3|^2} \sum_{m,n,i=0}^{N-1} (-1)^{ms} 
         q^{  N \left( \frac{i}{N} + m + \frac{r}{2N} \right)^2} (-1)^{-ns} 
    \bar{q}^{ N \left( \frac{i}{N} + n + \frac{r}{2N} \right)^2}.
\ee
With this the sums have been decoupled and can now be written as absolute values squared. Now we write $(-1)^{ms}=\exp{(2\pi i(m+i/N+r/2N)s/2)\exp{((-i/n-r/2N})s/2)}$, $(-1)^{-ns}=\exp{(-2\pi i(m+i/N+r/2N)s/2)\exp{((i/n+r/2N)}s/2)}$. From the general definition of theta-functions we get
\be
\langle\sigma_1 \sigma_r \sigma_{r+s} \sigma_s \rangle = 
   |z(1-z)|^{-1/4}\frac{1}{|\theta_3|^2} \sum_{i=0}^{N-1} = 
   \left| \theta{\left[ \begin{array}{c} 
      { \frac{1}{N}\left( i+{r}/{2} \right)} \\ 
      { {s}/{2}} 
   \end{array}\right]}(N\tau) \right|^2.
\ee
We now split the various cases $r,s$. 

\subsubsection{Case $\langle\sigma_1 \sigma_1 \sigma_1 \sigma_1\rangle$}
Take $r=0,s=0$. Here we simplify (\ref{TTTTgen}) further as follows. Note that $\theta_{[i/N,0]}(N\tau) = \theta_{[(N-i)/N,0]}(N\tau)$. With this we can write
\be
&& \langle\sigma_1 \sigma_1 \sigma_1 \sigma_1 \rangle =  
   |z(1-z)|^{-1/4}\frac{1}{|\theta_3|^2} \times \nn \\ && \times 
      \left( \left| \theta{ \left[ \begin{array}{c} 
         { 0} \\ 
         { 0} 
      \end{array}\right]}(N\tau) \right|^2 +
      \left| \theta\left[\begin{array}{c} 
         { \um} \\ 
         { 0} 
      \end{array}\right](N\tau) \right|^2 + 
   2\sum_{i=1}^{N/2-1}
      \left| \theta\left[\begin{array}{c} 
         {\frac{i}{N}} \\ 
         {0} 
      \end{array}\right](N\tau) \right|^2
      \right). 
\label{Nevenr0TTTT}
\ee
The three sets of theta-functions signal emergence of three types of conformal blocks. Let us check the asymptotics to see which fields are propagating in the $S$-channel. If $z\rightarrow 0$, $q\rightarrow z/16$, so we can expand (\ref{Nevenr0TTTT}) in $q$. We get
\be
{\cal F}_A^{r,s=0} \sim {\cal_N}_A \times z^{-{1}/{8} + 0},\;\;
{\cal F}_B^{r,s=0} \sim {\cal_N}_B \times z^{-{1}/{8} + {N}/{4}},\;\;
{\cal F}_C^{r,s=0} \sim {\cal_N}_C \times z^{-{1}/{8} + {(2i)^2}/{4N}}.
\ee
This is consistent with propagation of the fields $0$, $\Phi^i$ and $\Phi_{2i}$ ($\Phi^1$ for $N/2$ even, $\Phi^2$ for $N/2$ odd), the normalization factors being the inverse of the OPE coefficients. The fusion rules are $\sigma_1 \times \sigma_1 = 0 + \Phi^i + \sum \Phi_{2i}$, so asymptotics are consistent with what we expected to be propagating. Finally we can write an explicit expression for the conformal blocks of $N$ even, $r,s=0$
\be
&&{\cal F}_{0}^{\sigma_1 \sigma_1 \sigma_1 \sigma_1}(z) = 1 \times 
   \frac{(z(1-z))^{-1/8}} {\theta_3} \theta{ 
      \left[\begin{array}{c} 
         { 0} \\ 
         { 0} 
      \end{array}\right]}(N\tau) \nn \\
&&{\cal F}_{\Phi^i}^{\sigma_1 \sigma_1 \sigma_1 \sigma_1}(z) = 
   \um 16^{N/4}\times 
   \frac{(z(1-z))^{-1/8}}{\theta_3} \theta{ 
      \left[\begin{array}{c} 
         { 1/2} \\ 
         { 0} 
      \end{array}\right]}(N\tau) \;\;\;  
   \mbox{$\Phi^i = \Phi^1$ for $N/2$ even, else $\Phi^2$}   \nn \\
&&{\cal F}_{\Phi_{2i}}^{\sigma_1 \sigma_1 \sigma_1 \sigma_1}(z) = 
     16^{i^2/N} \times 
     \frac{(z(1-z))^{-1/8}}{\theta_3} \theta{ 
        \left[\begin{array}{c} 
        { i/N} \\ 
        { 0} 
      \end{array}\right]}(N\tau) \;\;\; i=1,\ldots, N/2-1.
\ee
The operator product coefficients follow from our choice of normalization of the conformal blocks. Since we chose to normalize conformal blocks to one, these turn out to be
\be
   C_{\sigma_1 \sigma_1}^{\ph{\sigma_1 \sigma_1}0} 
   C_{\sigma_1 \sigma_1 0} &=& 1, \nn \\
   C_{\sigma_1 \sigma_1}^{\ph{\sigma_1 \sigma_1}\Phi^i} 
   C_{\sigma_1 \sigma_1 \Phi^i} &=& (2\times 16^{-N/4})^2,  \\
   C_{\sigma_1 \sigma_1}^{\ph{\sigma_1 \sigma_1}\Phi_{2i}} 
   C_{\sigma_1 \sigma_1 \Phi_{2i}} &=& (\sqrt{2}\times 16^{-i^2/N})^2. \nn
\ee
We see that the operator product coefficients are powers of 16, as was also noted in \cite{dixon}.

The next step is to compare the blocks at $z$ and $1-z$ to extract the fusing matrices via ${\cal F}_p(z) = \sum_q F_{pq} {\cal F}_q(1-z)$. Going from $z$ to $1-z$ means going from $\tau$ to $-1/\tau$. This leads to three sets of equations, each one for $0$, $\Phi^i$ and $\Phi_{2i}$
\be
   \frac{(z(1-z))^{-1/8}}{\theta_3(\tau)}
      \theta \left[\begin{array}{c} { 0} \\ { 0} \end{array}\right]({ N\tau}) 
   &=& F_{00} \frac{((1-z)z)^{-1/8}}{\theta_3(-1/\tau)}
       \theta \left[\begin{array}{c} { 0} \\ { 0} \end{array}\right]({ -N/\tau}) +                    \nn \\
   && +F_{0\Phi^i} \um 16^{N/4} \frac{((1-z)z)^{-1/8}}{\theta_3(-1/\tau)}
       \theta \left[\begin{array}{c} { 1/2} \\ { 0} \end{array}\right]({ -N/\tau})                    \nn \\ 
   && +\sum_{i=1}^{N/2-1} F_{0\Phi_{2i}} 16^{i^2/N} \frac{((1-z)z)^{-1/8}}{\theta_3(-1/\tau)}
       \theta \left[\begin{array}{c} { i/N} \\ { 0} \end{array}\right]({ -N/\tau}),                   \nn
\ee
\be
      \um 16^{N/4}\frac{(z(1-z))^{-1/8}}{\theta_3(\tau)}
      \theta \left[\begin{array}{c} { 1/2} \\ { 0} \end{array}\right]({ N\tau}) 
   &=& F_{\Phi^i 0} \frac{((1-z)z)^{-1/8}}{\theta_3(-1/\tau)}
       \theta \left[\begin{array}{c} { 1/2} \\ { 0} \end{array}\right]({ -N/\tau}) +                  \nn \\ 
   &&+ F_{\Phi^i\Phi^i} \um 16^{N/4} \frac{((1-z)z)^{-1/8}}{\theta_3(-1/\tau)}
       \theta \left[\begin{array}{c} { 1/2} \\ { 0} \end{array}\right]({ -N/\tau})                    \nn \\ 
   &&+ \sum_{i=1}^{N/2-1} F_{\Phi^i \Phi_{2i}} 16^{i^2/N} \frac{((1-z)z)^{-1/8}}{\theta_3(-1/\tau)}
       \theta \left[\begin{array}{c} { i/N} \\ { 0} \end{array}\right]({ -N/\tau}),                   \nn 
\ee
\be
       16^{i^2/N}\frac{(z(1-z))^{-1/8}}{\theta_3(\tau)}
       \theta \left[\begin{array}{c} { i/N} \\ { 0} \end{array}\right]({ N\tau}) 
   &=& F_{\Phi_{2i} 0} \frac{((1-z)z)^{-1/8}}{\theta_3(-1/\tau)}
       \theta \left[\begin{array}{c} { 0} \\ { 0} \end{array}\right]({ -N/\tau}) +                    \nn \\ 
   &&+ F_{\Phi_{2i} \Phi^i} \um 16^{N/4} \frac{((1-z)z)^{-1/8}}{\theta_3(-1/\tau)}
       \theta \left[\begin{array}{c} { 1/2} \\ { 0} \end{array}\right]({ -N/\tau})                    \nn \\ 
   &&+ \sum_{j=1}^{N/2-1} F_{\Phi_{2i}\Phi_{2j}} 16^{j^2/N} \frac{((1-z)z)^{-1/8}}{\theta_3(-1/\tau)}
       \theta \left[\begin{array}{c} { j/N} \\ { 0} \end{array}\right]({ -N/\tau}).                   \nn \\
\label{Nevenr0s0fusing}
\ee
At first sight it seems rather difficult to find matrices $F_{pq}$ relating theta-functions at $N\tau$ and $-N/\tau$, but this can be done as follows. Note the following equality defining the $S$-matrix of a free boson theory
\be
\chi_i(\tau)=\sum_j S_{ij} \chi_j(-1/\tau) \Leftrightarrow 
\frac{\theta \left[\begin{array}{c} { i/2N} \\ { 0} \end{array}\right]({ 2N\tau})}{\eta(\tau)} = \sum_{j=0}^{2N-1}
\frac{e^{-2\pi \I \frac{ij}{2N}}}{\sqrt{2N}} \frac{\theta \left[\begin{array}{c} { j/2N} \\ { 0} \end{array}\right]({ -2 N/\tau})}{\eta(-1/\tau)},
\label{U(1)smatrix}
\ee
which can be proven taking a Fourier tranformation of the left-hand-side. With some reshuffling we can rewrite (\ref{U(1)smatrix}) as
\be
&& \sqrt{-2N \I \tau} \; 
   \theta \left[\begin{array}{c} 
      { k/2N} \\ 
      { 0} 
   \end{array}\right]({ 2N\tau}) = 
   \theta \left[\begin{array}{c} 
      { 0} \\ 
      { 0} 
   \end{array}\right]({ -2 N/\tau}) + 
      e^{-\pi \I k}\theta \left[\begin{array}{c} 
      { 1/2} \\ 
      { 0} 
   \end{array}\right]({ -2 N/\tau}) + \nn \\ && + 
   \sum_{k^\prime = 1}^{N-1} 
      2\cos{\left(\frac{2\pi k k^\prime}{2N}\right)}\theta 
   \left[\begin{array}{c} 
      { k^\prime/2N} \\ 
      { 0} 
   \end{array}\right]({ -2 N/\tau}).
\label{Nevenr0sofourier}
\ee
Since in (\ref{Nevenr0s0fusing}) $N$ is even, we can use (\ref{Nevenr0sofourier}) on it. The factors $(z(1-z))^{-1/8}$ drop out and using also $\theta_3({ -1/\tau}) = \sqrt{-\I \tau}\theta_3({ \tau})$, we simplify (\ref{Nevenr0s0fusing}) to
\be
&&\sqrt{-\I \tau}\; \theta \left[\begin{array}{c} { 0} \\ { 0} \end{array}\right]({ N\tau}) = 
F_{00} \theta \left[\begin{array}{c} { 0} \\ { 0} \end{array}\right]({ -N/\tau}) + \nn \\
&&+ F_{0\Phi^i} \um 16^{N/4} \theta \left[\begin{array}{c} { 1/2} \\ { 0} \end{array}\right]({ -N/\tau}) + \sum_{i=1}^{N/2-1}
F_{0\Phi_{2i}} 16^{i^2/N} \theta \left[\begin{array}{c} { i/N} \\ { 0} \end{array}\right]({ -N/\tau}), \nn \\ &&\nn \\&&\nn \\
&&\um 16^{N/4}\sqrt{-\I \tau} \; \theta \left[\begin{array}{c} { 1/2} \\ { 0} \end{array}\right]({ N\tau}) = 
F_{\Phi^i 0} \theta \left[\begin{array}{c} { 0} \\ { 0} \end{array}\right]({ -N/\tau}) + \nn \\ &&+ F_{\Phi^i\Phi^i} \um 16^{N/4} \theta \left[\begin{array}{c} { 1/2} \\ { 0} \end{array}\right]({ -N/\tau}) + \sum_{i=1}^{N/2-1} F_{\Phi^i \Phi_{2i}} 16^{i^2/N} \theta \left[\begin{array}{c} { i/N} \\ { 0} \end{array}\right]({ -N/\tau}), 
\label{Nevenr0s0fusingintermediate} \\ &&\nn \\&&\nn \\
&& 16^{i^2/N} \sqrt{-\I \tau} \; \theta \left[\begin{array}{c} { i/N} \\ { 0} \end{array}\right]({ N\tau}) = 
F_{\Phi_{2i} 0} \theta \left[\begin{array}{c} { 0} \\ { 0} \end{array}\right]({ -N/\tau}) + \nn \\ &&+ F_{\Phi_{2i} \Phi^i} \um 16^{N/4} \theta \left[\begin{array}{c} { 1/2} \\ { 0} \end{array}\right]({ -N/\tau}) + \sum_{j=1}^{N/2-1}
F_{\Phi_{2i}\Phi_{2j}} 16^{j^2/N} \theta \left[\begin{array}{c} { j/N} \\ { 0} \end{array}\right]({ -N/\tau}). \nn 
\ee
Due to (\ref{Nevenr0sofourier}) these equalities hold for a fusing matrix
\be
F\left[\begin{array}{cc} \sigma_1 & \sigma_1 \\ \sigma_1 & \sigma_1 \end{array}\right] = \frac{1}{\sqrt{N}}\times
\begin{array}{c|ccc} 
 & 0 & \Phi^i & \Phi_{2k} \\ \hline
0 & 1 & 2. 16^{-N/4} & 2. 16^{-k^2/N} \\
\Phi^i & \frac{1}{2} 16^{N/4} & {(-1)^{N/2}} & (-1)^k 16^{N/4-k^2/N} \\
\Phi_{2k^\prime} & 16^{k^{\prime 2}/N} & 2(-1)^{k^\prime} 16^{-N/4+k^{\prime 2}/N} & 2. 16^{(k^{\prime 2}-k^2)/N} {\cos{\left(\frac{2\pi k k^\prime}{N}\right)}}
\end{array}
\ee
We can check the first hexagon equation here. If we assume $\xi_{ij}^{k}=1$ the first hexagon reduces to $F^2=1$ for matrices of type $F[iiii]$. Since we conjecture $\xi_{TT}^U=1$, we have to check this matrix squares to one, which is indeed so.

\subsubsection{Case $\langle\sigma_1 \sigma_1\sigma_2 \sigma_2\rangle$}
Now let us look at the $r=0,s=1$ case. From the fusion rules, the same $0,\Phi^i,\Phi_{2k}$ should be propagating in the $S$-channel. However from (\ref{TTTTgen}), the block for $\Phi^i$ is $\theta_{[1/2,1/2]}=0$. The vanishiment of this block is needed, as we shall see. Using $|\theta_{[a,1/2]}|^2 = |\theta_{[-a,1/2]}|^2$, the $r=0,s=1$ correlator can then be rewritten as
\be
\langle\sigma_1 \sigma_1 \sigma_2 \sigma_2 \rangle =  
   |z(1-z)|^{-1/4}\frac{1}{|\theta_3|^2} \times \left(
   \left| \theta{ \left[\begin{array}{c} 
      { 0} \\ 
      { 1/2} 
   \end{array}\right]}(N\tau) \right|^2 + 2\sum_{i=1}^{N/2-1}
   \left| \theta\left[\begin{array}{c} 
      { i/N} \\ 
      { 1/2} 
   \end{array}\right](N\tau) \right|^2\right).
\label{Nevenr0s1TTTT}
\ee
Studying the asymptotics we come to
\be
&& {\cal F}_{0}^{\sigma_1 \sigma_1 \sigma_2 \sigma_2}(z)= 
      1 \times \frac{(z(1-z))^{-1/8}}{\theta_3} \theta{ 
      \left[\begin{array}{c} 
         { 0} \\ 
         { 1/2} 
      \end{array}\right]}(N\tau), \nn \\
&& {\cal F}_{\Phi^i}^{\sigma_1 \sigma_1 \sigma_2 \sigma_2}(z) = 0, \\
&& {\cal F}_{\Phi_{2i}}^{\sigma_1 \sigma_1 \sigma_2 \sigma_2}(z)= 
     16^{i^2/N} \times \frac{(z(1-z))^{-1/8}}{\theta_3} \theta{ 
     \left[\begin{array}{c} 
         { i/N} \\ 
         { 1/2} 
     \end{array}\right]}(N\tau) \;\;\; i=1,\ldots, N/2-1. \nn
\ee
The operator product coefficients are
\be
   C_{\sigma_1 \sigma_1}^{\ph{\sigma_1 \sigma_1}0}
   C_{\sigma_2 \sigma_2 0} &=& 1, \nn \\
   C_{\sigma_1 \sigma_1}^{\ph{\sigma_1 \sigma_1}\Phi_{2i}}
   C_{\sigma_2 \sigma_2 \Phi_{2i}} &=& (\sqrt{2}\times 16^{-i^2/N})^2.
\ee
(The coefficient $C_{\sigma_1 \sigma_1}^{\ph{\sigma_1 \sigma_1}\Phi^i}$ cannot be derived from $\langle 1122 \rangle$, but we already know it from $\langle 1111 \rangle$.)
The equation defining the fusing matrix for this case is
\be
{\cal F}_{p}^{\sigma_1\sigma_1\sigma_2\sigma_2}(z) = 
   F_{pq} \left[ \begin{array}{cc} 
      \sigma_1 & \sigma_2 \\ 
      \sigma_1 & \sigma_2
   \end{array}\right]
{\cal F}_{q}^{\sigma_1\sigma_2\sigma_2\sigma_1}({1-z}), \label{1122-1221}
\ee
so we need to study the $r=1,s=0$ case to proceed.

\subsubsection{Case $\langle\sigma_1 \sigma_2\sigma_2 \sigma_1\rangle$}
From the fusion rules, the fields propagating in $\langle1221\rangle$\footnote{We occasionally abbreviate $\sigma_{1,2}$ by 1,2, which should be clear from the context.} are $\Phi_{2i+1}$. Noting that $\theta_{[(i+1/2)N,0]}=\theta_{[(N-i-1+1/2)/N,0]}$ we can rewrite (\ref{TTTTgen}) as
\be
\langle\sigma_1 \sigma_2 \sigma_2 \sigma_1 \rangle = 
   |z(1-z)|^{-1/4}\frac{1}{|\theta_3|^2} \times 2\sum_{i=0}^{N/2-1}
      \left| \theta\left[ \begin{array}{c} 
         { \frac{1}{N}(i+1/2)} \\ 
         { 0} 
      \end{array}\right](N\tau) \right|^2.
\label{Nevenr1s0TTTT}
\ee
This has exactly the same number of fields propagating as $\langle1122\rangle$, namely $N/2$ of them. That is why the block ${\cal F}_{\Phi^i}^{1122}$ vanishes: if it were not so, then the dimension of the space of conformal blocks of $\langle1122\rangle$ would be bigger than that of $\langle1221\rangle$. The asymptotics yield
\be
{\cal F}_{\Phi_{2i+1}}^{\sigma_1 \sigma_1 \sigma_2 \sigma_2}(z) 
   = 16^{(2i+1)^2/4N} \times \frac{(z(1-z))^{-1/8}}{\theta_3} 
      \theta{ \left[ \begin{array}{c} 
         { \frac{1}{N}(i+1/2)} \\ 
         { 0} 
      \end{array}\right]}(N\tau), \;\;\; i=0,\ldots, N/2-1. \nn \\
\ee
with operator coefficients
\be
C_{\sigma_1 \sigma_2}^{\ph{\sigma_1 \sigma_2}\Phi_{2i+1}}
C_{\sigma_2 \sigma_1 \Phi_{2i+1}} = (\sqrt{2}\times 16^{-(2i+1)^2/4N})^2.
\ee
The fusing matrix equation is now known. It is
\be
&& \frac{(z(1-z))^{-1/8}}{\theta_3(\tau)}\theta 
   \left[\begin{array}{c} 
      { 0} \\ 
      { 1/2} 
   \end{array}\right]({ N\tau}) = \nn \\ 
&&=\frac{((1-z)z)^{-1/8}}{\theta_3(-1/\tau)} \sum_{q=0}^{N/2-1} 
   F_{0\Phi_{2q+1}} 16^{(2q+1)^2/4N} \theta 
   \left[\begin{array}{c} 
      { \frac{1}{N}(q+1/2)} \\ 
      { 0} 
   \end{array}\right]({ -N/\tau}), \nn\\ &&\nn\\ \nn \\
&& \frac{(z(1-z))^{-1/8}}{\theta_3(\tau)}16^{i^2/N} \theta 
   \left[\begin{array}{c} 
      { i/N} \\ 
      { 1/2} 
   \end{array}\right]({ N\tau}) = \nn \\ 
&&=\frac{((1-z)z)^{-1/8}}{\theta_3(-1/\tau)} \sum_{q=0}^{N/2-1} 
   F_{\Phi_{2i} \Phi_{2q+1}} 16^{(2q+1)^2/4N} \theta 
   \left[\begin{array}{c} 
     { \frac{1}{N}(q+1/2)} \\ 
     { 0} 
   \end{array}\right]({ -N/\tau}).
\label{1122-1221blocks}
\ee
To solve this, we can use the following identity, derived again from Fourier transformations
\be
\sqrt{-\I N\tau} \; \theta 
   \left[\begin{array}{c} 
      { k/N} \\ 
      { 1/2}
   \end{array}\right]({ N\tau}) = 
e^{2\pi \I k/2N} \sum_{g=0}^{N/2-1} 2 
   \cos{\left( \frac{2\pi k (g\!+\!1/2)}{N} \right)}\theta 
   \left[\begin{array}{c} 
      { \frac{1}{N}(g+1/2)} \\ 
      { 0} 
   \end{array}\right]({ -N/\tau}). \nn \\
\ee
Comparing this with (\ref{1122-1221blocks}) we conclude that the fusing matrix is
\be
F  \left[\begin{array}{cc} 
      \sigma_1 & \sigma_2 \\ 
      \sigma_1 & \sigma_2 
   \end{array}\right] = \frac{1}{\sqrt{N}} \times
   \begin{array}{c|c}  & \Phi_{2k^\prime+1} \\ \hline
      0 & 2.16^{-(2k^{\prime}+1)^2/4N} \\
      \Phi_{2k} & 2.e^{2\pi i k/2N} 
         16^{k^{2}/N-(2k^{\prime}+1)^2/4N} 
         {\cos{\left(\frac{2\pi k (k^\prime\!+\!1/2)}{N}\right)}}
\end{array} \label{F1122}
\ee
The fusing matrix $F[1221]$ is just the inverse of this matrix (due to the hexagon). We will however derive $F[1221]$ explicitly from the conformal blocks. This will serve as a check that we have the correct matrix.

The equation defining the fusing matrix for this case is
\be
{\cal F}_{p}^{\sigma_1\sigma_2\sigma_2\sigma_1}(z) = 
   F_{pq} \left[ \begin{array}{cc} 
      \sigma_2 & \sigma_2 \\ 
      \sigma_1 & \sigma_1
   \end{array}\right]
{\cal F}_{q}^{\sigma_1\sigma_1\sigma_2\sigma_2}({1-z}). \label{1221-1122}
\ee
This leads to
\be
&& \frac{(z(1-z))^{-1/8}}{\theta_3(\tau)}16^{(2p+1)^2/4N}\theta 
   \left[\begin{array}{c} 
      { \frac{1}{N}(p+1/2)} \\ 
      { 0} 
   \end{array}\right]({ N\tau}) =  \\ 
&&=\frac{((1-z)z)^{-1/8}}{\theta_3(-1/\tau)} \left( F_{\Phi_{2p}0}\theta 
   \left[\begin{array}{c} 
      { 0} \\ 
      { 1/2} 
   \end{array}\right]({ -N/\tau}) +\sum_{q=1}^{N/2-1} 
   F_{\Phi_{2p}\Phi_{2q+1}} 16^{q^2/N} \theta \left[\begin{array}{c} 
      { q/N} \\ 
      { 1/2} 
   \end{array}\right]({ -N/\tau}) \right). \nn
\label{1221-1122blocks}
\ee
We now use the (by now familiar) Fourier equality
\be
&& \sqrt{-\I N\tau}\; \theta \left[\begin{array}{c} 
      { \frac{1}{N}(k+1/2)} \\ 
      { 1/2} 
   \end{array}\right]({ N\tau}) = \nn \\
&&=\theta \left[\begin{array}{c} 
      { 0} \\ 
      { 1/2} 
   \end{array}\right]({ -N/\tau}) + \sum_{g=1}^{N/2-1} 
   e^{-2\pi \I g/2N} 2 \cos{ \left( \frac{2\pi g (k\!+\!1/2)}{N} \right)}
   \theta \left[\begin{array}{c} 
      { g/N} \\ 
      { 1/2} 
   \end{array}\right]({ -N/\tau}) \nn \\
\label{Fourier3}
\ee
Comparing the above with (\ref{1221-1122blocks}) we come to the fusing matrix
\be
F\left[\begin{array}{cc} \sigma_2 & \sigma_2 \\ \sigma_1 & \sigma_1 \end{array}\right] = \frac{1}{\sqrt{N}} \times
\begin{array}{c|cc} &
0 & \Phi_{2k^\prime} \\ \hline
\Phi_{2k+1} &  16^{(2k+1)^2/4N} &
2.e^{-2\pi i k^\prime/2N} 16^{k^{\prime 2}/N+(2k+1)^2/4N} {\cos{\left(\frac{2\pi k^\prime (k+1/2)}{N}\right)}}
\end{array} \nn\\ \label{F2211}
\ee
Because of the hexagon relation
\be
F\left[\begin{array}{cc} \sigma_1 & \sigma_2 \\ \sigma_1 & \sigma_2 \end{array}\right] F\left[\begin{array}{cc} \sigma_2 & \sigma_2 \\ \sigma_1 & \sigma_1 \end{array}\right] = 1, \nn
\ee
(\ref{F2211}) should be the inverse of (\ref{F1212}), which can checked to be so.

\subsubsection{Case $\langle\sigma_1 \sigma_2\sigma_1 \sigma_2\rangle$}
The final case is $r,s=1$, which will give us $F[1212]$. Here we expect the fields $\Phi_{2k+1}$ to propagate in the $S$-channel. Using $|\theta_{[a,1/2]}|^2=|\theta_{[-a,1/2]}|^2$ we get from (\ref{TTTTgen})
\be
\langle\sigma_1 \sigma_2 \sigma_1 \sigma_2 \rangle = 
   |z(1-z)|^{-1/4}\frac{1}{|\theta_3|^2} \times 2\sum_{i=0}^{N/2-1}
   \left| \theta \left[\begin{array}{c} 
      { \frac{1}{N}(i+1/2)} \\ 
      { 1/2} \end{array}
   \right](N\tau) \right|^2.
\label{Nevenr1s1TTTT}
\ee
The asymptotics as $z\rightarrow 0$ behave as expected. We get
\be
&& {\cal F}_{\Phi_{2p+1}}^{\sigma_1 \sigma_2 \sigma_1 \sigma_2}(z)
   = 16^{(2p+1)^2/4N} \times \frac{(z(1-z))^{-1/8}}{\theta_3}
   \theta{ \left[\begin{array}{c} 
      { \frac{1}{N}(p+1/2)} \\ 
      { 1/2} 
   \end{array}\right]}(N\tau), \;\;\; p=0,\ldots,N/2-1,\nn \\
&& C_{\sigma_1 \sigma_2}^{\ph{\sigma_1 \sigma_2}\Phi_{2p+1}} 
   C_{\sigma_1 \sigma_2 \Phi_{2p+1}} = (\sqrt{2} \times 16^{-(2p+1)^2/4N})^2.
\ee
The fusing matrix equation ${\cal F}^{1212}_p(z)=F_{pq}[1212]{\cal F}_{q}^{1212}(1-z)$ leads to
\be
&& \frac{(z(1-z))^{-1/8}}{\theta_3(\tau)}16^{(2p+1)^2/4N}
   \theta \left[\begin{array}{c} 
      { \frac{1}{N}(p+1/2)} \\ 
      { 1/2} 
   \end{array}\right]({ N\tau}) = \nn \\ 
&&=\frac{((1-z)z)^{-1/8}}{\theta_3(-1/\tau)} \sum_{q=0}^{N/2-1}
   F_{\Phi_{2p+1} \Phi_{2q+1}} 16^{(2q+1)^2/4N} \theta \left[\begin{array}{c} 
      { \frac{1}{N}(q+1/2)} \\ 
      { 1/2} 
   \end{array}\right]({ -N/\tau}).  
\label{1212-1212blocks}
\ee
The Fourier equality to use is
\be
&& \sqrt{-iN\tau}\; \theta \left[\begin{array}{c} 
      { 1/N(k+1/2)} \\ 
      { 1/2} 
   \end{array}\right]({ N\tau}) = \nn \\
&&=e^{ 2\pi \I \frac{k+1/2}{2N}} \sum_{g=0}^{N/2-1} 
   e^{-2\pi \I \frac{g+1/2}{2N}} 2 
   \cos{\left(\frac{2\pi (g\!+\!1/2) (k\!+\!1/2)}{N}
       \right)}\theta \left[\begin{array}{c} 
      { 1/N(g\!+\!1/2)} \\ 
      { 1/2}
   \end{array}\right]({ -N/\tau}). \nn \\ 
\label{Fourier4}
\ee
Comparing this with (\ref{1212-1212blocks}) gives us the fusing matrix
\be
F\left[\begin{array}{cc} \sigma_2 & \sigma_1 \\ \sigma_1 & \sigma_2 \end{array}\right] = \frac{1}{\sqrt{N}} \times
\begin{array}{c|c} 
& \Phi_{2k^\prime+1} \\ \hline
\Phi_{2k+1} &
2.e^{-2\pi \I \frac{k-k^\prime}{2N}} 16^{\frac{(2k+1)^2-(2k^\prime+1)^2}{4N}} {\cos{\left(\frac{2\pi (k+1/2) (k^\prime+1/2)}{N}\right)}}
\end{array}. \label{F1212}
\ee
The hexagon here is again $F^2=1$, which can be checked to be so.

\subsection{Case $N$ odd}
This case has subtlety that the true twist conformal field is not $\sigma_i$ but a complex combination of the two. One way to see this is to note that the correlator $\langle \sigma_1 \sigma_r \sigma_{r+s} \sigma_s \rangle$ for arbitrary $R$ is like a partition function, a real number. It is also an unextended theory object, which doesn't discriminate between $R$'s. Thus it is insensitive to the fact that the twist fields are complex for $R^2=2N$, $N$ odd. Therefore the $\sigma_i$ that enter the four-twist correlator are not the true conformal fields. As it was noted before in (\ref{Noddvertexes}), the true twist conformal field is the combination $ \hat{\sigma}_1 = \squm (\sigma_1 + \I \sigma_2),\;\hat{\sigma}_2 = \squm (\sigma_1 - \I \sigma_2)$.

Since we are after $\langle \hat{\sigma}_1 \hat{\sigma}_r\hat{\sigma}_{r+s}\hat{\sigma}_s
 \rangle$, we have to evaluate the appropriate linear combinations. Using the dihedral group symmetries to get correlators with $\sigma_1 \leftrightarrow \sigma_2$ we get
\be
\langle \hat{\sigma}_1\hat{\sigma}_1\hat{\sigma}_{1}\hat{\sigma}_1
 \rangle = \um \bigg(
\langle \sigma_1 \sigma_1 \sigma_{1} \sigma_1 \rangle - 
\langle \sigma_1 \sigma_1 \sigma_{2} \sigma_2 \rangle -
\langle \sigma_1 \sigma_2 \sigma_{2} \sigma_1 \rangle -
\langle \sigma_1 \sigma_2 \sigma_{1} \sigma_2 \rangle \bigg),\nn \\
\langle \hat{\sigma}_1\hat{\sigma}_1\hat{\sigma}_{2}\hat{\sigma}_2
 \rangle = \um \bigg(
\langle \sigma_1 \sigma_1 \sigma_{1} \sigma_1 \rangle -
\langle \sigma_1 \sigma_1 \sigma_{2} \sigma_2 \rangle +
\langle \sigma_1 \sigma_2 \sigma_{2} \sigma_1 \rangle +
\langle \sigma_1 \sigma_2 \sigma_{1} \sigma_2 \rangle \bigg),\nn \\\langle \hat{\sigma}_1\hat{\sigma}_2\hat{\sigma}_{2}\hat{\sigma}_1
 \rangle = \um \bigg(
\langle \sigma_1 \sigma_1 \sigma_{1} \sigma_1 \rangle + 
\langle \sigma_1 \sigma_1 \sigma_{2} \sigma_2 \rangle -
\langle \sigma_1 \sigma_2 \sigma_{2} \sigma_1 \rangle +
\langle \sigma_1 \sigma_2 \sigma_{1} \sigma_2 \rangle \bigg),\label{combos} \\
\langle \hat{\sigma}_1\hat{\sigma}_2\hat{\sigma}_{1}\hat{\sigma}_2
 \rangle = \um \bigg(
\langle \sigma_1 \sigma_1 \sigma_{1} \sigma_1 \rangle + 
\langle \sigma_1 \sigma_1 \sigma_{2} \sigma_2 \rangle +
\langle \sigma_1 \sigma_2 \sigma_{2} \sigma_1 \rangle -
\langle \sigma_1 \sigma_2 \sigma_{1} \sigma_2 \rangle \bigg). \nn
\ee
We now specify (\ref{TTTTgen}) to the $N$ odd case. To simplify this expression we do the following steps. Pull a factor $2/R$ out of the square, split $m$ into odd and even contributions, resum over $n\rightarrow n-mN$, split again $m$ into odd and even contributions and finally resum over $m-n\rightarrow m$. The result is
\be
&& \langle \sigma_1 \sigma_r \sigma_{r+s} \sigma_s \rangle = 
      \frac{|z(1-z)|^{-1/4}}{|\theta_3|^2} \times \nn \\
&&\mbox{\hspace{.2cm}} 
          \left( \sum_{g=0}^{N-1} \left| \theta \left[ \begin{array}{c} 
             { \frac{1}{2N}(g+r/2)} \\ 
             { 0} 
          \end{array} \right] { (4N\tau)} \right|^2
        + \left| \theta \left[ \begin{array}{c} 
             { \um+\frac{1}{2N}(g+r/2)} \\ 
             { 0} 
          \end{array} \right] { (4N\tau)} \right|^2 \right. \nn \\
&&+ \left. (-1)^s \theta \left[ \begin{array}{c}
             { \frac{1}{4} + \frac{1}{2N}(g+r/2)} \\ 
             { 0} 
          \end{array} \right] { (4N\tau)} \;\; 
          \overline{\theta} \left[  \begin{array}{c} 
             { -\frac{1}{4}+\frac{1}{2N}(g+r/2)} \\ 
             { 0} 
          \end{array} \right] { (4N\tau)} \right.  \label{NoddRawcorr} \\
&&+ \left. (-1)^s \theta \left[ \begin{array}{c} 
             { -\frac{1}{4} + \frac{1}{2N}(g+r/2)} \\ 
             { 0} 
          \end{array} \right] { (4N\tau)} \;\; 
          \overline{\theta} \left[\begin{array}{c} 
             { \frac{1}{4}+\frac{1}{2N}(g+r/2)} \\ 
             { 0} 
          \end{array} \right] { (4N\tau)} \right). \nn
\ee
Note that this isn't even an object of type $\sum_p |{\cal F}_p(z)|^2$, but that is because it is not a true correlator. Some manipulations are needed to rewrite this expression as a sum of squares of conformal blocks.
Taking specific $r,s$ and ignoring the conformal prefactor for the moment, we come to
\be
&& \langle \sigma_1 \sigma_1 \sigma_{1} \sigma_1 \rangle = 
   \sum_{g=0}^{N-1} |A|^2 + |B|^2 + C\bar{D} + D\bar{C}, \nn \\
&& \langle \sigma_1 \sigma_1 \sigma_{2} \sigma_2 \rangle = 
   \sum_{g=0}^{N-1} |C|^2 + |D|^2 + A\bar{B} + B\bar{A}, \nn \\
&& \langle \sigma_1 \sigma_2 \sigma_{2} \sigma_1 \rangle = 
   \sum_{g=0}^{N-1} |A|^2 + |B|^2 - C\bar{D} - D\bar{C}, \label{rsABCD} \\
&& \langle \sigma_1 \sigma_2 \sigma_{1} \sigma_2 \rangle = 
   \sum_{g=0}^{N-1} |C|^2 + |D|^2 - A\bar{B} - B\bar{A}, \nn
\ee
with
\be
A = \theta \left[ \begin{array}{c}
       { \frac{g}{2N}} \\ 
       { 0} 
    \end{array} \right] { (4N\tau)}, &
B = \theta \left[ \begin{array}{c}
       { \frac{1}{2} + \frac{g}{2N}} \\ 
       { 0} 
    \end{array} \right] { (4N\tau)}, \nn\\
C = \theta \left[ \begin{array}{c}
       { \frac{1}{4}+\frac{g}{2N}} \\ 
       { 0} 
    \end{array} \right] { (4N\tau)}, &
D = \theta \left[ \begin{array}{c}
       { \frac{1}{4} - \frac{g}{2N}} \\ 
       { 0} 
    \end{array} \right] { (4N\tau)}.
\label{defABCD}
\ee
It doesn't look like (\ref{NoddRawcorr}) leads to this result for $\langle \sigma_1 \sigma_2 \sigma_{2} \sigma_1 \rangle$ and $\langle \sigma_1 \sigma_2 \sigma_{1} \sigma_2 \rangle$, but one can prove it using the following equalities
\be
&& \sum_{g=0}^{N-1} \left| \theta \left[ \begin{array}{c} 
      { \frac{1}{2N}(g+\um)} \\ 
      { 0} 
   \end{array} \right] \right|^2 + \left| \theta \left[ \begin{array}{c} 
      { \um+\frac{1}{2N}(g+\um)} \\ 
      { 0} 
   \end{array} \right] \right|^2 =
   \sum_{g=0}^{N-1} \left| \theta \left[ \begin{array}{c} 
      { \frac{1}{4}+ \frac{g}{2N}} \\ 
      { 0}
   \end{array} \right] \right|^2 + \left| \theta \left[ \begin{array}{c} 
      { \frac{1}{4}-\frac{g}{2N}} \\ 
      { 0} 
   \end{array} \right] \right|^2, \nn \\
&& \sum_{g=0}^{N-1} \theta \left[ \begin{array}{c} 
      { \frac{1}{4} + \frac{1}{2N}(g+\um)} \\ 
      { 0}
   \end{array} \right] \; \bar{\theta} \left[ \begin{array}{c} 
      { -\frac{1}{4}+\frac{1}{2N}(g+\um)} \\ 
      { 0} 
   \end{array} \right] + CC = 
   \sum_{g=0}^{N-1} \theta \left[ \begin{array}{c} 
      { \frac{g}{2N}} \\ 
      { 0} 
   \end{array} \right] \; \bar{\theta} \left[ \begin{array}{c} 
      { \frac{1}{2}+\frac{g}{2N}} \\ 
      { 0} 
   \end{array} \right] + CC. \nn
\ee
Taking the appropriate linear combinations (\ref{combos}), the true correlators are
\be
\langle \hat{\sigma}_1 \hat{\sigma}_1 \hat{\sigma}_{1} \hat{\sigma}_1 \rangle =
\sum_{g=0}^{N-1} -|C-D|^2, &
\langle \hat{\sigma}_1 \hat{\sigma}_1 \hat{\sigma}_{2} \hat{\sigma}_2 \rangle =
\sum_{g=0}^{N-1} |C+D|^2, \nn \\
\langle \hat{\sigma}_1 \hat{\sigma}_2 \hat{\sigma}_{2} \hat{\sigma}_1 \rangle =\sum_{g=0}^{N-1} |A-B|^2, &
\langle \hat{\sigma}_1 \hat{\sigma}_2 \hat{\sigma}_{1} \hat{\sigma}_2 \rangle =
\sum_{g=0}^{N-1} |A+B|^2. \label{truersABCD}
\ee
The sums and differences of theta-functions look strange but can be simplified using the equalities
\be
\theta \left[ \begin{array}{c}
      { \frac{g}{2N}} \\ 
      { 0}
   \end{array} \right] { (4N\tau)} + 
\theta \left[ \begin{array}{c}
      { \frac{1}{2}-\frac{g}{2N}} \\ 
      { 0} 
   \end{array} \right] { (4N\tau)}
&=&\theta \left[ \begin{array}{c}
      { \frac{g}{N}} \\ 
      { 0} 
   \end{array} \right] { (N\tau)}, \nn \\
\theta \left[ \begin{array}{c}
      { \frac{g}{2N}} \\ 
      { 0} 
   \end{array} \right] { (4N\tau)} - 
\theta \left[ \begin{array}{c}
      { \frac{1}{2}-\frac{g}{2N}} \\ 
      { 0} \end{array} \right] { (4N\tau)}
&=& e^{-\I \pi g/N}\theta \left[ \begin{array}{c}
      { \frac{g}{N}} \\ 
      { 1/2} 
   \end{array} \right] { (N\tau)}, \nn \\
\theta \left[ \begin{array}{c}
      { \frac{1}{4}+\frac{g}{2N}} \\ 
      { 0} 
   \end{array} \right] { (4N\tau)} + 
\theta \left[ \begin{array}{c}
      { \frac{1}{4}-\frac{g}{2N}} \\ 
      { 0} 
   \end{array} \right] { (4N\tau)} 
&=& \theta \left[ \begin{array}{c} 
      { \frac{1}{2}+\frac{g}{2N}} \\ 
      { 0} 
   \end{array} \right] { (N\tau)}, \\
\theta \left[ \begin{array}{c}
      { \frac{1}{4}+\frac{g}{2N}} \\ 
      { 0} 
   \end{array} \right] { (4N\tau)} - 
\theta \left[ \begin{array}{c}
      { \frac{1}{4}-\frac{g}{2N}} \\ 
      { 0} 
   \end{array} \right] { (4N\tau)} 
&=& e^{-\I \pi(1/2+g/N)} \theta \left[ \begin{array}{c} 
      { \frac{1}{2}+\frac{g}{2N}} \\ 
      { 1/2} 
   \end{array} \right] { (N\tau)}. \nn 
\ee
(The phase factors above disappear in the absolute squares.) Restoring the conformal factors, the true correlators then simplify to
\be
&& \langle \hat{\sigma}_1 \hat{\sigma}_1 
           \hat{\sigma}_1 \hat{\sigma}_1 \rangle = 
   \frac{|z(1-z)|^{-1/4}}{|\theta_3|^2} \left(-2 \sum_{g=1}^{\frac{N-1}{2}} 
   \left| \theta \left[ \begin{array}{c}
      { \frac{1}{2}-\frac{g}{N}} \\ 
      { \um} 
   \end{array} \right] { (N\tau)} \right|^2 \right), \nn \\
&& \langle \hat{\sigma}_1 \hat{\sigma}_1 
           \hat{\sigma}_2 \hat{\sigma}_2 \rangle = 
   \frac{|z(1-z)|^{-1/4}}{|\theta_3|^2} \left( |\theta_2{(N\tau)}|^2 + 2 
   \sum_{g=1}^{\frac{N-1}{2}} \left| \theta \left[ \begin{array}{c}
      { \frac{1}{2}-\frac{g}{N}} \\ 
      { 0} 
   \end{array} \right] { (N\tau)} \right|^2 \right), \nn \\
&& \langle \hat{\sigma}_1 \hat{\sigma}_2 
           \hat{\sigma}_2 \hat{\sigma}_1 \rangle = 
   \frac{|z(1-z)|^{-1/4}}{|\theta_3|^2} \left( |\theta_4 {(N\tau)}|^2 + 2
   \sum_{g=1}^{\frac{N-1}{2}} \left| \theta \left[ \begin{array}{c} 
      { \frac{g}{N}} \\ 
      { \um} 
   \end{array} \right] { (N\tau)} \right|^2 \right), \label{truers} \\
&& \langle \hat{\sigma}_1 \hat{\sigma}_2 
           \hat{\sigma}_1 \hat{\sigma}_2 \rangle = 
   \frac{|z(1-z)|^{-1/4}}{|\theta_3|^2} \left( |\theta_3{(N\tau)}|^2 + 2
   \sum_{g=1}^{\frac{N-1}{2}} \left| \theta \left[ \begin{array}{c} 
      { \frac{g}{N}} \\ 
      { 0} 
   \end{array} \right] { (N\tau)} \right|^2 \right). \nn
\ee
And in this form we explicitly see the expected behaviour $\sum_p |{\cal F}_p(z)|^2$. Now the asymptotics can be checked. Let us first do $\langle \hat{\sigma}_1 \hat{\sigma}_1 \hat{\sigma}_{1} \hat{\sigma}_1 \rangle$. The Theta-functions behave as
\be
\theta \left[ \begin{array}{c}
      { \frac{1}{2}-\frac{g}{N}} \\ 
      { \um} 
   \end{array} \right] { (N\tau)} \sim q^{\frac{(N-2g)^2}{4N}}.
\ee
The fusion rules are $\hat{\sigma}_1 \times \hat{\sigma}_1 = \Phi^1 + \Phi_{2k+1}$, but the field $\Phi^1$ cannot propagate due to fusion at the right-hand vertex of the correlator, so only the $\Phi_{2k+1}$ fields propagate. The asymptotics are consistent with propagation of fields $\Phi_{N-2g}$, of which there are $(N-1)/2$ of, exactly as many terms as the corresponding expression (\ref{truers}) consists of. Evaluating the normalizations we get
\be
&& {\cal F}_{\Phi_{N-2g}}^{\hat{\sigma}_1\hat{\sigma}_1 
                           \hat{\sigma}_1\hat{\sigma}_1} (z) = 
   16^{\frac{(N-2g)^2}{4N}} \frac{(z(1-z))^{-1/8}}{\theta_3} \;
   \theta \left[ \begin{array}{c}
      { \frac{1}{2}-\frac{g}{N}} \\ 
      { \um} 
   \end{array} \right] { (N\tau)}, \nn \\
&& C_{\hat{\sigma}_1\hat{\sigma}_1}^{\ph{\hat{\sigma}_1\hat{\sigma}_1}
                                    \Phi_{N-2g}} 
   C_{\hat{\sigma}_1\hat{\sigma}_1\Phi_{N-2g}} = 
   \left(\I \sqrt{2} \times 16^{-\frac{(N-2g)^2}{4N}} \right)^2.
\ee
The blocks for $\langle \hat{\sigma}_1 \hat{\sigma}_1 \hat{\sigma}_{2} \hat{\sigma}_2 \rangle$ are similar, but in addition we have a block for $\Phi^1$, which can now propagate. The asymptotics and normalizations lead to
\be
&& {\cal F}_{\Phi^1}^{\hat{\sigma}_1\hat{\sigma}_1
                      \hat{\sigma}_2\hat{\sigma}_2} (z)= 
   \um 16^{N/4} \frac{(z(1-z))^{-1/8}}{\theta_3} \;
   \theta_2 { (N\tau)}, \nn \\
&& {\cal F}_{\Phi_{N-2g}}^{\hat{\sigma}_1\hat{\sigma}_1 
                           \hat{\sigma}_2\hat{\sigma}_2} (z)= 
   16^{\frac{(N-2g)^2}{4N}} \frac{(z(1-z))^{-1/8}}{\theta_3} \;
   \theta \left[ \begin{array}{c}
      { \frac{1}{2}-\frac{g}{N}} \\ 
      { 0} 
   \end{array} \right] { (N\tau)}, \nn \\
&& C_{\hat{\sigma}_1\hat{\sigma}_1}^{\ph{\hat{\sigma}_1\hat{\sigma}_1}\Phi^1}
   C_{\Phi^1\hat{\sigma}_2\hat{\sigma}_2} = 
      \left( 2 \times 16^{-N/4} \right)^2,  \\
&& C_{\hat{\sigma}_1\hat{\sigma}_1}^{\ph{\hat{\sigma}_1\hat{\sigma}_1}
                                     \Phi_{N-2g}} 
   C_{\hat{\sigma}_2\hat{\sigma}_2\Phi_{N-2g}} = 
      \left(\sqrt{2} \times 16^{-\frac{(N-2g)^2}{4N}}\right)^2. \nn
\ee
Now we look for the blocks of $\langle \sigma_1 \sigma_2 \sigma_{2} \sigma_1 \rangle$. The theta-functions behave as
\be
\theta \left[ \begin{array}{c} 
      { \frac{g}{N}} \\ 
      { \um} 
   \end{array} \right] { (N\tau)} \sim z^{\frac{(2g)^2}{4N}}.
\ee
The fusion rules are $\sigma_1 \times \sigma_2 = 0 + \Phi_{2k}$. The $g=0$ term is the identity block and other $g$-terms have the asymptotics for propagation of $\Phi_{2g}$. Checking the normalizations we get
\be
&& {\cal F}_{0}^{\hat{\sigma}_1\hat{\sigma}_2\hat{\sigma}_2\hat{\sigma}_1}(z)=
   \frac{(z(1-z))^{-1/8}}{\theta_3} \; \theta \left[ \begin{array}{c}       
      { 0} \\ 
      { \um}  
   \end{array} \right] { (N\tau)}, \nn \\
&& {\cal F}_{\Phi_{2g}}^{\hat{\sigma}_1\hat{\sigma}_2 
                         \hat{\sigma}_2\hat{\sigma}_1}(z)= 
   16^{\frac{(2g)^2}{4N}} \frac{(z(1-z))^{-1/8}}{\theta_3} \; 
   \theta \left[ \begin{array}{c} 
      { \frac{g}{N}} \\ 
      { \um} 
   \end{array} \right] { (N\tau)}, \nn \\
&& C_{\hat{\sigma}_1\hat{\sigma}_2}^{\ph{\hat{\sigma}_1\hat{\sigma}_2}0} 
   C_{\hat{\sigma}_2\hat{\sigma}_1 0} = 1, \\
&& C_{\hat{\sigma}_1\hat{\sigma}_2}^{\ph{\hat{\sigma}_1\hat{\sigma}_2}
                                     \Phi_{2g}} 
   C_{\hat{\sigma}_2\hat{\sigma}_1\Phi_{2g}} = 
      \left(\sqrt{2} \times 16^{-\frac{(2g)^2}{4N}}\right)^2. \nn
\ee
The blocks for $\langle \sigma_1 \sigma_2 \sigma_{1} \sigma_2 \rangle$ are similar, namely
\be
&& {\cal F}_{0}^{\hat{\sigma}_1\hat{\sigma}_2\hat{\sigma}_1\hat{\sigma}_2}(z)=
   \frac{(z(1-z))^{-1/8}}{\theta_3} \; \theta \left[ \begin{array}{c}       
      { 0} \\ 
      { 0} 
   \end{array} \right] { (N\tau)}, \nn \\
&& {\cal F}_{\Phi_{2g}}^{\hat{\sigma}_1\hat{\sigma}_2 
                         \hat{\sigma}_1\hat{\sigma}_2}(z)= 
   16^{\frac{(2g)^2}{4N}} \frac{(z(1-z))^{-1/8}}{\theta_3} \; 
   \theta \left[ \begin{array}{c} 
      { \frac{g}{N}} \\ 
      { 0} 
   \end{array} \right] { (N\tau)}, \nn \\
&& C_{\hat{\sigma}_1\hat{\sigma}_2}^{\ph{\hat{\sigma}_1\hat{\sigma}_2}0} 
   C_{\hat{\sigma}_1\hat{\sigma}_2 0} = 1, \\
&& C_{\hat{\sigma}_1\hat{\sigma}_2}^{\ph{\hat{\sigma}_1\hat{\sigma}_2}
                                     \Phi_{2g}} 
   C_{\hat{\sigma}_1\hat{\sigma}_2\Phi_{2g}} = 
      \left(\sqrt{2} \times 16^{-\frac{(2g)^2}{4N}}\right)^2. \nn
\ee
With all blocks under control, we can determine the fusing matrices for $N$ odd.

\subsubsection{Equivalence to the free boson}
We can check the conformal block formulas by comparing the orbifold result for $N\!=\!1$ with its equivalent theory, the extended free boson with $R^2=8$ theory. Fields in the circle theory are characterized by their momentum $p_n=n/\sqrt{8}$, $n=-3,\ldots,4$ with conformal weights $\um p_n^2$. The field correspondence is
\be
\begin{array}{cccc}
\mbox{Free boson labels} & \mbox{momentum} & \mbox{Orb $N\!=\!1$ labels} & \mbox{Weight} \\ \hline
0 & 0    & 0        & 0    \\
1 & 1/\sqrt{8}  & \sigma_1 & 1/16 \\
2 & 2/\sqrt{8}  & \Phi^1   & 1/4  \\
3 & 3/\sqrt{8}  & \tau_1   & 9/16 \\
4 & 4/\sqrt{8}  & J        & 1    \\
5 & -3/\sqrt{8} & \tau_2   & 9/16 \\
6 & -2/\sqrt{8} & \Phi^2   & 1/4  \\
7 & -1/\sqrt{8} & \sigma_2 & 1/16 \\
\end{array}\nn.
\ee
The free boson four-point correlators on the diagonal invariant are given by the Koba-Nielsen formula
\be
\langle \prod_{i} V_i(z_i,\bar{z}_i) \rangle = \prod_{i<j} |z_i-z_j|^{2p_i p_j}\times\delta_{\sum_i p_i,0}.
\ee
Taking canonical insertions this leads to
\be
&\langle 1111 \rangle_{\text{free boson}} = 0 
 \;\;\;\;\;\;\;\,\;\;\;\;\;\;\;\;\;\;\;\;\;\; & 
 \langle 1177 \rangle_{\text{free boson}} = |1-z|^{-1/4}|z|^{1/4} \nn \\ 
&\langle 1771 \rangle_{\text{free boson}} = |1-z|^{1/4}|z|^{-1/4} &
 \langle 1717 \rangle_{\text{free boson}} = |1-z|^{-1/4}|z|^{-1/4}.
\ee
The orbifold result should reproduce this. The $g$-terms in (\ref{truers}) all vanish and we are left with the classical theta-functions only. We have then
\be
&& \langle \hat{\sigma}_1 \hat{\sigma}_1 \hat{\sigma}_1 
           \hat{\sigma}_1 \rangle_{\text{Orb(2)}} = 
   \langle 1111 \rangle_{\text{free boson}} = 0, \nn \\
&& \langle \hat{\sigma}_1 \hat{\sigma}_1 
           \hat{\sigma}_2 \hat{\sigma}_2 \rangle_{\text{Orb(2)}} = 
   \langle 1177 \rangle_{\text{free boson}} = 
   \frac{|z(1-z)|^{-1/4}}{|\theta_3|^2}
      \left| \theta_2 { (\tau)} \right|^2 = |1-z|^{-1/4}|z|^{1/4}, \nn \\
&& \langle \hat{\sigma}_1 \hat{\sigma}_2 
           \hat{\sigma}_2 \hat{\sigma}_1 \rangle_{\text{Orb(2)}} = 
   \langle 1771 \rangle_{\text{free boson}} = 
   \frac{|z(1-z)|^{-1/4}}{|\theta_3|^2}
      \left| \theta_4 { (\tau)} \right|^2 = |1-z|^{1/4}|z|^{-1/4}, \nn \\
&& \langle \hat{\sigma}_1 \hat{\sigma}_2 
           \hat{\sigma}_1 \hat{\sigma}_2 \rangle_{\text{Orb(2)}} = 
   \langle 1717 \rangle_{\text{free boson}} = 
   \frac{|z(1-z)|^{-1/4}}{|\theta_3|^2}
      \left| \theta_3 { (\tau)} \right|^2 = |z(1-z)|^{-1/4}.
\ee
The results match, as expected. Had we not taken the complex combinations $\hat{\sigma}_{1,2}$ there would be no way to make the equivalence work. \newline

\noindent
The final simplification one can do is to substitute $N-2g \rightarrow 2k+1$ in the conformal blocks. Doing this leads to
\be
&& {\cal F}_{\Phi_{2k+1}}^{\hat{\sigma}_1\hat{\sigma}_1 
                           \hat{\sigma}_1\hat{\sigma}_1}(z)= 
   16^{\frac{(2k+1)^2}{4N}} \frac{(z(1-z))^{-1/8}}{\theta_3}
   \theta \left[ \begin{array}{c} 
      { \frac{2k+1}{2N}} \\ 
      { \um} 
   \end{array} \right] { (N\tau)},\nn \\
&& {\cal F}_{\Phi^1}^{\hat{\sigma}_1\hat{\sigma}_1
                      \hat{\sigma}_2\hat{\sigma}_2}(z)= 
   \um 16^{N/4} \frac{(z(1-z))^{-1/8}}{\theta_3} \;
   \theta_2 { (N\tau)}, \nn \\
&& {\cal F}_{\Phi_{2k+1}}^{\hat{\sigma}_1\hat{\sigma}_1 
                           \hat{\sigma}_2\hat{\sigma}_2}(z)= 
   16^{\frac{(2k+1)^2}{4N}} \frac{(z(1-z))^{-1/8}}{\theta_3}  \;
   \theta \left[ \begin{array}{c} 
      { \frac{2k+1}{2N}} \\ 
      { 0} 
   \end{array} \right] { (N\tau)}, \nn \\
&& {\cal F}_{0}^{\hat{\sigma}_1\hat{\sigma}_2
                 \hat{\sigma}_2\hat{\sigma}_1}(z)=
   \frac{(z(1-z))^{-1/8}}{\theta_3} \; \theta \left[ \begin{array}{c}       
      { 0} \\ 
      { \um}  
   \end{array} \right] { (N\tau)}, \nn \\
&& {\cal F}_{\Phi_{2g}}^{\hat{\sigma}_1\hat{\sigma}_2 
                         \hat{\sigma}_2\hat{\sigma}_1}(z)= 
   16^{\frac{(2g)^2}{4N}} \frac{(z(1-z))^{-1/8}}{\theta_3} \; 
   \theta \left[ \begin{array}{c} 
      { \frac{2g}{2N}} \\ 
      { \um} 
   \end{array} \right] { (N\tau)}, \label{Noddblocks} \\
&& {\cal F}_{0}^{\hat{\sigma}_1\hat{\sigma}_2
                 \hat{\sigma}_1\hat{\sigma}_2}(z)= 
   \frac{(z(1-z))^{-1/8}}{\theta_3} \; \theta \left[ \begin{array}{c}   
      { 0} \\ 
      { 0} 
   \end{array} \right] { (N\tau)}, \nn \\
&& {\cal F}_{\Phi_{2g}}^{\hat{\sigma}_1\hat{\sigma}_2 
                         \hat{\sigma}_1\hat{\sigma}_2} (z)= 
   16^{\frac{(2g)^2}{4N}} \frac{(z(1-z))^{-1/8}}{\theta_3} \; 
   \theta \left[ \begin{array}{c} 
      { \frac{2g}{2N}} \\ 
      { 0} 
   \end{array} \right] { (N\tau)},
      \;\;\; k,g=1,\ldots,(N-1)/{2}.\nn
\ee
These are precisely the same blocks of the $N$ even case, but with some labels swapped.

Having the conformal blocks, the next step is to extract the fusing matrices. Since the blocks are the same as $N$ even, the fusing matrices are also the same. The only subtlety is the zero-block of $\langle 1111\rangle$ of $N$ even, which is moved into a different correlator in $N$ odd.

\subsubsection{Case $\langle \hat{\sigma}_1 \hat{\sigma}_1 \hat{\sigma}_{1} \hat{\sigma}_1 \rangle$}
The fusion rules here are $\hat{\sigma_i} \times \hat{\sigma_i} = \Phi^{m} + \Phi_{2k-1}$, but, as we saw before, the $\Phi^m$ field cannot propagate. The fusing equation is ${\cal F}_p^{\hat{1}\hat{1}\hat{1}\hat{1}}(z) = \sum_q F[\hat{1}\hat{1}\hat{1}\hat{1}]_{pq} {\cal F}_q^{\hat{1}\hat{1}\hat{1}\hat{1}}(1-z)$, and leads to
\be
16^{\frac{(2p-1)^2}{4N}} \sqrt{-\I \tau} \theta \left[ \begin{array}{c} 
      { \frac{2p-1}{2N}} \\ 
      { \um} 
    \end{array} \right] { (N\tau)} = \sum_q F_{pq}\left[ \begin{array}{cc} 
       \hat{\sigma}_1 & \hat{\sigma}_1 \\ 
       \hat{\sigma}_1 & \hat{\sigma}_1 
    \end{array} \right] 16^{\frac{(2q-1)^2}{4N}} \theta 
    \left[ \begin{array}{c} 
       { \frac{2q-1}{2N}} \\ 
       { \um} 
    \end{array} \right] { (-N/\tau)}.
\ee
Using the Fourier equality
\be
&&\sqrt{-\I N\tau}
\theta \left[ \begin{array}{c} 
      { \frac{2k-1}{2N}} \\ 
      { \um} 
   \end{array} \right] { (N\tau)} = e^{2\pi \I (2k-1)/4N} \nn \\ &&\times  
   \sum_{g=1}^{(N-1)/2}2.e^{-2\pi \I (2g-1)/4N} 
      \cos{\left(2\pi\frac{{(2k-1)(2g-1)}}{{4N}}\right)} 
   \theta \left[ \begin{array}{c} 
      { \frac{2g-1}{2N}} \\ 
      { \um} 
   \end{array} \right] { (-N/\tau)},
\ee
we come to
\be
F  \left[ \begin{array}{cc} 
      \hat{\sigma}_1 & \hat{\sigma}_1 \\ 
      \hat{\sigma}_1 & \hat{\sigma}_1 
   \end{array} \right] =  \frac{1}{\sqrt{N}} \times \begin{array}{c|c} 
      & \Phi_{2q-1}  \\ \hline 
      \Phi_{2p-1} & 2.16^{\frac{(2p-1)^2-(2q-1)^2}{4N}} 
      e^{2\pi \I((2p-1)-(2q-1))/4N} 
      \cos{(2\pi \frac{(2p-1)(2q-1)}{4N})} 
   \end{array} \nn \\
\ee
The hexagon is $F^2=1$ which holds, since it is the same as the $N$ even case.

\subsubsection{Case $\langle \hat{\sigma}_1 \hat{\sigma}_2 \hat{\sigma}_{1} \hat{\sigma}_2 \rangle$}
The fusion rules are now $\hat{\sigma_i} \times \hat{\sigma_i} = 0 + \Phi_{2k}$. The equation for the fusing matrix is ${\cal F}_p^{\hat{1}\hat{2}\hat{1}\hat{2}}(z) = \sum_q F[\hat{1}\hat{2}\hat{1}\hat{2}]_{pq} {\cal F}_q^{\hat{1}\hat{2}\hat{1}\hat{2}}(1-z)$. This leads to
\be
&& \sqrt{-\I \tau} \theta \left[ \begin{array}{c} 
      { 0} \\ 
      { 0} 
   \end{array} \right] { (N\tau)} = F_{00} \theta \left[ \begin{array}{c} 
      { 0} \\ 
      { 0} 
   \end{array} \right] { (-N/\tau)}+ \sum_{q=0}^{\frac{N-1}{2}} 
F_{0\Phi_{2q}} 16^{\frac{(2q)^2}{4N}} \theta \left[ \begin{array}{c} 
      { \frac{2q}{2N}} \\ 
      { 0} \end{array} 
   \right] { (-N/\tau)}, \\
&& \sqrt{-\I \tau} \theta  \left[ \begin{array}{c} 
      { \frac{2k}{2N}} \\ 
      { 0} 
   \end{array} \right] { (N\tau)} = 
F_{0\Phi_{2k}} \theta \left[ \begin{array}{c} 
      { 0} \\ 
      { 0} 
   \end{array} \right] { (-N/\tau)}+ \sum_{q=0}^{\frac{N-1}{2}} 
F_{\Phi_{2k}\Phi_{2q}}  16^{\frac{(2q)^2}{4N}} \theta \left[ \begin{array}{c} 
      { \frac{2q}{2N}} \\ 
      { 0} 
   \end{array} \right] { (-N/\tau)}. \nn
\ee 
The Fourier equality to use is
\be
\sqrt{-\I N\tau} \theta \left[ \begin{array}{c} 
      { \frac{2k}{2N}} \\ 
      { 0} 
   \end{array} \right] { (N\tau)} = \theta \left[ \begin{array}{c} 
      { 0} \\ 
      { 0} 
   \end{array} \right] { (-N/\tau)} + \sum_{g=1}^{\frac{N-1}{2}}
      2\cos{\left(2\pi\frac{{2k.2g}}{4N}\right)} \label{NoddFourier2112}
   \theta \left[ \begin{array}{c} 
      { \frac{2g}{2N}} \\ 
      { 0} 
   \end{array} \right] { (-N/\tau)},
\ee
which gives a fusing matrix
\be
F  \left[\begin{array}{cc} 
      \hat{\sigma}_2 & \hat{\sigma}_1 \\ 
      \hat{\sigma}_1 & \hat{\sigma}_2 
   \end{array}\right] = \frac{1}{\sqrt{N}} \times \begin{array}{c|cc}
       & 0 & \Phi_{2k^\prime} \\ \hline
       0 & 1 & 2.16^{-(2k^{\prime})^2/4N} \\
       \Phi_{2k} & 16^{(2k)^2/4N} & 2.16^{((2k)^2-(2k^\prime))^2/4N} 
       \cos{(2\pi\frac{2k.2k^\prime}{4N})} 
   \end{array}. \label{FF1122}
\ee
The hexagon is again $F^2=1$, which holds.

\subsubsection{Case $\langle \hat{\sigma}_1 \hat{\sigma}_2 \hat{\sigma}_{2} \hat{\sigma}_1 \rangle$}
The fusing equation is ${\cal F}_p^{\hat{1}\hat{1}\hat{2}\hat{2}}(z) = \sum_q F[\hat{1}\hat{1}\hat{2}\hat{2}]_{pq} {\cal F}_q^{\hat{1}\hat{2}\hat{2}\hat{1}}(1-z)$, leading to
\be
\um 16^{N/4} \sqrt{-\I \tau} \theta \left[ \begin{array}{c} 
      { \um} \\ 
      { 0} 
   \end{array} \right] { (N\tau)} &=& 
F_{\Phi^1 0}  \theta \left[ \begin{array}{c} 
      { 0} \\ 
      { \um} 
   \end{array} \right] { (-N/\tau)} \nn \\
&&+ \sum_{q=0}^{\frac{N-1}{2}} 
F_{\Phi^i \Phi_{2q}}  16^{\frac{(2q)^2}{4N}} \theta \left[ \begin{array}{c} 
      { \frac{2q}{2N}} \\ 
      { \um} 
   \end{array} \right] { (-N/\tau)}, \nn \\
16^{(2k-1)^2/4N} \sqrt{-\I \tau} \theta \left[ \begin{array}{c} 
      { \frac{2k-1}{2N}} \\ 
      { 0} 
   \end{array} \right] { (N\tau)} &=& 
F_{\Phi_{2k-1}0} \theta \left[ \begin{array}{c} 
      { 0} \\ 
      { \um} 
   \end{array} \right] { (-N/\tau)} \label{NoddF1212}  \\ 
&&+ \sum_{q=0}^{\frac{N-1}{2}} 
F_{\Phi_{2k-1}\Phi_{2q}}  16^{\frac{(2q)^2}{4N}} 
   \theta \left[ \begin{array}{c} 
      { \frac{2q}{2N}} \\ 
      { \um} 
   \end{array} \right] { (-N/\tau)}. \nn
\ee
Using the Fourier equality
\be
\sqrt{-\I N\tau} \theta \left[ \begin{array}{c} 
      { \frac{2k+1}{2N}} \\ 
      { 0} 
   \end{array} \right] { (N\tau)} &=& \theta \left[ \begin{array}{c} 
      { 0} \\ 
      { \um} 
   \end{array} \right] { (-N/\tau)} \\ &&+ \sum_{g=1}^{(N-1)/2} 
   2.e^{-2\pi \I f/2N} \cos{\left(2\pi\frac{{(2k+1)2g}}{4N}\right)} 
\label{NoddFourier1212}
   \theta \left[ \begin{array}{c} 
      { \frac{2g}{2N}} \\ 
      { 0} 
   \end{array} \right] { (-N/\tau)}, \nn
\ee
(pick $k=(N-1)/2$ for the first line of (\ref{NoddF1212})) we get
\be
F  \left[\begin{array}{cc} 
      \hat{\sigma}_1 & \hat{\sigma}_2 \\ 
      \hat{\sigma}_1 & \hat{\sigma}_2 
   \end{array}\right] = \frac{1}{\sqrt{N}} \times \begin{array}{c|cc} 
      & 0 & \Phi_{2g} \\ \hline
      \Phi^i & \um 16^{N/4} & (-1)^g e^{-2\pi \I g/2N} 16^{N/4-(2g)^2/4N} \\
      \Phi_{2k+1} & 16^{(2k+1)^2/4N} & 2.16^{((2k+1)^2-(2g))^2/4N} 
      e^{-2\pi \I g/2N} \cos{(2\pi\frac{(2k+1)2g}{4N})}
   \end{array} \nn \\
\ee

\subsubsection{Case $\langle \hat{\sigma}_1 \hat{\sigma}_1 \hat{\sigma}_{1} \hat{\sigma}_1 \rangle$}
The last fusing equation is ${\cal F}_p^{\hat{1}\hat{2}\hat{2}\hat{1}}(z) = \sum_q F[\hat{1}\hat{2}\hat{2}\hat{1}]_{pq} {\cal F}_q^{\hat{1}\hat{2}\hat{2}\hat{1}}(1-z)$, which leads to
\be
\sqrt{-\I \tau} \theta \left[ \begin{array}{c} 
      { 0} \\ 
      { \um} 
   \end{array} \right] { (N\tau)} &=& 
F_{0 \Phi^1} \um 16^{N/4} \theta \left[ \begin{array}{c} 
      { \um} \\ 
      { 0} 
   \end{array} \right] { (-N/\tau)} \nn \\
   &&+ \sum_{q=0}^{\frac{N-1}{2}} 
F_{0 \Phi_{2q-1}}  16^{\frac{(2q-1)^2}{4N}} \theta \left[ \begin{array}{c} 
      { \frac{2q-1}{2N}} \\ 
      { 0} 
   \end{array} \right] { (-N/\tau)}, \\
16^{(2k)^2/4N} \sqrt{-\I \tau} \theta \left[ \begin{array}{c} 
      { \frac{2k}{2N}} \\ 
      { \um} 
   \end{array} \right] { (N\tau)} &=&  
F_{\Phi_{2k}\Phi^i} \um 16^{N/4} \theta \left[ \begin{array}{c} 
      { \um} \\ 
      { 0} 
   \end{array} \right] { (-N/\tau)} \nn \\
   &&+ \sum_{q=0}^{\frac{N-1}{2}} 
F_{\Phi_{2k}\Phi_{2q-1}}  16^{\frac{(2q-1)^2}{4N}} 
   \theta \left[ \begin{array}{c} 
      { \frac{2q-1}{2N}} \\ 
      { 0} 
   \end{array} \right] { (-N/\tau)}.
\label{NoddF2211} \nn
\ee
The Fourier equality is here
\be
&& \sqrt{-\I N\tau} \theta \left[ \begin{array}{c} 
      { \frac{2k}{2N}} \\ 
      { \um} 
   \end{array} \right] { (N\tau)} = e^{2\pi \I k/2N} (-1)^k \theta 
   \left[ \begin{array}{c} 
      { \um} \\ 
      { 0} 
   \end{array} \right] { (-N/\tau)} \nn \\&&+
      \sum_{g=1}^{(N-1)/2} 2.e^{2\pi \I k/2N} 
      \cos{\left(2\pi\frac{{2k(2g-1)}}{4N}\right)}
\label{NoddFourier2211}
   \theta \left[ \begin{array}{c} 
      { \frac{2g-1}{2N}} \\ 
      { 0} 
   \end{array} \right] { (-N/\tau)},
\ee
from which we get
\be
&& F\left[\begin{array}{cc} 
      \hat{\sigma}_2 & \hat{\sigma}_2 \\ 
      \hat{\sigma}_1 & \hat{\sigma}_1 
   \end{array}\right] = \frac{1}{\sqrt{N}} \times \\ && 
   \begin{array}{c|cc} 
      & \Phi^i & \Phi_{2k+1} \\ \hline
      0 & 2.16^{N/4} & 2.16^{-(2k+1)^2/4N} \\
      \Phi_{2g} & 2(-1)^g e^{2\pi \I g/2N} 16^{-N/4+(2g)^2/4N} & 
      2.16^{((2g)^2-(2k+1))^2/4N} 
      e^{2\pi \I g/2N} \cos{(2\pi\frac{(2k+1)2g}{4N})} 
\end{array} \nn
\ee
The hexagon is $F[1122]F[1221]=1$ and it holds. If one had the mixed correlators with $\tau_i$ fields, the hexagons could be used to determine fusing matrices $F[TTTT]$ with two or four $\tau_i$'s.

\chapter{Outlook}
\label{conclusion}
The quest of describing the macroscopic world and the microscopic world in a unified framework is perhaps the greatest theoretical puzzle since the events that preceded the discovery of Quantum Mechanics and Relativity. The challenge and renown of discovering a grand unified theory attracted many scientists to the field but the task turned out to be formidable indeed. There is not yet a convincing theory of quantum gravity, although many proposals have been advanced. String Theory is the most promising candidate for such a realization. A spectrum with gravitons and good ultraviolet behavior indicates it may very well be a step in the right direction.

String Theory is a fascinating subject. It is so comprehensive and widespread that it stimulates a lot of active research in many branches of Physics and Mathematics. Finding the path from the critical ten dimensions down to four is one of the main goals of string phenomenology and it should at least hint strongly towards correctness or not of String Theory itself. In this thesis we explored how one can start the reduction of dimensions in a consistent way by studying algebraic theories that are equivalent to compactification of one dimension. This was done in the context of open string theories, following the algebraic approach.

Open string theories have a much richer structure as compared to closed ones. The appearance of extra features on the world-sheet, such as boundaries and crosscaps, makes it an interesting challenge to try and understand the mechanics of the conformal field theory living on it. Geometrically, the boundaries introduce D-branes and crosscaps introduce orientifold O-planes. The presence of such objects in an open string theory allows for quite some flexibility in devising phenomenological scenarios. The brane world scenarios, for instance, are a very interesting idea based on D-branes whose test may even be within experimental reach. Black-hole geometries are another idea based on D-branes that is currently undergoing a lot of research. Open string theories are in any case part of the web of string dualities and therefore an essential piece fitting in the M-theory puzzle. They are by themselves a worthy object of pure research, and their relevance in various models is added motivation for their study.

Open strings come hand-to-hand with orientifold planes. The O-planes are necessary to restore consistency of the total theory, both at space-time and world-sheet level. Orientifolds project some states out of the theory and therefore gives rise to unoriented string theories. From the phenomenological side, O-planes have been used to construct novel cosmological scenarios \cite{mig}.

The open descendants construction allows one to build an open string theory starting from a closed string one. This requires us to first search for a parent closed string theory, from which the open strings will descend. In the algebraic approach, this amounts to finding a conformal field theory model with modular invariant torus partition function. Then we must define the states that describe a boundary or crosscap, which is done via linear combinations of closed string Ishibashi states. Closed string scattering between boundary and crosscap states can then be calculated and that provides us the Klein bottle, annulus and Moebius strip. After taking due care with orientation matters, transforming the result to the open string channel provides us with the unoriented closed and open string partition functions. Partition functions count states, so requiring positive and integer state degeneracies becomes the first consistency conditions on the open string theory.

Positivity and integrality are very restrictive requirements with which a lot can be done. They were applied in this thesis to extended free boson orbifold theories, which correspond to compactifications on line segments. The orbifold torus partition function changes according to the free boson compactification radius. When the radius is integer, the modular invariant is a simple current invariant, for which the boundary and crosscap coefficients were derived in \cite{foe}. If the radius is rational the invariant is exceptional, which was the case in this thesis. Nevertheless the positivity and integrality requirements turned out to be enough to postulate a consistent set of boundary and crosscap coefficients.

The positivity and integrality constraints on one-loop partition functions are a necessary consistency requirement for open string theories. Necessary, but not sufficient. Consistency of the string perturbative expansion requires a check at all orders, not just at one-loop. The sewing constraints' inductive mechanism gives us a set of equations which, when satisfied, guarantee that sewing two world-sheet topologies together will yield a consistent theory in the sewn surface, in the sense that it will be unambiguous and that it will have the correct factorization properties.

Checking the sewing constraints requires specific data from the conformal field theory on the world-sheet. This data includes model-dependent quantities like operator product coefficients, conformal block functions and duality matrices, which are often hard to get since they require solving complicated differential equations for correlation functions. In the free boson orbifold case the task of obtaining the correlation functions is tractable because it is a simple model. The untwisted sector of the orbifold has vertex operators that are combination of free boson ones and therefore the untwisted orbifold correlators are combinations of free boson correlators, which are known. Part of the twisted sector correlators are also known, and with all this one can derive some of the quantities needed to check the sewing constraints.

On this thesis, we concentrated on deriving the orbifold fusing and braiding matrices. These matrices have a pivotal importance in the sewing constraints, since all of the constraint equations depend on them explicitly. They are exchange matrices that interpolate between the various channel expansions of closed string correlation functions. The duality matrices are themselves also constrained by the pentagon and hexagon identities, two identities derived from alternative but equivalent ways of writing the conformal blocks the matrices relate. For conformal field theories with charge-conjugation modular invariant, the fusing matrices turn out to be equivalent to the boundary operator product coefficients, a simplification that leads to equivalence of the boundary four-point function sewing constraint to the pentagon identity. Finding a set of fusing matrices is then equivalent to solving one of the sewing constraints for a class of conformal field theories.

From the correlation functions for the orbifold untwisted sector one could extract the conformal blocks and from there the fusing and braiding matrices could be obtained. This procedure brought however some unexpected surprises. The orbifold theory has extended symmetries in its chiral algebra and this brings up a few features that are usually not present if one considers the Virasoro algebra alone. Existence of chiral fields of equal conformal weights is one of the possible complications at the level of conformal blocks and this does indeed happen for the orbifold. Because of it ambiguities arise in comparing conformal blocks to obtain the duality matrices. Understanding how can one relate the conformal blocks by means of duality matrices in the presence of extended chiral algebras is an interesting problem to be investigated.

In practice what happened in the orbifold case was that the naive calculations produced the correct result for duality matrices up to signs. It was then possible to resort to the pentagon and hexagon identities to find a solution for the fusing and braiding matrices with correct signs and to check it was correct up to $N=21$ and that it was essentially unique. For the mixed and twisted sector, fusing matrices were derived from the correlators involving twisted $\sigma_i$ fields, with which some of the hexagon identities could be solved. A full check of the whole orbifold pentagon and hexagon identities can be done once the correlators involving twisted $\tau_i$ fields are available. The set of untwisted fusing matrices is however already a good start for proving consistency of free boson orbifold theories since it is automatically the solution of one of the sewing constraints.

The stage is then set for the first thorough study of open string consistency to all orders in perturbation theory. When done, it will reassure us that compactifications are indeed consistent and that String Theory rests on solid grounds.

\chapter*{Samenvatting}
\addcontentsline{toc}{chapter}{Samenvatting}
Natuurkunde, in haar oneindige zoektocht om te kunnen begrijpen en verklaren hoe het universum werkt, komt vele fascinerende enigmas tegen. Tegenwoordig is de unificatie, het verenigen, van de macroscopische wereld met de microscopiche wereld het meest fascinerende onderdeel voor vele natuurkundigen. Einstein's beschrijving van de macroscopische wereld en de Standaard Model beschrijving van de microscopische wereld beschrijven hun desbetreffende wereld tot een verbazende precisie. Maar als we proberen een extrapolatie te maken van die theorie\"en buiten hun domeinen, zodat een verbinding ertussen gevonden kan worden, komen we tot de conclusie dat het niet lukt. Het lijkt alsof de twee werelden door verschillende wetten geregeerd worden, zonder een blijkbare connectie.

Het gaat onze logica te boven, te denken dat deze scheiding onoplosbaar is. De macroscopische en microscopische werelden maken deel uit van een en hetzelfde universum, hoewel op een verschillende schaal bekeken. Daarom denken wetenschappers dat algemene relativiteit en het Standard Model louter verschillende begrenzingen van dezelfde uniforme theorie zijn. Het vinden van een consistente theorie van quantum zwaartekracht zou ons begrip van de relatie tussen de verschillende fysieke schalen drastische vergroten en dit is de favoriete puzzel van de theoretici van de afgelopen 50 jaren geweest.

Snaartheorie is een veelbelovende kandidaat om de unificatie te realiseren. Haar principe is heel eenvoudig: in plaats van puntdeeltjes nemen we kleine snaartjes. Deze hoofdaanname, samen met enkele eenvoudige andere (zoals supersymmtrie) en de eis van mathematische consistentie is voldoende om ons een systeem op te leveren met een zeer complexe structuur en onvoorstelbare eigenschappen die (tot nu toe) vrij zijn van inconsistenties. Echter, volledige consistentie kan alleen worden bereikt in een ruimte-tijd met een onnatuurlijk aantal dimensies: $D=10$.

Het mysterische resultaat $D=10$ kan feitelijk in ons voordeel omgekeerd worden. Aangezien er geen expliciete eis wordt gelegd op geometrie van ruimte-tijd, kunnen de extra dimensies als klein en periodiek worden genomen. Het samenvoegen van zes dimensies kan op haast oneindig veel manieren gedaan worden, en elk van die manieren levert een verschillende vier-dimensionale snaartheorie op. Het is daarom een belangrijke taak om deze theori\"en te classificeren en op consistentie te controleren.

Snaren worden wiskundig beschreven door de conformele veld theori\"en die op hun wereld-oppervlakke bestaan. Omdat er gesloten en geopende snaren mogelijk zijn, moeten we de conformele veld theorie op gesloten oppervlakken en op oppervlakken met begrenzing bestuderen. Bovendien, hebben open snaren eindpunten. De objecten waar deze eindpunten liggen worden D-branes genoemd. Consistentie vereist dan het introduceren van een ander object, de O-planes, als D-branes voorkomen in een snaartheorie. De O-planes introduceren, op hun beurt, crosscaps op de wereld-oppervlakke en dit leidt tot het voorkomen van niet-georienteerde snaartheorien. In open snaartheorien komen uiteindelijk open of gesloten, geori\"enteerde of niet-geori\"enteerde snaren, voor waarvan hun wereld-oppervlakke uit oppervlakken kunnen opkomen als de torus, Klein vles, annulus of Moebius band. Dit draagt bij tot het bestuderen van de conformele veld theorie op oppervlakken met grenzen en crosscaps. Het feit dat sommige dimensies compact zijn kan ook door de conformele veld theorie beschreven worden. Dit is de zo genoemde `algebraische methode', een alternatief op compactificatie, en deze is hierop in verschillende aspecten superieur.

In dit boek hebben we op de algebraische manier open snaartheori\"en geconstrueerd op basis van gesloten snaren. Bij het analiseren van verschillende diagrammen, zoals de one-loop torus, Klein vles, annulus en Moebius band, hebben we positief en integraal eisen afgeleid die gebruikt kunnen worden om elk model te testen. Deze procedure werd daarna gebruikt om een klasse van conformele veld theori\"en te klassificeren en voor een expliciete berekening van enkele van de relevante quantities van een eenvoudig voorbeeld van vrije boson orbifold modellen.

Volledige consistentie controles gaan de positiviteit en integraliteit controles te boven. Consistentie van de verstorende expansie van de snaar maakt het noodzakleijk om de `sewing constraints' van het model na te lopen. Deze constraints zijn relaties tussen verschillende model afhankelijke hoeveelheden die, als er aan voldaan is, ons een correcte en onweerlegbare factorisatie van de correlatie functies op oppervlakken met grenzen en crosscaps van de conformele veld theorie verzekeren. Het laatste deel van dit boek behandelt de berekening van de vrije boson orbifold fusie matrices. Deze matrices zijn model-afhankelijke quantities die een cruciaal stuk informatie bevatten die nodig zijn bij het werken met sewing constraints. De berekening gaf opheldering over enkele subtiliteiten van de conformele veld theorie met uitgebreide symmetrie. Hoewel niet alle matrices gevonden werden, konden met de vele wel gevonden matrices een aantal sewing constraints geverifieerd worden.

Dit boek draagt, als een eerste stap, bij aan een volledige consistentie check van een van de eenvoudigste modellen waarop snaartheori\"en gebouwd kunnen worden, namelijk de vrije boson orbifolds. Snaartheorie is een zeer interessante theorie die momenteel intensief bestudeerd wordt, ook omdat haar eigenschappen breed toepasbaar zijn. Deze eigenschappen worden benut om een serie van modellen die gebruikt worden in andere takken van de natuurkunde te onderbouwen. Snaartheorie is daadwerkelijk een serieuze kandidaat voor een universele theorie.
\thispagestyle{plain}

\chapter*{Acknowledgements}
\addcontentsline{toc}{chapter}{Acknowledgements}
First of all I would like to thank my supervisor Bert Schellekens for all the time and trouble he had helping me. He was always there when I needed him, and that is the best thing one can say of a friend and supervisor.

From the Physics side I thank my office mate Lennaert Huiszoon, with whom I held many fruitful conversations, not only about work. Thanks also to Yassen Stanev and Augusto Sagnotti for discussions about open descendants. It was, after all, them who started the field and is always a privilege to learn the subject first hand from the masters themselves. It was also a pleasure to meet J\"urgen Fuchs, Christoph Schweigert and Ingo Runkel, who I thank for explaining me the more mathematical side of the subject. I wish them good luck in their attempt to give conformal field theory a solid mathematical basis. Thanks also to Robbert Dijkgraaf for explaining me his work on free boson orbifolds and to Jos Vermaseren for his suggestions to disentangle the mixed sector.

A big thanks also to the people in Coimbra, especially to Eef van Beveren, supervisor to my graduation thesis and a friend of Bert, to whom he recommended me. Regards to Nicholas Petropoulos for sharing his office and computer when I was writing up the manuscript and to Lino Tralh\~ao for explaining details on complex analysis.

For providing financial support I thank Portuguese Funda\cc \~ao para a Ci\^encia e Tecnologia which supported me for four years under the reference BD/13770/97. Honorable mention to Teus van Egdom for arranging for a scholarship extension, awarded by Stichting F.O.M. and to Justus Koch for encouragement.

Finally I would like to thank all of my family, especially my girlfriend and my mother for encouragement and to all my friends and colleagues in Portugal and in Holland for just being there.

\chapter*{Curriculum vit\ae}
\addcontentsline{toc}{chapter}{Curriculum vit\ae}
Nuno Miguel Marques de Sousa was born 17th of January 1973 in Coimbra, Portugal. He did his basic education in several schools throughout the center of Portugal. In 1995 he got his degree in Physics from the University of Coimbra. In 1998 and in the same university he defended his M.Sc. thesis in Theoretical Many-Body Physics entitled {\em Spin tunneling in the Lipkin model}. During his period in Coimbra he won various regional and national titles in Chess and Bridge. In 1995 he was a member of the Portuguese junior Bridge team that participated in the European Junior Bridge Championships, held in Cardiff, Wales.

From 1998 to present he did his Ph.D. research on open string conformal field theory at the NIKHEF institute in Amsterdam, Holland, under the supervision of Prof.Dr. A.N. Schellekens.

Nuno M.M. de Sousa has applied for a three-year mixed post-doctoral position in the Technical University of Lisbon and University of Porto, where he hopes to continue his research on String Theory.

\appendix
\chapter{Exceptional automorphisms of the free boson orbifold}
\label{EA}
In this appendix we describe the exceptional fusion rule automorphisms of the free boson orbifold theory. We denote the automorphism by $\omega$. It acts trivially on all chiral fields other than $\Phi_k$: 
\be
\omega(0)=0,\;\;\; 
\omega(J)=J ,\;\;\; 
\omega(\Phi^i)=\Phi^i,\;\;\; 
\omega(\sigma_i)=\sigma_i,\;\;\; 
\omega(\tau_i)=\tau_i.
\ee
On the $\Phi_k$ the action is as follows. Write $\Phi_k$ as
\be
\Phi_k \rightarrow 
   \left\{ 
      \begin{array}{cl} 
         \Phi_{N-2g} & \mbox{for $k$ odd} \\ 
         \Phi_{2g}   & \mbox{for $k$ even}
      \end{array} 
   \right.
\ee
with $g$ taking the values $1,\ldots,(N-1)/2$. Now look for the smallest integer $m>1$ such that $m^2=1 \; \mod \; N$. For any $x$ define the (unique) number $[x]_N$, $0 \leq [x]_N \leq N/2$, such that $x \equiv \pm [x]_N \; \mod \; N$ for some choice of sign and define the permutation 
\be
\pi_m: \pi_m(g)  = [mg]_N \label{appperm}.
\ee
The automorphism then acts on $\Phi_k$ as
\be
\omega(\Phi_{2g}) &= &\Phi_{2[mg]_N} \nn \\
\omega(\Phi_{N-2g}) &=&  \Phi_{N-2[mg]_N}. 
\ee
The D+A torus is then
\be
T=\sum_g \chi_{\Phi_{2g}} \overline{\chi}_{\Phi_{2[mg]_N}} + \chi_{\Phi_{N-2g}} \overline{\chi}_{\Phi_{N-2[mg]_N}} +  \mbox{diagonal in the other fields}.
\label{appTd+a}\ee
The C+A torus is obtained applying charge conjugation to (\ref{appTd+a}). One can verify that the two tori obey $[T,Z]=[S,Z]=0$ and are thus modular invariant.

\chapter{Duality matrices for free boson orbifolds}
\label{appappfusing}
First we review the chiral field content. The extended free boson orbifold has the following chiral labels

\begin{center}
\begin{tabular}{c|c|ccc|c|c}
   \multicolumn{3}{c}{\large $N$ even} & {\hspace{1cm}} & 
   \multicolumn{3}{c}{\large $N$ odd} \\
    Label & weight & conjugate & {\hspace{1cm}} & 
    Label & weight & conjugate \\
         \cline{1-3} \cline{5-7}
      0   &     0  &      0     & {\hspace{1cm}} & 
      0   &     0  &      0    \\
      $J$   &     1  &      $J$     & {\hspace{1cm}} & 
      $J$   &     1  &      $J$    \\
   $\Phi^i, \; i=1,2$ &          $N/4$ &   $\Phi^i$   & {\hspace{1cm}} & 
   $\hat{\Phi}^i, \; i=1,2$ &    $N/4$ &   $\hat{\Phi}^j, \; i\neq j$   \\
   $\Phi_k, \; k=1\ldots N-1$ & $k^2/4N$ &  $\Phi_k$     & {\hspace{1cm}} & 
   $\Phi_k, \; k=1\ldots N-1$ & $k^2/4N$ &  $\Phi_k$   \\
   $\sigma_i, \; i=1,2 $       & 1/16   & $\sigma_i$    & {\hspace{1cm}} & 
   $\hat{\sigma}_i, \; i=1,2$ & 1/16   & $\hat{\sigma}_i, \; i\neq j$   \\ 
   $\tau_      i, \; i=1,2$   & 9/16   & $\tau_i$      & {\hspace{1cm}} & 
   $\hat{\tau}_i, \; i=1,2$   & 9/16   & $\hat{\tau}_i, \; i\neq j$
\end{tabular}
\end{center}
The indices $i,j,k,l$ refer to a generic chiral label, but are also used in $\Phi^i$, $\sigma_i$ and $\tau_i$ (with and without hats). The difference should be clear from the context. The twisted sector fields are $\sigma_i$ and $\tau_i$, the remaining fields being untwisted sector fields. We designate a generic untwisted/twisted field by $U,T$ respectively.  

For convenience, we repeat the definition of $[a]$, which is as follows. Given an integer $a$, define $[a]$ as
\be
[a] = \left\{ \ba{cl}
         |a|   & \text{if $-N<a<N$} \\
         2N-a  & \text{if $a>N$}    \\
         a+2N  & \text{if $a<-N$}
      \ea \right. \nn
\ee
This function maps an index $k$ of $\Phi_k$ into the fundamental range $[0,N-1]$. As gauge choice for the fusing matrices we use (\ref{FRSgauge}), together with conformal block normalization of (\ref{blockdesc}).

\section{$N$ even}
\subsubsection{Untwisted sector}
\subsubsection{Braiding matrices}
The braiding matrices can be derived from the fusing matrix via (\ref{b->f}). For that we need the eigenvalues $\xi_{ij}^k$, which for $N$ even are
\be
\xi_{ij}^k = 1, \; \forall \; i,j,k.
\ee

\subsubsection{Fusing matrices}
\subsubsection{Case $F[ijk0]$} 
\be
 F\!\left[\begin{array}{cc} j & k \\ i & 0 \end{array}\right] = 1,
    \;\;\; \mbox{if $N_{ij}^{\ph{ij}{k}}=1$.} \label{appF[ijk0]even} 
\ee
Likewise for permutations $F[ij0k]=F[i0jk]=F[0ijk]=1$.

\subsubsection{Case $F[JJJJ]$}
\be 
 F\!\left[\begin{array}{cc} J & J \\ J & J \end{array}\right] = 1.
\ee 
 
\subsubsection{Case $F[JJ\Phi^i\Phi^i]$} 
\be 
&&   F\!\left[ \begin{array}{cc}  
        J & \Phi^{i} \\ 
        J & \Phi^{i} 
        \end{array}\right]= \frac{N}{2}\;\;\;
     F\!\left[ \begin{array}{cc}  
        \Phi^{i} & \Phi^{i} \\ 
           J     &    J 
        \end{array}\right]= \frac{2}{N}\;\;\;
     F\!\left[ \begin{array}{cc}  
        \Phi^{i} &    J \\ 
           J     & \Phi^{i} 
        \end{array}\right]=1  
\nn \\
&&   F\!\left[ \begin{array}{cc}  
           J     &    J \\ 
        \Phi^{i} & \Phi^{i} 
        \end{array}\right]= \frac{2}{N}\;\;\;
     F\!\left[ \begin{array}{cc}  
        \Phi^{i} & J \\ 
        \Phi^{i} & J 
        \end{array}\right]= \frac{N}{2}\;\;\;
     F\!\left[ \begin{array}{cc}  
           J     & \Phi^{i} \\ 
        \Phi^{i} &    J 
        \end{array}\right]= 1
\ee
 
\subsubsection{Case $F[JJ\Phi_k \Phi_k]$} 
\be 
&&   F\!\left[ \begin{array}{cc}  
        J & \Phi_k \\ 
        J & \Phi_k 
        \end{array}\right]= \frac{k^2}{2N}\;\;\;
     F\!\left[ \begin{array}{cc}  
        \Phi_k & \Phi_k \\ 
        J & J 
        \end{array}\right]= \frac{2N}{k^2}\;\;\;
     F\!\left[ \begin{array}{cc}  
        \Phi_k & J \\ 
        J & \Phi_k 
        \end{array}\right]= 1 \nn \\ 
&&   F\!\left[ \begin{array}{cc}  
        J & J \\ 
        \Phi_k & \Phi_k 
        \end{array}\right]= \frac{2N}{k^2}\;\;\;
     F\!\left[ \begin{array}{cc}  
        \Phi_k & J \\ 
        \Phi_k & J 
        \end{array}\right]=\frac{k^2}{2N}\;\;\;
     F\!\left[ \begin{array}{cc}  
        J & \Phi_k \\ 
        \Phi_k & J 
        \end{array}\right]= 1 
\ee 

\subsubsection{Case $F[J \Phi_k \Phi_a \Phi_b]$} 
\be
F\!\left[ \begin{array}{cc}  
      \Phi_k & \Phi_a \\ 
      J & \Phi_b 
   \end{array}\right]= 
      \left\{ \begin{array}{l} 
      \frac{b}{b-a},\; k=\pm(a-b)  \\ 
      \frac{b}{b+a},\; k=a+b       \\ 
      \frac{-b}{2N-a-b}, \; k=2N-a-b \end{array} 
      \right. &
F\!\left[ \begin{array}{cc}  
      \Phi_b & J \\ 
      \Phi_a & \Phi_k 
   \end{array}\right]= 
      \left\{ \begin{array}{l} 
      \frac{b}{b-a},\; k=\pm(a-b)  \\ 
      \frac{b}{b+a},\; k=a+b       \\ 
      \frac{-b}{2N-a-b},\; k=2N-a-b \end{array} 
   \right. \nn \\
F\!\left[ \begin{array}{cc}  
      J & \Phi_a \\ 
      \Phi_k & \Phi_b 
   \end{array}\right]= 
      \left\{ \begin{array}{l} 
      \frac{a}{a-b},\; k=\pm(a-b)  \\ 
      \frac{a}{a+b},\; k=a+b       \\ 
      \frac{-a}{2N-a-b}, k=2N-a-b \end{array} 
   \right. & 
F\!\left[ \begin{array}{cc}  
      \Phi_b & \Phi_k \\ 
      \Phi_a & J 
   \end{array}\right]=
      \left\{ \begin{array}{l} 
      \frac{a}{a-b},\; k=\pm(a-b)  \\ 
      \frac{a}{a+b},\; k=a+b       \\ 
      \frac{-a}{2N-a-b},\;k=2N-a-b \end{array} 
   \right. \nn \\
\ee 

\subsubsection{Case $F[J\Phi_k \Phi_{N-k} \Phi^i]$} 
\be
\ba{lll}
F\!\left[ \begin{array}{cc}  
      \Phi_k & \Phi_{N-k} \\ 
      J & \Phi^{i} 
   \end{array}\right]_{{}_{{}_{{}_{{}_{}}}}}
                    = \; {\displaystyle \frac{N}{k}}       &
F\!\left[ \begin{array}{cc} 
      \Phi^{i} & \Phi_{N-k} \\ 
      J & \Phi_k 
   \end{array}\right]= \; {\displaystyle \frac{k}{N}}      &
F\!\left[ \begin{array}{cc}  
      \Phi_{N-k} & \Phi^{i} \\ 
      J & \Phi_k 
   \end{array}\right]= \; {\displaystyle \frac{k}{k-N}}    \\ 
F\!\left[ \begin{array}{cc}  
      J & \Phi_{N-k} \\ 
      \Phi_k & \Phi^{i} 
   \end{array}\right]_{{}_{{}_{{}_{{}_{}}}}}
                     = \; {\displaystyle \frac{k-N}{k}}     &
F\!\left[ \begin{array}{cc} 
      J & \Phi_{k} \\ 
      \Phi^{i} & \Phi_{N-k} 
   \end{array}\right]= \; {\displaystyle \frac{k}{N}}       &
F\!\left[ \begin{array}{cc}  
      J & \Phi^{i} \\ 
      \Phi_k & \Phi_{N-k} 
   \end{array}\right]= \; {\displaystyle \frac{N}{k}}      \\
F\!\left[ \begin{array}{cc}  
      \Phi_k & J \\ 
      \Phi^{i} & \Phi_{N-k} 
   \end{array}\right]_{{}_{{}_{{}_{{}_{}}}}}
                     = \; {\displaystyle \frac{k}{k-N}}     &
F\!\left[ \begin{array}{cc} 
      \Phi_{N-k} & J \\ 
      \Phi_{k} & \Phi^{i} 
   \end{array}\right]= \; {\displaystyle \frac{N-k}{N}}     &
F\!\left[ \begin{array}{cc}  
      \Phi^{i} & J \\ 
      \Phi_k & \Phi_{N-k} 
   \end{array}\right]= \; {\displaystyle \frac{N}{N-k}}    \\ 
F\!\left[ \begin{array}{cc}  
      \Phi^{i} & \Phi_{N-k} \\ 
      \Phi_k & J 
   \end{array}\right]= \; {\displaystyle \frac{k}{k-N}}     &
F\!\left[ \begin{array}{cc} 
      \Phi_{N-k} & \Phi_{k} \\ 
      \Phi^{i} & J 
   \end{array}\right]= \; {\displaystyle \frac{N}{k}}       &
F\!\left[ \begin{array}{cc}  
      \Phi_{N-k} & \Phi^{i} \\ 
      \Phi_k & J 
   \end{array}\right]= \; {\displaystyle \frac{k}{N}}
\ea
\label{appF[Jk(N-k)i]even}
\ee 

\subsubsection{Case $F[\Phi_a \Phi_a \Phi^i \Phi^{j}]$}
\be
\ba{ll}
F\!\left[ \begin{array}{cc}  
      \Phi^i & \Phi_a \\ 
      \Phi_a & \Phi^{j} 
   \end{array}\right]= {\displaystyle (-1)^{i+j+a}}
& F\!\left[ \begin{array}{cc}  
      \Phi_a & \Phi^{j} \\ 
      \Phi^i & \Phi_a 
   \end{array}\right]_{{}_{{}_{{}_{{}_{}}}}}= {\displaystyle (-1)^{i+j+a}} 
\\ 
F\!\left[ \begin{array}{cc} 
      \Phi_a & \Phi^{i} \\ 
      \Phi_a & \Phi^{i} 
   \end{array}\right]= {\displaystyle \frac{1}{2}} \;\;\;
F\!\left[ \begin{array}{cc} 
      \Phi_a & \Phi^{i} \\ 
      \Phi_a & \Phi^{j} 
   \end{array}\right]= {\displaystyle \frac{1}{a}}  
& F\!\left[ \begin{array}{cc} 
      \Phi^i & \Phi^{i} \\ 
      \Phi_a & \Phi_{a} 
   \end{array}\right]= {\displaystyle 2} \;\;\;
F\!\left[ \begin{array}{cc} 
      \Phi^i & \Phi^{j} \\ 
      \Phi_a & \Phi_a 
   \end{array}\right]_{{}_{{}_{{}_{{}_{}}}}}= {\displaystyle a}
\\ 
F\!\left[ \begin{array}{cc} 
      \Phi^{i} & \Phi_a \\ 
      \Phi^{i} & \Phi_a 
   \end{array}\right]= {\displaystyle\frac{1}{2}} \;\;\;
F\!\left[ \begin{array}{cc} 
      \Phi^j & \Phi_a \\ 
      \Phi^i & \Phi_a 
   \end{array}\right]= {\displaystyle\frac{1}{a}} 
& F\!\left[ \begin{array}{cc} 
      \Phi_a & \Phi_a \\ 
      \Phi^{i} & \Phi^{i} 
   \end{array}\right]= {\displaystyle 2} \;\;\;
F\!\left[ \begin{array}{cc}
      \Phi_a & \Phi_a \\ 
      \Phi^i & \Phi^j 
   \end{array}\right]= {\displaystyle a}
\ea
\ee 

\subsubsection{Case $F[\Phi^i \Phi_k \Phi_a \Phi_b]$} 
\be 
F\!\left[ \begin{array}{cc}
\Phi_a & \Phi_b \\ 
\Phi^i & \Phi_c 
\end{array}\right]= \left\{ \ba{cl} -(-1)^{i+b+\frac{N}{2}} & \mbox{if $a=b+c-N$} \\ -(-1)^{i+b+\frac{N}{2}} & \mbox{if $c=a+b-N$} \\ 1 & \mbox{else}\ea \right. 
\ee 
The permutations $F[c\Phi^i ab]$, $F[bc\Phi^i a]$ and $F[abc\Phi^i]$ are similar, the field taking the role of $\Phi_b$ now being the one diagonally opposite to $\Phi^i$.

\subsubsection{Case $F[\Phi^i\Phi^j\Phi^k\Phi^l]$} 
\be 
F\!\left[ \begin{array}{cc} 
      \Phi^i & \Phi^i \\ 
      \Phi^i & \Phi^i 
   \end{array}\right]= 1 \;\;\; 
F\!\left[ \begin{array}{cc} 
      \Phi^j & \Phi^i \\ 
      \Phi^i & \Phi^j 
   \end{array}\right]= 1 \;\;\;
F\!\left[ \begin{array}{cc} 
      \Phi^j & \Phi^j \\ 
      \Phi^i & \Phi^i 
   \end{array}\right]= \frac{2}{N} \;\;\; 
F\!\left[ \begin{array}{cc} 
      \Phi^i & \Phi^j \\ 
      \Phi^i & \Phi^j 
   \end{array}\right]= \frac{N}{2}
\ee

\subsubsection{Case $F[\Phi_a \Phi_a \Phi_a \Phi_a]$}
\be
F\! \left[ \begin{array}{cc} 
       \Phi_a & \Phi_a \\ 
       \Phi_a & \Phi_a 
   \end{array} \right] = 
   \begin{array}{c|ccc} & 0 & J & \Phi_{[2a]} \\ \hline 
                        0 & \um & \frac{x}{2} & \um \\ 
                        J & \frac{1}{2x} & \um & -\frac{1}{2x} \\ 
                        \Phi_{[2a]} & 1 & -x & 0  
   \end{array}\;\;\; x=\frac{a^2}{2N}, \;\;\; a \neq N/2.
\ee
\be
F\! \left[ \begin{array}{cc} 
       \Phi_{\frac{N}{2}} & \Phi_{\frac{N}{2}} \\ 
       \Phi_{\frac{N}{2}} & \Phi_{\frac{N}{2}} 
    \end{array} \right] 
  = \begin{array}{c|cccc} 
       & 0 & J & \Phi^{1} & \Phi^2 \\ \hline 
       0 & \um & \frac{x}{2} & \frac{1}{4} & \frac{1}{4} \\ 
       J & \frac{1}{2x} & \um & -\frac{1}{4x} & -\frac{1}{4x} \\ 
       \Phi^{1} & 1 & -x & \um (-1)^{\frac{N}{2}} & 
          -\um (-1)^{\frac{N}{2}} \\ 
       \Phi^2 & 1 & -x & -\um (-1)^{\frac{N}{2}} & \um (-1)^{\frac{N}{2}}  
    \end{array} \;\;\; x=\frac{N}{8}. 
\ee

\subsubsection{Case  $F[\Phi_a \Phi_a \Phi_a \Phi_b]$} 
\be
F\!\left[ \begin{array}{cc} \Phi_a & \Phi_a \\ 
   \Phi_a & \Phi_b \end{array} \right]=
   \left\{ \ba{cl} 1 & \mbox{if $b=3a$} \\ (-1)^a & \mbox{else} 
   \ea \right. 
\ee

\subsubsection{Case  $F[\Phi_a \Phi_a \Phi_b \Phi_b]$} 
\be 
&& F\!\left[ \begin{array}{cc}  
      \Phi_b & \Phi_a \\ 
      \Phi_a & \Phi_b 
   \end{array}\right]= 
   \begin{array}{c|cc}  
      & \Phi_{[a+b]} & \Phi_{[a-b]} \\ \hline 
      \Phi_{[a+b]} & 0 & \epsilon \\ 
      \Phi_{[a-b]} & \epsilon & 0 
   \end{array} \;\;\; \epsilon=
      \left\{ \ba{cl} (-1)^{a+b} & \mbox{if $a+b>N$} \\ 1 & \mbox{else} \ea 
      \right. \\ 
&& F\!\left[ \begin{array}{cc}  
      \Phi_a & \Phi_c \\ 
      \Phi_a & \Phi_c 
      \end{array}\right]= 
   \begin{array}{c|cc}  
      & \Phi_{[a+c]} & \Phi_{[a-c]} \\ \hline 
      0 &     \um    &    \um \\ 
      J & \frac{-1}{2x} & \frac{1}{2x} 
   \end{array} \;\;\; 
F\!\left[ \begin{array}{cc}  
      \Phi_c & \Phi_c \\ 
      \Phi_a & \Phi_a 
   \end{array}\right]= 
   \begin{array}{c|cc}  
      &      0     &    J    \\ \hline 
      \Phi_{[a+c]}  &      1     &   -x   \\ 
      \Phi_{[a-c]}  &      1     &    x 
   \end{array} \;\;\; x'=\frac{ac}{2N} \nn
\ee
\be 
F\!\left[ \begin{array}{cc}  
      \Phi_{(N-a)} & \Phi_a \\ 
      \Phi_a  & \Phi_{(N-a)} 
   \end{array}\right]&=& 
   \begin{array}{c|ccc}  
                &   \Phi^1   &   \Phi^2 & \Phi_{[N-2a]} \\ \hline 
      \Phi^1    &  \um (-1)^a    &  -\um (-1)^a    &      1 \\ 
      \Phi^2    & -\um (-1)^a    &   \um (-1)^a    &      1 \\ 
      \Phi_{[N-2a]} &    \um     &    \um   &      0 
   \end{array} \nn \\
F\!\left[ \begin{array}{cc}  
      \Phi_{(N-a)} & \Phi_{(N-a)} \\ 
      \Phi_a & \Phi_a 
   \end{array}\right]&=&
   \begin{array}{c|ccc}  
                    &     0     &     J    & \Phi_{[2a]} \\ \hline  
      \Phi^1        &     1     &    -x   &     (-1)^{\frac{N}{2}}     \\ 
      \Phi^2        &     1     &    -x   &    -(-1)^{\frac{N}{2}}     \\ 
      \Phi_{[N-2a]} &     1     &     x   &     0 
   \end{array} \;\;\; x=\frac{a(N-a)}{2N} \\ 
F\!\left[ \begin{array}{cc}  
      \Phi_a & \Phi_{(N-a)} \\ 
      \Phi_a & \Phi_{(N-a)}
   \end{array}\right]&=& 
   \begin{array}{c|ccc}  
      &    \Phi^1  &   \Phi^2   &  \Phi_{[N-2a]}    \\ \hline 
      0  &      \frac{1}{4}     &   \frac{1}{4}       &  \um    \\ 
      J  &-\frac{1}{4x}         &  -\frac{1}{4x}      &  \frac{1}{2x} \\ 
      \Phi_{[2a]} & \um(-1)^{\frac{N}{2}} & -\um(-1)^{\frac{N}{2}} & 0 
\end{array} \nn 
\ee
\clearpage 

\subsubsection{Case $F[\Phi_a \Phi_a \Phi_b \Phi_c]$} 
\be 
F\!\left[ \begin{array}{cc}  
      \Phi_{\frac{N}{2}} & \Phi_b \\
      \Phi_{\frac{N}{2}} & \Phi_{(N-b)}
   \end{array}\right]&=&\begin{array}{c|cc}  
             & \Phi_{[\frac{N}{2}+b]} & \Phi_{[\frac{N}{2}-b]} \\ \hline
      \Phi^1 &   -(-1)^b   &    1   \\ 
      \Phi^2 &    (-1)^b   &    1 
   \end{array} \nn \\
F\!\left[ \begin{array}{cc}  
      \Phi_{(N-b)} &\Phi_b \\ 
      \Phi_{\frac{N}{2}} & \Phi_{\frac{N}{2}} 
   \end{array}\right]&=&\begin{array}{c|cc}  
           & \Phi^1 & \Phi^2 \\ \hline 
      \Phi_{[\frac{N}{2}+b]} &   -\um(-1)^b   &    \um(-1)^b   \\ 
      \Phi_{[\frac{N}{2}-b]} &    \um   &    \um
   \end{array} \nn \\
F\!\left[ \begin{array}{cc}  
      \Phi_{\frac{N}{2}} & \Phi_{\frac{N}{2}} \\ 
      \Phi_{(N-b)} & \Phi_b
   \end{array}\right]&=&\begin{array}{c|cc}  
           & \Phi^1 & \Phi^2 \\ \hline 
      \Phi_{[\frac{N}{2}+b]} &   -\um(-1)^b   &    \um(-1)^b   \\ 
      \Phi_{[\frac{N}{2}-b]} &    \um   &    \um 
   \end{array} \nn \\
F\!\left[ \begin{array}{cc}  
      \Phi_b & \Phi_{\frac{N}{2}} \\ 
      \Phi_{(N-b)} & \Phi_{\frac{N}{2}}
   \end{array}\right]&=&\begin{array}{c|cc}  
             & \Phi_{[\frac{N}{2}+b]} & \Phi_{[\frac{N}{2}-b]} \\ \hline 
      \Phi^1 &   -(-1)^b   &    1   \\ 
      \Phi^2 &    (-1)^b   &    1 
\end{array} \label{apptype6} \\ 
F\!\left[ \begin{array}{cc}  
      \Phi_{(N-b)} & \Phi_{\frac{N}{2}}\\ 
      \Phi_{\frac{N}{2}} & \Phi_b 
   \end{array}\right]&=&\begin{array}{c|cc}  
           & \Phi_{[\frac{N}{2}+b]} & \Phi_{[\frac{N}{2}-b]} \\ \hline 
      \Phi_{[\frac{N}{2}+b]} & (-1)^b   &    0   \\ 
      \Phi_{[\frac{N}{2}-b]} &    0     &    1 
   \end{array} \nn \\
F\!\left[ \begin{array}{cc}  
      \Phi_{\frac{N}{2}} & \Phi_{(N-b)} \\ 
      \Phi_b & \Phi_{\frac{N}{2}}  
   \end{array}\right]&=&\begin{array}{c|cc}  
           & \Phi_{[\frac{N}{2}+b]} & \Phi_{[\frac{N}{2}-b]} \\ \hline 
      \Phi_{[\frac{N}{2}+b]} &  (-1)^b   &    0   \\ 
      \Phi_{[\frac{N}{2}-b]} &    0      &    1 
\end{array} \nn
\ee
 
\subsubsection{Case $F[\Phi_a \Phi_b \Phi_c \Phi_d]$}
\be
F\!\left[ \begin{array}{cc}  
      \Phi_b & \Phi_c \\ 
      \Phi_a & \Phi_d 
   \end{array}\right]=\left\{ \ba{cl}  
      (-1)^{q}   &  \mbox{if couplings are of type DDPP} \\ 
      (-1)^{p}   &  \mbox{if couplings are of type PPDD} \\ 
      (-1)^{c}   &  \mbox{if couplings are of type DPDP} \\ 
      (-1)^{a}   &  \mbox{if couplings are of type PDPD} \\ 
      (-1)^{b}   &  \mbox{if couplings are of type PDDP} \\ 
      (-1)^{d}   &  \mbox{if couplings are of type DPPD} \\ 
      1 & \mbox{else} \ea 
   \right. \;\;\;a \neq N-d, \;\;\; c\neq N-b.
\label{appgenabcd}
\ee
See section \ref{sectionDDPP} for a definition of $D,P$. This formula is valid for any $a,b,c,d$ that leads to a 1x1 fusing matrix.

\be
F\!\left[ \begin{array}{cc}  
      \Phi_{(N-a)} & \Phi_c \\ 
      \Phi_a & \Phi_{(N-c)} 
   \end{array}\right]&=&\begin{array}{c|cc}  
             & \Phi_{[a+(N-c)]} & \Phi_{[a-(N-c)]} \\ \hline 
      \Phi^1 &   (-1)^{q+\frac{N}{2}}   &    1   \\ 
      \Phi^2 &  -(-1)^{q+\frac{N}{2}}   &    1 
   \end{array} \nn \\
F\!\left[ \begin{array}{cc}  
      \Phi_{(N-c)} & \Phi_c\\ 
      \Phi_a & \Phi_{(N-a)} 
   \end{array}\right]&=&\begin{array}{c|cc}  
              & \Phi^1 & \Phi^2 \\ \hline 
      \Phi_{[a+(N-c)]} &  \um(-1)^{p+\frac{N}{2}} 
                       & -\um(-1)^{p+\frac{N}{2}}   \\ 
      \Phi_{[a-(N-c)]} & \um  & \um
   \end{array} \\
F\!\left[ \begin{array}{cc}  
      \Phi_c & \Phi_{(N-a)} \\ 
      \Phi_a & \Phi_{(N-c)} 
   \end{array}\right]&=&\begin{array}{c|cc}  
             & \Phi_{[c+(N-a)]} & \Phi_{[c-(N-a)]} \\ \hline 
      \Phi_{[a+c]} &    (-1)^{\frac{p+q+N}{2}}   &    0   \\ 
      \Phi_{[a-c]} &    0   &    (-1)^{\frac{p+q+N}{2}} 
\end{array} \nn
\ee
with $p,q$ the charge of $\Phi_k$ fields propagating in the $S-$ and $T-$channel respectively.

\subsubsection{Mixed sector}
\be
B^+\left[ \begin{array}{cc} 
      \Phi_a & \Phi_b \\ 
      \sigma_i & \sigma_j
   \end{array} \right] = e^{\I \pi \frac{ab}{2N}} \times
   \begin{array}{c|cc} 
      & \sigma_m & \tau_m \\ \hline
    \sigma_m  &    
       \cos{(\frac{\pi ab}{2N} )} & 
       {\I \frac{ab}{N}} \sin{(\frac{\pi ab}{2N})} \\ 
    \tau_m   &  \frac{\I N}{{ab}} \sin{(\frac{\pi ab}{2N})} & 
        \cos{(\frac{\pi ab}{2N})} 
   \end{array}. \label{appTUUTB}
\ee

\subsubsection{Twisted sector}
\be
&& F\left[\begin{array}{cc} \sigma_1 & \sigma_1 \\ \sigma_1 & \sigma_1 \end{array}\right] = \frac{1}{\sqrt{N}} 
\begin{array}{c|ccc} 
 & 0 & \Phi^i & \Phi_{2k} \nn \\ \hline
0 & 1 & 2. 16^{-N/4} & 2. 16^{-k^2/N} \\
\Phi^i & \frac{1}{2} 16^{N/4} & {(-1)^{N/2}} & (-1)^k 16^{N/4-k^2/N} \\
\Phi_{2k^\prime} & 16^{k^{\prime 2}/N} & 2(-1)^{k^\prime} 16^{-N/4+k^{\prime 2}/N} & 2. 16^{(k^{\prime 2}-k^2)/N} {\cos{\left(\frac{2\pi k k^\prime}{N}\right)}}
\end{array} \\
&&F  \left[\begin{array}{cc} 
      \sigma_1 & \sigma_2 \\ 
      \sigma_1 & \sigma_2 
   \end{array}\right] = \frac{1}{\sqrt{N}} 
   \begin{array}{c|c}  & \Phi_{2k^\prime+1} \\ \hline
      0 & 2.16^{-(2k^{\prime}+1)^2/4N} \\
      \Phi_{2k} & 2.e^{2\pi i k/2N} 
         16^{k^{2}/N-(2k^{\prime}+1)^2/4N} 
         {\cos{\left(\frac{2\pi k (k^\prime\!+\!1/2)}{N}\right)}}
\end{array} \label{appF1122app} \\
&&F\left[\begin{array}{cc} \sigma_2 & \sigma_2 \\ \sigma_1 & \sigma_1 \end{array}\right] = \frac{1}{\sqrt{N}}
\begin{array}{c|cc} &
0 & \Phi_{2k^\prime} \\ \hline
\Phi_{2k+1} &  16^{(2k+1)^2/4N} &
2.e^{-2\pi i k^\prime/2N} 16^{k^{\prime 2}/N+(2k+1)^2/4N} {\cos{\left(\frac{2\pi k^\prime (k\!+\!1/2)}{N}\right)}}
\end{array} \nn\\ \label{appF2211} \nn \\
&&F\left[\begin{array}{cc} \sigma_2 & \sigma_1 \\ \sigma_1 & \sigma_2 \end{array}\right] = \frac{1}{\sqrt{N}}
\begin{array}{c|c} 
& \Phi_{2k^\prime+1} \\ \hline
\Phi_{2k+1} &
2.e^{-2\pi i \frac{k-k^\prime}{2N}} 16^{\frac{(2k+1)^2-(2k^\prime+1)^2}{4N}} {\cos{\left(\frac{2\pi (k\!+\!1/2) (k^\prime\!+\!1/2)}{N}\right)}}
\end{array} \label{appF1212} \nn
\ee

\section{$N$ odd}
\subsubsection{Braiding matrices}
The eigenvalues $\xi_{ij}^k$ needed for evaluating the braiding matrices are for $N$ odd
\be
\xi_{ij}^k = \left\{ \ba{cl} -1  &  \text{if $Q(i)+Q(j)+Q(k)=2N$} \\
                              1  &  \text{otherwise} \ea \right.
\ee
with $Q(0)=Q(J)=0,\; Q(\hat{\Phi}^i)=N,\; Q(\Phi_k)=k$.

\subsubsection{Fusing matrices}

\subsubsection{Case $F[ijk0]$} 
\be
 F\!\left[\begin{array}{cc} j & k \\ i & 0 \end{array}\right] = 1,
    \;\;\; \mbox{if $N_{ij}^{\ph{ij} k}=1$.} \label{appF[ijk0]} 
\ee
Likewise for permutations $F[ij0k]=F[i0jk]=F[0ijk]=1$.

\subsubsection{Case $F[JJJJ]$}
\be 
 F\!\left[\begin{array}{cc} J & J \\ J & J \end{array}\right] = 1.
\ee

\subsubsection{Case $F[JJ\Phi^i\Phi^i]$}
\be 
&&   F\!\left[ \begin{array}{cc}  
        J & {\Phi}^{i} \\ 
        J & {\Phi}^{j} 
        \end{array}\right]= \frac{N}{2}\;\;\; 
     F\!\left[ \begin{array}{cc}  
        {\Phi}^{i} & {\Phi}^{j} \\ 
            J      &     J 
        \end{array}\right]= \frac{2}{N} \nn \;\; 
     F\!\left[ \begin{array}{cc}  
        {\Phi}^{i} &     J \\ 
            J      & {\Phi}^{j} 
        \end{array}\right]= 1 \nn \\ 
&&   F\!\left[ \begin{array}{cc}  
            J      &     J \\ 
        {\Phi}^{i} & {\Phi}^{j} 
        \end{array}\right]= \frac{2}{N} \;\;\; 
     F\!\left[ \begin{array}{cc}  
        {\Phi}^{i} & J \\ 
        {\Phi}^{j} & J 
        \end{array}\right]= \frac{N}{2} \;\;\; 
     F\!\left[ \begin{array}{cc}  
            J      & {\Phi}^{j} \\ 
        {\Phi}^{i} &     J 
        \end{array}\right]= 1
\ee

\subsubsection{Case $F[JJ\Phi_k \Phi_k]$} 
\be 
&&   F\!\left[ \begin{array}{cc}  
        J & \Phi_k \\ 
        J & \Phi_k 
        \end{array}\right]= \frac{k^2}{2N} \;\;\;
     F\!\left[ \begin{array}{cc}  
        \Phi_k & \Phi_k \\ 
        J & J 
        \end{array}\right]= \frac{2N}{k^2} \;\;\;
     F\!\left[ \begin{array}{cc}  
        \Phi_k & J \\ 
        J & \Phi_k 
        \end{array}\right]= 1 \nn \\ 
&&   F\!\left[ \begin{array}{cc}  
        J & J \\ 
        \Phi_k & \Phi_k 
        \end{array}\right]= \frac{2N}{k^2} \;\;\;
     F\!\left[ \begin{array}{cc}  
        \Phi_k & J \\ 
        \Phi_k & J 
        \end{array}\right]=\frac{k^2}{2N} \;\;\;
     F\!\left[ \begin{array}{cc}  
        J & \Phi_k \\ 
        \Phi_k & J 
        \end{array}\right]= 1 
\ee

\subsubsection{Case $F[J \Phi_k \Phi_a \Phi_b]$} 
\be
F\!\left[ \begin{array}{cc}  
      \Phi_k & \Phi_a \\ 
      J & \Phi_b 
   \end{array}\right]= 
      \left\{ \begin{array}{l} 
      \frac{b}{b-a},\; k=\pm(a-b)  \\ 
      \frac{b}{b+a},\; k=a+b       \\ 
      \frac{-b}{2N-a-b}, \; k=2N-a-b \end{array} 
      \right. &
F\!\left[ \begin{array}{cc}  
      \Phi_b & J \\ 
      \Phi_a & \Phi_k 
   \end{array}\right]= 
      \left\{ \begin{array}{l} 
      \frac{b}{b-a},\; k=\pm(a-b)  \\ 
      \frac{b}{b+a},\; k=a+b       \\ 
      \frac{-b}{2N-a-b},\; k=2N-a-b \end{array} 
   \right. \nn \\
F\!\left[ \begin{array}{cc}  
      J & \Phi_a \\ 
      \Phi_k & \Phi_b 
   \end{array}\right]= 
      \left\{ \begin{array}{l} 
      \frac{a}{a-b},\; k=\pm(a-b)  \\ 
      \frac{a}{a+b},\; k=a+b       \\ 
      \frac{-a}{2N-a-b}, k=2N-a-b \end{array} 
   \right. & 
F\!\left[ \begin{array}{cc}  
      \Phi_b & \Phi_k \\ 
      \Phi_a & J 
   \end{array}\right]=
      \left\{ \begin{array}{l} 
      \frac{a}{a-b},\; k=\pm(a-b)  \\ 
      \frac{a}{a+b},\; k=a+b       \\ 
      \frac{-a}{2N-a-b},\;k=2N-a-b \end{array} 
   \right. \nn \\
\ee

\subsubsection{Case $F[J\Phi_k \Phi_{N-k} \Phi^i]$} 
\be
\ba{lll}
F\!\left[ \begin{array}{cc}  
      \Phi_k & \Phi_{N-k} \\ 
      J & \Phi^{i} 
   \end{array}\right]_{{}_{{}_{{}_{{}_{}}}}}
                    = \; {\displaystyle \frac{N}{k}}       &
F\!\left[ \begin{array}{cc} 
      \Phi^{i} & \Phi_{N-k} \\ 
      J & \Phi_k 
   \end{array}\right]= \; {\displaystyle \frac{k}{N}}      &
F\!\left[ \begin{array}{cc}  
      \Phi_{N-k} & \Phi^{i} \\ 
      J & \Phi_k 
   \end{array}\right]= \; {\displaystyle \frac{k}{k-N}}    \\ 
F\!\left[ \begin{array}{cc}  
      J & \Phi_{N-k} \\ 
      \Phi_k & \Phi^{i} 
   \end{array}\right]_{{}_{{}_{{}_{{}_{}}}}}
                     = \; {\displaystyle \frac{k-N}{k}}     &
F\!\left[ \begin{array}{cc} 
      J & \Phi_{k} \\ 
      \Phi^{i} & \Phi_{N-k} 
   \end{array}\right]= \; {\displaystyle \frac{k}{N}}       &
F\!\left[ \begin{array}{cc}  
      J & \Phi^{i} \\ 
      \Phi_k & \Phi_{N-k} 
   \end{array}\right]= \; {\displaystyle \frac{N}{k}}      \\
F\!\left[ \begin{array}{cc}  
      \Phi_k & J \\ 
      \Phi^{i} & \Phi_{N-k} 
   \end{array}\right]_{{}_{{}_{{}_{{}_{}}}}}
                     = \; {\displaystyle \frac{k}{k-N}}     &
F\!\left[ \begin{array}{cc} 
      \Phi_{N-k} & J \\ 
      \Phi_{k} & \Phi^{i} 
   \end{array}\right]= \; {\displaystyle \frac{N-k}{N}}     &
F\!\left[ \begin{array}{cc}  
      \Phi^{i} & J \\ 
      \Phi_k & \Phi_{N-k} 
   \end{array}\right]= \; {\displaystyle \frac{N}{N-k}}    \\ 
F\!\left[ \begin{array}{cc}  
      \Phi^{i} & \Phi_{N-k} \\ 
      \Phi_k & J 
   \end{array}\right]= \; {\displaystyle \frac{k}{k-N}}     &
F\!\left[ \begin{array}{cc} 
      \Phi_{N-k} & \Phi_{k} \\ 
      \Phi^{i} & J 
   \end{array}\right]= \; {\displaystyle \frac{N}{k}}       &
F\!\left[ \begin{array}{cc}  
      \Phi_{N-k} & \Phi^{i} \\ 
      \Phi_k & J 
   \end{array}\right]= \; {\displaystyle \frac{k}{N}}
\ea
\label{appF[Jk(N-k)i]odd}
\ee

\subsubsection{Case $F[\Phi_a \Phi_a \Phi^i \Phi^{j}]$}
\be
\ba{ll}
F\!\left[ \begin{array}{cc}  
      \Phi^i & \Phi_a \\ 
      \Phi_a & \Phi^{j} 
   \end{array}\right]= {\displaystyle -(-1)^{i+j+a}}
& F\!\left[ \begin{array}{cc}  
      \Phi_a & \Phi^{j} \\ 
      \Phi^i & \Phi_a 
   \end{array}\right]_{{}_{{}_{{}_{{}_{}}}}}= {\displaystyle -(-1)^{i+j+a}} 
\\ 
F\!\left[ \begin{array}{cc} 
      \Phi_a & \Phi^{i} \\ 
      \Phi_a & \Phi^{j} 
   \end{array}\right]= {\displaystyle \frac{1}{2}} \;\;\;
F\!\left[ \begin{array}{cc} 
      \Phi_a & \Phi^{i} \\ 
      \Phi_a & \Phi^{i} 
   \end{array}\right]= {\displaystyle \frac{1}{a}}  
& F\!\left[ \begin{array}{cc} 
      \Phi^i & \Phi^{j} \\ 
      \Phi_a & \Phi_{a} 
   \end{array}\right]= {\displaystyle 2} \;\;\;
F\!\left[ \begin{array}{cc} 
      \Phi^i & \Phi^{i} \\ 
      \Phi_a & \Phi_a 
   \end{array}\right]_{{}_{{}_{{}_{{}_{}}}}}= {\displaystyle a}
\\ 
F\!\left[ \begin{array}{cc} 
      \Phi^{j} & \Phi_a \\ 
      \Phi^{i} & \Phi_a 
   \end{array}\right]= {\displaystyle \frac{1}{2}} \;\;\;
F\!\left[ \begin{array}{cc} 
      \Phi^i & \Phi_a \\ 
      \Phi^i & \Phi_a 
   \end{array}\right]= {\displaystyle \frac{1}{a}} 
& F\!\left[ \begin{array}{cc} 
      \Phi_a & \Phi_a \\ 
      \Phi^{i} & \Phi^{j} 
   \end{array}\right]= {\displaystyle 2} \;\;\;
F\!\left[ \begin{array}{cc}
      \Phi_a & \Phi_a \\ 
      \Phi^i & \Phi^i 
   \end{array}\right]= {\displaystyle a}
\ea
\ee

\subsubsection{Case $F[\Phi^i \Phi_k \Phi_a \Phi_b]$} 
\be 
F\!\left[ \begin{array}{cc} 
\Phi_a & \Phi_b \\ 
\Phi^i & \Phi_c 
\end{array}\right]= \left\{ \ba{cl} -(-1)^{i+b+\frac{N-1}{2}} & \mbox{if $a=b+c-N$} \\ (-1)^{i+b+\frac{N-1}{2}} & \mbox{if $c=a+b-N$} 
\\  1 & \mbox{else}\ea \right. 
\ee 
The permutations $F[c\Phi^i ab]$, $F[bc\Phi^i a]$ and $F[abc\Phi^i]$ are similar, the field taking the role of $\Phi_b$ now being the one diagonally opposite $\Phi^i$.

\subsubsection{Case $F[\Phi^i\Phi^j\Phi^k\Phi^l]$}
\be 
F\!\left[ \begin{array}{cc} 
      \Phi^i & \Phi^i \\ 
      \Phi^i & \Phi^i 
   \end{array}\right]= 1 \;\;\;
F\!\left[ \begin{array}{cc}
      \Phi^j & \Phi^i \\ 
      \Phi^i & \Phi^j 
   \end{array}\right]= 1 \;\;\; 
F\!\left[ \begin{array}{cc} 
      \Phi^i & \Phi^j \\ 
      \Phi^i & \Phi^j 
   \end{array}\right]= \frac{2}{N} \;\;\; 
F\!\left[ \begin{array}{cc} 
      \Phi^j & \Phi^j \\ 
      \Phi^i & \Phi^i 
   \end{array}\right]= \frac{N}{2}
\ee

\subsubsection{Case $F[\Phi_a \Phi_a \Phi_a \Phi_a]$} 
\be
F\!\left[ \begin{array}{cc} 
       \Phi_a & \Phi_a \\ 
       \Phi_a & \Phi_a 
   \end{array} \right] = 
   \begin{array}{c|ccc} & 0 & J & \Phi_{[2a]} \\ \hline 
                        0 & \um & \frac{x}{2} & \um \\ 
                        J & \frac{1}{2x} & \um & -\frac{1}{2x} \\ 
                        \Phi_{[2a]} & 1 & -x & 0  
   \end{array}\;\;\; x=\frac{a^2}{2N}
\ee

\subsubsection{Case $F[\Phi_a \Phi_a \Phi_a \Phi_b]$} 
\be
F\!\left[ \begin{array}{cc} 
      \Phi_a & \Phi_a \\ 
      \Phi_a & \Phi_b 
   \end{array} \right]=
   \left\{ \ba{cl} 1 & \mbox{if $b=3a$} \\ (-1)^a & \mbox{else} \ea 
   \right. 
\ee

\subsubsection{Case $F[\Phi_a \Phi_a \Phi_b \Phi_b]$} 
\be 
&& F\!\left[ \begin{array}{cc}  
      \Phi_b & \Phi_a \\ 
      \Phi_a & \Phi_b 
      \end{array}\right]= 
   \begin{array}{c|cc}  
      & \Phi_{[a+b]} & \Phi_{[a-b]} \\ \hline 
      \Phi_{[a+b]} & 0 & \epsilon \\ 
      \Phi_{[a-b]} & \epsilon & 0 
   \end{array} \;\;\; \epsilon=
      \left\{ \ba{cl} (-1)^{a+b} & \mbox{if $a+b>N$} \\ 1 & \mbox{else} \ea 
      \right. \\ 
&& F\!\left[ \begin{array}{cc}  
      \Phi_a & \Phi_c \\ 
      \Phi_a & \Phi_c 
      \end{array}\right]= 
   \begin{array}{c|cc}  
      & \Phi_{[a+c]} & \Phi_{[a-c]} \\ \hline 
      0 &     \um    &    \um \\ 
      J & \frac{-1}{2x} & \frac{1}{2x} 
   \end{array} \;\;\; 
F\!\left[ \begin{array}{cc}  
      \Phi_c & \Phi_c \\ 
      \Phi_a & \Phi_a 
   \end{array}\right]= 
   \begin{array}{c|cc}  
      &      0     &    J    \\ \hline 
      \Phi_{[a+c]}  &      1     &   -x   \\ 
      \Phi_{[a-c]}  &      1     &    x 
   \end{array} \;\;\; x=\frac{ac}{2N} \nn
\ee

\be 
F\!\left[ \begin{array}{cc}  
      \Phi_{(N-a)} &  \Phi_a \\ 
      \Phi_a  & \Phi_{(N-a)} 
   \end{array}\right]&=& 
   \begin{array}{c|ccc}  
                &   \Phi^1   &   \Phi^2 & \Phi_{[N-2a]} \\ \hline 
      \Phi^1    &  \um (-1)^a    &  -\um (-1)^a    &      1 \\ 
      \Phi^2    & -\um (-1)^a    &   \um (-1)^a    &      1 \\ 
      \Phi_{[N-2a]} &    \um     &    \um   &      0 
   \end{array} \nn
\ee

\be 
F\!\left[ \begin{array}{cc}  
      \Phi_{(N-a)} & \Phi_{(N-a)} \\ 
      \Phi_a & \Phi_a 
   \end{array}\right]&=& 
   \begin{array}{c|ccc}  
                    &   0   &     J   & \Phi_{[2a]} \\ \hline  
      \Phi^1        &   1   &    -x   & -\epsilon(-1)^{\frac{N-1}{2}}   \\ 
      \Phi^2        &   1   &    -x   &  \epsilon(-1)^{\frac{N-1}{2}}   \\ 
      \Phi_{[N-2a]} &   1   &     x   &     0 
   \end{array} \;\;\; \epsilon=-1\;\mbox{if $2a>N$} \nn \\
F\!\left[ \begin{array}{cc}  
      \Phi_a & \Phi_{(N-a)} \\ 
      \Phi_a & \Phi_{(N-a)}
   \end{array}\right]&=&
   \begin{array}{c|ccc}  
         &    \Phi^1  &   \Phi^2   &  \Phi_{[N-2a]}    \\ \hline 
      0  &  \frac{1}{4}       &   \frac{1}{4}     &  \um    \\ 
      J  & -\frac{1}{4x}      &  -\frac{1}{4x}    &  \frac{1}{2x} \\ 
      \Phi_{[2a]} & -\um\epsilon(-1)^{\frac{N-1}{2}} & 
                     \um\epsilon(-1)^{\frac{N-1}{2}} & 0 
   \end{array}
\ee

\subsubsection{Case $F[\Phi_a \Phi_b \Phi_c \Phi_d]$} 
\be
F\!\left[ \begin{array}{cc}  
      \Phi_b & \Phi_c \\ 
      \Phi_a & \Phi_d 
   \end{array}\right]=\left\{ \ba{cl}  
      (-1)^{q}   &  \mbox{if couplings are of type DDPP} \\ 
      (-1)^{p}   &  \mbox{if couplings are of type PPDD} \\ 
      (-1)^{c}   &  \mbox{if couplings are of type DPDP} \\ 
      (-1)^{a}   &  \mbox{if couplings are of type PDPD} \\ 
      (-1)^{b}   &  \mbox{if couplings are of type PDDP} \\ 
      (-1)^{d}   &  \mbox{if couplings are of type DPPD} \\ 
      1 & \mbox{else} \ea 
   \right. \;\;\;a \neq N-d, \;\;\; c\neq N-b.
\label{appgenabcdodd}
\ee
See section \ref{sectionDDPP} for a definition of $D,P$. Valid for any $a,b,c,d$ that leads to a 1x1 fusing matrix.

\be 
F\!\left[ \begin{array}{cc}  
      \Phi_{(N-a)} & \Phi_c\\ 
      \Phi_a & \Phi_{(N-c)} 
   \end{array}\right]&=&\begin{array}{c|cc}  
             & \Phi_{[a+(N-c)]} & \Phi_{[a-(N-c)]} \\ \hline 
      \Phi^1 &   \eta(-1)^{q+\frac{N-1}{2}}   &    1   \\ 
      \Phi^2 &  -\eta(-1)^{q+\frac{N-1}{2}}   &    1 
   \end{array} \;\;\; \eta=\left\{ \ba{cl} -1  
      & \mbox{if $a<c$}  \\  1 
      & \mbox{if $a>c$} \ea \right.\nn \\
F\!\left[ \begin{array}{cc}  
      \Phi_{(N-c)} & \Phi_c\\ 
      \Phi_a & \Phi_{(N-a)} 
   \end{array}\right]&=&\begin{array}{c|cc}  
             & \Phi^1 & \Phi^2 \\ \hline 
      \Phi_{[a+(N-c)]} &    \um\eta(-1)^{p+\frac{N-1}{2}}  
                       &   -\um\eta(-1)^{p+\frac{N-1}{2}}   \\ 
      \Phi_{[a-(N-c)]} &    \um   &    \um 
   \end{array} \\
F\!\left[ \begin{array}{cc}  
      \Phi_c & \Phi_{(N-a)} \\ 
      \Phi_a & \Phi_{(N-c)} 
   \end{array}\right]&=&\begin{array}{c|cc}  
             & \Phi_{[c+(N-a)]} & \Phi_{[c-(N-a)]} \\ \hline 
      \Phi_{[a+c]} &    -(-1)^{\frac{p+q+N}{2}}   &    0   \\ 
      \Phi_{[a-c]} &    0   &    -(-1)^{\frac{p+q+N}{2}} 
   \end{array} \nn  \label{appac(N-a)(N-c)odd} 
\ee
with $p,q$ the charge of $\Phi_k$ fields propagating in the $S-$ and $T-$channel respectively.

\subsubsection{Twisted sector}
\be
&&F  \left[ \begin{array}{cc} 
      \hat{\sigma}_1 & \hat{\sigma}_1 \\ 
      \hat{\sigma}_1 & \hat{\sigma}_1 
   \end{array} \right] =  \frac{1}{\sqrt{N}} \begin{array}{c|c} 
      & \Phi_{2q-1}  \\ \hline 
      \Phi_{2p-1} & 2.16^{\frac{(2p-1)^2-(2q-1)^2}{4N}} 
      e^{2\pi \I((2p-1)-(2q-1))/4N} 
      \cos{(2\pi \frac{(2p-1)(2q-1)}{4N})} 
   \end{array} \nn \\
&&F  \left[\begin{array}{cc} 
      \hat{\sigma}_2 & \hat{\sigma}_1 \\ 
      \hat{\sigma}_1 & \hat{\sigma}_2 
   \end{array}\right] = \frac{1}{\sqrt{N}} \begin{array}{c|cc}
       & 0 & \Phi_{2k^\prime} \\ \hline
       0 & 1 & 2.16^{-(2k^{\prime})^2/4N} \\
       \Phi_{2k} & 16^{(2k)^2/4N} & 2.16^{((2k)^2-(2k^\prime))^2/4N} 
       \cos{(2\pi\frac{2k.2k^\prime}{4N})} 
   \end{array} \label{appFF1122} \\
&&F  \left[\begin{array}{cc} 
      \hat{\sigma}_1 & \hat{\sigma}_2 \\ 
      \hat{\sigma}_1 & \hat{\sigma}_2 
   \end{array}\right] = \frac{1}{\sqrt{N}} \begin{array}{c|cc} 
      & 0 & \Phi_{2g} \\ \hline
      \Phi^i & \um 16^{N/4} & (-1)^g e^{-2\pi \I g/2N} 16^{N/4-(2g)^2/4N} \\
      \Phi_{2k+1} & 16^{(2k+1)^2/4N} & 2.16^{((2k+1)^2-(2g))^2/4N} 
      e^{-2\pi \I g/2N} \cos{(2\pi\frac{(2k+1)2g}{4N})}
   \end{array} \nn
\ee
\be
&& F\left[\begin{array}{cc} 
      \hat{\sigma}_2 & \hat{\sigma}_2 \\ 
      \hat{\sigma}_1 & \hat{\sigma}_1 
   \end{array}\right] = \frac{1}{\sqrt{N}} \times \nn \\ && 
   \begin{array}{c|cc} 
      & \Phi^i & \Phi_{2k+1} \\ \hline
      0 & 2.16^{N/4} & 2.16^{-(2k+1)^2/4N} \\
      \Phi_{2g} & 2(-1)^g e^{2\pi \I g/2N} 16^{-N/4+(2g)^2/4N} & 
      2.16^{((2g)^2-(2k+1))^2/4N} 
      e^{2\pi \I g/2N} \cos{(2\pi\frac{(2k+1)2g}{4N})} 
\end{array} \nn
\ee

\end{document}